\documentclass[a4paper,11pt]{article}
\pdfoutput=1
\usepackage{jheppub}
\usepackage[utf8]{inputenc}

\usepackage[vcentermath]{youngtab}
\usepackage{pdflscape}
\usepackage{tikz}
\usepackage{marvosym}
\usetikzlibrary{positioning}
\usetikzlibrary{arrows}
\usetikzlibrary{arrows.meta}
\usetikzlibrary{decorations.pathreplacing}
\usepackage{subcaption}
\usepackage{xcolor,colortbl}
\usepackage{mathtools}
\usepackage{ragged2e}
\newlength\ubwidth

\usepackage{paralist}
\usetikzlibrary{arrows,fit,decorations.pathreplacing, shapes.misc,decorations.pathmorphing}
\usepackage[noabbrev,capitalise]{cleveref}
\usepackage{xcolor}
\usepackage{soul}
\usepackage{colortbl}
\definecolor{Gray}{gray}{0.9}
\usepackage{multirow}
\usepackage[colorinlistoftodos]{todonotes}
\usepackage{amsthm}
\usepackage{amsmath}
\usepackage{amsthm}
\usepackage{changepage}
\usepackage{tabularx}
\usepackage{tikz}
\usepackage{multirow}
\usepackage{longtable}
\usepackage{multirow}
\usepackage{amsmath}
\usepackage{amssymb}
\usepackage{mathtools}
\usepackage{flafter}
\usepackage{adjustbox}
\usetikzlibrary{shapes,arrows}
\usepackage[english]{babel}
\newcolumntype{P}[1]{>{\centering\arraybackslash}p{#1}}
\usepackage{tabularx}

\usepackage{color}
\usepackage{subcaption}
\usepackage{ytableau}
\usepackage{setspace}
\usepackage{gensymb}
\usepackage{pdflscape}

\usepackage[all]{xy}
\newcommand{\slice}[1]{\mathcal{S}_{\mathcal N,#1} }
\newcommand{\orbit}[1]{\overline{\mathcal{O}}_{#1} }
\newcommand{\ghs}[1]{g_{HS}^{#1}}
\newcommand{\bv}[1]{d_{BV}(#1)}
\newcommand{\pal}{\ldots \text{palindrome}\ldots}

\usetikzlibrary{shapes.misc}
\tikzset{gauge1/.style={draw=none,minimum size=0.6cm,fill=white,circle, draw}}
\tikzset{gauge3/.style={draw=none,minimum size=0.4cm,fill=white,circle, draw}}
\tikzset{gauge33/.style={draw=none,minimum size=0.35cm,fill=white,circle, draw}}
\tikzset{crosses/.style={cross out, draw=black, minimum size=0.3cm, inner sep=0pt, outer sep=0pt},
cross/.default={1pt}}
\tikzset{blank/.style={draw=none,minimum size=0.4cm,fill=none,circle, draw}}
\tikzset{flavor2/.style={draw=none,minimum size=0.4cm,fill=white,regular polygon sides=4,draw}}
\tikzset{flavorBlue/.style={draw=none,minimum size=0.4cm,fill=blue,regular polygon sides=4,draw}}
\tikzset{flavorRed/.style={draw=none,minimum size=0.4cm,fill=red,regular polygon sides=4,draw}}
\tikzset{none/.style={draw=none}}
\tikzset{redgauge/.style={draw=none,minimum size=0.4cm,fill=red,circle, draw}}
\tikzset{miniU/.style={draw=none,minimum size=0.1cm,fill=red,circle, draw}}
\tikzset{smallgauge1/.style={draw=none,minimum size=0.1cm,fill=white,circle, draw}}
\tikzset{miniBlue/.style={draw=none,minimum size=0.1cm,fill=blue,circle, draw}}
\tikzset{gauge2/.style={draw=none,minimum size=0.35mm,fill=red,circle, draw}}
\tikzset{bluegauge/.style={draw=none,minimum size=0.4cm,fill=blue,circle, draw}}
\tikzset{flavor1/.style={draw=none,minimum size=0.35mm,fill=blue, regular polygon,regular polygon sides=4,draw}}
\tikzset{flavor0/.style={draw=none,minimum size=0.35mm,fill=white, regular polygon,regular polygon sides=4,draw}}
\tikzset{smalldot/.style={draw=none,minimum size=0.1mm,fill=black, circle,draw}}
\tikzset{dotsize/.style={circle,fill,inner sep=1.5pt,draw}}
\tikzset{doubleguys/.style={double, double distance = 3pt}}
\tikzset{tripleguys/.style={triple}}
\tikzset{new edge style 1/.style={dashed}}
\tikzset{thickline/.style={line width=0.06cm}}
\tikzset{darke/.style={line width=0.3mm,black}}
\tikzset{brace/.style={decorate,decoration={brace,amplitude=10pt}}}
\pgfdeclarelayer{edgelayer}
\pgfdeclarelayer{nodelayer}
\usetikzlibrary{decorations.pathreplacing}
\tikzset{blankflavor/.style={draw=none,minimum size=0.8mm,fill=none, regular polygon,regular polygon sides=4,draw}}
\pgfsetlayers{edgelayer,nodelayer,main}

\tikzset{wd/.style={circle, draw,inner sep=3.5pt}}
\tikzset{bd/.style={circle, draw,inner sep=3.5pt, fill=black}}

\DeclareFontFamily{U}{rcjhbltx}{}
\DeclareFontShape{U}{rcjhbltx}{m}{n}{<->rcjhbltx}{}
\DeclareSymbolFont{hebrewletters}{U}{rcjhbltx}{m}{n}

\let\aleph\relax\let\beth\relax
\let\gimel\relax\let\daleth\relax

\DeclareMathSymbol{\aleph}{\mathord}{hebrewletters}{39}
\DeclareMathSymbol{\beth}{\mathord}{hebrewletters}{98}
\DeclareMathSymbol{\gimel}{\mathord}{hebrewletters}{103}
\DeclareMathSymbol{\daleth}{\mathord}{hebrewletters}{100}

\DeclareMathSymbol{\lamed}{\mathord}{hebrewletters}{108}
\DeclareMathSymbol{\mem}{\mathord}{hebrewletters}{109}
\DeclareMathSymbol{\ayin}{\mathord}{hebrewletters}{96}
\DeclareMathSymbol{\tsadi}{\mathord}{hebrewletters}{118}
\DeclareMathSymbol{\qof}{\mathord}{hebrewletters}{114}
\DeclareMathSymbol{\shin}{\mathord}{hebrewletters}{152}

\allowdisplaybreaks

\preprint{Imperial/TP/20/AH/07}

\title{Magnetic Lattices for Orthosymplectic Quivers}

\author[\lamed]{Antoine Bourget,}
\author[\lamed]{Julius F. Grimminger,}
\author[\lamed]{ Amihay Hanany,}
\author[\lamed]{ Rudolph Kalveks,}
\author[\beth]{ Marcus Sperling}
\author[\lamed]{and Zhenghao Zhong}
\affiliation[\lamed]{Theoretical Physics Group, The Blackett Laboratory, Imperial College London, Prince Consort Road
London, SW7 2AZ, UK}
\affiliation[\beth]{Yau Mathematical Sciences Center, Tsinghua University, Haidian District, Beijing, 100084, China}
\emailAdd{a.bourget@imperial.ac.uk}
\emailAdd{julius.grimminger17@imperial.ac.uk}
\emailAdd{a.hanany@imperial.ac.uk}
\emailAdd{rudolph.kalveks09@imperial.ac.uk}
\emailAdd{marcus.sperling@univie.ac.at}
\emailAdd{zhenghao.zhong14@imperial.ac.uk}

\abstract{For any gauge theory, there may be a subgroup of the gauge group which acts trivially on the matter content. While many physical observables are not sensitive to this fact, the choice of the precise gauge group becomes crucial when the magnetic lattice of the theory is considered. This question is addressed in the context of Coulomb branches for $3$d $\mathcal{N}=4$ quiver gauge theories, which are moduli spaces of dressed monopole operators. We compute the Coulomb branch Hilbert series of many unitary-orthosymplectic quivers for different choices of gauge groups, including diagonal quotients of the product gauge group of individual factors, where the quotient is by a trivially acting subgroup. Choosing different such diagonal groups results in distinct Coulomb branches, related as orbifolds. Examples include nilpotent orbit closures of the exceptional E-type algebras and magnetic quivers that arise from brane physics. This includes Higgs branches of theories with 8 supercharges in dimensions $4$, $5$, and $6$. 
A crucial ingredient in the calculation of exact refined Hilbert series is the alternative construction of unframed magnetic quivers from resolved Slodowy slices, whose Hilbert series can be derived from Hall-Littlewood polynomials.}

\begin{document} 
\maketitle

\section{Introduction}
Supersymmetric gauge theories with 8 supercharges are central objects in  string theory and quantum field theory. Due to the high amount of supersymmetry (SUSY), there are some families in which the field theory ingredients such as the gauge groups, matter content, and the superpotential interactions can be neatly encoded in so-called \emph{quiver} diagrams \cite{Kronheimer1990,Douglas:1996sw}. 
The moduli space of a $3$d $\mathcal{N}=4$ quiver gauge theory at the IR fixed point contains a Higgs branch and a Coulomb branch, which are both (singular) hyper-K\"ahler spaces. The (classical) Higgs branch is a hyper-K\"ahler quotient \cite{Hitchin:1986ea,Argyres:1996eh,Antoniadis:1996ra}. The Coulomb branch is the space of dressed monopole operators \cite{monopole}. Monopole operators \cite{Borokhov:2002ib,Borokhov:2002cg,Borokhov:2003yu} are, among others, characterized by their magnetic charges which form the magnetic lattice. By definition \cite{Englert:1976ng,Goddard:1976qe}, the magnetic lattice is the weight lattice of the dual group of the gauge group, sometimes called GNO-dual. For unframed unitary quivers
it is understood that 
a $\mathrm{U}(1)^{\textrm{diag}}$ has to be \emph{ungauged} to compute the correct magnetic lattice. However, for unframed orthosymplectic quivers there has been no satisfactory determination of the magnetic lattice.
For a given quiver $\mathsf{Q}$, we aim to clarify how the magnetic lattice of the corresponding theory is determined.
The first step is the identification of the precise gauge group associated to the quiver.

Given a quiver $\mathsf{Q}$, the product group $G$ of the individual gauge group factors may have a subgroup $H$ acting trivially on the matter content. 
For any such $H$, one can construct the group $G_H=G/H$,
and ask if there is a physical difference between the choices of $G_H$ and of $G$. While some observables are insensitive to this choice, others are not.
For instance, consider the Higgs branch of the theory. Taking the hyper-K\"ahler quotient with respect to $G_H$ or with respect to $G$ results in the same moduli space; therefore, the Higgs branch is insensitive to the difference between the two gauge groups. 
However, the GNO-duals of the groups $G_H$ and $G$ are different. Hence the computation of the Coulomb branch is sensitive to this distinction. In fact, $H$ now becomes part of the global symmetry $F$ of the Coulomb branch.

Pictorially, the effect of removing $H$ from the gauge group can be summarized as follows:
\begin{equation}
\label{eq:effect_ungauge}
    \raisebox{-.5\height}{
\begin{tikzpicture}
	\begin{pgfonlayer}{nodelayer}
		\node [style=none] (0) at (-3, -1) {$G_H$};
		\node [style=none] (1) at (3, -1) {$G$};
		\node [style=none] (2) at (-3, 0.25) {$F_H$};
		\node [style=none] (3) at (3, 0.25) {$F$};
		\node [style=none] (4) at (-1.5, 0.25) {};
		\node [style=none] (5) at (1.5, 0.25) {};
		\node [style=none] (6) at (-1.5, -1) {};
		\node [style=none] (7) at (1.75, -1) {};
		\node [style=none] (8) at (0, -0.75) {ungauge $H$};
		\node [style=none] (9) at (0, 0.5) {gauge $H$};
		\node [style=none] (10) at (-7, 0.25) {Global Symmetry:};
		\node [style=none] (11) at (-7, -1) {Gauge Symmetry:};
		\node [style=none] (12) at (-3.75, 0.5) {};
		\node [style=none] (13) at (-3.75, -1.25) {};
		\node [style=none] (14) at (3.75, 0.5) {};
		\node [style=none] (15) at (3.75, -1.25) {};
	\end{pgfonlayer}
	\begin{pgfonlayer}{edgelayer}
		\draw [->] (4.center) to (5.center);
		\draw [->] (7.center) to (6.center);
	\end{pgfonlayer}
\end{tikzpicture}}
\end{equation}
Note that the action of $H$ affects the gauge and flavor groups in opposite ways, while $G_H$ is `smaller' than $G$, $F_H$ is `bigger' than $F$.

The central question addressed in this paper is the following: for a given quiver $\mathsf{Q}$ what are the choices for the subgroup $H$ and their associated magnetic lattice? 
It is important to stress that this question is not restricted to physics in 3 dimensions because recent developments around \emph{magnetic quivers} \cite{TropicalSanti,6dmagnetic,newMarcus,5dweb}, see also \cite{DelZotto:2014kka,GiuilaInf,Ferlito:2017xdq,Mekareeya:2017jgc,NoppadolE8,Hanany:2018vph}, have shown that Coulomb branches of $3$d $\mathcal{N}=4$ quivers do describe finite and infinite gauge coupling Higgs branches of theories with 8 supercharges in $d=4,5,6$ dimensions.
A similar question is relevant for the Standard Model gauge group $(\mathrm{SU}(3)\times \mathrm{SU}(2) \times \mathrm{U}(1) )\slash H$ where $H\in \{\{1\}, \mathbb{Z}_2, \mathbb{Z}_3, \mathbb{Z}_6\}$. Analogously, the choice of gauge group manifests itself in the spectrum of electric, magnetic and dyonic line operators and higher form symmetries \cite{Aharony:2013hda,Aharony:2013kma,Gaiotto:2014kfa,Tong:2017oea,Albertini:2020mdx,Morrison:2020ool}.

Focusing on $3$d $\mathcal{N}=4$ quiver gauge theories, these can be roughly classified according to the type of gauge group and presences of flavor nodes.
To begin with, quivers that only contain unitary gauge groups are usually dubbed \textit{unitary quivers} and are well-known. 
Another family are the  \emph{orthosymplectic quivers}, which contain (special) orthogonal and symplectic gauge groups. 
In contrast to the unitary quivers, the details of orthosymplectic quivers are more subtle. 
For instance, if the $3$d $\mathcal{N}=4$ quiver gauge theory originates from a Type IIB brane configuration with orientifolds as in \cite{Feng:2000eq}, it is not a priori clear if the gauge groups associated to a B or D type Lie algebra are orthogonal or special orthogonal, see \cite{CabreraZhong,Cabrera:2017njm} for a detailed study.
In addition, the existence of disconnected groups $\mathrm{O}(n)$ imposes computational challenges \cite{Hanany:2012dm, rudolph1601,Antoinefirstquiverpaper}. Nevertheless, recent studies  \cite{Feng:2000eq,GaiottoWitten,tsigmarho,Sicilian,CabreraZhong,Cabrera:2017njm,NoppadolE8,newMarcus,RudolphSlodowy,5dweb} in orthosymplectic quivers, especially in relation to electric-magnetic quiver pairs, help in building the tools necessary to develop the understanding further. 

Alternatively, quivers can be further classified into \emph{framed/flavored} and 
\emph{unframed/ flavorless} quivers, indicated by the presence or absence of flavor nodes. When studying the moduli space of unitary quivers derived from brane configurations \cite{Hanany:1996ie}, there always exists a trivially acting diagonal $\mathrm{U}(1)^{\textrm{diag}}$ which may or may not be a subgroup of the gauge group. This $\mathrm{U}(1)$ describes the center of mass.
Hence, if the quiver is flavorless the physical gauge group is $G_{\mathrm{U}(1)} = G \slash \mathrm{U}(1)^{\textrm{diag}}$. 
Studies involving flavorless orthosymplectic quivers are scarce. These can arise as class $\mathcal{S}$ theories \cite{Benini:2010uu,Sicilian} or as magnetic quivers for $6$d $\mathcal{N}=(1,0)$ theories \cite{NoppadolE8,newMarcus} as well as $5$d $\mathcal{N}=1$ theories \cite{5dweb}. From the Type II brane perspective \cite{Feng:2000eq}, the existence of orientifold planes fixes the center of mass, making it unnecessary to quotient  $\mathrm{U}(1)^{\textrm{diag}}$. 
However, in some cases there exists a remaining diagonal $\mathbb{Z}_2\subset G$, which we call $\mathbb{Z}_2^{\textrm{diag}}$, that acts trivially on the matter multiplets and one may choose to quotient it out as detailed below. 
For theories originating from brane systems, it is natural to quotient by $\mathbb{Z}_2^{\textrm{diag}}$; even though the background without the quotient seems to be a consistent background. This is supported by the field theory perspective where one does not necessarily have to quotient by $\mathbb{Z}_2^{\textrm{diag}}$, because both theories are valid in their own right.

To study the Higgs branch of a quiver gauge theory in $d=3,4,5,6$ with 8 supercharges or the Coulomb branch of a $3$d $\mathcal{N}=4$ quiver gauge theory, we compute
the Hilbert series \cite{PlethysticProgram}, which is a generating function counting holomorphic functions on the moduli spaces graded by their dimension.
This approach allows one to study the moduli space from a geometric perspective as an affine variety. Explicit computations of the Hilbert series enable us to check proposed electric-magnetic quiver pairs, where the Higgs branch Hilbert series of the electric theory is the same as the $3$d $\mathcal{N}=4$ Coulomb branch Hilbert series of the magnetic quiver. In addition, it allows for a whole set of predictions for moduli spaces which were not computable beforehand.
\begin{figure}[t]
    \centering
    \begin{tabular}{|c|c|}
        \hline
         Unframed Quiver Type & $H$ \\
         \hline
         Unitary & $\mathrm{U}(1)^{\textrm{diag}}$\\
         Unitary plus one $\mathrm{SU}(k)$ node & $\mathbb{Z}_k^{\mathrm{ diag}}$\\
         (Unitary-)Orthosymplectic without $\mathrm{SO}(2n+1)$ gauge groups & $\mathbb{Z}_2^{\textrm{diag}}$\\
         (Unitary-)Orthosymplectic with  $\mathrm{SO}(2n+1)$ gauge groups & $\{1\}$\\
         \hline
    \end{tabular}
    \caption[Unframed quivers and trivially acting subgroups]{Maximal subgroups $H$ of the product group $G$ of different types of unframed simply laced quivers.}
    \label{tab:trivials}
\end{figure}

In recent years, a series of papers \cite{tsigmarho,Sicilian,CremonesiHall, CabreraZhong} emerged which present the $3$d $\mathcal{N}=4$ Coulomb branch Hilbert series of orthosymplectic quivers using the monopole formula \cite{monopole}. 
If there exists a diagonal $\mathbb{Z}_2^{\textrm{diag}} \subset G$ that acts trivially on the matter content, then the choice of $G_H$ or $G$ affects the magnetic lattice and hence the moduli space. This is supported by explicit Hilbert series computations. Therefore, the monopole formula offers a perfect opportunity to study this $\mathbb{Z}_2^{\textrm{diag}}$  action and the choice of the gauge group. As derived below,   $\mathbb{Z}_2^{\textrm{diag}}$ is crucial in orthosymplectic quivers composed of $\mathrm{SO}(2m)$ and $\mathrm{USp}(2n)$ gauge groups.

The family of flavorless quivers can be enlarged by including a mixture of orthosymplectic gauge groups \emph{and} unitary gauge groups. We shall dub these quivers \emph{unitary-orthosymplectic} quivers, see also the companion paper \cite{5dweb}. Most notably, the Coulomb branch of some of its members are closures of minimal nilpotent orbits of the exceptional $\mathfrak{e}_n$ algebras. As elaborated in the main text, in order to compute the Coulomb branch Hilbert series of such quivers, one needs to take care of a $\mathbb{Z}_2^{\textrm{diag}}\subset G$.
In particular, the $\mathbb{Z}_2^{\textrm{diag}}$ crucially affects the magnetic lattice for unitary-orthosymplectic quivers composed of $\mathrm{U}(k)$, $\mathrm{SO}(2l)$, and $\mathrm{USp}(2m)$ gauge groups. In Figure \ref{tab:trivials} the maximal subgroups $H$ for various types of unframed simply laced quivers are tabulated.

The computations presented in this paper rely on the precise identification of the magnetic lattice. In some cases, these computations can be cross-checked by an independent calculation of the Hilbert series by gluing together \emph{resolved Slodowy slices} computed using the Hall-Littlewood formula (for details see \cite{CremonesiHall, RudolphSlodowy} or Appendix \ref{app:HL}).

The paper is organized as follows: Section \ref{monopole} details our main proposal: for a quiver $\mathsf{Q}$ a choice of a physical gauge group $G_H$ is derived from the simple product gauge group $G$ modded out by $H$, which is a subgroup of the kernel of $G$. In the example of flavorless unitary-orthosymplectic quivers, we extensively discuss how the magnetic lattice is affected by this quotient.
In the remaining sections, we present Coulomb branch Hilbert series of various families of unitary-orthosymplectic quiver theories where $H$ equals the kernel of $G$. In Section \ref{5dexamples}, we focus on the $E_n$ exceptional theories: unitary-orthosymplectic quivers whose Coulomb branches are closures of minimal nilpotent orbits of the exceptional $\mathfrak{e}_n$ algebras. In Section \ref{general5dfamilysection}, we generalize the exceptional theories to families whose Coulomb branch corresponds to the Higgs branch of certain $5$d $\mathcal{N}=1$ theories at infinite gauge coupling. In Section \ref{classSguys}, we study the $3$d $\mathcal{N}=4$ Coulomb branches of three-legged unitary-orthosymplectic quivers, which are magnetic quivers corresponding to $4$d $\mathcal{N}=2$ class $\mathcal{S}$ theories. In particular, we focus on twisted A-type fixtures and untwisted D-type fixtures. In Section \ref{bouquets}, we consider magnetic quivers, which contain a bouquet of gauge nodes, corresponding to certain $6$d $\mathcal{N}=(1,0)$ theories. In Section \ref{sodd} we consider orthosymplectic quivers with of $\mathrm{SO(odd)}$ gauge groups. 
Section \ref{sec:conclusion} concludes and provides an outlook.
The appendices provide background material and computational results.
%
%
\section{Coulomb branch and magnetic lattice}{\label{monopole}}
The Coulomb branch of a $3$d $\mathcal{N}=4$ gauge theory or a magnetic quiver is the fundamental object of this work. Understood as space of dressed monopole operators, for which the magnetic lattice needs to be specified, this geometric space can be characterized by a Hilbert series. This section details how the magnetic lattice is defined for a given quiver and how this enters the Hilbert series computation.
\subsection{Main idea}
In this paper, we consider quiver gauge theories, which are gauge theories encoded by a labeled graph called a quiver. Here we specify \emph{how the gauge theory is defined by the quiver}. We first give this recipe in an abstract way, and then illustrate on several examples. 

A quiver $\mathsf{Q}$ can be seen as a way of encoding a representation $\phi : G \rightarrow GL(V)$ of a group $G$ into a finite dimensional vector space $V$. We follow standard conventions, associating a \emph{group} to each vertex of the quiver, and a bifundamental representation to a link between two vertices. Importantly, it is not enough to specify a Lie algebra at each vertex of the quiver; in general several groups correspond to the same algebra, and give distinct gauge theories. Other kinds of representations, beyond bifundamentals, are also allowed. 
For a given quiver $\mathsf{Q}$ with gauge group $G$ and symmetry group $\mathrm{ker}\phi$, there is a choice which $3$d $\mathcal{N}=4$ gauge theory one likes to consider. For any normal subgroup $H \trianglelefteq\mathrm{ker}\phi$, there exists a theory defined by a Lagrangian with matter content represented by $V$ and with a gauge group $G_H$, where
\begin{equation}
\label{defGaugeGroup}
\boxed{
    \frac{G}{\mathrm{ker} \, \phi} \subseteq
     G_H \coloneqq  \frac{G}{H} 
     \subseteq G 
     } \, , 
\end{equation}
as well as interactions dictated by supersymmetry. 
Given a quiver $\mathsf{Q}$ and the choice of $H$, we define the \emph{Coulomb branch} of the pair $(\mathsf{Q},H)$, denoted $\mathcal{C}_H(\mathsf{Q})$ as being the Coulomb branch of the $3$d $\mathcal{N}=4$ gauge theory defined in the previous sentence. It is important to notice that this is a definition of what we mean by the Coulomb branch of a quiver with a choice of group $H$. Furthermore, the different choices of subgroups of $\mathrm{ker} \phi$ lead to an orbifold relation among the Coulomb branches:  
\begin{subequations}
    \label{orbifoldddddrelation}
\begin{equation}
 \mathcal{C}_{\{1\}}(\mathsf{Q}) 
 = \frac{\mathcal{C}_{H}(\mathsf{Q})}{H}  
 \qquad \text{for any}
 \quad H \trianglelefteq\mathrm{ker} \phi \,.
\end{equation}
Alternatively, using the quotient $N_H = \frac{\mathrm{ker} \phi }{ H} $ , one arrives at
\begin{align}
\mathcal{C}_{H}(\mathsf{Q}) = \frac{\mathcal{C}_{\mathrm{ker} \phi}(\mathsf{Q}) }{ N_H } 
\,.
\end{align}
\end{subequations}
Given a unitary quiver with nodes $\mathrm{U}(k)$, for instance, then each node has a subgroup $\mathbb{Z}_k$ which one can choose to ungauge or not. Hence, there exists multitude of \emph{different} orbifold moduli spaces that one can construct from a single quiver simply by choosing different nodes to ungauge a discrete group.

 The $3$d theory is only an intermediate tool, and indeed $\mathcal{C}_{H}(\mathsf{Q})$ can be used in a large variety of contexts, in particular in higher dimensional SCFTs, as illustrated in the present paper, and as a combinatorial way of describing a symplectic singularity. In order to learn about $\mathcal{C}_{H}(\mathsf{Q})$, an important computational tool is the Hilbert series, which can be computed using the \emph{monopole formula}, in which the precise definition of the gauge group $G_H$ in (\ref{defGaugeGroup}) plays a central role. 

In the case where the kernel is trivial $\mathrm{ker} \, \phi = \{1\}$, the group $H$ is necessarily trivial as well, and the Coulomb branch of the quiver is denoted simply $\mathcal{C}(\mathsf{Q})$. 

As a remark, the choice $H=\mathrm{ker} \phi$ seems to be suitable for the Coulomb branches that describe the Higgs branches of the 6d theories \cite{6dmagnetic,newMarcus} of Section \ref{bouquets}, the 5d theories \cite{5dweb} of the Sections \ref{5dexamples} - \ref{general5dfamilysection}, and the class $\mathcal{S}$ theories of Section \ref{classSguys}.
Unless stated otherwise, $H=\rm{ker} \phi$ will be the standard convention in this paper.
 Nonetheless, the choice $H=\{1\}$ for orthosymplectic quivers is elaborated on in Section \ref{5dexamples}, because it defines a viable model as well.
\subsection{Monopole formula}
\label{subsectionMonopoleFormula}
The monopole formula proposed in \cite{monopole} is designed to compute the Coulomb branch Hilbert series of a good or ugly\footnote{For good or ugly theories, all operators have $\Delta \geq \frac{1}{2}$ and the only operator with $\Delta =0$ is the identity operator. Theories containing operators other than the identity with $\Delta<\frac{1}{2}$ are called bad and the monopole formula diverges.} $3$d $\mathcal{N}=4$ theory by counting \textit{dressed monopole operators}. As such, we can also view the Coulomb branch as the moduli space of dressed monopole operators.

\begin{figure}[t]
    \centering
    \begin{subtable}[t]{1\textwidth}
    \centering
    \begin{tabular}{|c|c|}  \hline 
        Group & $\Delta_{\text{vec}}$ \\  \hline 
        $\mathrm{U}(r)$ & $-\sum_{i<j}^r |m_i - m_j| $\\ 
        $\mathrm{SO}(2r)$ & $-\sum_{i<j}^r (|m_i+m_j| +|m_i - m_j|) $ \\ 
        $\mathrm{O}(2r)$ & $-\sum_{i<j}^r (|m_i+m_j| +|m_i - m_j|) $ \\ 
        $\mathrm{SO}(2r+1)$ & $-\sum_{i<j}^r (|m_i+m_j| +|m_i - m_j|)-\sum_{i=1}^r|m_i|$ \\ 
        $\mathrm{USp}(2r)$ & $-\sum_{i<j}^r (|m_i+m_j| +|m_i - m_j|) -2\sum_{i=1}^r|m_i|  $ \\ \hline 
    \end{tabular}
    \caption{The contribution of vector multiplets for various gauge groups.}
    \label{subtab:1}
    \end{subtable}
    
    \vspace{0.5cm}
    \begin{subtable}[t]{1\textwidth}
    \centering
    \begin{tabular}{|c|c|}  \hline 
        Representation & $\Delta_{\text{hyp}}$ \\  \hline 
        $\mathrm{SO}(2r)_{m} \times \mathrm{USp}(2k)_{n}$ bifundamental & $\frac{1}{2}  \sum_{i=1}^k \sum_{j=1}^r (|n_i-m_j| + |n_i + m_j|) $ \\
        $\mathrm{SO}(2r)_{m} \times \mathrm{U}(k)_{n}$ bifundamental & $\frac{1}{2} \sum_{i=1}^k \sum_{j=1}^r (|n_i-m_j| + |n_i + m_j|)$  \\  
        $ \mathrm{U}(r)_{m} \times \mathrm{USp}(2k)_{n}$ bifundamental & $\frac{1}{2}  \sum_{i=1}^k \sum_{j=1}^r (|n_i-m_j| + |n_i + m_j|) $ \\ 
        $\mathrm{SO}(2r+1)_{m} \times \mathrm{USp}(2k)_{n}$ bifundamental & $\frac{1}{2} \sum_{i=1}^k \sum_{j=1}^r (|n_i-m_j| + |n_i + m_j|) +\frac{1}{2} \sum_{i=1}^k |n_i| $  \\ \hline 
    \end{tabular} 
    \caption{The contribution of hypermultiplets in bifundamental representations.}
    \label{subtab:2}
    \end{subtable}
    
        \vspace{0.5cm}
      \begin{subtable}[t]{1\textwidth}
    \centering
    \begin{tabular}{|c|c|}  \hline 
        Representation & $\Delta_{\text{vec}}+\Delta_{\text{hyp}}$ \\  \hline 
        $\mathrm{SO}(2r)_{m}$ with antisymmetric $\Lambda^2$ & $0 $ \\ 
        $\mathrm{USp}(2k)_{m} $ with antisymmetric $\Lambda^2$ &$-2\sum_{i=1}^k |m_i| $  \\
        $\mathrm{U}(k)_{m}$  with $l$ charge 2 hypermultiplets & $l\sum_{i=1}^k |m_i| -\sum_{i<j}^k |m_i - m_j| $  \\   \hline 
    \end{tabular}
     \caption{The combined contribution of the vector and hypermultiplets for special representations.
     }
    \label{subtab:3}
    \end{subtable}
    \caption[Contributions to conformal dimension $\Delta$]{Contributions to the conformal dimension $\Delta$ that appear in the monopole formula \eqref{mono}. (\subref{subtab:1}) summarises the vector mulitplet contributions, (\subref{subtab:2}) collects the hypermultiplet contributions, and (\subref{subtab:3}) provides the combined parts for certain special representations.
    Note in particular that a bifundamental of  $\mathrm{SO}(2r)_{m} \times \mathrm{U}(k)_{n}$ contributes exactly the same as a bifundamental of $\mathrm{SO}(2r)_{m} \times \mathrm{USp}(2k)_{n}$ and of $\mathrm{USp}(2r)_{m} \times \mathrm{U}(k)_{n}$, as required by the fact that $\mathrm{U}(k)$ should be seen as a subgroup of $\mathrm{USp}(2k)$.}
    \label{tableDefinitions}
\end{figure}

For a theory with gauge group $G$, the unrefined\footnote{For every $\mathrm{U}(1)$ factor in the center $Z(G)$ of the gauge group $G$ there is a topological $\mathrm{U}(1)$ symmetry which can be used to refine the monopole formula \cite{monopole}.} monopole formula is given as: 
\begin{equation}{\label{mono}}
\mathrm{HS}(t)=\sum\limits_{ \textbf{m}\; \in \Lambda^{G^\vee}_{m}/\mathcal{W}_{G^\vee} } t^{2\Delta(\textbf{m})}P_G(\textbf{m},t) \; 
\end{equation}
where $\textbf{m}=(m_1,m_2,\ldots,m_{\text{rk(G)}})$ are the magnetic charges of the dressed monopole operators associated with the gauge group. To sum over the monopole operators, the formula works by summing over the magnetic charges in the weight lattice of the GNO dual group $\Lambda^{G^\vee}_{m}$. By taking the quotient of the weight lattice by the Weyl group $\mathcal{W}_{G^\vee}$, we ensure only gauge invariant monopole operators are counted \cite{GNO}. We briefly discuss the ingredients of (\ref{mono}), referring to \cite{monopole} for details. 
\begin{itemize}
    \item The conformal dimension $\Delta$ can be decomposed into two parts: 
\begin{equation}
    \Delta = \Delta_{\text{hyp}} +\Delta_{\text{vec}} 
\label{eq:cdimvec}
\end{equation}
where $\Delta_{\text{vec}}$ is the contribution from the vector multiplets and  $\Delta_{\text{hyp}}$ is the contribution from the hypermultiplets. The formulas for $\Delta_{\text{hyp}}$ and $\Delta_{\text{vec}}$ are given in Figure \ref{tableDefinitions}. 
We note the conformal dimension is the same if the groups are $\mathrm{O}(2r)$ and $\mathrm{O}(2r+1)$ \cite{tsigmarho,CabreraZhong}.
\item The classical factors $P_{\mathrm{SO}(n)}$ and $P_{\mathrm{USp}(2n)}$ are given in \cite[Appendix A]{monopole}. 
\item The summation range $\textbf{m}\; \in \Lambda^{G^\vee}_{m}/\mathcal{W}_{G^\vee} $ is discussed in the next subsection. 
\end{itemize}
Solely based on the conformal dimension, the Weyl group, and the dressing factors one can conjecture that
\begin{equation}
    \mathcal{C}\left(
    \raisebox{-.5\height}{
    \begin{tikzpicture}
        \node[gauge3,label=below:{$\mathrm{O}(2r)$}] (1) at (0,-0.5) {};
        \node[flavor0,label=above:{$\mathrm{C}_N$}] (2) at (0,0.5) {};
        \draw (1)--(2);
    \end{tikzpicture}
    }
    \right)
    =\mathcal{C}\left(
    \raisebox{-.5\height}{\begin{tikzpicture}
        \node[gauge3,label=below:{$\mathrm{O}(2r{+}1)$}] (1) at (0,0) {};
        \node[flavor0,label=above:{$\mathrm{C}_{N{+}1}$}] (2) at (0,1) {};
        \draw (1)--(2);
    \end{tikzpicture}}
    \right)
    =\mathcal{C}\left(
    \raisebox{-.5\height}{
    \begin{tikzpicture}
        \node[gauge3,label=below:{$\mathrm{SO}(2r{+}1)$}] (1) at (0,0) {};
        \node[flavor0,label=above:{$\mathrm{C}_{N{+}1}$}] (2) at (0,1) {};
        \draw (1)--(2);
    \end{tikzpicture}
    }
    \right)
    =\mathcal{C}\left(
    \raisebox{-.5\height}{
    \begin{tikzpicture}
        \node[gauge3,label=below:{$\mathrm{Sp}(r)$}] (1) at (0,0) {};
        \node[flavor0,label=above:{$\mathrm{D}_{N{+}2}$}] (2) at (0,1) {};
        \draw (1)--(2);
    \end{tikzpicture}}
    \right)
\end{equation}
holds for $N\geq2r$, as their Hilbert series agree.

\subsection{Generalities on lattices}

\renewcommand{\arraystretch}{1.5}
\begin{figure}[t]
    \centering
        \begin{subtable}[t]{1\textwidth}
    \centering
        \vspace*{2cm}\scalebox{0.9}{
    \begin{tabular}{|c|c|c|c|c|c|} \hline 
     $G$   &  \begin{tabular}{c}
         $\mathrm{U}(n)$ \\
        ($n \geq 1$)
     \end{tabular}   & \begin{tabular}{c}
         $\mathrm{SU}(n)$ \\
        ($n \geq 1$)
     \end{tabular}  & \begin{tabular}{c}
        $\mathrm{SO}(2n{+}1)$ \\
       ($n \geq 1$)
     \end{tabular}  & \begin{tabular}{c}
        $\mathrm{USp}(2n)$ \\
       ($n \geq 1$)
     \end{tabular}  & \begin{tabular}{c}
        $\mathrm{SO}(2n)$ \\
       ($n \geq 2$)
     \end{tabular}    \\  \hline 
       $\Lambda^{\mathfrak{g}}_r$ & $\mathbb{Z}^n_{\Sigma=0}$ & $\mathbb{Z}^{n}_{\Sigma=0}$ & $\mathbb{Z}^n$ & 
$\mathbb{Z}^n_{2|\Sigma}$
         & $\mathbb{Z}^n_{2|\Sigma}$ \\  \hline 
                \begin{tabular}{c}
         Weight Lattice \\ $\Lambda^{G}_w$
        \end{tabular}   & $\mathbb{Z}^n$ & $\mathbb{Z}^{n}/\delta$ &   $\mathbb{Z}^n$   &    $\mathbb{Z}^n$       &  $\mathbb{Z}^n$   \\  \hline 
$\Lambda^{\mathfrak{g}}_w$ & NA  &  $\mathbb{Z}^{n}/\delta$   &    $\mathbb{Z}^n_{\frac{1}{2}}$    & $\mathbb{Z}^n$   &   $\mathbb{Z}^n_{\frac{1}{2}}$ \\  \hline 
$ \Lambda^{\mathfrak{g}}_{cw}  = \left( \Lambda^{\mathfrak{g}}_{r} \right)^\ast $ & $\mathbb{Z}^{n}/\delta$   & $\mathbb{Z}^{n}/\delta$    & $\mathbb{Z}^n$ &  $\mathbb{Z}^n_{\frac{1}{2}}$  &$\mathbb{Z}^n_{\frac{1}{2}}$  \\  \hline 
\begin{tabular}{c}
         Magnetic Lattice \\
         $\Lambda^{G^{\vee}}_w = \left( \Lambda^{G}_{w} \right)^\ast $
        \end{tabular}  &  $\mathbb{Z}^n$  &  $\mathbb{Z}^{n}_{\Sigma=0}$  &    $\mathbb{Z}^n$   &    $\mathbb{Z}^n$       &  $\mathbb{Z}^n$  \\   \hline 
                 $  \Lambda^{\mathfrak{g}}_{cr} = \left( \Lambda^{\mathfrak{g}}_{w} \right)^\ast $ &  NA &   $\mathbb{Z}^{n}_{\Sigma=0}$ &   $\mathbb{Z}^n_{2|\Sigma}$ & 
          $\mathbb{Z}^n$
         &  $\mathbb{Z}^n_{2|\Sigma}$ \\  \hline              
        $\frac{ \Lambda^{\mathfrak{g}}_w}{ \Lambda^{\mathfrak{g}}_r} = \frac{  \Lambda^{\mathfrak{g}}_{cw}  }{  \Lambda^{\mathfrak{g}}_{cr} }  $ & NA & $\mathbb{Z}_n$  &     $\mathbb{Z}_2$ &  $\mathbb{Z}_2$   &  
       \begin{tabular}{cc}
          $\mathbb{Z}_4$ & ($n$ odd)\\
          $\mathbb{Z}_2 \times \mathbb{Z}_2$ & ($n$ even) 
        \end{tabular} \\  \hline             
       \begin{tabular}{c}
         Center \\
        $ Z(G) = \frac{\Lambda^{\mathfrak{g}}_{cw}}{\Lambda^{G^{\vee}}_w } = \frac{\Lambda^G_w}{\Lambda^{\mathfrak{g}}_r}$
        \end{tabular}  & $\mathrm{U}(1)$  &  $\mathbb{Z}_n$  &    $1$  &  $\mathbb{Z}_2$  &  $\mathbb{Z}_2$ \\  \hline 
    \end{tabular}
    }
    \caption{The various lattices for the groups used in this paper. The explanation of the notation is given in the Figure \ref{subtab:new2} below.}
    \label{subtab:new1}
    \end{subtable}
    
    \vspace{0.5cm}
        \begin{subtable}[t]{1\textwidth}
    \centering
    \begin{tabular}{|c|c|} \hline
    Set & Abbreviation \\ \hline
         \begin{tabular}{c} 
          $(a_1,\ldots,a_n) \in \mathbb{Z}^n$ \\
          $\sum a_i = 0$ 
        \end{tabular}          & $\mathbb{Z}^n_{\Sigma=0}$ \\ \hline
        $\mathbb{Z}^n \cup \left( \mathbb{Z}  + \frac{1}{2}\right)^n$ & $\mathbb{Z}^n_{\frac{1}{2}}$ \\ \hline
        \begin{tabular}{c}
          $(a_1,\ldots,a_n) \in \mathbb{Z}^n$ \\
          $\sum a_i$ even 
        \end{tabular} & $\mathbb{Z}^n_{2|\Sigma}$ \\ \hline 
   $\mathbb{Z}^{n}/\mathbb{Z}(1,\ldots , 1)$    & $\mathbb{Z}^{n}/\delta$  \\ \hline 
    \end{tabular}
    \caption{Explanation of the notation used in Figure \ref{subtab:new1}.}
    \label{subtab:new2}
    \end{subtable}
    \caption[Various lattices for groups]{The different lattices used in this paper are collected in (\subref{subtab:new1}); while the notation is clarified in (\subref{subtab:new2}). 
    The lattices $\mathbb{Z}^n_{\Sigma=0}$ and $\mathbb{Z}^{n}/\delta$ (of rank $n-1$) are dual, as are the lattices $\mathbb{Z}^n_{2|\Sigma}$ and $\mathbb{Z}^n_{\frac{1}{2}}$ (of rank $n$). Note that, as explained in the text, $\Lambda^{\mathfrak{g}}_w$ and $\Lambda^{\mathfrak{g}}_{cw}$ are not lattices for the non semisimple group $\mathrm{U}(n)$, whose center is not discrete, so they are not shown in the figure. In this case, the root lattice and the weight lattice have ranks differing by one. 
    } 
    \label{tableLattices}
\end{figure}
\renewcommand{\arraystretch}{1}

We now review some classical aspects of lattices associated to Lie groups and Lie algebras (see \cite[Chapters 15 and 18]{bump2004lie}, \cite[Chapter 4]{onishchik2012lie} and \cite{MarcusLattice} in the context of the monopole formula). 

Let $G$ be a connected compact Lie group, and let $T$ be a maximal compact torus. 
A \emph{character} of $T$ is a continuous homomorphism $T \rightarrow \mathrm{U}(1)$. The characters of $T$ form an additive group that we denote by $\Lambda_w (G)$, and that we call the \emph{weight lattice} of the group $G$. 

Assume now that $G$ is semisimple. Let $\mathfrak{g}$ be the (semisimple) Lie algebra of $G$, with Cartan subalgebra $\mathfrak{h}$. We can define in the standard way the \emph{weight lattice} $\Lambda^{\mathfrak{g}}_w$ and the \emph{root lattice} $\Lambda^{\mathfrak{g}}_r$ of $\mathfrak{g}$. We have the inclusions $\Lambda^{\mathfrak{g}}_r \subseteq \Lambda_w (G) \subseteq \Lambda^{\mathfrak{g}}_w \subseteq \mathfrak{h}^{\ast}$. 
For any lattice $\Lambda$, we call $\Lambda^{\ast} =  \mathrm{Hom}(\Lambda ,\mathbb{Z})$ the dual lattice. Taking the dual of all three lattices in the above inclusion, we obtain three more lattices, with reversed inclusions: 
\begin{equation}
\label{inclusions}
\begin{tabular}{cccccccccc}
 & & $\Lambda^{\mathfrak{g}}_r$ & $\subseteq $ & $\Lambda^G_w$ &  $\subseteq $ &  $\Lambda^{\mathfrak{g}}_w$ & $\subset$ & $\mathfrak{h}^*$\\
 & & $\updownarrow $ & & $\updownarrow $ & & $\updownarrow $ \\
$\mathfrak{h}$ & $\supset$ & $\Lambda^{\mathfrak{g}}_{cw} \coloneqq \left( \Lambda^{\mathfrak{g}}_{r} \right)^{\ast}$ & $\supseteq$ & $\Lambda^{G^{\vee}}_w \coloneqq \left( \Lambda^G_w \right)^{\ast} $ & $\supseteq$ & $\Lambda^{\mathfrak{g}}_{cr} \coloneqq \left( \Lambda^{\mathfrak{g}}_{w} \right)^{\ast} $.
\end{tabular}
\end{equation}
The lattice $\Lambda^{G^{\vee}}_w$ is called the \emph{magnetic lattice}. The center of $G$ is 
\begin{equation}
    Z(G) = \frac{\Lambda^{\mathfrak{g}}_{cw}}{\Lambda^{G^{\vee}}_w } = \frac{\Lambda^G_w}{\Lambda^{\mathfrak{g}}_r} \, . 
\end{equation}
Note specifically, that for $H$ a subgroup of $Z(G)$ we have:
\begin{equation}
\label{inclusions2}
\begin{tabular}{ccc}
 $\Lambda^{G/H}_w$ & $\subseteq $ & $\Lambda^G_w$\\
 $\updownarrow $ & & $\updownarrow $ \\
$\Lambda^{(G/H)^{\vee}}_w \coloneqq \left( \Lambda^{G/H}_w \right)^{\ast} $  & $\supseteq$ & $\Lambda^{G^{\vee}}_w \coloneqq \left( \Lambda^G_w \right)^{\ast} $ .
\end{tabular}
\end{equation}
We can now turn to the case of a reductive, but non semisimple group $G$. In this case, one can still define all the objects of (\ref{inclusions}), satisfying the same relations of inclusion and duality; however in this case they are not necessarily lattices \cite{KIRILLOV2006576}. This is in particular the case for $\mathrm{U}(n)$. Since we only need the $\mathrm{U}(n)$ case in this paper, we refrain from spelling out the entire theory and refer instead to Figure \ref{tableLattices}, where we gather the various lattices and spaces involved in (\ref{inclusions}) for several important choices of $G$.

For a direct product of several groups in Figure \ref{tableLattices}, one just takes the direct sum of the corresponding lattices. The magnetic lattice for $G$ in \eqref{defGaugeGroup} is obtained this way. Then to obtain the magnetic lattice of the quotient group $G_H$ of $G$ by $H \trianglelefteq \mathrm{ker} \, \phi$, the $H$ quotient has to be taken over this whole summed lattice. Examples are presented in the next subsection. 

In the following and throughout the rest of the paper, we will use the notation $\mathrm{HS}_{\Lambda}$ to denote Hilbert series where the monopole formula sums over the magnetic lattice $\Lambda$. For two lattices $\Lambda_1$ and $\Lambda_2$ such that $\Lambda_1 \cap \Lambda_2 = \emptyset$, we then have 
\begin{equation}
    \mathrm{HS}_{\Lambda_1 \cup \Lambda_2} = \mathrm{HS}_{\Lambda_1} + \mathrm{HS}_{ \Lambda_2}  \, . 
\end{equation}
Note that if $\Lambda$ does not contain the origin, the corresponding Hilbert series has a zero constant term.

\subsection{Examples}
The choice of $H$ changes the magnetic lattice. For $H=\{1\}$, the magnetic lattice is the direct sum of the individual magnetic lattices. In this section, we focus on some general families and their precise magnetic lattice for different choices of $H \trianglelefteq\mathrm{ker}{\phi}$. 
\subsubsection{Framed quivers with fundamental hypermultiplets \texorpdfstring{-- $\mathrm{ker}\phi=\{1\}$}{}}
If the quiver under consideration is framed, i.e.\ it contains fields in the fundamental representation of one of the factors of the gauge group, then $\mathrm{ker} \, \phi$ is trivial. This applies both to unitary, orthosymplectic, or mixed quivers. For those quivers, the magnetic lattice is just the direct sum of the magnetic lattices of the various simple groups that appear as nodes of the quiver. 
\subsubsection{QED with charge \texorpdfstring{$q$}{q} hypermultiplets \texorpdfstring{-- $\mathrm{ker}\phi=\mathbb{Z}_q$}{}}
\label{subsubsectionEX2}
We now consider the following quiver, which contains only one $\mathrm{U}(1)$ gauge node with $l$ charge 2 hypermultiplets: 
\begin{equation}
\label{Uwithcharge2}
\raisebox{-.5\height}{
    \begin{tikzpicture}
	\begin{pgfonlayer}{nodelayer}
		\node [style=gauge3] (41) at (-1, 1) {};
 		\node [style=none] (43) at (-0.5, 1) {$1$};
		\node [style=none] (44) at (-0.5, 2.3) {$l$};
		\node [style=flavor2] (48) at (-1, 2.3) {};
	\end{pgfonlayer}
	\begin{pgfonlayer}{edgelayer}
		\draw [line join=round,decorate, decoration={zigzag, segment length=4,amplitude=.9,post=lineto,post length=2pt}]  (48) -- (41);
	\end{pgfonlayer}
\end{tikzpicture}
}
\end{equation}
We use a wiggly line to denote specifically charge 2 hypermultiplets throughout the paper, as done in \cite{5dweb}. 
he initial gauge group is $G=\mathrm{U}(1)$, and since the charge of the hypermultiplets is 2, $\mathrm{ker}\phi =\mathbb{Z}_2$. Hence, there exist two choices of gauge groups: $G$ and $G_{\mathbb{Z}_2}=\mathrm{U}(1)/\mathbb{Z}_2$ with magnetic lattices $\mathbb{Z}$ and $\mathbb{Z}\cup(\mathbb{Z}+\frac{1}{2})$, respectively. We will now discuss each case.
\paragraph{Gauge group $G$.}
For this choice, the Hilbert series for the Coulomb branch is obtained from the formulas of Section \ref{subsectionMonopoleFormula}. One finds
\begin{equation}
    \mathrm{HS}_{\mathbb{Z}}(t) = \frac{1-t^{4l}}{(1-t^2)(1-t^{2l})^2} \, . 
    \label{nomodding}
\end{equation}
This corresponds to the Hilbert series of the Coulomb branch of a $\mathrm{U}(1)$ gauge theory with $2l$ charge $1$ hypermultiplets, which is known to be $\mathbb{C}^2/\mathbb{Z}_{2l}$.
\paragraph{Gauge group $G_{\mathbb{Z}_2}$.}
Modding out the kernel, the relevant monopole formula is 
\begin{equation}
  \mathrm{HS}_{\mathbb{Z}\; \cup  \;\mathbb{Z} + \frac{1}{2}}(t)=  \sum\limits_{ \mathbb{Z}\; \cup\; (\mathbb{Z}+\frac{1}{2})} \frac{t^{2l |m|}}{1-t^2} \, .
  \label{modding}
\end{equation}
This sum can be decomposed into a sum over the integers and a sum over the integers-plus-half. The first one $\mathrm{HS}_{\mathbb{Z}}$ is computed above in (\ref{nomodding}). The contribution of the integers-plus-half is 
\begin{equation}
    \mathrm{HS}_{\mathbb{Z} + \frac{1}{2}}(t) = \frac{2 t^{l}}{(1-t^2)(1-t^{2l})} \, . 
\end{equation}
Combining both contributions, we obtain 
\begin{equation}
  \mathrm{HS}_{\mathbb{Z}\; \cup  \;(\mathbb{Z} + \frac{1}{2})}(t)=  \mathrm{HS}_{ \mathbb{Z}}(t) +  \mathrm{HS}_{ \mathbb{Z} + \frac{1}{2}}(t) = \frac{1 - t^{2l}}{(1-t^2)(1-t^l)^2} \, . 
\end{equation}
The geometry is now $\mathbb{C}^2/\mathbb{Z}_{l}$. This shows clearly that the distinction between a $\mathrm{U}(1)$ gauge theory with $2l$ charge $1$ hypermultiplets and $l$ charge $2$ hypermultiplets comes from the magnetic lattice. Furthermore, the space $ \mathcal{C}_{\mathbb{Z}_2}\eqref{Uwithcharge2}$ with Hilbert series  \eqref{modding}  and the space $ \mathcal{C}_{\{1\}}\eqref{Uwithcharge2}$ with Hilbert series \eqref{nomodding}  are related by the orbifold relation:
\begin{equation}
    \mathcal{C}_{\{1\}}\eqref{Uwithcharge2} = \frac{   \mathcal{C}_{\mathbb{Z}_2}\eqref{Uwithcharge2}}{\mathbb{Z}_2} \, , 
\end{equation}
verifying the general relations in \eqref{orbifoldddddrelation}. 
In particular, for $l=1$ we find the following moduli space: 
\begin{equation}
\mathcal{C} \left( 
   \raisebox{-0.5\height}{ \begin{tikzpicture}
	\begin{pgfonlayer}{nodelayer}
		\node [style=gauge3] (41) at (-1, 1) {};
 		\node [style=none] (43) at (-0.5, 1) {$1$};
		\node [style=none] (44) at (-0.5, 2.3) {$1$};
		\node [style=flavor2] (48) at (-1, 2.3) {};
	\end{pgfonlayer}
	\begin{pgfonlayer}{edgelayer}
		\draw [line join=round,decorate, decoration={zigzag, segment length=4,amplitude=.9,post=lineto,post length=2pt}]  (48) -- (41);
	\end{pgfonlayer}
\end{tikzpicture} } \right) = \mathbb{H} \, ,
 \end{equation}
 which appears in Sections \ref{5dexamples} and  \ref{general5dfamilysection}.  
This argument generalizes without difficulty to $l$ hypermultiplets of charge $q$, where $q$ is any positive integer. In this case $\mathrm{ker} \phi=\mathbb{Z}_q$ and the magnetic lattice for this choice is $\bigcup_{i=0}^{q-1}(\mathbb{Z}+\frac{i}{q})$.
For the choice $H=\mathbb{Z}_d \trianglelefteq \mathbb{Z}_q$, with $d$ being a divisor of $q$, the Coulomb branches satisfy
\begin{subequations}
 \begin{align}
    \mathcal{C}_{\{1\}} &= \mathcal{C}_{\mathbb{Z}_d}\slash \mathbb{Z}_d
    = \mathbb{C}^2 \slash \mathbb{Z}_{q\cdot l} \; \, , \qquad 
\mathcal{C}_{\mathbb{Z}_d} =\mathbb{C}^2 \slash \mathbb{Z}_{\frac{ q\cdot l}{d}} \,, \\
\mathcal{C}_{\mathbb{Z}_d} &=
\mathcal{C}_{\mathbb{Z}_q} \slash \mathbb{Z}_{\frac{q}{d}} 
\quad \text{with} \quad 
\mathcal{C}_{\mathbb{Z}_q} = \mathbb{C}^2 \slash \mathbb{Z}_{ l}
\end{align}
\end{subequations}
which exemplifies \eqref{orbifoldddddrelation}.
 
If we take $\mathrm{U}(1)$ with $l$ hypermultiplets of possibly different charges $q_i$, then the kernel is $\mathbb{Z}_{q}$ with $q=\mathrm{gcd}(q_1,\dots,q_l)$.  
For an  $H=\mathbb{Z}_d \trianglelefteq \mathbb{Z}_{q}$, where $d$ is a divisor of $q$, the orbifold relations \eqref{orbifoldddddrelation} become
\begin{subequations}
\begin{align}
 \mathcal{C}_{\{1\}} &= \mathcal{C}_{\mathbb{Z}_d}\slash \mathbb{Z}_d
 = \mathbb{C}^2 \slash \mathbb{Z}_{\sum_{i}q_i} 
 \; , \qquad \mathcal{C}_{\mathbb{Z}_d} = \mathbb{C}^2 \slash \mathbb{Z}_{\frac{\sum_{i}q_i}{d}}\,, \\
\mathcal{C}_{\mathbb{Z}_d} &=
\mathcal{C}_{\mathbb{Z}_q} \slash \mathbb{Z}_{\frac{q}{d}} 
\quad \text{with} \quad 
\mathcal{C}_{\mathbb{Z}_q} = \mathbb{C}^2 \slash \mathbb{Z}_{ \frac{\sum_{i}q_i}{q}}
 \,.
\end{align}
 \end{subequations}
\subsubsection{Unframed unitary quivers \texorpdfstring{-- $\mathrm{ker}\phi=\mathrm{U}(1)$}{}}
\label{sec:uni}

Consider now a quiver with only unitary gauge nodes, and with only bifundamental matter (unframed). In this case $\mathrm{ker} \, \phi = \mathrm{U}(1)^{\textrm{diag}}$, the diagonal abelian part of all the unitary groups. The gauge group is therefore of the form 
\begin{equation}
\label{Uquotient}
   \left( \prod_i \mathrm{U}(n_i)  \right) / \mathrm{U}(1) \, . 
\end{equation}
Figure \ref{tableLattices} can be used to determine the magnetic lattice. Each $\mathrm{U}(n_i)$ contributes a $\mathbb{Z}^{n_i}$ lattice. The action of the $\mathrm{U}(1)^{\textrm{diag}}$ quotient can be read in the $\Lambda_{cw}^{\mathfrak{g}}$ line of Figure \ref{tableLattices}, as this is the magnetic lattice for the centerless group $G/Z(G)$. In the present case, the lattice is computed by taking a quotient by the line spanned by $(1,\ldots,1)$. Setting $N=\sum_i n_i$, the lattice becomes
\begin{equation}
    \mathbb{Z}^N / \delta \cong \{(0,a_2 , \cdots , a_N) | a_i \in \mathbb{Z} \} \, ,
\end{equation}
using the notation of Figure \ref{tableLattices}.
In other words, any one of the magnetic weights in $\mathbb{Z}^N$ can be set equal to $0$ in order to take the modding by $\delta$ into account, and the other $(N-1)$ magnetic weights belong to $\mathbb{Z}^{N-1}$. Physically, this is equivalent to setting an origin to the brane system. Note that the gauge group (\ref{Uquotient}) \emph{cannot} be equivalently realized by just replacing one of the groups in the quiver by an $\mathrm{SU}(n)$ group (in this case, there is still a $\mathbb{Z}_n$ kernel to gauge out), nor by a $\mathrm{PSU}(n)$ group (which does not admit the fundamental representation of $\mathrm{U}(n)$). See the example in Sections  \ref{sec:D4ex} -- \ref{sec:E6ex} for an explicit analysis.

\subsubsection{Unframed orthosymplectic quivers \texorpdfstring{-- $\mathrm{ker}\phi=\mathbb{Z}_2$}{}}
\label{subsubsectionEX1}
The Coulomb branch Hilbert series of orthosymplectic quivers had previously been studied and computed in \cite{tsigmarho,CabreraZhong,RudolphSlodowy}, where most of the Coulomb branches are either closures of nilpotent orbits or their intersections with Slodowy slices. A ubiquitous feature of these quivers is that they all contain flavor nodes\footnote{We note that correct Coulomb branch Hilbert series computations of flavorless orthosymplectic quivers had been given in \cite{Sicilian}. However, the approach used the Hall-Littlewood formula \cite{CremonesiHall} rather than the monopole formula. As a result, it does not use the explicit magnetic lattice of the gauge groups (aside from the central node of the three-legged theories).}. In this paper, we wish to extend our understanding by investigating orthosymplectic quivers that \textit{do not} have flavor nodes. In this case, $G$ is a product of special orthogonal and symplectic groups, and the representation $\phi$ is a direct sum of bifundamental representations. As a consequence, we distinguish two situations: 
\begin{itemize}
    \item If there is at least one $\mathrm{SO}(2r+1)$ node in the quiver, then $\mathrm{ker}\phi$ is trivial and $G_H=G$. 
    \item If there is no node of type $\mathrm{SO}(2r+1)$ in the quiver, then $\mathrm{ker}\phi=\mathbb{Z}_2^{\textrm{diag}}$, and we have two choices for the gauge group. $G_{\{1\}}=G$, which is the product of orthosymplectic gauge groups, and  $G_{\mathbb{Z}_2}=G /\mathbb{Z}_2^{\textrm{diag}}$, which is the product of orthosymplectic groups divided by $\mathbb{Z}_2^{\textrm{diag}}$. 
\end{itemize} 

To see the effect on magnetic lattices, consider a product of two groups, $\mathrm{SO}(2r)$ and $\mathrm{USp}(2k)$. The magnetic lattice for the product $G=\mathrm{SO}(2r) \times \mathrm{USp}(2k)$ is $\mathbb{Z}^r \oplus \mathbb{Z}^k$, see Figure \ref{tableLattices}. For $\mathrm{SO}(2r)/\mathbb{Z}_2 \times \mathrm{USp}(2k)/\mathbb{Z}_2$ we see in Figure \ref{tableLattices} that the magnetic lattice is 
\begin{equation}
    \left[  \mathbb{Z}^r \cup \left( \mathbb{Z}  + \frac{1}{2}\right)^r  \right] \oplus  \left[  \mathbb{Z}^k \cup \left( \mathbb{Z}  + \frac{1}{2}\right)^k  \right] \, . 
\end{equation}
Finally, for $(\mathrm{SO}(2r) \times \mathrm{USp}(2k))/\mathbb{Z}_2^{\textrm{diag}}$ where we quotient by a diagonal $\mathbb{Z}_2^{\textrm{diag}}$ subgroup, we obtain the magnetic lattice 
\begin{equation}
\label{latticeZ2}
    \left[  \mathbb{Z}^r \oplus \mathbb{Z}^k  \right] \cup  \left[  \left( \mathbb{Z}  + \frac{1}{2}\right)^r  \oplus \left( \mathbb{Z}  + \frac{1}{2}\right)^k  \right] = \mathbb{Z}^{r+k} \cup  \left( \mathbb{Z}  + \frac{1}{2}\right)^{r+k} \, . 
\end{equation}

\begin{figure}[t]
    \centering
    \begin{tabular}{|ccc|ccc|ccc|} \hline 
    \multicolumn{3}{|c}{sympletic} & \multicolumn{3}{|c}{orthogonal} & \multicolumn{3}{|c|}{unitary}  \\ \hline 
           \raisebox{-.5\height}{ \begin{tikzpicture}
	\begin{pgfonlayer}{nodelayer}
		\node [style=bluegauge] (20) at (0, 0) {};
		\node [style=none] (24) at (0, -0.5) {$2n$};
		\node [style=none] at (0,0.5) {};
	\end{pgfonlayer}
\end{tikzpicture} } & $\longleftrightarrow$ & $\mathrm{USp}(2n)$  &  
         \raisebox{-.5\height}{   \begin{tikzpicture}
	\begin{pgfonlayer}{nodelayer}
		\node [style=redgauge] (20) at (0, 0) {};
		\node [style=none] (24) at (0, -0.5) {$m$};
		\node [style=none] at (0,0.5) {};
	\end{pgfonlayer}
\end{tikzpicture} } & $\longleftrightarrow$ & $\mathrm{SO}(m)$  &  
        \raisebox{-.5\height}{    \begin{tikzpicture}
	\begin{pgfonlayer}{nodelayer}
		\node [style=gauge3] (20) at (0, 0) {};
		\node [style=none] (24) at (0, -0.5) {$k$};
		\node [style=none] at (0,0.5) {};
	\end{pgfonlayer}
\end{tikzpicture} }  & $\longleftrightarrow$ & $\mathrm{U}(k)$  \\ \hline 
    \end{tabular}
    \caption[Notation on colored nodes]{Colored nodes used in the paper. }
    \label{tabColoredNodes}
\end{figure}

As an illustration, we consider the simplest orthosymplectic quiver:
\begin{equation}
\label{exampleQuiver}
   \raisebox{-0.5\height}{
    \begin{tikzpicture}
	\begin{pgfonlayer}{nodelayer}
		\node [style=bluegauge] (20) at (-2.75, 0) {};
		\node [style=redgauge] (23) at (-4.5, 0) {};
		\node [style=none] (24) at (-2.75, -0.5) {2};
		\node [style=none] (25) at (-4.5, -0.5) {2};
		\node [style=none] (26) at (-3.625, 0.25) {$\tsadi$};
	\end{pgfonlayer}
	\begin{pgfonlayer}{edgelayer}
		\draw (20) to (23);
	\end{pgfonlayer}
\end{tikzpicture}
}
\end{equation}
where the color code for the nodes is summarized in Figure \ref{tabColoredNodes}. $\tsadi$ is the number of copies of hypermultiplets between the two gauge groups (the edge multiplicity). For $\tsadi \geq 3$, the $\mathrm{USp}(2)$ node is not bad in the sense of \cite{GaiottoWitten} and hence the  Coulomb branch Hilbert series does not diverge. The presence of bad gauge nodes occurs frequently in constructing orthosymplectic quivers which limits our ability to study its Coulomb branch.

\begin{figure}[t]
    \centering
    \hspace{-4cm}
    \scalebox{.8}{
       \begin{subfigure}{.1\textwidth}
       \centering
    \begin{tikzpicture}
\tikzstyle{wei}=[circle,inner sep=1pt,fill=red!80!black];
\tikzstyle{cowei}=[draw,circle,inner sep=2pt];
\foreach \x in {-2,-1,...,2}{\foreach \y in {-2,-1,...,2}{\node[wei] at (\x,\y) {};}}
\foreach \x in {-2,-1,...,2}{\foreach \y in {-2,-1,...,2}{\node[cowei] at (\x,\y) {};}}
\foreach \y in {-1,0,1}{\node[draw,star,inner sep=2pt] at (0,2*\y) {};}
\draw[->] (0,-2.5) -- (0,2.5);
\draw[->] (-2.5,0) -- (2.5,0);
\node at (0,3) {$\mathrm{USp}(2)=\mathrm{SU}(2)$};
\node at (3,0) {$\mathrm{SO}(2)$};
\draw[latex-latex] (-2.6,0.5) to[out=-135,in=135] (-2.6,-0.5);  
\node at (-3.123,-3.123) {};
\node at (3.123,3.123) {};
\end{tikzpicture}
\caption{}
\label{subfig:a}
\end{subfigure}
}
\begin{subfigure}{0.325\textwidth}
    \centering
    $\quad  $
\end{subfigure}
    \scalebox{.8}{
           \begin{subfigure}{.1\textwidth}
           \centering
    \begin{tikzpicture}
\tikzstyle{wei}=[circle,inner sep=1pt,fill=red!80!black];
\tikzstyle{cowei}=[draw,circle,inner sep=2pt];
\foreach \x in {-2,-1,...,2}{\foreach \y in {-1,0,1}{\node[wei] at (\x,2*\y) {};}}
\foreach \x in {-2,-1,...,2}{\foreach \y in {-4,-3,...,4}{\node[cowei] at (\x,0.5*\y) {};}}
\draw[->] (0,-2.5) -- (0,2.5);
\draw[->] (-2.5,0) -- (2.5,0);
\node at (0,3) {$\frac{\mathrm{USp}(2)}{\mathbb{Z}_2} \simeq \mathrm{SO}(3)$};
\node at (3,0) {$\mathrm{SO}(2)$};
\draw[latex-latex] (-2.6,0.5) to[out=-135,in=135] (-2.6,-0.5);  
\node at (-3.123,-3.123) {};
\node at (3.123,3.123) {};
\end{tikzpicture}
\caption{}
\label{subfig:b}
\end{subfigure}
}
\begin{subfigure}{0.325\textwidth}
    \centering
    $\quad  $
\end{subfigure}
    \scalebox{.8}{
           \begin{subfigure}{.1\textwidth}
           \centering
    \begin{tikzpicture}
\tikzstyle{wei}=[circle,inner sep=1pt,fill=red!80!black];
\tikzstyle{cowei}=[draw,circle,inner sep=2pt];
\foreach \y in {-2,-1,...,2}{\foreach \x in {-1,0,1}{\node[wei] at (2*\x,\y) {};}}
\foreach \y in {-2,-1,...,2}{\foreach \x in {-4,-3,...,4}{\node[cowei] at (0.5*\x,\y) {};}}
\draw[->] (0,-2.5) -- (0,2.5);
\draw[->] (-2.5,0) -- (2.5,0);
\node at (0,3) {$\mathrm{USp}(2)=\mathrm{SU}(2)$};
\node at (3,0) {$\frac{\mathrm{SO}(2)}{\mathbb{Z}_2}$};
\draw[latex-latex] (-2.6,0.5) to[out=-135,in=135] (-2.6,-0.5);  
\node at (-3.123,-3.123) {};
\node at (3.123,3.123) {};
\end{tikzpicture}   
\caption{}
\label{subfig:c}
\end{subfigure}
}
\\
\hspace{-4cm}
    \scalebox{.8}{
           \begin{subfigure}{.1\textwidth}
           \centering
\begin{tikzpicture}
\tikzstyle{wei}=[circle,inner sep=1pt,fill=red!80!black];
\tikzstyle{cowei}=[draw,circle,inner sep=2pt];
\foreach \y in {-1,0,1}{\foreach \x in {-1,0,1}{\node[wei] at (2*\x,2*\y) {};}}
\foreach \y in {-4,-3,...,4}{\foreach \x in {-4,-3,...,4}{\node[cowei] at (0.5*\x,0.5*\y) {};}}
\draw[->] (0,-2.5) -- (0,2.5);
\draw[->] (-2.5,0) -- (2.5,0);
\node at (0,3) {$\frac{\mathrm{USp}(2)}{\mathbb{Z}_2} \simeq \mathrm{SO}(3)$};
\node at (3,0) {$\frac{\mathrm{SO}(2)}{\mathbb{Z}_2}$};
\draw[latex-latex] (-2.6,0.5) to[out=-135,in=135] (-2.6,-0.5);  
\node at (-3.123,-3.123) {};
\node at (3.123,3.123) {};
\end{tikzpicture}
\caption{}
\label{subfig:d}
\end{subfigure}
}
\begin{subfigure}{0.325\textwidth}
    \centering
    $\quad  $
\end{subfigure}
\scalebox{.8}{
       \begin{subfigure}{.1\textwidth}
       \centering
\begin{tikzpicture}
\tikzstyle{wei}=[circle,inner sep=1pt,fill=red!80!black];
\tikzstyle{cowei}=[draw,circle,inner sep=2pt];
\foreach \y in {-1,0,1}{\foreach \x in {-1,0,1}{\node[wei] at (2*\x,2*\y) {};}}
\foreach \y in {0,1}{\foreach \x in {0,1}{\node[wei] at (2*\x-1,2*\y-1) {};}}
\foreach \y in {-2,-1,...,2}{\foreach \x in {-2,-1,...,2}{\node[cowei] at (\x,\y) {};}}
\foreach \y in {-1,...,2}{\foreach \x in {-1,...,2}{\node[cowei] at (\x-.5,\y-.5) {};}}
\draw[->] (0,-2.5) -- (0,2.5);
\draw[->] (-2.5,0) -- (2.5,0);
\draw[latex-latex] (-2.6,0.5) to[out=-135,in=135] (-2.6,-0.5);  
\node at (-3.123,-3.123) {};
\node at (3.123,3.123) {};
\node at (5,0) {$\mathrm{coker} \, \phi \simeq \frac{\mathrm{SO}(2) \times \mathrm{USp}(2)}{\mathbb{Z}_2}$};
\end{tikzpicture}
\caption{}
\label{subfig:e}
\end{subfigure}
}
    \caption[Lattices for SO(2) x USp(2) example]{ In all the diagrams, the red dots show the weight lattice, and the black circles show the dual lattice, which is the magnetic lattice involved in the monopole formula. The arrow denotes the action of the Weyl group.  
    \subref{subfig:a}: The stars show the root lattice of $\mathrm{USp}(2)$. This notion does not extend to the full group $\mathrm{SO}(2) \times \mathrm{USp}(2)$ because of the Abelian factor. We do not show the roots on the other diagrams. \subref{subfig:b}: $\mathrm{USp}(2)$ is replaced by $\mathrm{SO}(3)$. \subref{subfig:c}: $\mathrm{SO}(2)$ is replaced by $\mathrm{SO}(2)/\mathbb{Z}_2 \simeq \mathrm{SO}(2)$, which rescales the weights. \subref{subfig:d}: Combinations of the two $\mathbb{Z}_2$ modifications of \subref{subfig:b} and \subref{subfig:c}. The weight lattice has index $4$ compared to \subref{subfig:a}. \subref{subfig:e}: Finally this is the weight and coweight lattices for the quiver group. }
    \label{figLattices}
\end{figure}
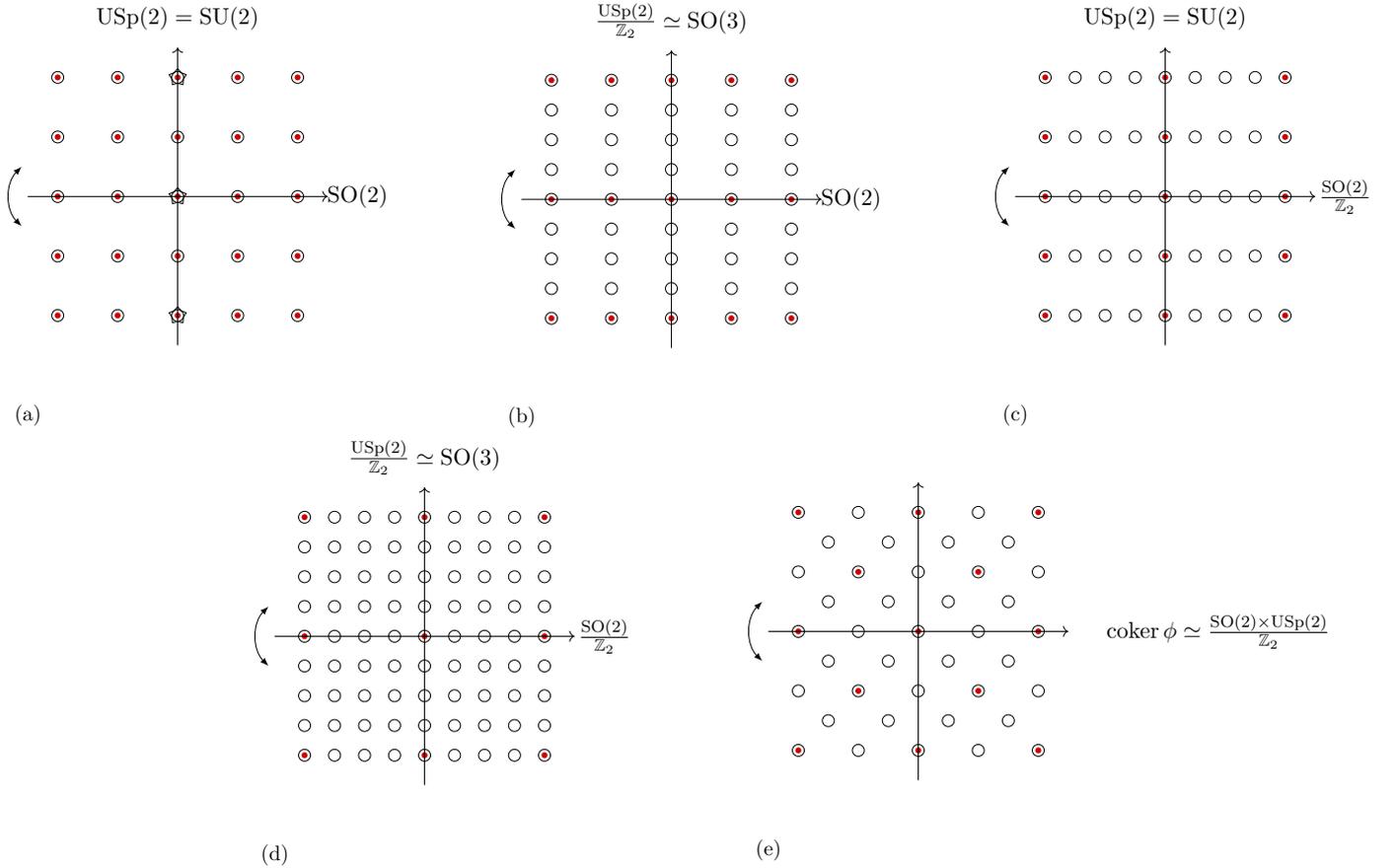

We can now present the different choices of discrete quotients in terms of Hilbert series: 
\begin{subequations}
    \label{proof}
\begin{eqnarray}
    \mathrm{HS}\Big(\mathrm{SO}(2) \times \mathrm{USp}(2)\Big) &=& \mathrm{HS}_{\mathbb{Z}^2}  \label{a}\\
       \mathrm{HS}\left(\frac{\mathrm{SO}(2)}{\mathbb{Z}_2} \times \mathrm{USp}(2)\right) &=& \mathrm{HS}_{\mathbb{Z}^2}+\mathrm{HS}_{(\mathbb{Z}+\frac{1}{2}) \oplus\mathbb{Z}} \\
          \mathrm{HS}\left(\mathrm{SO}(2)\times \frac{\mathrm{USp}(2)}{\mathbb{Z}_2}\right) &=& \mathrm{HS}_{\mathbb{Z}^2}+\mathrm{HS}_{\mathbb{Z} \oplus (\mathbb{Z}+\frac{1}{2})} \\ 
           \mathrm{HS}\left(\frac{\mathrm{SO}(2)}{\mathbb{Z}_2}\times \frac{\mathrm{USp}(2)}{\mathbb{Z}_2}\right) &=& \mathrm{HS}_{\mathbb{Z}^2}+\mathrm{HS}_{\mathbb{Z} \oplus( \mathbb{Z}+\frac{1}{2})} \nonumber\\ & & +\mathrm{HS}_{ \mathbb{Z} \oplus ( \mathbb{Z}+\frac{1}{2})}+\mathrm{HS}_{( \mathbb{Z}+\frac{1}{2})^2} \\
           \mathrm{HS}\left(\frac{\mathrm{SO}(2)\times \mathrm{USp}(2)}{\mathbb{Z}_2^{\textrm{diag}}}\right) &=& \mathrm{HS}_{\mathbb{Z}^2}+\mathrm{HS}_{( \mathbb{Z}+\frac{1}{2})^2} \label{e}
\end{eqnarray} 
\end{subequations}
where the Hilbert series subscripts $\Lambda_{w}^{\mathrm{SO}(2)^\vee} \oplus \Lambda_{w}^{\mathrm{USp}(2)^\vee}$ explicitly denotes the lattice of the magnetic charges of $\mathrm{SO}(2)$ and $\mathrm{USp}(2)$ respectively\footnote{ Note, this is the only time we explicitly label the Hilbert series with their respective magnetic charges. For the rest of the paper, we continue to use the notation $\mathrm{HS}_{\mathbb{Z}}$ to denote the Hilbert series where all charges are integer values and $\mathrm{HS}_{\mathbb{Z}+\frac{1}{2}}$ for half-plus-integer value charges and $\mathrm{HS}_{\mathbb{Z}\; \cup  \;\mathbb{Z}+\frac{1}{2}}$ for their sum.}.  
The Hilbert series takes the form above because, for example, if $m$ takes all integers and $n$ takes all positive integers and integers-plus-half, we can decompose them into two separate Hilbert series. For the quiver (\ref{exampleQuiver}) the gauge group is the quotient $G_{\mathbb{Z}_2} = \frac{\mathrm{SO}(2)\times \mathrm{USp}(2)}{\mathbb{Z}_2^{\textrm{diag}}}$ of $G=\mathrm{SO}(2) \times \mathrm{USp}(2)$ by the kernel $\mathbb{Z}_2^{\textrm{diag}}$. This means that the Coulomb branch Hilbert series of (\ref{exampleQuiver}) is given by \eqref{e}. Note that the GNO dual of $G_{\mathbb{Z}_2}$ is not a product of classical groups. The electric and magnetic lattices for the five cases of equations (\ref{proof}) are represented in Figure \ref{figLattices}. 

\subsubsection{Unframed unitary orthosymplectic quivers \texorpdfstring{-- $\mathrm{ker}\phi=\mathbb{Z}_2$}{}} 
Finally, we can consider quivers which contain both orthosymplectic gauge nodes and unitary nodes, with or without charge 2 hypermultiplets. When this is the case, the same analysis is valid, with $\mathrm{ker} \phi = \mathbb{Z}_2^{\textrm{diag}}$, and the magnetic lattice is a direct sum of a $\mathbb{Z}^r$ component and a $\left(\mathbb{Z} + \frac{1}{2}  \right)^r$ component. A myriad of examples is given in the remaining sections of the paper. 
As a remark, the quiver (\ref{Uwithcharge2}) can be attached to an unflavored unitary-orthosymplectic quiver, so that the kernel of the matter representation is the diagonal $\mathbb{Z}_2^{\textrm{diag}}$ coming from the two $\mathbb{Z}_2$ factors discussed in Sections \ref{subsubsectionEX1} and \ref{subsubsectionEX2}. In these cases, we still refer to them as \textit{unframed} for convenience despite the presence of fundamental hypermultiplets.

\paragraph{Notation.}
The lattice \eqref{latticeZ2}, or its generalizations to arbitrary unframed orthosymplectic or unitary orthosymplectic quivers, 
are relevant for the Coulomb branch Hilbert series computations for the choice $H=\mathrm{ker}\phi$.
To lighten the notation, the following convention is adopted. For the choice $H=\mathrm{ker}\phi$, the Hilbert series $\mathrm{HS}(t)$ can be decomposed as a sum of two pieces, which are symbolically called $\mathrm{HS}_{ \mathbb{Z}} (t)$ and $\mathrm{HS}_{ \mathbb{Z}+{\frac{1}{2}}} (t)$. 
For the choice $H=\{1\}$, the total Hilbert series is just $\mathrm{HS}_{ \mathbb{Z}} (t)$.

\subsection{Sums over magnetic sublattices}
Above, we see how the magnetic lattice of $G_{\mathbb{Z}_2}$ can be split into the lattice containing integer magnetic charges  $ \mathbb{Z}$ and the lattice containing magnetic charges shifted by a half, $ \mathbb{Z} + \frac{1}{2}$.  This can be generalized to $G_{\mathbb{Z}_k}$ for some $k$ where the lattice is the sum $\bigcup_{i=0}^{k-1}(\mathbb{Z}+\frac{i}{k})$. The Hilbert series can therefore be decomposed as:
\begin{equation}
    \mathrm{HS}(t) = \sum_{i=0}^{k-1}\mathrm{HS}_{ \mathbb{Z} + \frac{i}{k}}(t)
    \label{decomposition}
\end{equation}
where $\mathrm{HS}_{ \mathbb{Z} + \frac{i}{k}}(t)$ is the Hilbert series obtained by summing magnetic lattices with $\mathbb{Z} + \frac{i}{k}$ magnetic charges. 

It is currently unknown how to compute the refined Hilbert series for orthosymplectic quivers using the monopole formula. However, the refined Hilbert series can in some cases be inferred from a unitary quiver counterpart, or alternatively, in many cases (including star shaped orthosymplectic quivers), be computed using the Hall-Littlewood method (see Appendix \ref{app:HL}). 
An exact refined Hilbert series can be concisely encapsulated in the form of a highest weight generating function ($\mathrm{HWG}$) \cite{RudolphHWG}. 
Given a refined Hilbert series, the characters of the irreducible representations $\chi_{[n_1,n_2,\dots,n_k]}(x_i)$, where $x_i$ are the fundamental weight fugacities, are labeled by Dynkin labels $[n_1,n_2,\dots,n_k]$. These can be mapped into highest weight fugacities $\mu_i$ in the following way:
\begin{equation}
    \chi_{[n_1,n_2,\dots,n_k]} \to \prod_{i=1}^{k}\mu_i^{n_i} \,.
\end{equation}
The HWG is then a power series in the highest weight fugacities $\mu_i$ and $t$:
\begin{equation}
    \mathrm{HWG} (\mu_i,t)= \sum_{n_1,n_2,\dots, n_k, n \in \mathbb{N}}g_{n_1,n_2,\dots,n_k,n}\mu_1^{n_1}\mu_2^{n_2}\dots\mu_k^{n_k}t^n
\end{equation}
where $g_{n_1,n_2,\dots,n_k,n}$ count multiplicities.
The expression of the HWG follows the same decomposition as \eqref{decomposition}:
\begin{equation}
    \mathrm{HWG}(\mu_i,t) = \sum_{i=0}^{k-1}\mathrm{HWG}_{ \mathbb{Z} + \frac{i}{k}}(\mu_i,t)
    \label{decompositionHWG}
\end{equation}
For the remainder of the paper, we will often provide the $\mathrm{HWG}$ whenever it can be written as a rational function. These expressions are often provided in the form of a \emph{Plethystic Exponential} (PE) or a \emph{Plethystic Logarithm} (PL), see for instance \cite{PlethysticProgram}. 
\subsection{Affine Dynkin diagram examples}
In this section, we illustrate the points made above using a selection of examples in the form of affine Dynkin diagrams.
\subsubsection{\texorpdfstring{$D_4$}{D4} example}
\label{sec:D4ex}
In Section \ref{sec:uni} the $\mathrm{U}(1)^{\mathrm{diag}}$ quotient for an unframed unitary quiver was discussed, while the $\mathbb{Z}_2^{\mathrm{diag}}$ quotient for an unframed orthosymplectic quiver has been detailed in Section \ref{subsubsectionEX1}. Here we address the explicit computation for the $D_4$ quiver:
\begin{equation}
   \raisebox{-0.5\height}{
\begin{tikzpicture}
    \node[gauge3,label=left:{$1$}] (1) at (-1,1) {};
    \node[gauge3,label=left:{$1$}] (2) at (-1,-1) {};
    \node[gauge3,label=right:{$1$}] (3) at (1,1) {};
    \node[gauge3,label=right:{$1$}] (4) at (1,-1) {};
    \node[gauge3,label=below:{$2$}] (5) at (0,0) {};
    \draw (1)--(5)--(2) (3)--(5)--(4);
    \label{eq:D4unitary}
\end{tikzpicture}
}
\end{equation}
The gauge group is $G_{\mathrm{U}(1)}=\left( \mathrm{U}(1)^4\times \mathrm{U}(2)\right) \slash \mathrm{U}(1)^{\mathrm{diag}}$. The magnetic lattice can be computed in two different ways:
\begin{enumerate}
    \item Consider the magnetic lattice for $G=\mathrm{U}(1)^4\times \mathrm{U}(2)$ and set one of the magnetic charges to $0$.
    \item Assign one of the unitary nodes with a special unitary gauge group and then quotient by the trivially acting subgroup that remains. 
\end{enumerate}
\paragraph{Focusing on a $\mathrm{U}(1)$ node.}
For a $\mathrm{U}(1)$ gauge group, both procedures yield the same result. Either the associated magnetic charge is set to zero such that the gauge node becomes a single flavour or one could naively replace one of the $\mathrm{U}(1)$ nodes with $\mathrm{SU}(1)$, which results in the framed quiver:
\begin{equation}
   \raisebox{-0.5\height}{
\begin{tikzpicture}
    \node[flavor2,label=left:{$1$}] (1) at (-1,1) {};
    \node[gauge3,label=left:{$1$}] (2) at (-1,-1) {};
    \node[gauge3,label=right:{$1$}] (3) at (1,1) {};
    \node[gauge3,label=right:{$1$}] (4) at (1,-1) {};
    \node[gauge3,label=below:{$2$}] (5) at (0,0) {};
    \draw (1)--(5)--(2) (3)--(5)--(4);
\end{tikzpicture}
} \,.
\label{eq:D4flav}
\end{equation}
The gauge group is $\mathrm{U}(1)^3\times \mathrm{U}(2)$, which acts freely on the matter content (since the quiver is framed), and the magnetic lattice is straightforwardly computed.
\paragraph{Replacing a $\mathrm{U}(2)$ with an $\mathrm{SU}(2)$.}
On the other hand, if we replace the $\mathrm{U}(2)$ node with $\mathrm{SU}(2)$, we obtain the quiver
\begin{equation}  
\raisebox{-0.5\height}{
\begin{tikzpicture}
    \node[gauge3,label=left:{$1$}] (1) at (-1,1) {};
    \node[gauge3,label=left:{$1$}] (2) at (-1,-1) {};
    \node[gauge3,label=right:{$1$}] (3) at (1,1) {};
    \node[gauge3,label=right:{$1$}] (4) at (1,-1) {};
    \node[gauge3,label=right:{$\mathrm{SU}(2)$}] (5) at (0,0) {};
    \draw (1)--(5)--(2) (3)--(5)--(4);
\end{tikzpicture}
}
\label{eq:D4SU}
\end{equation}
and there exists a $\mathbb{Z}_2^{\rm diag}$ subgroup of $\mathrm{U}(1)^4\times \mathrm{SU}(2)$ which acts trivially on the matter representation. 
To see the action of $\mathbb{Z}_2^{\mathrm{diag}}$, we first decompose $\mathfrak{so}(8) \rightarrow \mathfrak{su}(2)^4$ with fugacities $\mu,\nu,\rho,\sigma$ for the four $\mathfrak{su}(2)$s. The adjoint representation $(\mu_2)_{\mathrm{so}(8)}$ decomposes as follows:
\begin{equation}
    (\mu_2)_{\mathrm{so}(8)} \rightarrow (\mu^2)_{\mathfrak{su}(2)} + (\nu^2)_{\mathfrak{su}(2)} + (\rho^2)_{\mathfrak{su}(2)} + (\sigma^2)_{\mathfrak{su}(2)} + (\mu)_{\mathfrak{su}(2)}(\nu)_{\mathfrak{su}(2)}(\rho)_{\mathfrak{su}(2)}(\sigma)_{\mathfrak{su}(2)} 
\end{equation}
The full highest weight generating function $\mathrm{HWG}$ takes the form:
\begin{equation}
   \mathrm{ HWG}=\mathrm{PE}
   \left[(\nu^2+\mu^2+\rho^2+\sigma^2+\nu\mu\rho\sigma)t^2+(1+\nu\mu\rho\sigma)t^4-\nu^2\mu^2\rho^2\sigma^2t^8
   \right]\;.
\end{equation}
which is the $\mathrm{HWG}$ of $\overline{\mathcal{O}}^{\mathfrak{so}(8)}_{\text{min}} $.
The $\mathbb{Z}_2^{\mathrm{diag}}$ then acts by sending the fundamental of one of the $\mathfrak{su}(2)$, say $\mu$ to $-\mu$ and $ \mathrm{HWG}_{ \mathbb{Z}}$ can be computed by taking the Molien sum:
\begin{align}
    \mathrm{HWG}_{ \mathbb{Z}}&= \frac{\mathrm{HWG}(\mu)+\mathrm{HWG}(-\mu)}{2} 
    \\
    &=\mathrm{PE}
   \left[(\nu^2+\mu^2+\rho^2+\sigma^2)t^2+(1+\nu^2\mu^2\rho^2\sigma^2)t^4+\nu^2\mu^2\rho^2\sigma^2t^6-\nu^4\mu^4\rho^4\sigma^4t^{10}
   \right]\;.\notag
\end{align}
This projection is confirmed by an explicit computation of the Hilbert series summing over the integer lattice $\mathrm{HS}_{ \mathbb{Z}}(t)$. For example, at order $t^2$, we have:
\begin{equation}
    \mathrm{HWG}_{ \mathbb{Z}} =  1 + (\mu^2 + \nu^2 + \rho^2 + \sigma^2)t^2 + \mathcal{O}(t^4)
\end{equation}
 This moduli space, with $\mathrm{HWG}_{\mathbb{Z}}$, is the $5$ dimensional orbifold $\overline{\mathcal{O}}^{\mathfrak{so}(8)}_{\text{min}}/\mathbb{Z}_2$ with the $\mathfrak{su}(2)^4$ global symmetry, which commutes with the $\mathbb{Z}_2$.  For $\mathrm{HWG}_{ \mathbb{Z}+1/2} $, the result can either be obtained by subtracting $\mathrm{HWG}_{ \mathbb{Z}} $ from $\mathrm{HWG}$ or through an explicit evaluation from the monopole formula. 

An interesting feature of this quiver is the isomorphism of the gauge group in \eqref{eq:D4SU}: 
\begin{align}
\left( \mathrm{U}(1)^4\times \mathrm{SU}(2) \right) \slash \mathbb{Z}_2^{\rm diag}=\left( \mathrm{SO}(2)^4\times \mathrm{USp}(2) \right) \slash \mathbb{Z}_2^{\rm diag}   \, , 
\end{align}
and in fact \eqref{eq:D4SU} turns into orthosymplectic quiver
\begin{equation}
   \raisebox{-0.5\height}{
\begin{tikzpicture}
    \node[redgauge,label=left:{$2$}] (1) at (-1,1) {};
    \node[redgauge,label=left:{$2$}] (2) at (-1,-1) {};
    \node[redgauge,label=right:{$2$}] (3) at (1,1) {};
    \node[redgauge,label=right:{$2$}] (4) at (1,-1) {};
    \node[bluegauge,label=below:{$2$}] (5) at (0,0) {};
    \draw (1)--(5)--(2) (3)--(5)--(4);
\end{tikzpicture}
}
\end{equation}
Note that this form of discrete gauging is different from the discrete gauging \cite{Hanany:2018vph,MarcusSymmetric,AntonBouquet} or \emph{wreathing} of \cite{Bourget:2020bxh}. When one gauges a $\mathbb{Z}_2$ subgroup of the automorphism of \eqref{eq:D4unitary} then the Coulomb branch of the new theory is $\overline{\mathcal{O}}^{\mathfrak{so}(7)}_{\text{n.min}}=\overline{\mathcal{O}}^{\mathfrak{so}(8)}_{\text{min}}/\mathbb{Z}_2$ and the Higgs branch is $\mathbb{C}^2/D_6=(\mathbb{C}^2/D_4)/\mathbb{Z}_2$.
\subsubsection{\texorpdfstring{$E_6$}{E6} example}
\label{sec:E6ex}
Consider the unitary quiver whose Coulomb branch is $\overline{\mathcal{O}}^{\mathfrak{e}_6}_{\text{min}}$:
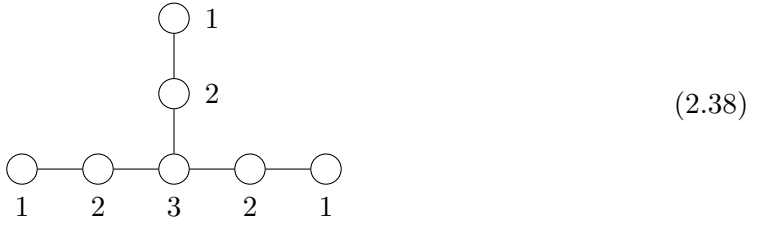
\begin{equation}
 \raisebox{-0.5\height}{
    \begin{tikzpicture}
	\begin{pgfonlayer}{nodelayer}
		\node [style=gauge3] (0) at (0, 0) {};
		\node [style=gauge3] (1) at (-1, 0) {};
		\node [style=gauge3] (2) at (-2, 0) {};
		\node [style=gauge3] (3) at (0, 1) {};
		\node [style=gauge3] (4) at (0, 2) {};
		\node [style=gauge3] (5) at (1, 0) {};
		\node [style=gauge3] (6) at (2, 0) {};
		\node [style=none] (7) at (0, -0.5) {3};
		\node [style=none] (8) at (-1, -0.5) {2};
		\node [style=none] (9) at (-2, -0.5) {1};
		\node [style=none] (10) at (1, -0.5) {2};
		\node [style=none] (11) at (2, -0.5) {1};
		\node [style=none] (12) at (0.5, 1) {2};
		\node [style=none] (13) at (0.5, 2) {1};
	\end{pgfonlayer}
	\begin{pgfonlayer}{edgelayer}
		\draw (4) to (3);
		\draw (3) to (0);
		\draw (0) to (5);
		\draw (5) to (6);
		\draw (2) to (1);
		\draw (1) to (0);
	\end{pgfonlayer}
\end{tikzpicture}
}
\label{E6_simply_laced}
\end{equation}
where all the gauge nodes are unitary. As in the previous subsection, the process of ungauging an overall $\mathrm{U}(1)^{\mathrm{diag}}$ can be executed by replacing a single $\mathrm{U}(n)$ node with $\mathrm{SU}(n)$ and quotienting the product group of all nodes by $\mathbb{Z}_{n}^{\mathrm{diag}}$.

\paragraph{Replacing a $\mathrm{U}(1)$ with $\mathrm{SU}(1)$.} 
The  simplest option is to replace one of the $\mathrm{U}(1)$ gauge nodes with $\mathrm{SU}(1)$. This is equivalent to replacing the $\mathrm{U}(1)$ gauge node with a flavour group: 
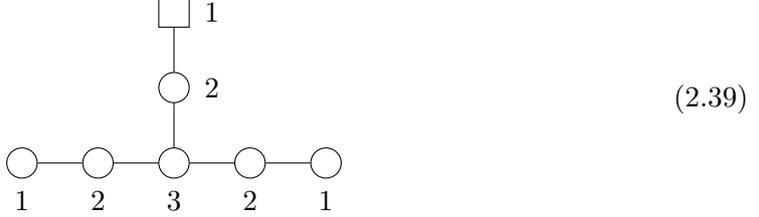
\begin{equation}
 \raisebox{-0.5\height}{
\begin{tikzpicture}
	\begin{pgfonlayer}{nodelayer}
		\node [style=gauge3] (0) at (0, 0) {};
		\node [style=gauge3] (1) at (-1, 0) {};
		\node [style=gauge3] (2) at (-2, 0) {};
		\node [style=gauge3] (3) at (0, 1) {};
		\node [style=gauge3] (5) at (1, 0) {};
		\node [style=none] (7) at (0, -0.5) {$3$};
		\node [style=none] (8) at (-1, -0.5) {2};
		\node [style=none] (9) at (-2, -0.5) {1};
		\node [style=none] (10) at (1, -0.5) {2};
		\node [style=none] (11) at (2, -0.5) {1};
		\node [style=none] (12) at (0.5, 1) {2};
		\node [style=none] (13) at (0.5, 2) {1};
		\node [style=gauge3] (14) at (2, 0) {};
		\node [style=flavor2] (15) at (0, 2) {};
	\end{pgfonlayer}
	\begin{pgfonlayer}{edgelayer}
		\draw (3) to (0);
		\draw (0) to (5);
		\draw (2) to (1);
		\draw (1) to (0);
		\draw (14) to (5);
		\draw (15) to (3);
	\end{pgfonlayer}
\end{tikzpicture}
}
\label{unitaryE6SU1}
\end{equation}
The resulting quiver is framed, hence the group $G=\mathrm{U}(3)\times \mathrm{U}(2)^3\times \mathrm{U}(1)^2$ acts freely on the matter representation. There is no need for a discrete quotient by $\mathbb{Z}_1^{\rm diag}=\{1\}$. The magnetic lattice of $G_{\{1\}}=G$ only includes integer charges. 

\paragraph{Replacing a $\mathrm{U}(2)$ with $\mathrm{SU}(2)$.} 
We now choose to replace one of the $\mathrm{U}(2)$ with an $\mathrm{SU}(2)$: 
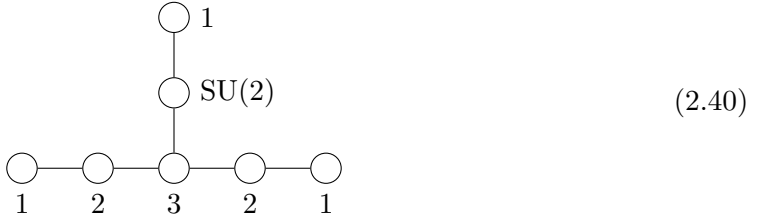
\begin{equation}
 \raisebox{-0.5\height}{
    \begin{tikzpicture}
		\node [style=gauge3] (0) at (0, 0) [label=below:3] {};
		\node [style=gauge3] (1) at (-1, 0) [label=below:2] {};
		\node [style=gauge3] (2) at (-2, 0) [label=below:1] {};
		\node [style=gauge3] (3) at (0, 1) [label=right:$\mathrm{SU}(2)$] {};
		\node [style=gauge3] (4) at (0, 2) [label=right:1] {};
		\node [style=gauge3] (5) at (1, 0) [label=below:2] {};
		\node [style=gauge3] (6) at (2, 0) [label=below:1] {};
		\draw (2)--(1)--(0)--(3)--(4) (0)--(5)--(6);
\end{tikzpicture}
\label{unitaryE6Z2}}
\end{equation}
There is a $\mathbb{Z}_2^{\rm diag}$ subgroup of $G=\mathrm{U}(3)\times \mathrm{SU}(2) \times \mathrm{U}(2)^2\times \mathrm{U}(1)^3$ acting trivially on the matter representation. The magnetic lattice of $G_{\mathbb{Z}_2}=G/\mathbb{Z}_2^{\rm diag}$ 
is the union of the integer and half-integer lattice. The Hilbert series is:
\begin{equation}
    \mathrm{HS}(t) = \mathrm{HS}_{\mathbb{Z}}(t) + \mathrm{HS}_{\mathbb{Z}+\frac{1}{2}}(t) \,.
    \label{HSE6Z2}
\end{equation} 
    Here again, the lattice summed in $\mathrm{HS}_{\mathbb{Z}}(t)$ is the direct sum of the individual magnetic lattices associated with the gauge groups in \eqref{unitaryE6Z2}. The subscript $ \mathbb{Z}$ indicates that all the charges are integer valued. Whereas the lattice in $\mathrm{HS}_{\mathbb{Z}+\frac{1}{2}}(t)$ is the summed lattice shifted by $\frac{1}{2}$, which is symbolically denote as $ \mathbb{Z}+\tfrac{1}{2} $. 
    As above, the HWG for the magnetic lattice of $G_{\mathbb{Z}_2}$ can be decomposed to obtain the HWG of the two lattices  $\mathrm{HWG}_{ \mathbb{Z}} $ and  $\mathrm{HWG}_{ \mathbb{Z}+1/2} $. This can be evaluated explicitly using the monopole formula for each of the lattices. Here, we offer an alternative method by first doing the branching $\mathfrak{e}_6 \rightarrow  \mathfrak{su}(6) \times \mathfrak{su}(2)$ with fugacities $\mu_i$ and $\nu$ respectively. For example, the decomposition of the  adjoint representation of $\mathfrak{e}_6$ gives: 
   \begin{equation}
    (\mu_6)_{\mathfrak{e}_6} \rightarrow (\nu^2)_{\mathfrak{su}(2)}+(\mu_1\mu_5)_{\mathfrak{su}(6)}+(\nu)_{\mathfrak{su}(2)}(\mu_3)_{\mathfrak{su}(6)}\;,
\end{equation} 
 and the $\mathrm{HWG}$ takes the form:
 \begin{equation}
   \mathrm{HWG}=\mathrm{PE}
   \left[(\nu^2+\mu_1\mu_5+\nu\mu_3)t^2+(1+\nu\mu_3+\mu_2\mu_4)t^4+\mu_3^2t^6-\nu^2\mu_3^2t^8
   \right]\;,
\end{equation}
which is the $\mathrm{HWG}$ of $\overline{\mathcal{O}}^{\mathfrak{e}_6}_{\text{min}} $. 
    The $\mathbb{Z}_{2}^{\mathrm{diag}}$ acts by sending $\nu$ to $-\nu$ whilst keeping $\mu_i$ the same. $\mathrm{HWG}_{ \mathbb{Z}}$ takes the form: 
\begin{equation}
\begin{split}
    \mathrm{HWG}_{ \mathbb{Z}}&= \frac{\mathrm{HWG}(\nu)+\mathrm{HWG}(-\nu)}{2} 
    \\&=\mathrm{PE} \left[ (\nu^2+\mu_1\mu_5)t^2+(1+\nu^2\mu_3^2+\mu_2\mu_4)t^4+(\mu_3^2\nu^2+\mu_3^2)t^6-\nu^4\mu_3^4t^{12} 
    \right]\;.
   \end{split}
   \label{E6Z2embed1}
\end{equation}
where at order $t^2$, the  $\mathbb{Z}_{2}^{\mathrm{diag}}$ projection yields: 
\begin{equation}
    \mathrm{HWG}_{ \mathbb{Z}} = 1 + (\nu^2 + \mu_1 \mu_5 )t^2 + \mathcal{O}(t^4) \,.
\end{equation}
The moduli space with $ \mathrm{HWG}_{ \mathbb{Z}}$ is the orbifold $\overline{\mathcal{O}}^{\mathfrak{e}_6}_{\text{min}}/\mathbb{Z}_2 $ with $\mathfrak{su}(6)\times \mathfrak{su}(2)$ global symmetry. 

For $\mathrm{HWG}_{ \mathbb{Z}+1/2} $, the result can be obtained by subtracting $\mathrm{HWG}_{ \mathbb{Z}} $ from $\mathrm{HWG}$.

\paragraph{Replacing $\mathrm{U}(3)$ with $\mathrm{SU}(3)$.} 
The quiver now takes the following form: 
\begin{equation}
 \raisebox{-0.5\height}{
    \begin{tikzpicture}
	\begin{pgfonlayer}{nodelayer}
		\node [style=gauge3] (0) at (0, 0) {};
		\node [style=gauge3] (1) at (-1, 0) {};
		\node [style=gauge3] (2) at (-2, 0) {};
		\node [style=gauge3] (3) at (0, 1) {};
		\node [style=gauge3] (4) at (0, 2) {};
		\node [style=gauge3] (5) at (1, 0) {};
		\node [style=gauge3] (6) at (2, 0) {};
		\node [style=none] (7) at (0, -0.5) {$\mathrm{SU}(3)$};
		\node [style=none] (8) at (-1, -0.5) {2};
		\node [style=none] (9) at (-2, -0.5) {1};
		\node [style=none] (10) at (1, -0.5) {2};
		\node [style=none] (11) at (2, -0.5) {1};
		\node [style=none] (12) at (0.5, 1) {2};
		\node [style=none] (13) at (0.5, 2) {1};
	\end{pgfonlayer}
	\begin{pgfonlayer}{edgelayer}
		\draw (4) to (3);
		\draw (3) to (0);
		\draw (0) to (5);
		\draw (5) to (6);
		\draw (2) to (1);
		\draw (1) to (0);
	\end{pgfonlayer}
\end{tikzpicture}
}
\label{E6quiver-SU3}
\end{equation}
There is a $\mathbb{Z}_3^{\rm diag}$ subgroup of $G=\mathrm{SU}(3)\times \mathrm{U}(2)^3\times \mathrm{U}(1)^3$ acting trivially on the matter representation. The magnetic lattice of $G_{\mathbb{Z}_3}=G/\mathbb{Z}_3^{\rm diag}$ is the union of three lattices, which are
the integer monopole lattice, and the same lattice shifted by $\frac{1}{3}$ and $\frac{2}{3}$, respectively. This is checked through an explicit unrefined Hilbert series computation: 
\begin{equation}
    \mathrm{HS}(t) = \mathrm{HS}_{\mathbb{Z}}(t) + \mathrm{HS}_{\mathbb{Z}+\frac{1}{3}}(t)+\mathrm{HS}_{\mathbb{Z}+\frac{2}{3}}(t)
\end{equation}
Similarly, the highest weight generating function (HWG) corresponding to the magnetic lattice of $G_{\mathbb{Z}_3}$ can be decomposed into the HWG of the three individual lattices $\mathrm{HWG}_{ \mathbb{Z}} $,  $\mathrm{HWG}_{ \mathbb{Z}+\frac{1}{3}} $ and  $\mathrm{HWG}_{ \mathbb{Z}+\frac{2}{3}} $.  
To do this, we first decompose $\mathfrak{e}_6 \rightarrow \mathfrak{su}(3)\times \mathfrak{su}(3)\times \mathfrak{su}(3)$ with highest weight fugacities $\mu_i$, $\nu_i$ and $\rho_i$ where ($i=1,2$).
\begin{equation}
    (\mu_2)_{\mathfrak{e}_6}=\mu_1\mu_2+\nu_1 \nu_2+\rho_1 \rho_2+\mu_1 \nu_1 \rho_1 + \mu_2 \nu_2 \rho_2\;.
\end{equation}
For the integer lattice we find
\begin{equation}
 \mathrm{HWG}_{ \mathbb{Z}} =1 +   (\mu_1 \mu_2+\nu_1 \nu_2 + \rho_1 \rho_2)t^2 +{\cal O}(t^4),
\end{equation}
The moduli space with $ \mathrm{HWG}_{ \mathbb{Z}}$ is the orbifold $\overline{\mathcal{O}}^{\mathfrak{e}_6}_{\text{min}}/\mathbb{Z}_3 $ with $\mathfrak{su}(3)^3$ global symmetry. For the integer $+\frac{1}{3}$ lattice we find
\begin{equation}
\mathrm{HWG}_{ \mathbb{Z}+\frac{1}{3}} =\mu_1 \nu_1 \rho_1 t^2 +{\cal O}(t^4),
\end{equation}
for the integer $+\frac{2}{3}$ lattice we find
\begin{equation}
\mathrm{HWG}_{ \mathbb{Z}+\frac{2}{3}} =\mu_2 \nu_2 \rho_2 t^2 + {\cal O}(t^4).
\end{equation}
The Dynkin labels $[n_1,n_2]$ for each of the three $\mathfrak{su}(3)$ global symmetries on the three $\mathrm{integer} + \frac{k}{3}$ lattices satisfy $(n_1 + 2 n_2)= k {~}\mathrm{mod}{~}3$. This is a manifestation of triality (or more generally $n$-ality for ${\rm SU}(n)$).

\subsubsection{\texorpdfstring{$E_6$}{E6} example (non-simply laced)}
\label{sec:E6ex2}
Consider the following non-simply laced unitary quiver: 
\begin{equation}
 \raisebox{-0.5\height}{
    \begin{tikzpicture}
	\begin{pgfonlayer}{nodelayer}
		\node [style=gauge3] (2) at (-0.925, 0) {};
		\node [style=none] (6) at (-0.925, -0.5) {3};
		\node [style=gauge3] (12) at (-1.925, 0) {};
		\node [style=none] (15) at (-1.925, -0.5) {2};
		\node [style=gauge3] (24) at (0.25, 0) {};
		\node [style=none] (29) at (0.25, -0.5) {4};
		\node [style=none] (30) at (-0.75, 0.075) {};
		\node [style=none] (31) at (0.25, 0.075) {};
		\node [style=none] (32) at (-0.75, -0.075) {};
		\node [style=none] (33) at (0.25, -0.075) {};
		\node [style=none] (34) at (-0.5, 0) {};
		\node [style=none] (35) at (-0.125, 0.375) {};
		\node [style=none] (36) at (-0.125, -0.375) {};
		\node [style=gauge3] (37) at (1, 0) {};
		\node [style=none] (38) at (1, -0.5) { $\mathrm{SU}(2)$};
		\node [style=gauge3] (39) at (-2.925, 0) {};
		\node [style=none] (42) at (-2.925, -0.5) {1};
	\end{pgfonlayer}
	\begin{pgfonlayer}{edgelayer}
		\draw (30.center) to (31.center);
		\draw (33.center) to (32.center);
		\draw (35.center) to (34.center);
		\draw (34.center) to (36.center);
		\draw (37) to (24);
		\draw (12) to (2);
		\draw (39) to (12);
	\end{pgfonlayer}
\end{tikzpicture}
} \, ,
\label{E6non-simply-laced}
\end{equation}
which appeared in \cite{Cabrera:2018uvz,Bourget:2019aer,foldinganton,Grimminger:2020dmg,Bourget:2020asf}.
%
%
The magnetic lattice of $G_{\mathbb{Z}_2} = G/\mathbb{Z}_2^{\mathrm{diag}}$ is again the union of integer and half-integer lattice with $\mathrm{HWG}_{\mathbb{Z}}$ and $\mathrm{HWG}_{\mathbb{Z}+\frac{1}{2}}$ respectively. The sum of both parts $\mathrm{HWG}=\mathrm{HWG}_{\mathbb{Z}}+\mathrm{HWG}_{\mathbb{Z}+\frac{1}{2}}$ takes the form:
\begin{equation}
    \mathrm{HWG}=\mathrm{PE} \left[\mu_1^2t^2 +\mu_2^2t^4 +\mu_3^2t^6+\mu_4t^2 + t^4 +\mu_4t^4 \right] \,,
\end{equation}
where $\mu_i$ are the fugacities for $\mathfrak{usp}(8)$.
This is again the $\mathrm{HWG}$ of $\overline{\mathcal{O}}^{\mathfrak{e}_6}_{\text{min}} $ obtained using the decomposition $\mathfrak{e}_6 \rightarrow \mathfrak{usp}(8)$ \cite{foldinganton}. 
The $\mathbb{Z}_2^{\mathrm{diag}}$ acts by sending $\mu_4$ to $-\mu_4$ and $\mathrm{HWG}_{\mathbb{Z}}$ takes the form:
\begin{equation}
      \mathrm{HWG}_{\mathbb{Z}}=\mathrm{PE} \left[\mu_1^2t^2 +\mu_2^2t^4 +\mu_3^2t^6+  t^4+\mu_4^2t^4 +\mu_4^2t^6 + \mu_4^2t^8 -\mu_4^4t^{12} \right]  \,.
\end{equation}
The moduli space with $ \mathrm{HWG}_{\mathbb{Z}}$ is the orbifold $\overline{\mathcal{O}}^{\mathfrak{e}_6}_{\text{min}} /\mathbb{Z}_2$, where the $\mathbb{Z}_2$ action preserves a residual $\mathrm{Sp}(4)$ global symmetry. 

This computation can be generalized to the whole $C_n$ family:
\begin{equation}
 \raisebox{-0.5\height}{
\begin{tikzpicture}
	\begin{pgfonlayer}{nodelayer}
		\node [style=gauge3] (2) at (-1, 0) {};
		\node [style=none] (6) at (-1, -0.5) {$n{-}1$};
		\node [style=none] (8) at (-1.75, 0) {\ldots};
		\node [style=gauge3] (12) at (-2.5, 0) {};
		\node [style=none] (15) at (-2.5, -0.5) {1};
		\node [style=none] (22) at (-2.125, 0) {};
		\node [style=none] (23) at (-1.35, 0) {};
		\node [style=gauge3] (24) at (0.25, 0) {};
		\node [style=none] (29) at (0.25, -0.5) {$n$};
		\node [style=none] (30) at (-1, 0.075) {};
		\node [style=none] (31) at (0.25, 0.075) {};
		\node [style=none] (32) at (-1, -0.075) {};
		\node [style=none] (33) at (0.25, -0.075) {};
		\node [style=none] (34) at (-0.5, 0) {};
		\node [style=none] (35) at (-0.125, 0.375) {};
		\node [style=none] (36) at (-0.125, -0.375) {};
		\node [style=gauge3] (37) at (1, 0) {};
		\node [style=none] (38) at (1, -0.5) {$\mathrm{SU}(2)$};
	\end{pgfonlayer}
	\begin{pgfonlayer}{edgelayer}
		\draw (23.center) to (2);
		\draw (22.center) to (12);
		\draw (30.center) to (31.center);
		\draw (33.center) to (32.center);
		\draw (35.center) to (34.center);
		\draw (34.center) to (36.center);
		\draw (37) to (24);
	\end{pgfonlayer}
\end{tikzpicture}
} \,.
\label{E6non-simply-laced2}
\end{equation}
The HWG takes the form
\begin{equation}
\mathrm{HWG} =\mathrm{PE} \left[\sum_{i=1}^{n-1}\mu_i^{2}t^{2i}+t^4+\mu_n(t^{n-2}+t^n)\right],
\end{equation}
and the sum over the integer lattice gives the HWG for the $\mathbb{Z}_2$ orbifold of the moduli space with a $\mathbb{Z}_2$ action that preserves an $\mathrm{Sp}(n)$ global symmetry and acts as $-1$ on $\mu_n$,
\begin{equation}
      \mathrm{HWG}_{\mathbb{Z}}=\mathrm{PE} \left[\sum_{i=1}^{n}\mu_i^{2}t^{2i}+  t^4+\mu_n^2\left(t^{2n-4} +t^{2n-2}\right) -\mu_n^4t^{4n-4} \right].
      \label{foldedE6HWG}
\end{equation}

\section{Orthosymplectic quivers for minimal \texorpdfstring{$E_n$}{En} orbits}
\label{5dexamples}

In this section, we focus on the closures of minimal nilpotent orbits of exceptional type $\overline{\mathcal{O}}^{\mathfrak{e}_n}_{\text{min}}$ for $n=1,\ldots,8$. These arise as moduli spaces of 5d $\mathcal{N}=1$ $\mathrm{USp}(2)$ gauge theories at infinite coupling. They are crucial in understanding theories with higher rank gauge groups, since their Higgs branches are possible elementary slices in other moduli spaces studied in this paper. In particular, magnetic quivers for rank 1 gauge theories and SCFTs constitute a basic input for the algorithm of quiver subtraction \cite{Cabrera:2018ann}. These 5d theories and their generalizations to higher rank gauge groups are studied further in Section \ref{general5dfamilysection}.

The algebras and moduli spaces are summarized in Figure \ref{exceptionallfamily}. \emph{Unitary quivers} for these orbits are well known as weighted affine Dynkin diagrams of the respective algebras. 
In this section, we list the (unitary)-orthosymplectic quivers obtained in \cite{5dweb}; The gauge group of the quiver has two choices, depending on the discrete group $H$ which is discussed in Section \ref{monopole}. For both these choices the Coulomb branch Hilbert series are computed and compared with known Hilbert series of the closures of minimal nilpotent orbits \cite{exceptionalRudolph}, as well as new $\mathbb{Z}_2$ orbifolds of these moduli spaces.

\subsection{A tale of two Coulomb branches}
As discussed above, given an unframed quiver $\mathsf{Q}$ containing $\mathrm{SO}(2n)$, $\mathrm{USp}(2n)$ and $\mathrm{U}(n)$ gauge groups, there are two associated Coulomb branch moduli spaces one can compute: $\mathcal{C}_{\mathbb{Z}_2}( \mathsf{Q})$ (with freely acting gauge group $G_{\mathbb{Z}_2}=G/\mathbb{Z}_2^{\mathrm{diag}}$) and $\mathcal{C}_{\{1\}}( \mathsf{Q})=\mathcal{C}_{\mathbb{Z}_2}( \mathsf{Q})/\mathbb{Z}_2^{\mathrm{diag}}$ (with the gauge group $G$).\footnote{Here there is a tricky point with the labeling of the two moduli space. If the gauge group is smaller by $\mathbb{Z}_2$ the flavor symmetry is bigger by $\mathbb{Z}_2$ and the resulting bigger space seems to be counter intuitive to the labeling. The notation of \eqref{eq:effect_ungauge} is chosen to reflect the $\mathbb{Z}_2$ in the gauge group, but an equivalent labeling can follow the global symmetry.} 
In this section, we focus on (unitary)-orthosymplectic quivers for which $\mathcal{C}_{\mathbb{Z}_2}( \mathsf{Q})=\overline{\mathcal{O}}^{\mathfrak{e}_n}_{\text{min}}$. In other words, the Coulomb branches are closures of the minimal nilpotent orbits of $\mathfrak{e}_n$ algebras and their $\mathbb{Z}_2$ orbifolds. 
The global symmetry of $\overline{\mathcal{O}}^{\mathfrak{e}_n}_{\text{min}}$ is $F_{\mathbb{Z}_2}=E_n$ whereas the global symmetry of $\overline{\mathcal{O}}^{\mathfrak{e}_n}_{\text{min}}/\mathbb{Z}_2$ is the subgroup $F$ which is the residual global symmetry after gauging $\mathbb{Z}_2^{\mathrm{diag}}$, as detailed in Figure \ref{integerHS}.
In these cases, one can always decompose representations of $\mathfrak{f}_{\mathbb{Z}_2}=\mathfrak{e}_n$ into representations of $\mathfrak{f}$ which has a subalgebra $\mathfrak{so}(2r) \subset \mathfrak{f}$, where $r=n-1$ for $n<8$ and $r=n$ for $n=8$. Taking $[0\dots,n_{r-1},0]_{\mathfrak{so}(2r)}$ and $[0\dots,0,n_{r}]_{\mathfrak{so}(2r)}$ to be the two Dynkin labels for the spinor nodes of $\mathfrak{so}(2r)$, and $\mathcal{F}=n_{r-1}+n_{r}$ to be the spinor number, 
The $\mathbb{Z}_2^{\mathrm{diag}}\subset \mathfrak{g}^{\rm global}$ turns out to be generated by $(-1)^\mathcal{F}$.
For $\overline{\mathcal{O}}^{\mathfrak{e}_n}_{\text{min}}$, all the chiral ring generators transform in the adjoint representation of 
$\mathfrak{e}_n$; decomposing this representation with respect to 
$\mathfrak{so}(2r)$, there are chiral ring generators both in the adjoint representation and in the spinor representations of 
$\mathfrak{so}(2r)$. Considering the orbifold $\overline{\mathcal{O}}^{\mathfrak{e}_n}_{\text{min}}/\mathbb{Z}_2$, only elements that are even under $\mathcal{F}$ are kept. As the spinor generators are odd under $\mathcal{F}$, they are no longer chiral ring generators, but even powers of them are still valid operators in the moduli space.

We assign the unrefined Hilbert series $\mathrm{HS}(t)$ and the refined Hilbert series, in the form of a $\mathrm{HWG}$ in the representations of $\mathfrak{f}$, to $\overline{\mathcal{O}}^{\mathfrak{e}_n}_{\text{min}}$, which includes summing over magnetic charges both in the integer and integers-plus-half lattice. And we assign $\mathrm{HS}(t)_{ \mathbb{Z}}$ and $\mathrm{HWG}_{ \mathbb{Z}} $ to $\overline{\mathcal{O}}^{\mathfrak{e}_n}_{\text{min}}/\mathbb{Z}_2$ which only includes magnetic charges in the integer lattice.
It follows from this discussion that $\mathrm{HWG}_{ \mathbb{Z}} $ can be obtained by taking the Molien sum over the spinor representation:
\begin{equation}
 \mathrm{HWG}_{ \mathbb{Z}}= \frac{\mathrm{HWG}(\mu_{\mathrm{spinor}})+\mathrm{HWG}(-\mu_{\mathrm{spinor}})}{2}    \,.
\end{equation}
We further note, both Hilbert series $\mathrm{HS}(t)$ and $\mathrm{HS}_{ \mathbb{Z}} (t) $ have palindromic numerators indicating the Coulomb branches are hyper-K\"ahler. Dividing their volumes 
\begin{equation}
    \frac{\text{HS}(t)}{\text{HS}_{ \mathbb{Z}} (t) }\bigg \rvert_{t \rightarrow 1}  = \left| \mathbb{Z}_2^{\textrm{diag}} \right| =2
    \label{quotientHSX}
\end{equation} 
gives the order of the $\mathbb{Z}_2^{\textrm{diag}}$ discrete group which is gauged. 

The Hilbert series $\mathrm{HS}_{ \mathbb{Z}} (t) $ and  $\mathrm{HWG}_{ \mathbb{Z}} $ are listed in the first row of \crefrange{E8HS}{E2HS}. Whereas, the Hilbert series $\mathrm{HS}(t) $ and  $\mathrm{HWG} $ are listed in the third row. 
Although the spaces $\overline{\mathcal{O}}^{\mathfrak{e}_n}_{\text{min}}/\mathbb{Z}_2$ are not the main focus of this section, the well behaved Hilbert series as well as $\mathrm{HWG}_{ \mathbb{Z}}$ shows they are interesting moduli spaces which deserve future studies. The algebra $\mathfrak{f}$ of the global symmetry group  can be obtained by looking at the balance on the nodes \cite{GaiottoWitten} (see also \cite{newMarcus}); whereas the group $F$ can be obtained by studying $\mathrm{HWG}_{ \mathbb{Z}}$ as listed in Figure \ref{integerHS}.

We argue that these statements generalize to all unframed quivers containing $\mathrm{SO}(2n)$, $\mathrm{USp}(2n)$ and $\mathrm{U}(n)$ gauge groups where the $\mathbb{Z}_2^{\mathrm{diag}}$ acts on the spinor representation of a suitable special orthogonal Lie subalgebra of $\mathfrak{f}$.

\begin{figure}[t]
\begin{adjustbox}{center}
	\begin{tabular}{|l|l|l|}
		\hline
	Theory&	  Exceptional algebra & Moduli space	\\ \hline
		$E_8$&	$\mathfrak{e}_8$ &$\overline{\mathcal{O}}^{\mathfrak{e}_8}_{\text{min}}$	\\ \hline
		$E_7$&	   	$\mathfrak{e}_7$&$\overline{\mathcal{O}}^{\mathfrak{e}_7}_{\text{min}}$\\ \hline
		$E_6$&	 	$\mathfrak{e}_6$&$\overline{\mathcal{O}}^{\mathfrak{e}_6}_{\text{min}}$	\\ \hline
		$E_5$&		$\mathfrak{e}_5\cong \mathfrak{so}(10)$	&$\overline{\mathcal{O}}^{\mathfrak{so}(10)}_{\text{min}}$\\ \hline
		$E_4$&	  	$\mathfrak{e}_4\cong \mathfrak{su}(5)$	&$\overline{\mathcal{O}}^{\mathfrak{su}(5)}_{\text{min}}$\\ \hline
		$E_3$&	   	$\mathfrak{e}_3\cong \mathfrak{su}(3)\times \mathfrak{su}(2)$	&$\overline{\mathcal{O}}^{\mathfrak{su}(3)}_{\text{min}}\cup \overline{\mathcal{O}}^{\mathfrak{su}(2)}_{\text{min}}$\\ \hline
		$E_2$&	   	$\mathfrak{e}_2\cong \mathfrak{su}(2)\times \mathfrak{u}(1)$	&$\overline{\mathcal{O}}^{\mathfrak{su}(2)}_{\text{min}}\cup \{ \bullet \bullet  \}$\\ \hline
		$E_1$&	   	$\mathfrak{e}_1\cong \mathfrak{su}(2)$ & $\overline{\mathcal{O}}^{\mathfrak{su}(2)}_{\text{min}}$	\\ \hline
	\end{tabular}
\end{adjustbox}
\caption[Exceptional $E_n$ theories]{The rank-1 exceptional $E_n$ theories have their Higgs branch equal to the closure of the minimal nilpotent orbit of the $\mathfrak{e}_n$ algebras. For lower ranks, the respective moduli spaces are identified in \cite{GiuilaInf}. }
\label{exceptionallfamily}
\end{figure}

\begin{figure}[t]
\begin{adjustbox}{center}
	\begin{tabular}{|l|l|l|}
		\hline
		  $\begin{matrix} \text{Global} \\ \text{symmetry} \end{matrix} \; F_{\mathbb{Z}_2} \supset \text{Subgroup}  $
		  & Dynkin diagrams &
		  $\begin{matrix} \text{Residual global} \\ \text{symmetry }F  \end{matrix} $ \\
		 \hline
$E_8\supset \mathrm{Ss}(16)$	& \raisebox{-.5\height}{\scalebox{.7}{\begin{tikzpicture}
	\node (1) at (0,0) [wd] {};
	\node (2) at (1,0) [wd] {};
	\node (3) at (2,0) [wd] {};
	\node (4) at (3,0) [wd] {};
	\node (5) at (4,0) [wd] {};
	\node (6) at (5,0) [wd] {};
	\node (7) at (6,0) [wd] {};
	\node (8) at (7,0) [bd] {};
	\node (0) at (5,1) [wd] {};
	\draw (1)--(2)--(3)--(4)--(5)--(6)--(7)--(8);
	\draw (6)--(0);
	\node at (0,1.5) {};
	\node at (0,-.5) {};
	\end{tikzpicture}}} & 	 $\mathrm{PSO}(16)$\\  \hline
$E_7/\mathbb{Z}_2\supset (\mathrm{Ss}(12)\times \mathrm{SU}(2))/\mathbb{Z}_2$	& \raisebox{-.5\height}{\scalebox{.7}{\begin{tikzpicture}
	\node (1) at (0,0) [wd] {};
	\node (2) at (1,0) [wd] {};
	\node (3) at (2,0) [wd] {};
	\node (4) at (3,0) [wd] {};
	\node (5) at (4,0) [bd] {};
	\node (6) at (3,1) [wd] {};
	\node (7) at (-1,0) [wd] {};
	\node (0) at (2,1) [wd] {};
	\draw (7)--(1)--(2)--(3)--(4)--(5)--(6);
	\draw (3)--(0);
	\node at (0,1.5) {};
	\node at (-3,-.5) {};
	\end{tikzpicture}}} & 	 $\mathrm{PSO}(12) \times \mathrm{PSU}(2)$\\  \hline
$E_6/\mathbb{Z}_3\supset \mathrm{Spin}(10)\times \mathrm{U}(1)$	& \raisebox{-.5\height}{\scalebox{.7}{\begin{tikzpicture}
	\node (1) at (0,0) [wd] {};
	\node (2) at (1,0) [wd] {};
	\node (3) at (2,0) [wd] {};
	\node (4) at (3,0) [wd] {};
	\node (5) at (4,0) [bd] {};
	\node (6) at (3,1) [bd] {};
	\node (0) at (2,1) [wd] {};
	\draw (1)--(2)--(3)--(4)--(5);
	\draw (3)--(0)--(6);
	\node at (0,1.5) {};
	\node at (-3,-.5) {};
	\end{tikzpicture}}}  & 	 $\mathrm{SO}(10)\times \mathrm{U}(1)$\\  \hline
$E_5/\mathbb{Z}_4=\mathrm{PSO}(10)\supset \mathrm{Ss}(8)\times \mathrm{U}(1)$	& \raisebox{-.5\height}{\scalebox{.7}{\begin{tikzpicture}
	\node (1) at (0,0) [wd] {};
	\node (2) at (1,0) [wd] {};
	\node (3) at (2,0) [wd] {};
	\node (4) at (3,0) [bd] {};
	\node (5) at (1,1) [wd] {};
	\node (0) at (2,1) [bd] {};
	\draw (1)--(2)--(3)--(4);
	\draw (3)--(0);
	\draw (2)--(5);
	\node at (0,1.5) {};
	\node at (-4,-.5) {};
	\end{tikzpicture}}}  & 	 $\mathrm{PSO}(8) \times \mathrm{U}(1)$\\  \hline
$E_4/\mathbb{Z}_5=\mathrm{PSU}(5)\supset \mathrm{Spin}(6)\times \mathrm{U}(1)$	& \raisebox{-.5\height}{\scalebox{.7}{\begin{tikzpicture}
	\node (1) at (0,0) [wd] {};
	\node (2) at (1,0) [wd] {};
	\node (3) at (2,0) [bd] {};
	\node (4) at (0,1) [wd] {};
	\node (0) at (1,1) [bd] {};
	\draw (1)--(2)--(3)--(0)--(4)--(1);
	\node at (0,1.5) {};
	\node at (-5,-.5) {};
	\end{tikzpicture}}} & 	 $\mathrm{SO}(6)\times \mathrm{U}(1)$\\  \hline
	\end{tabular}
\end{adjustbox}
\caption[Exceptional symmetries and Dynkin diagrams]{Exceptional group $F_{\mathbb{Z}_2}=E_n/\mathbb{Z}_{9-n}$ and its subgroup whose $\mathbb{Z}_2$ quotient is $F$. The number of $\mathrm{U}(1)$ factors is the number of black dots minus 1. Only the cases with connected Dynkin diagrams are presented. $\mathrm{Ss}$ is the semispin group, $\mathrm{PSO}$ is the projective special orthogonal group and $\mathrm{PSU}$ is the projective special unitary group (See Table \ref{tableGroupsD} for more details). The $\mathbb{Z}_2$ in row two acts as $-1$ simultaneously on the spinors of $\mathrm{SU}(2)$ and $\mathrm{Ss}(12)$. } 
\label{integerHS}
\end{figure}

\subsection{\texorpdfstring{$E_8$}{E8} quiver}
We start with the orthosymplectic quiver whose Coulomb branch is the closure of the $E_8$ minimal nilpotent orbit $\overline{\mathcal{O}}^{\mathfrak{e}_8}_{\text{min}}$. 
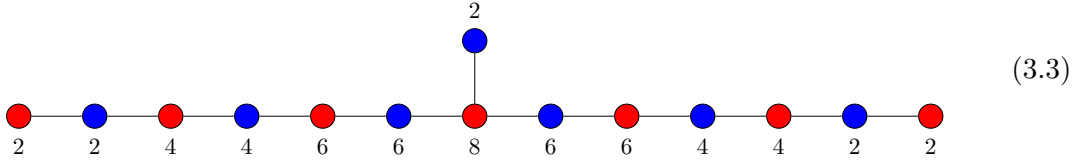
\begin{equation}
  \scalebox{.8}{ \raisebox{-.5\height}{ \begin{tikzpicture}
	\begin{pgfonlayer}{nodelayer}
		\node [style=bluegauge] (0) at (0, 0) {};
		\node [style=redgauge] (1) at (1.25, 0) {};
		\node [style=bluegauge] (2) at (2.5, 0) {};
		\node [style=redgauge] (3) at (3.75, 0) {};
		\node [style=redgauge] (4) at (-1.25, 0) {};
		\node [style=redgauge] (5) at (-3.75, 0) {};
		\node [style=bluegauge] (6) at (-2.5, 0) {};
		\node [style=none] (8) at (0, -0.5) {4};
		\node [style=none] (9) at (1.25, -0.5) {6};
		\node [style=none] (10) at (2.5, -0.5) {6};
		\node [style=none] (11) at (3.75, -0.5) {8};
		\node [style=none] (12) at (-1.25, -0.5) {4};
		\node [style=none] (13) at (-2.5, -0.5) {2};
		\node [style=none] (14) at (-3.75, -0.5) {2};
		\node [style=bluegauge] (23) at (5, 0) {};
		\node [style=none] (25) at (5, -0.5) {6};
		\node [style=bluegauge] (27) at (10, 0) {};
		\node [style=redgauge] (28) at (11.25, 0) {};
		\node [style=redgauge] (30) at (8.75, 0) {};
		\node [style=redgauge] (31) at (6.25, 0) {};
		\node [style=bluegauge] (32) at (7.5, 0) {};
		\node [style=none] (33) at (10, -0.5) {2};
		\node [style=none] (34) at (11.25, -0.5) {2};
		\node [style=none] (36) at (8.75, -0.5) {4};
		\node [style=none] (37) at (7.5, -0.5) {4};
		\node [style=none] (38) at (6.25, -0.5) {6};
		\node [style=bluegauge] (39) at (3.75, 1.25) {};
		\node [style=none] (40) at (3.75, 1.75) {2};
	\end{pgfonlayer}
	\begin{pgfonlayer}{edgelayer}
		\draw (5) to (6);
		\draw (6) to (4);
		\draw (4) to (0);
		\draw (0) to (1);
		\draw (1) to (2);
		\draw (2) to (3);
		\draw (3) to (23);
		\draw (31) to (32);
		\draw (32) to (30);
		\draw (30) to (27);
		\draw (27) to (28);
		\draw (23) to (31);
		\draw (3) to (39);
	\end{pgfonlayer}
\end{tikzpicture}}}
\label{quiverE8}
\end{equation}
The Coulomb branch Hilbert series is given in Figure \ref{E8HS} which matches the computation of the one $E_8$ instanton in \cite{Benvenuti:2010pq}. In addition to the unrefined Hilbert series, the exact refined Hilbert series is also given in the form of a highest weight generating function (HWG) \cite{RudolphHWG}. 

In Figure \ref{E8HS}, the first line shows the Hilbert series obtained by summing over the integer lattice $\mathrm{HS}_{ \mathbb{Z}} (t) $. The resulting Hilbert series is that of a symplectic singularity with global symmetry $\mathrm{PSO}(16)$, about which more is said at the end of this section. Note that the global symmetry is a strict subalgebra of the exceptional $\mathfrak{e}_8$, which is reflected in the decomposition of the 248 dimensional adjoint representation of $\mathfrak{e}_8$ into the $120$ dimensional adjoint and the $128$ dimensional spinor of $\mathfrak{so}(16)$. The integers-plus-half lattice $\mathrm{HS}_{ \mathbb{Z}+\frac{1}{2}} (t)$ adds precisely the missing $\mathfrak{so}(16)$ spinor at order $t^2$ in the Hilbert series, so that the full result displays $\mathfrak{e}_8$ global symmetry. These facts can be deduced directly from the HWGs, where $\mu_2$ stands for the $\mathfrak{so}(16)$ adjoint and $\mu_8$ for the spinor. 

As can be read from the table, the HWG of the minimal $\mathfrak{e}_8$ nilpotent orbit, which is $\mathrm{PE}[\mu_7 t^2]$ in terms of $\mathfrak{e}_8$ fugacities, reads $ \mathrm{PE}[(\mu_2+\mu_8)t^2 +  (1+\mu_4 + \mu_8)t^4  + \mu_6 t^6 ]$ in terms of $\mathfrak{so}(16)$ fugacities. A $\mathbb{Z}_2^{\textrm{diag}}$ quotient can be performed, in which the $\mathbb{Z}_2^{\textrm{diag}}$ acts non-trivially on the spinor terms $\mu_8 t^2$ and $\mu_8 t^4$. As explained in \cite{Hanany:2018vph}, this is accounted for by replacing $\mu_8 t^2 + \mu_8 t^4$ by $\mu_8^2 t^4 + \mu_8^2 t^6 + \mu_8^2 t^8 - \mu_8^4 t^{12}$, which gives exactly the HWG for the $\mathrm{PSO}(16)$ space from the integer lattice. 

In the following subsections, similar comments and observations can be made. 


\begin{figure}[tp]
\small
\begin{adjustbox}{center}
	\begin{tabular}{|c|c|c|}
\hline
		  & Hilbert Series  & $\mathrm{HWG}$\\ 
\hline
    $	\mathrm{HS}_{ \mathbb{Z}} (t) $	& 		
    $\begin{array}{l}
	    \dfrac{{\cal P}_{116}(t)}{(1 - t^2)^{29} (1 - t^4)^{29}} \\ \\
    = { 1 + 120 t^2 + 13560 t^4 + 881205 t^6}\\{+ 39574360 t^8 + 1321374912 t^{10} + O\left(t^{12}\right)} \\  
    \end{array}$ &
   $\begin{array}{l} \mathrm{PE}\big[
       \mu_2t^2 + (1+\mu_4 + \mu_8^2)t^4 \\\qquad + (\mu_6 + \mu_8^2) t^6 + \mu_8^2 t^8 \\\qquad - \mu_8^4 t^{12}\big]\\
    \end{array}$\\ 
\hline		
    $ \mathrm{HS}_{ \mathbb{Z}+\frac{1}{2}} (t) $	& 		
    $\begin{array}{l}
		\dfrac{128 t^2 {~} {\cal P}_{112}(t)}{(1 - t^2)^{29} (1 - t^4)^{29}}\\ \\
		=	128 t^2 + 13440 t^4 + 881920 t^6 \\+ 39568640 t^8 + 1321402368 t^{10} + O\left(t^{12}\right) \\ 
	\end{array}	$
    & $\begin{array}{l}\mu_8t^2\;\mathrm{PE} \big[t^2+\mu_2t^2+(\mu_4+\mu_8^2)t^4 \\\qquad \qquad +\mu_6t^6+\mu_8^2t^8\big]\end{array}$\\	 
\hline			
	$\mathrm{HS}(t)$	& 		
    $\begin{array}{l}
      \dfrac{(1 + t^2){~}{\cal P}_{56}(t)}{(1 - t^2)^{58} }\\ \\
      =   1 + 248 t^2 + 27000 t^4 + 1763125 t^6\\ + 79143000 t^8 + 2642777280 t^{10}+ O\left(t^{12}\right)
		\end{array}	$ 
& 
        $\begin{array}{l} \mathrm{PE}\big[(\mu_2+\mu_8)t^2\\\qquad +  (1+\mu_4 + \mu_8)t^4 \\\qquad + \mu_6 t^6  \big]
        \end{array}$\\
\hline
    \end{tabular}
\end{adjustbox}
\caption[Hilbert series for $E_8$ magnetic quiver]{Hilbert series for the $E_8$ magnetic quiver (\ref{quiverE8}). The first line of the table provides the Hilbert series when the GNO magnetic lattice is integer $\mathrm{HS}_{ \mathbb{Z}} (t) $, the second line shows the sum over the integers-plus-half lattice $\mathrm{HS}_{ \mathbb{Z}+\frac{1}{2}} (t) $ and the last line is their sum $\mathrm{HS}(t)$. The palindromic numerator terms ${\cal P}_k(t)$ are given in Appendix \ref{apx:PP}. In the third column, we show the HWG in terms of fugacities for $\mathfrak{so}(16) $ (denoted $\mu_1 , \ldots , \mu_8$). Note that we pick the branching of $\mathfrak{e}_8\rightarrow\mathfrak{so}(16)$ where $(\mu_7)_{\mathfrak{e}_8}\mapsto(\mu_2+\mu_8)_{\mathfrak{so}(16)}$.}
\label{E8HS}
\end{figure}

\subsection{\texorpdfstring{$E_7$}{E7} quiver}
We move on to the unitary-orthosymplectic quiver whose Coulomb branch is the closure of the $E_7$ minimal nilpotent orbit $\overline{\mathcal{O}}^{\mathfrak{e}_7}_{\text{min}}$. 
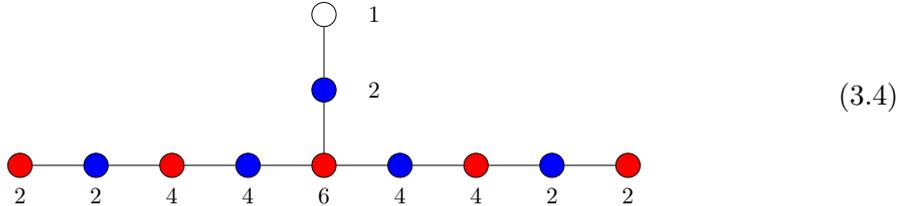
\begin{equation}
\scalebox{.8}{\raisebox{-.5\height}{
\begin{tikzpicture}
	\begin{pgfonlayer}{nodelayer}
		\node [style=bluegauge] (0) at (0, 0) {};
		\node [style=redgauge] (1) at (1.25, 0) {};
		\node [style=bluegauge] (2) at (2.5, 0) {};
		\node [style=redgauge] (3) at (3.75, 0) {};
		\node [style=redgauge] (4) at (-1.25, 0) {};
		\node [style=redgauge] (5) at (-3.75, 0) {};
		\node [style=bluegauge] (6) at (-2.5, 0) {};
		\node [style=none] (8) at (0, -0.5) {4};
		\node [style=none] (9) at (1.25, -0.5) {6};
		\node [style=none] (10) at (2.5, -0.5) {4};
		\node [style=none] (11) at (3.75, -0.5) {4};
		\node [style=none] (12) at (-1.25, -0.5) {4};
		\node [style=none] (13) at (-2.5, -0.5) {2};
		\node [style=none] (14) at (-3.75, -0.5) {2};
		\node [style=bluegauge] (23) at (5, 0) {};
		\node [style=redgauge] (24) at (6.25, 0) {};
		\node [style=none] (25) at (5, -0.5) {2};
		\node [style=none] (26) at (6.25, -0.5) {2};
		\node [style=bluegauge] (27) at (1.25, 1.25) {};
		\node [style=gauge3] (28) at (1.25, 2.5) {};
		\node [style=none] (29) at (2.075, 1.25) {2};
		\node [style=none] (30) at (2.075, 2.5) {1};
	\end{pgfonlayer}
	\begin{pgfonlayer}{edgelayer}
		\draw (5) to (6);
		\draw (6) to (4);
		\draw (4) to (0);
		\draw (0) to (1);
		\draw (1) to (2);
		\draw (2) to (3);
		\draw (3) to (23);
		\draw (23) to (24);
		\draw (28) to (27);
		\draw (27) to (1);
	\end{pgfonlayer}
\end{tikzpicture}
}}
\label{quiverE7}
\end{equation}
where the white node represents a $\mathrm{U}(1)$ gauge group. As discussed in Section \ref{monopole}, under the diagonal $\mathbb{Z}_2^{\textrm{diag}}$ quotient, we take integer and integers-plus-half magnetic charges for the unitary gauge groups as well. 

The $\mathrm{U}(1)$ gauge node can equivalently be expressed as an $\mathrm{SO}(2)$ gauge node and this quiver thereby acts as a bridge between different families of quivers. As an $\mathrm{SO}(2)$, it comes more naturally from D-type punctures of $4$d $\mathcal{N}=2$ class $\mathcal{S}$ theories. As $\mathrm{U}(1)$  it comes more naturally from a brane construction with O5 planes \cite{5dweb}. 
The Coulomb branch Hilbert series is given in Figure \ref{E7HS}, which matches the computation of the one $E_7$ instanton in \cite{Benvenuti:2010pq}, alongside with the Hilbert series for the orbifold moduli space with global symmetry $\mathrm{PSO}(12)\times \mathrm{PSU}(2)$.


\begin{figure}[tp]
\small
\begin{adjustbox}{center}
	\begin{tabular}{|c|c|c|}
	\hline
		  & Hilbert Series  & $\mathrm{HWG}$\\ 
     \hline
$	\mathrm{HS}_{ \mathbb{Z}} (t) $	
& 		
$\begin{array}{l}
	\dfrac{{\cal P}_{68}(t)}{(1 - t^2)^{17} (1 - t^4)^{17}}\\ \\= 1 + 69 t^2 + 3723 t^4 + 119434 t^6 \\+ 2625390 t^8 + 42857892 t^{10}+O\left(t^{12}\right) \\ 
\end{array}$ 
&
   $\begin{array}{l}\mathrm{PE}\big[
       (\mu_2 + \nu^2) t^2 \\+ (1+\mu_4 + \nu^2 \mu_6^2) t^4 \\+ (\mu_6^2 + \nu^2 \mu_6^2) t^6 \\- \nu^4 \mu_6^4 t^{12} \big]
       \end{array}$\\ 
\hline		
$ \mathrm{HS}_{ \mathbb{Z}+\frac{1}{2}} (t) $	& 
 $\begin{array}{l}
		\dfrac{64 t^2 {~} {\cal P}_{64}(t)}{(1 - t^2)^{17} (1 - t^4)^{17}} \\ \\=
			64 t^2 + 3648 t^4 + 119168 t^6 \\+ 2623360 t^8 + 42852096 t^{10} +O\left(t^{12}\right) \\ 
		\end{array}	$
& $ \begin{array}{l}\mu_6\nu t^2\; \mathrm{PE} \big[t^2 + \nu^2t^2 + \mu_2t^2  \\ \mu_4t^4 + \mu_6^2t^6 + \nu^2\mu_6^2t^4  \big]\end{array}$\\	 
\hline			
	$\mathrm{HS}(t)$	
& 		
$\begin{array}{l}
      \dfrac{(1 + t^2){~}{\cal P}_{32}(t)}{(1 - t^2)^{34} }\\ \\
      =     1 + 133 t^2 + 7371 t^4 + 238602 t^6 \\+ 5248750 t^8 + 85709988 t^{10}+O\left(t^{12}\right)
		\end{array}	$ 
& 
        $\begin{array}{l}\mathrm{PE}\big[ (\mu_2 + \nu^2 + \nu \mu_6) t^2 \\+ (1+\mu_4 + \nu \mu_6) t^4 \\+ \mu_6^2 t^6 - \nu^2 \mu_6^2 t^8 \big]
        \end{array}$\\
        \hline
    \end{tabular}
    \end{adjustbox}
    \caption[Hilbert series for $E_7$ magnetic quiver]{Hilbert series for the $E_7$ magnetic quiver (\ref{quiverE7}). The first line of the table provides the Hilbert series when the GNO magnetic lattice is integer $\mathrm{HS}_{ \mathbb{Z}} (t) $, the second line shows the sum over the integers-plus-half lattice $\mathrm{HS}_{ \mathbb{Z}+\frac{1}{2}} (t) $ and the last line is their sum $\mathrm{HS}(t)$. The palindromic numerator terms ${\cal P}_k(t)$ are given in Appendix \ref{apx:PP}. In the third column, we show the HWG in terms of fugacities for $\mathfrak{su}(2)$ and   $\mathfrak{so}(12)$ (denoted $\nu$ and $\mu_1 , \ldots , \mu_6$ respectively). Note that we pick the branching of $\mathfrak{e}_7\rightarrow\mathfrak{su}(2)\times\mathfrak{so}(12)$ where $(\mu_1)_{\mathfrak{e}_7}\mapsto(\nu\mu_6+\nu^2+\mu_2)_{\mathfrak{su}(2)\times\mathfrak{so}(12)}$.}
\label{E7HS}
\end{figure}

\subsection{\texorpdfstring{$E_6$}{E6} quiver}
The unitary-orthosymplectic quiver whose Coulomb branch is the closure of the $E_6$ minimal nilpotent orbit $\overline{\mathcal{O}}^{\mathfrak{e}_6}_{\text{min}}$ takes the following form:
\begin{eqnarray}
\scalebox{.8}{\raisebox{-.5\height}{
\begin{tikzpicture}
	\begin{pgfonlayer}{nodelayer}
		\node [style=bluegauge] (0) at (0, 0) {};
		\node [style=redgauge] (1) at (1.25, 0) {};
		\node [style=bluegauge] (2) at (2.5, 0) {};
		\node [style=redgauge] (3) at (3.75, 0) {};
		\node [style=redgauge] (4) at (-1.25, 0) {};
		\node [style=redgauge] (5) at (-3.75, 0) {};
		\node [style=bluegauge] (6) at (-2.5, 0) {};
		\node [style=gauge3] (7) at (0, 1) {};
		\node [style=none] (8) at (0, -0.5) {4};
		\node [style=none] (9) at (1.25, -0.5) {4};
		\node [style=none] (10) at (2.5, -0.5) {2};
		\node [style=none] (11) at (3.75, -0.5) {2};
		\node [style=none] (12) at (-1.25, -0.5) {4};
		\node [style=none] (13) at (-2.5, -0.5) {2};
		\node [style=none] (14) at (-3.75, -0.5) {2};
		\node [style=none] (15) at (0, 1.5) {1};
	\end{pgfonlayer}
	\begin{pgfonlayer}{edgelayer}
		\draw (5) to (6);
		\draw (6) to (4);
		\draw (4) to (0);
		\draw (0) to (7);
		\draw (0) to (1);
		\draw (1) to (2);
		\draw (2) to (3);
	\end{pgfonlayer}
\end{tikzpicture}}
} {~}={~} 
\scalebox{.8}{\raisebox{-.5\height}{
\begin{tikzpicture}
	\begin{pgfonlayer}{nodelayer}
		\node [style=bluegauge] (0) at (0, 0) {};
		\node [style=redgauge] (1) at (1.25, 0) {};
		\node [style=bluegauge] (2) at (2.5, 0) {};
		\node [style=redgauge] (3) at (3.75, 0) {};
		\node [style=redgauge] (4) at (-1.25, 0) {};
		\node [style=redgauge] (5) at (-3.75, 0) {};
		\node [style=bluegauge] (6) at (-2.5, 0) {};
		\node [style=redgauge] (7) at (0, 1) {};
		\node [style=none] (8) at (0, -0.5) {4};
		\node [style=none] (9) at (1.25, -0.5) {4};
		\node [style=none] (10) at (2.5, -0.5) {2};
		\node [style=none] (11) at (3.75, -0.5) {2};
		\node [style=none] (12) at (-1.25, -0.5) {4};
		\node [style=none] (13) at (-2.5, -0.5) {2};
		\node [style=none] (14) at (-3.75, -0.5) {2};
		\node [style=none] (15) at (0, 1.5) {2};
	\end{pgfonlayer}
	\begin{pgfonlayer}{edgelayer}
		\draw (5) to (6);
		\draw (6) to (4);
		\draw (4) to (0);
		\draw (0) to (7);
		\draw (0) to (1);
		\draw (1) to (2);
		\draw (2) to (3);
	\end{pgfonlayer}
\end{tikzpicture}}
} 
\label{quiverE6}
\end{eqnarray}
Again, the $\mathrm{U}(1)$ gauge node can equivalently be expressed as an $\mathrm{SO}(2)$ gauge node.
The Coulomb branch Hilbert series is given in Figure \ref{E6HS} which matches the computation of the one $E_6$ instanton in \cite{Benvenuti:2010pq}, along side with the Hilbert series for the orbifold moduli space with global symmetry $\mathrm{SO}(10)\times \mathrm{U}(1)$ . 

In Subsection \ref{sec:E6ex} the $F_{\mathbb{Z}_2}$ subgroup is $\mathrm{SU}(6)\times \mathrm{SU}(2)$ while in this subsection the $F_{\mathbb{Z}_2}$ subgroup is $\mathrm{SO}(10)\times \mathrm{U}(1)$, giving two different $\mathbb{Z}_2$ orbifolds of $\overline{\mathcal{O}}_{\min}^{\mathfrak{e}_6}$.
This is a nice demonstration that one can use different embeddings of $\mathbb{Z}_2$ inside $E_6$ which yield different orbifolds of the same space. 
An explicit demonstration of this fact is the splitting of the Coulomb branch Hilbert series into $\mathrm{HS}_{ \mathbb{Z}}$ and $\mathrm{HS}_{ \mathbb{Z}+1/2}$. Even though, the Coulomb branch Hilbert series of  \eqref{quiverE6} and \eqref{unitaryE6Z2} agree, the individual components need not. This follows because the $\mathrm{HS}_{ \mathbb{Z}}$ is the Hilbert series of different $\mathbb{Z}_2$ orbifolds.


\begin{figure}[tp]
\small
\begin{adjustbox}{center}
	\begin{tabular}{|c|c|c|}
	\hline
		  & Hilbert Series  & $\mathrm{HWG}$\\ 
     \hline
$	\mathrm{HS}_{ \mathbb{Z}} (t) $	& 
    $\begin{array}{l}
    	\dfrac
    	{{\cal P}_{44}(t)}
        { {(1 - t^2)^{11}(1 - t^4)^{11}}}\\\\
        = 1 + 46 t^2 + 1278 t^4 + 22254 t^6 \\+ 270798 t^8 + 2491731 t^{10} + O\left(t^{12}\right) \\  
	\end{array}$	&
    $\begin{array}{l} \mathrm{PE}\big[
       \mu_2 t^2 +t^2 \\+ (\mu_4^2 q^6 + \mu_5^2 q^{-6} + \mu_4 \mu_5 ) t^4 \\-\mu_4^2 \mu_5^2 t^8 \big]\\
    \end{array} $\\  
\hline		
	$\mathrm{HS}_{ \mathbb{Z}+\frac{1}{2}} (t) $	
& 		
    $\begin{array}{l}
    	 \dfrac
    	 {32 t^2{~}{\cal P}_{40}(t)}
    	{(1 - t^2)^{11}(1 - t^4)^{11}}\\\\
    	= 32 t^2 + 1152 t^4 + 21504 t^6 \\+ 267168 t^8 + 2477376 t^{10}+ O\left(t^{12}\right) \\ 
    \end{array}	$
& $ \begin{array}{l}(\mu_4 q^3+ \frac{\mu_5}{q^3})t^2\; \mathrm{PE} \big[t^2 + \mu_2t^2 +\mu_4^2q^6t^4+\frac{\mu_5^2}{q^6}t^4  \big]\end{array}$\\	 
\hline			
	$\mathrm{HS}(t)$	
	& 		
	$\begin{array}{l}
        \dfrac
        {(1 + t^2){~} {\cal P}_{20}(t)}
        {(1 - t^2)^{22}}\\ \\
        = 1 + 78 t^2 + 2430 t^4 + 43758 t^6 \\ + 537966 t^8 + 4969107 t^{10} + O\left(t^{12}\right) 
		\end{array}	$
		& 
     $\begin{array}{l} \mathrm{PE}\big[
     \mu_2 t^2 + t^2 \\+ (\mu_4 q^3 + \mu_5 q^{-3}) t^2    \big]
    \end{array}$\\ 	
\hline	
	\end{tabular}
\end{adjustbox}
\caption[Hilbert series for $E_6$ magnetic quiver]{Hilbert series for the $E_6$ magnetic quiver (\ref{quiverE6}). The first line of the table provides the Hilbert series when the GNO magnetic lattice is integer $\mathrm{HS}_{ \mathbb{Z}} (t) $, the second line shows the sum over the integers-plus-half lattice $\mathrm{HS}_{ \mathbb{Z}+\frac{1}{2}} (t) $ and the last line is their sum $\mathrm{HS}(t)$. The palindromic numerator terms ${\cal P}_k(t)$ are given in Appendix \ref{apx:PP}. In the third column, we show the HWGs in terms of fugacities for $\mathfrak{so}(10)$ (denoted $\mu_1 , \ldots , \mu_5$) and $\mathfrak{u}(1)$ (denoted $q$). Note that the $q$-charge is normalized, such that the branching of $\mathfrak{e}_6\rightarrow\mathfrak{so}(10)\times \mathfrak{ u}(1)$ is $(\mu_6)_{\mathfrak{e}_6} \mapsto q^{+2}(\mu_1)_{\mathfrak{so}(10)}+q^{-1}(\mu_4)_{\mathfrak{so}(10)}+q^{-4}$.}
\vspace{0.5cm}
\label{E6HS}
\end{figure}

\subsection{\texorpdfstring{$E_5$}{E5} quiver}
Using the isomorphism $\mathfrak{e}_5 \cong \mathfrak{so}(10)$, the unitary-orthosymplectic quiver whose Coulomb branch is the closure of the $\mathfrak{so}(10)$ minimal nilpotent orbit $\overline{\mathcal{O}}^{\mathfrak{so}(10)}_{\text{min}}$ takes the following form:
\begin{equation}
\scalebox{.8}{\raisebox{-.5\height}{
\begin{tikzpicture}
	\begin{pgfonlayer}{nodelayer}
		\node [style=bluegauge] (0) at (0, 0) {};
		\node [style=redgauge] (1) at (1.25, 0) {};
		\node [style=bluegauge] (2) at (2.5, 0) {};
		\node [style=redgauge] (3) at (3.75, 0) {};
		\node [style=redgauge] (4) at (-1.25, 0) {};
		\node [style=none] (8) at (0, -0.5) {2};
		\node [style=none] (9) at (1.25, -0.5) {4};
		\node [style=none] (10) at (2.5, -0.5) {2};
		\node [style=none] (11) at (3.75, -0.5) {2};
		\node [style=none] (12) at (-1.25, -0.5) {2};
		\node [style=none] (29) at (1.75, 1.25) {1};
		\node [style=gauge3] (30) at (1.25, 1.25) {};
	\end{pgfonlayer}
	\begin{pgfonlayer}{edgelayer}
		\draw (4) to (0);
		\draw (0) to (1);
		\draw (1) to (2);
		\draw (2) to (3);
		\draw (30) to (1);
	\end{pgfonlayer}
\end{tikzpicture}}
}
\label{quiverE5}
\end{equation}
In this case, as opposed to equation (\ref{quiverE6}), the $\mathrm{U}(1)$ node can not be replaced by an $\mathrm{SO}(2)$ node, as the central node of the quiver is already of orthogonal type. Instead, one may view the $\mathrm{U}(1)$ as gauging a subgroup of a $\mathrm{USp}(2)$ flavour symmetry. 
The Hilbert series is given in Figure \ref{E5HS} and matches the moduli space of $\overline{\mathcal{O}}_{\text{min}}^{\mathfrak{so}(10)}$ \cite{CabreraZhong}, along side with the Hilbert series for the orbifold moduli space with global symmetry $\mathrm{PSO}(8)\times \mathrm{U}(1)$ .


\begin{figure}[tp]
\small
\begin{adjustbox}{center}
	\begin{tabular}{|c|c|c|}
		\hline
		  & Hilbert Series  & $\mathrm{HWG}$\\ 
    \hline
$	\mathrm{HS}_{ \mathbb{Z}} (t) $&
	\begin{tabular}{l}
    	\parbox[t]{8cm}{{\footnotesize
    	{$ \dfrac
    	{\left(\begin{array}{c}1 + 22 t^2 + 245 t^4 + 1442 t^6 + 5355 t^8 + 12978 t^{10} \\+ 21919 t^{12} + 25900 t^{14} +\pal + t^{28}\end{array}\right)}
    	{(1 - t^2)^{7}(1 - t^4)^7}$}}\\\\
         $ = 1 + 29 t^2 + 434 t^4 + 4060 t^6 + 27384 t^8 + 144312 t^{10}+ O\left(t^{12}\right)$} \\  
	\end{tabular} &
    \begin{tabular}{l}
       \parbox[t]{3cm}{$\mathrm{PE}\big[ \mu_2 t^2 +t^2\\
       +(\mu_4^2 q^4 + \mu_4^2 q^{-4})t^4\\ + \mu_4^2 t^4 - \mu_4^4 t^8  \big]$}\\
       \end{tabular}\\
    \hline		
	$\mathrm{HS}_{ \mathbb{Z}+\frac{1}{2}} (t) $	& 		
	\begin{tabular}{l}
	    \parbox[t]{8cm}{{\footnotesize 
	    $ \dfrac
	    { 16 t^2 \left(\begin{array}{c} 1 + 14 t^2 + 91 t^4 + 336 t^6 + 819 t^8 + 1362 t^{10} \\+ 1618 t^{12}+\pal + t^{24}\end{array}\right)}
	    {(1 - t^2)^{7}(1 - t^4)^7}$} \\\\
	    $ = 16 t^2 + 336 t^4 + 3584 t^6 + 25536 t^8 + 138432 t^{10} + O\left(t^{12}\right)$} \\ 
		\end{tabular}	&$ \begin{array}{l}\mu_4(q^2+\frac{1}{q^2}) t^2\; \mathrm{PE} \big[t^2 + \mu_2t^2  \\ +\frac{\mu_4^2}{q^2}t^4 + \mu_4^2q^4t^4 \big]\end{array}$\\	 
	\hline			
	$\mathrm{HS}(t)$	& 		
	\begin{tabular}{l}
	    \parbox[t]{8cm}{{\footnotesize $\dfrac
	    {\left(\begin{array}{c}1 + 30t^2 + 201t^4+394t^6 + 201t^8 + 30t^{10} + t^{12}\end{array}\right)}
	    {(1- t^2)^{14}(1 + t^2)^{-1}} $}\\\\
	    $ = 1 + 45 t^2 + 770 t^4 + 7644 t^6 + 52920 t^8 + 282744 t^{10}+ O\left(t^{12}\right)$} 
	\end{tabular}	& 
    \begin{tabular}{l} 
    \parbox[t]{3cm}{$\mathrm{PE}\big[\mu_2 t^2+t^2\\ +(\mu_4 q^2 + \mu_4 q^{-2})t^2 \big]$ } 
    \end{tabular}\\
    \hline	
	\end{tabular}
\end{adjustbox}
\caption[Hilbert series for $E_5$ magnetic quiver]{Hilbert series for the $E_5$ magnetic quiver (\ref{quiverE5}). The first line of the table provides the Hilbert series when the GNO magnetic lattice is integer $\mathrm{HS}_{ \mathbb{Z}} (t) $, the second line shows the sum over the integers-plus-half lattice $\mathrm{HS}_{ \mathbb{Z}+\frac{1}{2}} (t) $ and the last line is their sum $\mathrm{HS}(t)$. In the third column, we show the HWG in terms of fugacities for $\mathfrak{so}(8)$ (denoted $\mu_1 , \ldots , \mu_4$) and $\mathfrak{u}(1)$ (denoted $q$).  Note that the $q$-charge is normalized, such that the branching of $\mathfrak{so}(10)\rightarrow\mathfrak{so}(8)\times \mathfrak{ u}(1)$ is $(\mu_5)_{\mathfrak{so}(10)} \mapsto q^{+1}(\mu_1)_{\mathfrak{so}(8)}+q^{-1}(\mu_3)_{\mathfrak{so}(8)}$.}
\label{E5HS}
\end{figure}

\subsection{\texorpdfstring{$E_4$}{E4} quiver}
For the exceptional theory of $E_4$, the algebra has the isomorphism $\mathfrak{e}_4 \cong \mathfrak{su}(5)$. Hence, we expect to find a quiver whose Coulomb branch is the closure of the $\mathfrak{su}(5)$ minimal nilpotent orbit $\overline{\mathcal{O}}^{\mathfrak{su}(5)}_{\text{min}}$. Here, there is another novel feature in the quiver, the existence of a \textit{charge 2  hypermultiplet}  between the $\mathrm{U}(1)$ gauge group and the $\mathrm{U}(1)$ flavor group. The quiver takes the form:
\begin{equation}
\scalebox{.8}{\raisebox{-.5\height}{
    \begin{tikzpicture}
	\begin{pgfonlayer}{nodelayer}
		\node [style=redgauge] (3) at (-2, 0) {};
		\node [style=bluegauge] (9) at (-1, 0) {};
		\node [style=redgauge] (16) at (0, 0) {};
		\node [style=none] (29) at (-1, -0.5) {2};
		\node [style=none] (35) at (0, -0.5) {2};
		\node [style=none] (39) at (-2, -0.5) {2};
		\node [style=gauge3] (41) at (-1, 1) {};
 		\node [style=none] (43) at (-0.5, 1) {1};
		\node [style=none] (44) at (-1, 2.8) {$1$};
		\node [style=flavor2] (48) at (-1, 2.3) {};
	\end{pgfonlayer}
	\begin{pgfonlayer}{edgelayer}
		\draw (3) to (9);
		\draw (9) to (41);
		\draw (9) to (16);
		\draw [line join=round,decorate, decoration={zigzag, segment length=4,amplitude=.9,post=lineto,post length=2pt}]  (48) -- (41);
	\end{pgfonlayer}
\end{tikzpicture}}
}
\label{quiverE4}
\end{equation}
where the wiggly line represents the charge 2 hypermultiplet, see Section \ref{subsubsectionEX2}. 
The Hilbert series is given in Figure \ref{E4HS} and matches the moduli space of $\overline{\mathcal{O}}^{\mathfrak{su}(5)}_{\text{min}}$, along side with the Hilbert series for the orbifold moduli space with global symmetry $\mathrm{SO}(6)\times \mathrm{U}(1)$ .

\begin{figure}[t]
\small
\begin{adjustbox}{center}
	\begin{tabular}{|c|c|c|}
\hline
		  & Hilbert Series  & $\mathrm{HWG}$\\ 
\hline
    $	\mathrm{HS}_{ \mathbb{Z}} (t) $	& 		
	\begin{tabular}{l}
	\parbox[t]{8cm}{{\footnotesize
	$\dfrac
	{1+12t^2+58t^4+124t^6+170t^8+
	\pal t^{16}}
	{(1-t^2)^4(1-t^4)^4}$}\\ \\
	$ = 1 + 16 t^2 + 120 t^4 + 560 t^6 + 1995 t^8 + 5824 t^{10}+ O\left(t^{12}\right)$} \\  
	\end{tabular}	&
    \begin{tabular}{l}
       \parbox[t]{4cm}{
       $ \mathrm{PE}\big[\mu_1 \mu_3 t^2 +t^2 \\ + (\mu_1^2 q^{10} + \frac{\mu_3^2}{q^{10}})t^4 - \mu_1^2 \mu_3^2 t^8 \big]$}\\
    \end{tabular}\\  
\hline		
	$\mathrm{HS}_{ \mathbb{Z}+\frac{1}{2}} (t) $	& 		
	\begin{tabular}{l}
	\parbox[t]{8cm}{{\footnotesize 
	$\dfrac
	{8t^2(1+6t^2+17t^4+22t^6+17t^8+6t^{10}+t^{12})}
	{(1-t^2)^4(1-t^4)^4}$}\\ \\
	$= 8 t^2 + 80 t^4 + 440 t^6 + 1680 t^8 + 5152 t^{10}+O\left(t^{12}\right)$} \\ 
	\end{tabular}
	&$ \begin{array}{l}(\mu_1q^5 + \frac{\mu_3}{q^5})\; \mathrm{PE}\big[ \mu_1\mu_3t^2 + t^2\\+ \mu_1^2 q^{10}t^4+ \frac{\mu_3^2}{q^{10}}t^4  \big]\end{array}$\\	 
\hline			
	$\mathrm{HS}(t)$	& 		
	\begin{tabular}{l}
	\parbox[t]{8cm}{{\footnotesize 
	$\dfrac
	{1+16t^2+36t^4+16t^6+t^8}
	{(1-t^2)^8}$}\\ \\
	$ = 1 + 24 t^2 + 200 t^4 + 1000 t^6 + 3675 t^8 + 10976 t^{10}+O\left(t^{12}\right)$} 
	\end{tabular}
	& 
    \begin{tabular}{l} 
    \parbox[t]{4cm}{$\mathrm{PE}\big[\mu_1 \mu_3 t^2 +t^2 \\ +(\mu_1 q^{5} + \frac{\mu_3}{q^5})t^2 -  \mu_1 \mu_3 t^4 \big]$ }   
    \end{tabular}\\ 	 
\hline	
	\end{tabular}
\end{adjustbox}
\caption[Hilbert series for $E_4$ magnetic quiver]{Hilbert series for the $E_4$ magnetic quiver (\ref{quiverE4}). The first line of the table provides the Hilbert series when the GNO magnetic lattice is integer $\mathrm{HS}_{ \mathbb{Z}} (t) $, the second line shows the sum over the integers-plus-half lattice $\mathrm{HS}_{ \mathbb{Z}+\frac{1}{2}} (t) $ and the last line is their sum $\mathrm{HS}(t)$. In the third column, we show the HWG in terms of fugacities for $\mathfrak{su}(4)$ (denoted $\mu_1 , \ldots , \mu_3$) and $\mathfrak{u}(1)$ (denoted $q$). Note that the $q$-charge is normalized, such that the branching of $\mathfrak{su}(5)\rightarrow\mathfrak{su}(4)\times \mathfrak{ u}(1)$ is $(\mu_1)_{\mathfrak{su}(5)} \mapsto q^{+1}(\mu_1)_{\mathfrak{su}(4)}+q^{-4}$.}
\label{E4HS}
\end{figure}

\subsection{\texorpdfstring{$E_3$}{E3} quiver}
For the exceptional theory of $E_3$, the algebra has the isomorphism $\mathfrak{e}_3 \cong \mathfrak{su}(3)\times \mathfrak{su}(2)$. The moduli space is a union of two hyper-K\"ahler cones, the closure of the $\mathfrak{su}(3)$ minimal nilpotent orbit $\overline{\mathcal{O}}^{\mathfrak{su}(3)}_{\text{min}}$ and the closure of the $\mathfrak{su}(2)$ orbit $\overline{\mathcal{O}}^{\mathfrak{su}(2)}_{\text{min}}$. We therefore expect two unitary-orthosymplectic quivers whose Coulomb branches are the two cones. 
The quiver for $\overline{\mathcal{O}}^{\mathfrak{su}(3)}_{\text{min}}$ cone is:
\begin{equation}
\scalebox{0.8}{\raisebox{-.5\height}{
    \begin{tikzpicture}
	\begin{pgfonlayer}{nodelayer}
		\node [style=redgauge] (9) at (-1, 0) {};
		\node [style=none] (29) at (-1, -0.5) {2};
		\node [style=gauge3] (41) at (-1, 1) {};
 		\node [style=none] (43) at (-0.5, 1) {1};
		\node [style=none] (44) at (-1, 2.8) {$1$};
		\node [style=flavor2] (48) at (-1, 2.3) {};
	\end{pgfonlayer}
	\begin{pgfonlayer}{edgelayer}
		\draw (9) to (41);
				\draw [line join=round,decorate, decoration={zigzag, segment length=4,amplitude=.9,post=lineto,post length=2pt}]  (48) -- (41);
	\end{pgfonlayer}
\end{tikzpicture}
}}
\label{E3quiver}
\end{equation}
where the wiggly line is a charge 2 hypermultiplet and the $\overline{\mathcal{O}}^{\mathfrak{su}(2)}_{\text{min}}$ cone is the Coulomb branch of an $\mathrm{SO}(2)$ gauge theory with 1 flavor\footnote{For this quiver, the flavor is a usual charge 1 hypermultiplet and therefore one does not need to ungauge an overall $\mathbb{Z}_2$. Only the integer lattice needs to be summed.}. 
The Hilbert series is given in Figure \ref{E3HS} and matches the moduli space of $\overline{\mathcal{O}}^{\mathfrak{su}(3)}_{\text{min}}$, along side with the Hilbert series for the orbifold moduli space with global symmetry $\mathrm{Spin}(3)\times \mathrm{U}(1)$ . 

As discussed in Section \ref{general5dfamilysection}, the exceptional theories can arise as the infinite coupling limit of the Higgs branch of certain $5$d $\mathcal{N}=1$ theories. For the $E_3$ quiver, this is the Higgs branch of $\mathrm{USp}(2)$ gauge theory with 2 flavors at infinite coupling. At finite coupling, the two cones are the same and the moduli space is their union $\overline{\mathcal{O}}_{[2,2]}^{\mathfrak{so}(4)}\cong\overline{\mathcal{O}}^{\mathfrak{su}(2)}_{\text{min}}\cup \overline{\mathcal{O}}^{\mathfrak{su}(2)}_{\text{min}}$, where we recall that $\overline{\mathcal{O}}^{\mathfrak{su}(2)}_{\text{min}}\cong \mathbb{C}^2/\mathbb{Z}_2$. At infinite coupling, one of the two cones is enhanced to $\overline{\mathcal{O}}^{\mathfrak{su}(3)}_{\text{min}}$ whereas the other cone remains as $\overline{\mathcal{O}}^{\mathfrak{su}(2)}_{\text{min}}$.


\begin{figure}[t]
\small
\begin{adjustbox}{center}
	\begin{tabular}{|c|c|c|}
\hline
	& Hilbert Series  &$\mathrm{HWG}$\\ 
\hline
    $	\mathrm{HS}_{ \mathbb{Z}} (t) $	& 		
	\begin{tabular}{l}
	    \parbox[t]{8cm}{{\footnotesize 
	    $\dfrac
    	{1+2t^2+6t^4+2t^6+t^8}
    	{(1-t^2)^4(1+t^2)^2}$}\\ \\
    	$= 1 + 4 t^2 + 15 t^4 + 32 t^6 + 65 t^8 + 108 t^{10} + O\left(t^{12}\right)$} \\  
	\end{tabular}	&
    \begin{tabular}{l}
        \parbox[t]{4cm}{$\mathrm{PE} \big[\mu^2t^2 +t^2 \\+\mu^2(q^6+\frac{1}{q^6})t^4  -\mu^4t^8\big]$}\\
       \end{tabular}\\
\hline		
	$\mathrm{HS}_{ \mathbb{Z}+\frac{1}{2}} (t) $	& 		
	\begin{tabular}{l}
	\parbox[t]{8cm}{{\footnotesize
	$\dfrac
	{4t^2(1-t+t^2)(1+t+t^2)}
	{(1-t^2)^4(1+t^2)^2}$} \\ \\
	$= 4 t^2 + 12 t^4 + 32 t^6 + 60 t^8 + 108 t^{10} +O\left(t^{12}\right)$} \\ 
	\end{tabular}
	&$ \begin{array}{l}\mu (q^3+\frac{1}{q^3}) t^2\; \mathrm{PE} \big[ \mu^2t^2 +t^2\\  + \mu^2 (q^6 + \frac{1}{q^6})t^4 -\mu^2 t^4 \big]\end{array}$\\ 
\hline			
	$\mathrm{HS}(t)$	& 		
	\begin{tabular}{l}
    	\parbox[t]{8cm}{{\footnotesize
    	$\dfrac
    	{1+4t^2+t^4}
    	{(1-t^2)^4} $} \\ \\
    	$=1 + 8 t^2 + 27 t^4 + 64 t^6 + 125 t^8 + 216 t^{10}+ O\left(t^{12}\right)$} 
	\end{tabular} & 
    \begin{tabular}{l} 
         \parbox[t]{4cm}{$\mathrm{PE} \big[\mu^2t^2 +t^2 \\ +\mu( q^3+ \frac{1}{q^3})t^2  -\mu^2t^4 \big]$ }   
    \end{tabular}\\
\hline	
	\end{tabular}
\end{adjustbox}
\caption[Hilbert series for $E_3$ magnetic quiver]{Hilbert series for the $E_3$ magnetic quiver (\ref{E3quiver}), representing the $\overline{\mathcal{O}}^{\mathfrak{su}(3)}_{\text{min}}$ cone in $\overline{\mathcal{O}}^{\mathfrak{e}_3}_{\text{min}}$. The first line of the table provides the Hilbert series when the GNO magnetic lattice is integer $\mathrm{HS}_{ \mathbb{Z}} (t) $, the second line shows the sum over the integers-plus-half lattice $\mathrm{HS}_{ \mathbb{Z}+\frac{1}{2}} (t) $ and the last line is their sum $\mathrm{HS}(t)$. In the third column, we show the HWG in terms of fugacities for $\mathfrak{su}(2)$ (denoted $\mu$) and $\mathfrak{u}(1)$ (denoted $q$). Note that the $q$-charge is normalized, such that the branching of $\mathfrak{su}(3)\rightarrow\mathfrak{su}(2)\times \mathfrak{u}(1)$ is $(\mu_1)_{\mathfrak{su}(3)} \mapsto q^{+1}(\mu)_{\mathfrak{su}(2)}+q^{-2}$. }
\label{E3HS}
\end{figure}

\subsection{\texorpdfstring{$E_2$}{E2} and \texorpdfstring{$E_1$}{E1} quivers}
For the $E_2$ theory, the moduli space is $\overline{\mathcal{O}}^{\mathfrak{su}(2)}_{\text{min}} \cup \{\bullet \bullet\} \cong \mathbb{C}^2/\mathbb{Z}_2 \cup \{\bullet \bullet\}$ \cite{GiuilaInf}. The discrete moduli space $\{\bullet \bullet\}$ is generated by a nilpotent element,\footnote{Specifically, the gaugino bilinear $S$, see \cite{GiuilaInf}.} and our quivers, which ultimately arise from brane configurations, are insensitive to it. The moduli space for $E_1$ is simply $\overline{\mathcal{O}}^{\mathfrak{su}(2)}_{\text{min}} \cong \mathbb{C}^2/\mathbb{Z}_2 $. Therefore, for both $E_2$ and $E_1$ theories, we find a unitary-orthosymplectic quiver that has the moduli space  $\mathbb{C}^2/\mathbb{Z}_2 $:

\begin{equation}
\scalebox{.8}{\raisebox{-.5\height}{
    \begin{tikzpicture}
	\begin{pgfonlayer}{nodelayer}
		\node [style=gauge3] (41) at (-1, 1) {};
 		\node [style=none] (43) at (-1, 0.5) {1};
		\node [style=none] (44) at (-1, 3) {$2$};
		\node [style=flavor2] (48) at (-1, 2.5) {};
	\end{pgfonlayer}
	\begin{pgfonlayer}{edgelayer}
		\draw [line join=round,decorate, decoration={zigzag, segment length=4,amplitude=.9,post=lineto,post length=2pt}]  (48) -- (41);
	\end{pgfonlayer}
\end{tikzpicture}
}}
\label{E2quiver}
\end{equation}
where the wiggly line is a charge 2 hypermultiplet. The Hilbert series is given in Figure \ref{E2HS} alongside with the Hilbert series for the orbifold moduli space $\mathbb{C}^2/\mathbb{Z}_4$ with global symmetry $\mathrm{U}(1)$.


\begin{figure}[t]
\small
\begin{adjustbox}{center}
	\begin{tabular}{|c|c|c|}
\hline
	& Hilbert Series  & $\mathrm{HWG}$\\
\hline
    $\mathrm{HS}_{ \mathbb{Z}} (t) $	& 		
	\begin{tabular}{l}
	    \parbox[t]{6.5cm}{{\footnotesize
	    $\dfrac
	    {(1+t^4)}
	    {(1-t^2)^2(1+t^2)}$}\\  \\
	    $ = 1 + t^2 + 3 t^4 + 3 t^6 + 5 t^8 + 5 t^{10} + O\left(t^{12}\right)$} \\  
	\end{tabular}	&
    \begin{tabular}{l}
  $\mathrm{PE} \big[t^2+(q^4+\frac{1}{q^4})t^4-t^8 \big]$
    \end{tabular}\\  
\hline		
	$\mathrm{HS}_{ \mathbb{Z}+\frac{1}{2}} (t) $	& 		
	\begin{tabular}{l}
    	\parbox[t]{6.5cm}{{\footnotesize
    	$\dfrac
    	{2t^2}
    	{(1-t^2)^2(1+t^2)}$}\\ \\
    	$ = 2 t^2 + 2 t^4 + 4 t^6 + 4 t^8 + 6 t^{10}+ O\left(t^{12}\right)$} \\ 
	\end{tabular}
	&$ (q^2+\frac{1}{q^2})t^2{~}\mathrm{PE} \big[t^2 +(q^4 +\frac{1}{q^4})t^4 -t^4 \big]$\\	
\hline			
	$\mathrm{HS}(t)$	& 		
	\begin{tabular}{l}
    	\parbox[t]{6.5cm}{{\footnotesize
    	$\dfrac
    	{\left(\begin{array}{c}1+t^2\end{array}\right)}
    	{(1-t^2)^2}$}\\ \\
    	$ = 1 + 3 t^2 + 5 t^4 + 7 t^6 + 9 t^8 + 11 t^{10}+ O\left(t^{12}\right)$} 
	\end{tabular}	& 
    \begin{tabular}{l}
 $\mathrm{PE}\big[t^2+(q^2+\frac{1}{q^2})t^2-t^4 \big]$
    \end{tabular}\\
\hline	
	\end{tabular}
\end{adjustbox}
\caption[Hilbert series for $E_2$ magnetic quiver]{Hilbert series for the $E_2$ magnetic quiver (\ref{E2quiver}). The first line of the table provides the Hilbert series when the GNO magnetic lattice is integer $\mathrm{HS}_{ \mathbb{Z}} (t) $, the second line shows the sum over the integers-plus-half lattice $\mathrm{HS}_{ \mathbb{Z}+\frac{1}{2}} (t) $ and the last line is their sum $\mathrm{HS}(t)$. In the third column, we show the HWG in terms of fugacities for $\mathfrak{u}(1)$ (denoted $q$). Note that the $q$-charge is normalized such that the branching of $\mathfrak{su}(2)\rightarrow \mathfrak{u}(1)$ is  $(\mu)_{\mathfrak{su}(2)} \mapsto q^{+1}+q^{-1}$.}
\label{E2HS}
\end{figure}

\subsection{\texorpdfstring{$\tilde{E_1}$}{tilde E1} and \texorpdfstring{$E_0$}{E0} quivers}
 In \cite{GiuilaInf}, we note two more members of the exceptional family: $\tilde{E}_1$ and $E_0$. The moduli space for $\tilde{E}_1$ is $\{\bullet \bullet\}$ and is generated by nilpotent elements that our quivers are not sensitive to. For $E_0$ the moduli space is trivial.

\subsection{Higgs branch}
So far we only computed the Coulomb branch of the $E_n$ theories. We can also go ahead and compute the Higgs branch using the hyper-K\"ahler quotient construction, which by definition is not affected by the quotient in \eqref{defGaugeGroup}. 
Before any computation, one verifies that the quaternionic dimension of the Higgs branch of the unitary-orthosymplectic $E_n$ theories are all 1. Hence, one may expect that these Higgs branches are Kleinian singularities.

The Higgs branch Hilbert series are obtained using the Molien-Weyl formula with the individual gauge groups exactly as given in the quiver, see Figure \ref{tabColoredNodes} for conventions. The results of the explicit Hilbert series computations are tabulated in Figure \ref{higgs}. The Higgs branches are indeed the expected $\mathbb{C}^2/\Gamma_{E_n}$ Kleinian singularities, where $\Gamma_{E_n}$ are the binary polyhedral groups and are discrete subgroups of $\mathrm{SU}(2)$.

As some of the quivers involve $\mathrm{U}(1)$ with charge 2 hypermultiplets, let us see how this is computed in the Molien-Weyl formula. For $\mathrm{U}(1)$ with $n$ charge 1 hypermultiplets, the hyper-K\"ahler quotient takes the following form:
\begin{equation}
    \mathrm{HS}(t)=\oint_{\mathrm{U}(1)} d\mu_{\mathrm{U}(1)} \dfrac{\text{PE}[n(x+\frac{1}{x})t]}{\text{PE}[t^2]} =\oint_{\mathrm{U}(1)} d\mu_{\mathrm{U}(1)} \dfrac{(1-t^2)}{(1-xt)^n(1-\frac{1}{x}t)^n} 
\end{equation}
where $d\mu_{\mathrm{U}(1)}=\frac{dx}{x}$ is the $\mathrm{U}(1)$ Haar measure and $x$ is the  $\mathrm{U}(1)$ character. The contribution in the numerator comes from the hypermultiplets and the denominator from the adjoint of $\mathrm{U}(1)$. 

For a $\mathrm{U}(1)$ and $n$ charge 2 hypermultiplets, the formula take the following form:
\begin{equation}
    \mathrm{HS}(t)=\oint_{\mathrm{U}(1)} d\mu_{\mathrm{U}(1)} \dfrac{\text{PE}[n(x^2+\frac{1}{x^2})t]}{\text{PE}[t^2]}=\oint_{\mathrm{U}(1)} d\mu_{\mathrm{U}(1)} \dfrac{(1-t^2)}{(1-x^2t)^n(1-\frac{1}{x^2}t)^n} \,.
    \label{charge2}
\end{equation}
Both results give the moduli space of $\mathbb{C}^2/\mathbb{Z}_n$. 

\eqref{charge2} is the only modification we need when computing the Higgs branch Hilbert series of a quiver containing a $\mathrm{U}(1)$ with $n$ charge 2 hypermultiplets as a subquiver. This is done for the computation of the $E_4$ and $A_2 \subset E_3$ Higgs branches. 

\begin{figure}[t]
\begin{adjustbox}{center}
	\begin{tabular}{|c|c|c|}
		\hline
		  Exceptional quiver &Higgs branch Hilbert series & Higgs branch moduli space \\ \hline
$E_8$	& 	 	 $\frac{1 - t^{60}}{(1 - t^{12}) (1 - t^{20}) (1 - t^{30})}$&$\mathbb{C}^2/\mathbb{B}\mathbb{I}$ \\  \hline
$E_7$	& 	  $\frac{1 - t^{36}}{(1 - t^8) (1 - t^{12}) (1 - t^{18})}$&$\mathbb{C}^2/\mathbb{B}\mathbb{O}$ \\  \hline
$E_6$	& $\frac{1 - t^{24}}{(1 - t^6) (1 - t^8) (1 - t^{12})}$&$\mathbb{C}^2/\mathbb{B}\mathbb{T}$\\  \hline
$E_5$	& $\frac{1 - t^{16}}{(1 - t^4) (1 - t^6) (1 - t^{8})}$	 &$\mathbb{C}^2/\mathrm{Dic}_5$\\  \hline
$E_4$	& $\frac{1 - t^{10}}{(1 - t^2) (1 - t^5)^2 }$&	$\mathbb{C}^2/\mathbb{Z}_5$\\  \hline
$A_2 \subset E_3$	& $\frac{1 - t^{6}}{(1 - t^2) (1 - t^3)^2 }$	& 	$\mathbb{C}^2/\mathbb{Z}_3$\\  \hline
$A_1 \subset E_2$	& $\frac{1 - t^{4}}{(1 - t^2)^3 }$	 &	$\mathbb{C}^2/\mathbb{Z}_2$\\  \hline
$E_1$	& $\frac{1 - t^{4}}{(1 - t^2)^3 }$	 &	$\mathbb{C}^2/\mathbb{Z}_2$\\  \hline
	\end{tabular}
\end{adjustbox}
\caption[Higgs branch Hilbert series]{We take the unitary-orthosymplectic quivers in this section and computed their Higgs branch Hilbert series. $\mathbb{B}\mathbb{I}$ is the binary icosahedral group, $\mathbb{B}\mathbb{O}$ is the binary octahedral group, $\mathbb{B}\mathbb{T}$ is the binary tetrahedral group and $\mathrm{Dic}_5$ is the binary dihedral group. }
\label{higgs}
\end{figure}

\subsection{Inequivalent embeddings}
Given a simple Lie algebra $\mathfrak{g}$ with Dynkin diagram, the regular embeddings of $\mathbb{Z}_m \hookrightarrow \mathfrak{g}$ are classified by Kac \cite{Kac:1994}. Considering the unitary affine Dynkin quivers for the minimal $E_n$ nilpotent orbit closures, the prescription of turning a single $\mathrm{U}(k)$ gauge node into an $\mathrm{SU}(k)$ gauge node leaves a $\mathbb{Z}_k$ discrete group. Gauging this $\mathbb{Z}_k$, i.e.\ taking the choice $H=\{1\}$, results in an orbifold moduli space with global symmetry prescribed by the Kac classification. However, only the cases where flux is given to a single node is covered by this approach.

Considering the ortho-symplectic quivers of this section, the $\mathbb{Z}_2$ orbifold moduli spaces have global symmetry detailed in Figure \ref{integerHS}.

Focusing, for instance, on the $\mathbb{Z}_2$ orbifolds of the $E_6$ case, the unitary Dynkin quiver prescription of Section \ref{sec:E6ex} yields 
\begin{compactitem}
    \item A $\mathbb{Z}_2$ ``orbifold" ($\mathbb{Z}_2$ acts trivially) which preserves $E_6$ via \eqref{unitaryE6SU1}, 
    \item A $\mathbb{Z}_2$ orbifold which preserves $\mathrm{SU}(2) \times \mathrm{SU}(6)$ via \eqref{unitaryE6Z2}.
\end{compactitem}
In addition, the orthosymplectic quiver \eqref{quiverE6} yields 
\begin{compactitem}
    \item A $\mathbb{Z}_2$ orbifold which preserves $\mathrm{SO}(10) \times \mathrm{U}(1)$. 
\end{compactitem}
These cases complete the regular embeddings $\mathbb{Z}_2 \hookrightarrow E_6$ of Kac. 
In addition, other irregular embeddings are known via non-simply laced unitary quivers 
\begin{compactitem}
    \item $\mathbb{Z}_2$ orbifold which preserves $\mathrm{USp}(8) $ via \eqref{E6non-simply-laced},  
    \item $\mathbb{Z}_2$ orbifold which preserves $F_4$ via  \cite[Eq.\ (4.1)]{foldinganton}.
\end{compactitem}

To complete the discussion, the remaining subgroups of $E_6$ are 
\begin{compactitem}
\item Regular subgroup $\mathrm{SU}(3)\times \mathrm{SU}(3) \times \mathrm{SU}(3)$ which is realised as global symmetry of the $\mathbb{Z}_3$ orbifold defined by \eqref{E6quiver-SU3}.
\item Irregular subgroups $\mathrm{SU}(3)\times G_2$ ,  $\mathrm{SU}(3)$, and  $ G_2$ for which no quiver realisation is known yet.
\end{compactitem}

The analysis of this $E_6$ example generalises straightforwardly to other algebras.
%
%
\section{5d \texorpdfstring{$\mathcal{N}=1$}{N=1} theories }
\label{general5dfamilysection}
Next, the general exceptional sequences  $E_n$ for $-\infty < n \leq 8$ are analyzed. Here the negative label $n$ may come as a surprise, but it comes as a natural extension when one studies $\mathrm{USp}(2k)$ gauge theories with $N_f$ fundamental flavours. The relation is $n =N_f -2k+3 $. The index $n$ labels the moduli space global symmetry for $k=1$ and is then used to denote the entire sequence. Within each fixed $n$ sequence, the members are distinguished by the rank $k$ of the 5d $\mathcal{N}=1$ electric gauge group together with the restriction $N_f \leq 2k+5$ for the existence of a 5d fixed point \cite{Bergman:2015dpa}.
\subsection{\texorpdfstring{$E_n$}{En} sequences of 5d \texorpdfstring{$\mathcal{N}=1$}{N=1} theories ($H=\mathrm{ker}\phi = \mathbb{Z}_2^{\mathrm{diag}}$)}
Using recently developed concepts of magnetic quivers \cite{TropicalSanti,newMarcus,6dmagnetic,5dweb}, we can view the $E_n$ unitary-orthosymplectic quivers as \textit{magnetic quivers} of certain $5$d $\mathcal{N}=1$ \textit{electric theories}. To be more precise, the $3$d $\mathcal{N}=4$ Coulomb branch of the $E_n$ unitary-orthosymplectic quivers (with $H=\mathbb{Z}_2^{\mathrm{diag}}$) is the $5$d $\mathcal{N}=1$ Higgs branch of $\mathrm{USp}(2k)$ gauge theory with $N_f$ flavors at infinite gauge coupling :
\begin{equation} \mathcal{H}_{\infty}^{5d} \left( 
\scalebox{.8}{\raisebox{-.5\height}{
\begin{tikzpicture}
	\begin{pgfonlayer}{nodelayer}
		\node [style=flavorRed] (1) at (3.25, 0) {};
		\node [style=none] (2) at (1.5, -0.5) {$2k$};
		\node [style=none] (3) at (3.25, -0.5) {$N_f$};
		\node [style=bluegauge] (22) at (1.5, 0) {};
		\draw (22) to (1);
	\end{pgfonlayer}
\end{tikzpicture}}} \right) = \mathcal{C}^{3d} \left( \begin{array}{c}
    E_n \textrm{ unitary-} \\ \textrm{orthosymplectic} \\ \textrm{quivers}
\end{array} \right) 
\,,
\label{electricquivers}
\end{equation}
for $n=N_f -2k+3$.
Such a relation is arrived by first constructing the electric gauge theories using brane webs and O5 orientifold planes \cite{Zafrir:2015ftn}. Upon setting the mass parameters to zero and taking the infinite coupling limit, the magnetic quiver can be read off from the brane configuration following the techniques developed in the companion paper \cite{5dweb}. 
The respective magnetic quivers are 
shown in Figure \ref{generalOrtho}. 
For the $E_n$ sequences with $n\leq 4$, we observe charge two hypermultiplets transforming under the bifundamental of a $\mathrm{U}(1)$ gauge group and a $\mathrm{U}(l+1)$ flavor group, with $l=k-\lfloor \frac{N_f}{2} \rfloor$. 
In fact, for $n\leq 4$ one can divide the exceptional $E_n$ families into two groups, one for $n$ even such that the moduli space is a single cone and one for $n$ odd where the moduli space is a union of two cones.

\paragraph{$E_{4-2l}$ family.} 
For $n$ even, the families are $E_4$, $E_2$, $E_0$, $E_{-2} $, $E_{-4}$, $\ldots$. The parameter $l$, introduced above, characterizes the group $E_{4-2l}$, where $l+1$ is the number of charge 2 hypermultiplets.

\paragraph{$E_{3-2l}$ family.} Similarly, the group where $n$ is odd consists of $E_3$, $E_1$, $E_{-1}$, $E_{-3}, \ldots$  and we can characterize them by $E_{3-2l}$.  Again, the number of charge 2 hypermultiplets equals $l+1$.
The Higgs branch of $\mathrm{USp}(2k)$ gauge group with $2k-2l$ flavors at finite gauge coupling is  $\overline{\mathcal{O}}_{[2^{2k-2l}]}^{\mathfrak{so}(4k-4l)}$. This is known in the literature as \textit{very even} D-type orbits and the space is the union of two identical hyper-K\"ahler cones. Each cone is given by the Coulomb branch of the flavored orthosymplectic quiver in Figure \ref{secondcone}. The intersection between both cones is non-trivial and equals a nilpotent orbit of type $\overline{\mathcal{O}}_{[2^{2k-2l-2},1^4]}^{\mathfrak{so}(4k-4l)}$. In the infinite coupling limit, one of the two cones gets enhanced and is given by the Coulomb branch of the unitary-orthosymplectic quiver listed in the last row of Figure \ref{generalOrtho}. Importantly, the intersection between the two cones at infinite coupling is the same as the intersection at finite coupling.

Due to a $5$d duality \cite{Gaiotto:2015una,Hayashi:2015zka}, the unitary-orthosymplectic quivers coincide with families of unitary quivers as detailed in \cite{5dweb}. These are the magnetic quivers associated with $5$d $\mathcal{N}=1$ SQCD theories at infinite gauge coupling given in \cite{TropicalSanti}. Specifically, the associated electric theory \cite{Gaiotto:2015una,Hayashi:2015zka} is a $\mathrm{SU}(k+1)$ gauge theory with $N_f$ flavors and $|\text{CS}| = k-(N_f)/2+3$ where CS is the Chern-Simons level. As a result, the exact refined Hilbert series, expressed in terms of the HWG, are already known \cite{pini,tropicalHWG}. We carried out explicit Coulomb branch Hilbert series computations for several members of families of orthosymplectic quiver and checked that they are indeed consistent with the expected HWG.

\subsection{Rank 0 limit}
When $k=1$, the quiver families return the $E_n$ theories. If we go further down to $k=0$, the resulting theories become free. This is clear from the electric quiver where the Higgs branch at infinite gauge coupling consists of free hypermultiplets. We tabulate all the non-trivial $k=0$ cases in Figure \ref{free}.  
Due to the fact that the moduli space is $\mathbb{H}^l$, for some $l$, the global symmetry is enhanced to $\mathrm{USp}(2l)$. The branching rule of the enhanced global symmetry is also tabulated.

\subsection{Global symmetry}
Following the prescription in Appendix \ref{originalGF}, we can determine exactly the global symmetry group and not just its algebra by looking at the HWG. These are provided in Figure \ref{generalOrtho} and \ref{Enorbifolds}.

\subsection{Dimension of Higgs branch}
As in the end of Section \ref{5dexamples}, we now turn our attention to the Higgs branch of the families of unitary-orthosymplectic quivers in Figure \ref{generalOrtho}. The quaternionic dimension of the Higgs branch of \textit{all} the quivers in the families is $k$. This is not surprising, as they are magnetic quivers for a theory with rank $k$ gauge group. 
This offers another non-trivial check for the existence of charge 2 hypermultiplets in $E_{4-2l}$ and $E_{3-2l}$ which gives the correct Higgs branch dimension.

\subsection{\texorpdfstring{$E_n$}{En} sequences of 5d \texorpdfstring{$\mathcal{N}=1$}{N=1} theories ($H=\{1\}$)}
Now, we can investigate the orthosymplectic quivers of the $E_n$ sequences where the discrete group is $H=\{1\}$. The Coulomb branches here are $\mathbb{Z}_2$ orbifolds of those in Figure \ref{generalOrtho}. The $\mathrm{HWG}_{\mathbb{Z}}$ of the orbifolds are easily obtained with the $\mathbb{Z}_2$ action $-1$ on the spinors. The results are tabulated in Figure \ref{Enorbifolds}. 
\begin{landscape}
\small
\begin{figure}[!ht]
\centering
\vspace{-1cm}
\begin{adjustbox}{center}
	\begin{tabular}{|c|c|c|}
\hline
		 Orthosymplectic Quiver & Global Symmetry & $\mathrm{PL}[\mathrm{HWG}]$  \\ 
\hline
\begin{tabular}{c}
\scalebox{0.80}{	
\begin{tikzpicture}
	\begin{pgfonlayer}{nodelayer}
		\node [style=miniU] (0) at (-3.7, 0) {};
		\node [style=miniBlue] (1) at (-2.7, 0) {};
		\node [style=miniBlue] (12) at (2.675, 0) {};
		\node [style=miniU] (19) at (3.675, 0) {};
		\node [style=none] (21) at (-3.7, -0.5) {2};
		\node [style=none] (22) at (-2.7, -0.5) {2};
		\node [style=none] (27) at (0, -0.5) {\small$2k+6$};
		\node [style=none] (28) at (1.15, -0.5) {\small$2k+4$};
		\node [style=none] (32) at (2.675, -0.5) {2};
		\node [style=none] (33) at (3.675, -0.5) {2};
		\node [style=miniU] (42) at (0, 0) {};
		\node [style=miniBlue] (43) at (-1.15, 0) {};
		\node [style=miniBlue] (44) at (1.15, 0) {};
		\node [style=none] (46) at (-1.9, 0) {$\cdots$};
		\node [style=none] (47) at (1.9, 0) {$\cdots$};
		\node [style=none] (48) at (-2.25, 0) {};
		\node [style=none] (49) at (-1.5, 0) {};
		\node [style=none] (50) at (1.5, 0) {};
		\node [style=none] (51) at (2.25, 0) {};
		\node [style=miniBlue] (52) at (0, 1) {};
		\node [style=none] (53) at (0, 1.5) {2};
		\node [style=none] (54) at (-1.15, -0.5) {\small$2k+4$};
	\end{pgfonlayer}
	\begin{pgfonlayer}{edgelayer}
		\draw (0) to (1);
		\draw (12) to (19);
		\draw (43) to (42);
		\draw (42) to (44);
		\draw (1) to (48.center);
		\draw (49.center) to (43);
		\draw (44) to (50.center);
		\draw (51.center) to (12);
		\draw (52) to (42);
	\end{pgfonlayer}
\end{tikzpicture}
}
\end{tabular}
	&
    \begin{tabular}{c}
       $E_8$ \; $(k=1)$ \\ $\mathrm{Ss}(4k+12)$ \; $(k > 1)$
    \end{tabular}& 
    \begin{tabular}{c}{
    $ \sum\limits_{i=1}^{k+2} \mu_{2i} t^{2i} + t^4 + \mu_{2k+6} \left( t^{k+1}+t^{k+3} \right)$}
    \end{tabular}\\
    \hline		
    \begin{tabular}{c}
    \scalebox{0.80}{	
\begin{tikzpicture}
	\begin{pgfonlayer}{nodelayer}
		\node [style=miniU] (0) at (-3.7, 0) {};
		\node [style=miniBlue] (1) at (-2.7, 0) {};
		\node [style=miniBlue] (12) at (2.675, 0) {};
		\node [style=miniU] (19) at (3.675, 0) {};
		\node [style=none] (21) at (-3.7, -0.5) {2};
		\node [style=none] (22) at (-2.7, -0.5) {2};
		\node [style=none] (27) at (0, -0.5) {\small$2k+4$};
		\node [style=none] (28) at (1.15, -0.5) {\small$2k+2$};
		\node [style=none] (32) at (2.675, -0.5) {2};
		\node [style=none] (33) at (3.675, -0.5) {2};
		\node [style=miniU] (42) at (0, 0) {};
		\node [style=miniBlue] (43) at (-1.15, 0) {};
		\node [style=miniBlue] (44) at (1.15, 0) {};
		\node [style=none] (46) at (-1.9, 0) {$\cdots$};
		\node [style=none] (47) at (1.9, 0) {$\cdots$};
		\node [style=none] (48) at (-2.25, 0) {};
		\node [style=none] (49) at (-1.5, 0) {};
		\node [style=none] (50) at (1.5, 0) {};
		\node [style=none] (51) at (2.25, 0) {};
		\node [style=miniBlue] (52) at (0, 1) {};
		\node [style=none] (53) at (0.5, 1) {2};
		\node [style=none] (54) at (-1.15, -0.5) {\small$2k+2$};
		\node [style=miniU] (55) at (0, 2) {};
		\node [style=none] (56) at (0.5, 2) {2};
	\end{pgfonlayer}
	\begin{pgfonlayer}{edgelayer}
		\draw (0) to (1);
		\draw (12) to (19);
		\draw (43) to (42);
		\draw (42) to (44);
		\draw (1) to (48.center);
		\draw (49.center) to (43);
		\draw (44) to (50.center);
		\draw (51.center) to (12);
		\draw (52) to (42);
		\draw (55) to (52);
	\end{pgfonlayer}
\end{tikzpicture}
}
\end{tabular}&	
    \begin{tabular}{c}    $E_7 / \mathbb{Z}_2$ \; $(k=1)$ \\ $(\mathrm{Ss}(4k+8)\times \mathrm{Spin}(3))/\mathbb{Z}_2$ \; $(k >1)$\end{tabular}
		&
	\begin{tabular}{c}
	\parbox[t]{7cm}{
	$ \sum\limits_{i=1}^{k+1} \mu_{2i} t^{2i}  + \nu^2 t^2 + t^4 + \nu \mu_{2k+4} \left(t^{k+1}+t^{k+3}\right) \\ + \mu_{2k+4}^2 t^{2k+4} - \nu^2 \mu_{2k+4}^2  t^{2k+6}$}
	\end{tabular}\\ 
\hline
    \begin{tabular}{c}
    \scalebox{0.80}{	
\begin{tikzpicture}
	\begin{pgfonlayer}{nodelayer}
		\node [style=miniU] (0) at (-3.7, 0) {};
		\node [style=miniBlue] (1) at (-2.7, 0) {};
		\node [style=miniBlue] (12) at (2.675, 0) {};
		\node [style=miniU] (19) at (3.675, 0) {};
		\node [style=none] (21) at (-3.7, -0.5) {2};
		\node [style=none] (22) at (-2.7, -0.5) {2};
		\node [style=none] (27) at (0, -0.5) {\small$2k+2$};
		\node [style=none] (28) at (1.15, -0.5) {\small$2k+2$};
		\node [style=none] (32) at (2.675, -0.5) {2};
		\node [style=none] (33) at (3.675, -0.5) {2};
		\node [style=miniU] (42) at (0, 0) {};
		\node [style=miniBlue] (43) at (-1.15, 0) {};
		\node [style=miniBlue] (44) at (1.15, 0) {};
		\node [style=none] (46) at (-1.9, 0) {$\cdots$};
		\node [style=none] (47) at (1.9, 0) {$\cdots$};
		\node [style=none] (48) at (-2.25, 0) {};
		\node [style=none] (49) at (-1.5, 0) {};
		\node [style=none] (50) at (1.5, 0) {};
		\node [style=none] (51) at (2.25, 0) {};
		\node [style=none] (54) at (-1.15, -0.5) {\small$2k+2$};
		\node [style=gauge3] (55) at (0, 1) {};
		\node [style=none] (56) at (0, 1.5) {1};
		\node [style=miniBlue] (57) at (0, 0) {};
		\node [style=miniU] (58) at (1.15, 0) {};
		\node [style=miniU] (59) at (-1.15, 0) {};
	\end{pgfonlayer}
	\begin{pgfonlayer}{edgelayer}
		\draw (0) to (1);
		\draw (12) to (19);
		\draw (43) to (42);
		\draw (42) to (44);
		\draw (1) to (48.center);
		\draw (49.center) to (43);
		\draw (44) to (50.center);
		\draw (51.center) to (12);
		\draw (55) to (42);
	\end{pgfonlayer}
\end{tikzpicture}
}
\end{tabular}	& 
  \begin{tabular}{c}    $E_6  / \mathbb{Z}_3$ \; $(k=1)$ \\ $\mathrm{Spin}(4k+6)\times \mathrm{U}(1)$ \; $(k > 1)$\end{tabular}&
  \begin{tabular}{c}{
  $ \sum\limits_{i=1}^{k} \mu_{2i} t^{2i} + t^2 + \left(\mu_{2k+2} q + \mu_{2k+3} \frac{1}{q} \right) t^{k+1}$} 
  \end{tabular}\\ 
 \hline	
  	\begin{tabular}{c}
  	\scalebox{0.80}{	
\begin{tikzpicture}
	\begin{pgfonlayer}{nodelayer}
		\node [style=miniU] (0) at (-3.7, 0) {};
		\node [style=miniBlue] (1) at (-2.7, 0) {};
		\node [style=miniBlue] (12) at (2.675, 0) {};
		\node [style=miniU] (19) at (3.675, 0) {};
		\node [style=none] (21) at (-3.7, -0.5) {2};
		\node [style=none] (22) at (-2.7, -0.5) {2};
		\node [style=none] (27) at (0, -0.5) {\small$2k+2$};
		\node [style=none] (28) at (1.15, -0.5) {\small$2k$};
		\node [style=none] (32) at (2.675, -0.5) {2};
		\node [style=none] (33) at (3.675, -0.5) {2};
		\node [style=miniU] (42) at (0, 0) {};
		\node [style=miniBlue] (43) at (-1.15, 0) {};
		\node [style=miniBlue] (44) at (1.15, 0) {};
		\node [style=none] (46) at (-1.9, 0) {$\cdots$};
		\node [style=none] (47) at (1.9, 0) {$\cdots$};
		\node [style=none] (48) at (-2.25, 0) {};
		\node [style=none] (49) at (-1.5, 0) {};
		\node [style=none] (50) at (1.5, 0) {};
		\node [style=none] (51) at (2.25, 0) {};
		\node [style=none] (54) at (-1.15, -0.5) {\small$2k$};
		\node [style=gauge3] (55) at (0, 1) {};
		\node [style=none] (56) at (0, 1.5) {1};
	\end{pgfonlayer}
	\begin{pgfonlayer}{edgelayer}
		\draw (0) to (1);
		\draw (12) to (19);
		\draw (43) to (42);
		\draw (42) to (44);
		\draw (1) to (48.center);
		\draw (49.center) to (43);
		\draw (44) to (50.center);
		\draw (51.center) to (12);
		\draw (55) to (42);
	\end{pgfonlayer}
\end{tikzpicture}
}
\end{tabular}	& 
  \begin{tabular}{c}    $E_5   / \mathbb{Z}_4 \cong \mathrm{Spin}(10)  / \mathbb{Z}_4 $ \; $(k=1)$ \\ $\mathrm{Ss}(4k+4)\times \mathrm{U}(1)$ \; $(k > 1)$\end{tabular}&
  \begin{tabular}{c}{
  $ \sum\limits_{i=1}^{k} \mu_{2i} t^{2i} + t^2 +\mu_{2k+2}  \left( q + \frac{1}{q} \right) t^{k+1}$} 
  \end{tabular} \\ 
\hline
  	\begin{tabular}{c}
  	\scalebox{0.80}{	
\begin{tikzpicture}
	\begin{pgfonlayer}{nodelayer}
		\node [style=miniU] (0) at (-3.725, 0) {};
		\node [style=miniBlue] (1) at (-2.725, 0) {};
		\node [style=miniBlue] (12) at (3.05, 0) {};
		\node [style=miniU] (19) at (4.05, 0) {};
		\node [style=none] (21) at (-3.725, -0.5) {2};
		\node [style=none] (22) at (-2.725, -0.5) {2};
		\node [style=none] (26) at (-1, -0.5) {\small $2k-2l$};
		\node [style=none] (27) at (0.15, -0.5) {\small $2k-2l$};
		\node [style=none] (28) at (1.3, -0.5) {\small $2k-2l$};
		\node [style=none] (32) at (3.05, -0.5) {2};
		\node [style=none] (33) at (4.05, -0.5) {2};
		\node [style=miniU] (34) at (-1, 0) {};
		\node [style=miniU] (35) at (1.3, 0) {};
		\node [style=gauge3] (36) at (0.15, 1) {};
		\node [style=miniBlue] (38) at (0.15, 0) {};
		\node [style=none] (41) at (0.8, 1) {1};
		\node [style=gauge3] (41) at (0.15, 1) {};
		\node [style=none] (44) at (1, 2) {$l+1$};
		\node [style=flavor2] (48) at (0.15, 2) {};
		\node [style=none] (49) at (-1.8, 0) {$\cdots$};
		\node [style=none] (50) at (2.125, 0) {$\cdots$};
		\node [style=none] (51) at (2.5, 0) {};
		\node [style=none] (52) at (1.725, 0) {};
		\node [style=none] (53) at (-1.4, 0) {};
		\node [style=none] (54) at (-2.225, 0) {};
	\end{pgfonlayer}
	\begin{pgfonlayer}{edgelayer}
		\draw (0) to (1);
		\draw (12) to (19);
		\draw (34) to (38);
		\draw (38) to (36);
		\draw (38) to (35);
		\draw (53.center) to (34);
		\draw (1) to (54.center);
		\draw (35) to (52.center);
		\draw (51.center) to (12);
		\draw [line join=round,decorate, decoration={zigzag, segment length=4,amplitude=.9,post=lineto,post length=2pt}]  (48) -- (41);
	\end{pgfonlayer}
\end{tikzpicture}
}
\end{tabular}	& 
  \begin{tabular}{c}    $E_4   / \mathbb{Z}_5 \cong \mathrm{PSU}(5)$ \; $(k=1 \;\& \;l=0)$ \\ $\mathrm{SU}(2)\subset E_2$ \; $(k=l=1)$ \\ $\mathrm{U}(1)$\; $(k=l \neq 1)$ \\ $\mathrm{Spin}(4k-4l+2)\times \mathrm{U}(1)$ \; otherwise
  \end{tabular}&	
  \begin{tabular}{c}
    \parbox{7cm}{
    $ \sum\limits_{i=1}^{k-l-1} \mu_{2i} t^{2i} + t^2 + \mu_{2k-2l}\mu_{2k-2l+1}t^{2k-2l} \\- \mu_{2k-2l}\mu_{2k-2l+1}t^{2k+2} \\+ \left( \mu_{2k-2l} q + \mu_{2k-2l+1} \frac{1}{q} \right) t^{k+1} $}
  \end{tabular}\\ 
\hline	
  	\begin{tabular}{c}
  	\scalebox{0.80}{	
\begin{tikzpicture}
	\begin{pgfonlayer}{nodelayer}
		\node [style=miniU] (0) at (-3.725, 0) {};
		\node [style=miniBlue] (1) at (-2.725, 0) {};
		\node [style=miniBlue] (12) at (3.05, 0) {};
		\node [style=miniU] (19) at (4.05, 0) {};
		\node [style=none] (21) at (-3.725, -0.5) {2};
		\node [style=none] (22) at (-2.725, -0.5) {2};
		\node [style=none] (26) at (-1, -1) {\small $2k-2l-2$};
		\node [style=none] (27) at (0.15, -0.5) {\small $2k-2l$};
		\node [style=none] (28) at (1.3, -1) {\small $2k-2l-2$};
		\node [style=none] (32) at (3.05, -0.5) {2};
		\node [style=none] (33) at (4.05, -0.5) {2};
		\node [style=miniU] (34) at (-1, 0) {};
		\node [style=miniU] (35) at (1.3, 0) {};
		\node [style=gauge3] (36) at (0.15, 1) {};
		\node [style=miniBlue] (38) at (0.15, 0) {};
		\node [style=none] (410) at (0.8, 1) {1};
		\node [style=gauge3] (41) at (0.15, 1) {};
		\node [style=none] (44) at (1, 2) {$l+1$};
		\node [style=flavor2] (48) at (0.15, 2) {};
		\node [style=none] (49) at (-1.8, 0) {$\cdots$};
		\node [style=none] (50) at (2.125, 0) {$\cdots$};
		\node [style=none] (51) at (2.5, 0) {};
		\node [style=none] (52) at (1.725, 0) {};
		\node [style=none] (53) at (-1.4, 0) {};
		\node [style=none] (54) at (-2.225, 0) {};
		\node [style=miniU] (55) at (0.15, 0) {};
		\node [style=miniBlue] (56) at (1.3, 0) {};
		\node [style=miniBlue] (57) at (-1, 0) {};
	\end{pgfonlayer}
	\begin{pgfonlayer}{edgelayer}
		\draw (0) to (1);
		\draw (12) to (19);
		\draw (34) to (38);
		\draw (38) to (36);
		\draw (38) to (35);
		\draw (53.center) to (34);
		\draw (1) to (54.center);
		\draw (35) to (52.center);
		\draw (51.center) to (12);
		\draw [line join=round,decorate, decoration={zigzag, segment length=4,amplitude=.9,post=lineto,post length=2pt}]  (48) -- (41);
	\end{pgfonlayer}
\end{tikzpicture}
}
\end{tabular}	& 
  \begin{tabular}{c}    $\mathrm{SU}(3) \subset E_3$ \; $(k=1 \;\& \;l=0)$ \\  $E_1\cong \mathrm{SU}(2)$ \; $(k=l=1)$ \\ $\mathrm{U}(1)$\; $(k=l \neq 1)$\\$\mathrm{Ss}(4k-4l)\times \mathrm{U}(1)$ $\;$ otherwise
  \end{tabular}&	
  \begin{tabular}{c}
     \parbox{7cm}{
     $\sum\limits_{i=1}^{k-l-1} \mu_{2i}t^{2i} + t^2 +\mu_{2k-2l}^2t^{2k-2l} -\mu_{2k-2l}^2t^{2k+2} +\mu_{2k-2l} \left( q+\dfrac{1}{q} \right) t^{k+1}$}
  \end{tabular}\\ 
  \hline
	\end{tabular}
\end{adjustbox}
\caption[$E_n$ unitary-orthsymplectic quivers]{The unitary-orthosymplectic $E_n$ families. The discrete group here is chosen to be $H=\mathrm{ker}\phi =\mathbb{Z}_2^{\mathrm{diag}}$. We also provide the Hilbert series in the form of Highest Weight Generating (HWG) functions where $\mu_i$, $\nu$ and $q$ are the Dynkin fugacities of the (non-exceptional) global symmetry.  In the second row, the $\mathbb{Z}_2$ acts as $-1$ simultaneously on $\nu$ and $\mu_{2k+4}$. The wiggly line represents a hypermultiplet of charge 2. The last two rows are the $E_{4-2l}$ and $E_{3-2l}  (\textrm{larger cone})$ families. }
\label{generalOrtho}
\end{figure}
\end{landscape}

\begin{figure}[ht]
\small
\centering
\begin{adjustbox}{center}
	\begin{tabular}{|c|c|c|c|}
\hline
		Family & Orthosymplectic Quiver &
		\begin{tabular}{c} Global\\ Symmetry\end{tabular} & $\mathrm{PL}[\mathrm{HWG}]$  \\ 
\hline
  \begin{tabular}{c}
  $E_{3-2l}$\\(\scriptsize{smaller cone})
  \end{tabular} &		
    \begin{tabular}{c}
    \scalebox{0.70}{\begin{tikzpicture}
	\begin{pgfonlayer}{nodelayer}
		\node [style=miniU] (0) at (-3.7, 0) {};
		\node [style=miniBlue] (1) at (-2.7, 0) {};
		\node [style=miniBlue] (12) at (2.675, 0) {};
		\node [style=miniU] (19) at (3.675, 0) {};
		\node [style=none] (21) at (-3.7, -0.5) {2};
		\node [style=none] (22) at (-2.7, -0.5) {2};
		\node [style=none] (26) at (-1.15, -0.8) {\small$2k-2l-2$};
		\node [style=none] (27) at (0, -0.5) {\small$2k-2l$};
		\node [style=none] (28) at (1.15, -0.8) {\small$2k-2l-2$};
		\node [style=none] (32) at (2.675, -0.5) {2};
		\node [style=none] (33) at (3.675, -0.5) {2};
		\node [style=none] (41) at (0.65, 1) {2};
		\node [style=miniU] (42) at (0, 0) {};
		\node [style=miniBlue] (43) at (-1.15, 0) {};
		\node [style=miniBlue] (44) at (1.15, 0) {};
		\node [style=flavorBlue] (45) at (0, 1) {};
		\node [style=none] (46) at (-1.9, 0) {$\cdots$};
		\node [style=none] (47) at (1.9, 0) {$\cdots$};
		\node [style=none] (48) at (-2.25, 0) {};
		\node [style=none] (49) at (-1.5, 0) {};
		\node [style=none] (50) at (1.5, 0) {};
		\node [style=none] (51) at (2.25, 0) {};
	\end{pgfonlayer}
	\begin{pgfonlayer}{edgelayer}
		\draw (0) to (1);
		\draw (12) to (19);
		\draw (43) to (42);
		\draw (42) to (45);
		\draw (42) to (44);
		\draw (1) to (48.center);
		\draw (49.center) to (43);
		\draw (44) to (50.center);
		\draw (51.center) to (12);
	\end{pgfonlayer}
\end{tikzpicture}
}
    \end{tabular}	& 
    $\mathrm{SO}(4k-4l)$ &
    \begin{tabular}{c}
        \parbox{4cm}{\footnotesize $\sum\limits_{i=1}^{k-l-1} \mu_{2i}t^{2i} +\mu_{2k-2l-1}^2t^{2k-2l}$}
    \end{tabular}\\ 
    \hline
	\end{tabular}
\end{adjustbox}
\caption[Small cone of $E_{3-2l}$ family]{The smaller cone of the $E_{3-2l}$ family along with the global symmetry and HWG. The HWG contains the spinor fugacity $\mu_{2k-2l-1}$ because the larger cone in Figure \ref{generalOrtho}  contains the other spinor fugacity $\mu_{2k-2l}$. This originates from the finite coupling case where the two cones are the same, but the HWG for each of them carries one of the two spinor fugacities. Their union then contains both spinors fugacities and the intersection contains neither spinors.   }
\label{secondcone}
\end{figure}

\begin{figure}[ht]
\small
\centering
\begin{adjustbox}{center}
	\begin{tabular}{|c|c|c|c|c|}
\hline
		Family & $k=0$ Orthosymplectic Quiver &
		\begin{tabular}{c} Global\\ Symmetry\end{tabular} &\begin{tabular}{c} Hilbert\\ Series \end{tabular}& Branching rules \\ 
\hline
  $E_8$ &		
    \begin{tabular}{c}
    \scalebox{0.60}{
\begin{tikzpicture}
	\begin{pgfonlayer}{nodelayer}
		\node [style=miniBlue] (0) at (0, 1) {};
		\node [style=miniU] (1) at (0, 0) {};
		\node [style=miniBlue] (2) at (1, 0) {};
		\node [style=miniBlue] (3) at (-1, 0) {};
		\node [style=miniBlue] (4) at (3, 0) {};
		\node [style=miniBlue] (5) at (-3, 0) {};
		\node [style=miniU] (6) at (2, 0) {};
		\node [style=miniU] (7) at (4, 0) {};
		\node [style=miniU] (8) at (-2, 0) {};
		\node [style=miniU] (9) at (-4, 0) {};
		\node [style=none] (10) at (0.5, 1) {2};
		\node [style=none] (11) at (0, -0.5) {6};
		\node [style=none] (12) at (1, -0.5) {4};
		\node [style=none] (13) at (-1, -0.5) {4};
		\node [style=none] (14) at (-2, -0.5) {4};
		\node [style=none] (15) at (2, -0.5) {4};
		\node [style=none] (16) at (3, -0.5) {2};
		\node [style=none] (17) at (4, -0.5) {2};
		\node [style=none] (18) at (-3, -0.5) {2};
		\node [style=none] (19) at (-4, -0.5) {2};
	\end{pgfonlayer}
	\begin{pgfonlayer}{edgelayer}
		\draw (9) to (5);
		\draw (5) to (8);
		\draw (8) to (3);
		\draw (3) to (1);
		\draw (1) to (0);
		\draw (1) to (2);
		\draw (2) to (6);
		\draw (6) to (4);
		\draw (4) to (7);
	\end{pgfonlayer}
\end{tikzpicture}
}
    \end{tabular}	& 
    $\mathrm{USp}(32)$ &
    \begin{tabular}{c}
        \parbox{1.2cm}{\footnotesize $\dfrac{1}{(1-t)^{32}}$}
    \end{tabular} & \begin{tabular}{l}
    $[1,0,\ldots,0]_{\mathfrak{usp}(32)}\rightarrow $\\ \\
    $ [0,0,0,0,0,1]_{\mathfrak{so}(12)} $  
    \end{tabular} \\ 
    \hline
      $E_7$ &		
    \begin{tabular}{c}
    \scalebox{0.70}{
\begin{tikzpicture}
	\begin{pgfonlayer}{nodelayer}
		\node [style=miniBlue] (0) at (0, 1) {};
		\node [style=miniU] (1) at (0, 0) {};
		\node [style=miniBlue] (2) at (1, 0) {};
		\node [style=miniBlue] (3) at (-1, 0) {};
		\node [style=miniU] (6) at (2, 0) {};
		\node [style=miniU] (8) at (-2, 0) {};
		\node [style=none] (10) at (0.5, 1) {2};
		\node [style=none] (11) at (0, -0.5) {4};
		\node [style=none] (12) at (1, -0.5) {2};
		\node [style=none] (13) at (-1, -0.5) {2};
		\node [style=none] (14) at (-2, -0.5) {2};
		\node [style=none] (15) at (2, -0.5) {2};
		\node [style=miniU] (20) at (0, 2) {};
		\node [style=none] (21) at (0.5, 2) {2};
	\end{pgfonlayer}
	\begin{pgfonlayer}{edgelayer}
		\draw (8) to (3);
		\draw (3) to (1);
		\draw (1) to (0);
		\draw (1) to (2);
		\draw (2) to (6);
		\draw (20) to (0);
	\end{pgfonlayer}
\end{tikzpicture}
}
    \end{tabular}	& 
    $\mathrm{USp}(16)$ &
    \begin{tabular}{c}
        \parbox{1.2cm}
        {\footnotesize $\dfrac{1}{(1-t)^{16}}$}
    \end{tabular}&
    \begin{tabular}{l}
    $[1,0,\ldots,0]_{\mathfrak{usp}(16)}\rightarrow$\\ \\ $[0,0,0,1;1]_{\mathfrak{so}(8)\times \mathfrak{su}(2)}$
    \end{tabular}\\ 
    \hline
      $E_6$ &		
    \begin{tabular}{c}
    \scalebox{0.70}{
\begin{tikzpicture}
	\begin{pgfonlayer}{nodelayer}
		\node [style=miniU] (6) at (1, 0) {};
		\node [style=miniU] (8) at (-1, 0) {};
		\node [style=none] (10) at (0.5, 1) {1};
		\node [style=none] (11) at (0, -0.5) {2};
		\node [style=none] (12) at (1, -0.5) {2};
		\node [style=none] (13) at (-1, -0.5) {2};
		\node [style=miniBlue] (22) at (0, 0) {};
		\node [style=gauge3] (23) at (0, 1) {};
	\end{pgfonlayer}
	\begin{pgfonlayer}{edgelayer}
		\draw (8) to (22);
		\draw (22) to (23);
		\draw (22) to (6);
	\end{pgfonlayer}
\end{tikzpicture}}
    \end{tabular}	& 
    $\mathrm{USp}(8)$ &
    \begin{tabular}{c}
        \parbox{1.2cm}{\footnotesize $\dfrac{1}{(1-t)^{8}}$}
    \end{tabular} &
    \begin{tabular}{l}
    $[1,0,0,0]_{\mathfrak{usp}(8)} \rightarrow $ \\ \\ $(q[0,1,0]+\frac{1}{q}[0,0,1])_{\mathfrak{so}(6)\times \mathfrak{u}(1)} $
    \end{tabular}\\ 
    \hline
      $E_5$ &		
    \begin{tabular}{c}
    \scalebox{0.70}{
\begin{tikzpicture}
	\begin{pgfonlayer}{nodelayer}
		\node [style=none] (10) at (0, 1.5) {1};
		\node [style=gauge3] (23) at (0, 1) {};
		\node [style=miniU] (24) at (0, 0) {};
		\node [style=none] (25) at (0, -0.5) {2};
	\end{pgfonlayer}
	\begin{pgfonlayer}{edgelayer}
		\draw (23) to (24);
	\end{pgfonlayer}
\end{tikzpicture}
}
    \end{tabular}	& 
    $\mathrm{USp}(4)$ &
    \begin{tabular}{l}
        \parbox{1.2cm}{\footnotesize $\dfrac{1}{(1-t)^{4}}$}
    \end{tabular}&
     \begin{tabular}{l}
    $[1,0]_{\mathfrak{usp}(4)}\rightarrow$ \\ \\ $(q+\frac{1}{q})[1]_{\mathfrak{su}(2)\times \mathfrak{u}(1)} $ 
     \end{tabular}\\
    \hline
      $E_4$ &		
    \begin{tabular}{c}
    \scalebox{0.70}{
\begin{tikzpicture}
	\begin{pgfonlayer}{nodelayer}
		\node [style=none] (10) at (0, 0.5) {1};
		\node [style=gauge3] (23) at (0, 1) {};
		\node [style=flavor2] (24) at (0, 2) {};
		\node [style=none] (25) at (0, 2.5) {1};
	\end{pgfonlayer}
	\begin{pgfonlayer}{edgelayer}
	\draw [line join=round,decorate, decoration={zigzag, segment length=4,amplitude=.9,post=lineto,post length=2pt}]  (24) -- (23);
	\end{pgfonlayer}
\end{tikzpicture}}
    \end{tabular}	& 
    $\mathrm{USp}(2)$ &
    \begin{tabular}{c}
        \parbox{1.2cm}{\footnotesize $\dfrac{1}{(1-t)^{2}}$}
    \end{tabular}& Trivial\\ 
    \hline
   $E_{3}$ &		
    \begin{tabular}{c}
    \scalebox{0.70}{
\begin{tikzpicture}
	\begin{pgfonlayer}{nodelayer}
		\node [style=none] (10) at (0, 0.5) {1};
		\node [style=gauge3] (23) at (0, 1) {};
		\node [style=flavor2] (24) at (0, 2) {};
		\node [style=none] (25) at (0, 2.5) {1};
	\end{pgfonlayer}
	\begin{pgfonlayer}{edgelayer}
			\draw [line join=round,decorate, decoration={zigzag, segment length=4,amplitude=.9,post=lineto,post length=2pt}]  (24) -- (23);
	\end{pgfonlayer}
\end{tikzpicture}
}
    \end{tabular}	& 
    $\mathrm{USp}(2)$ &
    \begin{tabular}{c}
        \parbox{1.2cm}{\footnotesize $\dfrac{1}{(1-t)^{2}}$}
    \end{tabular}&Trivial\\ 
    \hline
	\end{tabular}
\end{adjustbox}
\caption[Case $k=0$ of the $E_n$ quivers]{The $k=0$ limit of the family of $E_n$ quivers in Figure \ref{generalOrtho}. The choice of discrete group is $H=\mathbb{Z}_2$. For $n <3$, the $k=0$ limit is trivial. The last column lists the branching rules for the fundamental representation of the $\mathrm{USp}$ global symmetry group of the free theory to the global symmetry group in the $k=0$ limit of Figure \ref{generalOrtho}. Note the nice pattern of power of 2.}
\label{free}
\end{figure}

\begin{landscape}
\small
\begin{figure}[!ht]
\centering
\vspace{-1cm}
\begin{adjustbox}{center}
	\begin{tabular}{|c|c|c|}
\hline
		 Orthosymplectic Quiver & Global Symmetry & $\mathrm{PL}[\mathrm{HWG}]$  \\ 
\hline
\begin{tabular}{c}
\scalebox{0.80}{	
\begin{tikzpicture}
	\begin{pgfonlayer}{nodelayer}
		\node [style=miniU] (0) at (-3.7, 0) {};
		\node [style=miniBlue] (1) at (-2.7, 0) {};
		\node [style=miniBlue] (12) at (2.675, 0) {};
		\node [style=miniU] (19) at (3.675, 0) {};
		\node [style=none] (21) at (-3.7, -0.5) {2};
		\node [style=none] (22) at (-2.7, -0.5) {2};
		\node [style=none] (27) at (0, -0.5) {\small$2k+6$};
		\node [style=none] (28) at (1.15, -0.5) {\small$2k+4$};
		\node [style=none] (32) at (2.675, -0.5) {2};
		\node [style=none] (33) at (3.675, -0.5) {2};
		\node [style=miniU] (42) at (0, 0) {};
		\node [style=miniBlue] (43) at (-1.15, 0) {};
		\node [style=miniBlue] (44) at (1.15, 0) {};
		\node [style=none] (46) at (-1.9, 0) {$\cdots$};
		\node [style=none] (47) at (1.9, 0) {$\cdots$};
		\node [style=none] (48) at (-2.25, 0) {};
		\node [style=none] (49) at (-1.5, 0) {};
		\node [style=none] (50) at (1.5, 0) {};
		\node [style=none] (51) at (2.25, 0) {};
		\node [style=miniBlue] (52) at (0, 1) {};
		\node [style=none] (53) at (0, 1.5) {2};
		\node [style=none] (54) at (-1.15, -0.5) {\small$2k+4$};
	\end{pgfonlayer}
	\begin{pgfonlayer}{edgelayer}
		\draw (0) to (1);
		\draw (12) to (19);
		\draw (43) to (42);
		\draw (42) to (44);
		\draw (1) to (48.center);
		\draw (49.center) to (43);
		\draw (44) to (50.center);
		\draw (51.center) to (12);
		\draw (52) to (42);
	\end{pgfonlayer}
\end{tikzpicture}
}
\end{tabular}
	&
    \begin{tabular}{c}
        $\mathrm{PSO}(4k+12)$ \; $(k \geq 1)$
    \end{tabular}& 
    \begin{tabular}{c}{
    $ \sum\limits_{i=1}^{k+2} \mu_{2i} t^{2i} + t^4 + \mu_{2k+6}^2 \left( t^{2k+2} +t^{2k+4}+t^{2k+6}  \right) -\mu_{2k+6}^4t^{4k+8}$}
    \end{tabular}\\
    \hline		
    \begin{tabular}{c}
    \scalebox{0.80}{	
\begin{tikzpicture}
	\begin{pgfonlayer}{nodelayer}
		\node [style=miniU] (0) at (-3.7, 0) {};
		\node [style=miniBlue] (1) at (-2.7, 0) {};
		\node [style=miniBlue] (12) at (2.675, 0) {};
		\node [style=miniU] (19) at (3.675, 0) {};
		\node [style=none] (21) at (-3.7, -0.5) {2};
		\node [style=none] (22) at (-2.7, -0.5) {2};
		\node [style=none] (27) at (0, -0.5) {\small$2k+4$};
		\node [style=none] (28) at (1.15, -0.5) {\small$2k+2$};
		\node [style=none] (32) at (2.675, -0.5) {2};
		\node [style=none] (33) at (3.675, -0.5) {2};
		\node [style=miniU] (42) at (0, 0) {};
		\node [style=miniBlue] (43) at (-1.15, 0) {};
		\node [style=miniBlue] (44) at (1.15, 0) {};
		\node [style=none] (46) at (-1.9, 0) {$\cdots$};
		\node [style=none] (47) at (1.9, 0) {$\cdots$};
		\node [style=none] (48) at (-2.25, 0) {};
		\node [style=none] (49) at (-1.5, 0) {};
		\node [style=none] (50) at (1.5, 0) {};
		\node [style=none] (51) at (2.25, 0) {};
		\node [style=miniBlue] (52) at (0, 1) {};
		\node [style=none] (53) at (0.5, 1) {2};
		\node [style=none] (54) at (-1.15, -0.5) {\small$2k+2$};
		\node [style=miniU] (55) at (0, 2) {};
		\node [style=none] (56) at (0.5, 2) {2};
	\end{pgfonlayer}
	\begin{pgfonlayer}{edgelayer}
		\draw (0) to (1);
		\draw (12) to (19);
		\draw (43) to (42);
		\draw (42) to (44);
		\draw (1) to (48.center);
		\draw (49.center) to (43);
		\draw (44) to (50.center);
		\draw (51.center) to (12);
		\draw (52) to (42);
		\draw (55) to (52);
	\end{pgfonlayer}
\end{tikzpicture}
}
\end{tabular}&	
    \begin{tabular}{c}    $\mathrm{PSO}(4k+8)\times \mathrm{PSU}(2)$ \; $(k \geq1)$\end{tabular}
		&
	\begin{tabular}{c}
	\parbox[t]{5cm}{
	$ \sum\limits_{i=1}^{k+1} \mu_{2i} t^{2i}  + \nu^2 t^2 + t^4 + \nu^2 \mu_{2k+4}^2 \left(t^{2k+2}+t^{2k+4}\right) \\+ \mu_{2k+4}^2 t^{2k+4} - \nu^4 \mu_{2k+4}^4  t^{4k+8}$}
	\end{tabular}\\ 
\hline
    \begin{tabular}{c}
    \scalebox{0.80}{	
\begin{tikzpicture}
	\begin{pgfonlayer}{nodelayer}
		\node [style=miniU] (0) at (-3.7, 0) {};
		\node [style=miniBlue] (1) at (-2.7, 0) {};
		\node [style=miniBlue] (12) at (2.675, 0) {};
		\node [style=miniU] (19) at (3.675, 0) {};
		\node [style=none] (21) at (-3.7, -0.5) {2};
		\node [style=none] (22) at (-2.7, -0.5) {2};
		\node [style=none] (27) at (0, -0.5) {\small$2k+2$};
		\node [style=none] (28) at (1.15, -0.5) {\small$2k+2$};
		\node [style=none] (32) at (2.675, -0.5) {2};
		\node [style=none] (33) at (3.675, -0.5) {2};
		\node [style=miniU] (42) at (0, 0) {};
		\node [style=miniBlue] (43) at (-1.15, 0) {};
		\node [style=miniBlue] (44) at (1.15, 0) {};
		\node [style=none] (46) at (-1.9, 0) {$\cdots$};
		\node [style=none] (47) at (1.9, 0) {$\cdots$};
		\node [style=none] (48) at (-2.25, 0) {};
		\node [style=none] (49) at (-1.5, 0) {};
		\node [style=none] (50) at (1.5, 0) {};
		\node [style=none] (51) at (2.25, 0) {};
		\node [style=none] (54) at (-1.15, -0.5) {\small$2k+2$};
		\node [style=gauge3] (55) at (0, 1) {};
		\node [style=none] (56) at (0, 1.5) {1};
		\node [style=miniBlue] (57) at (0, 0) {};
		\node [style=miniU] (58) at (1.15, 0) {};
		\node [style=miniU] (59) at (-1.15, 0) {};
	\end{pgfonlayer}
	\begin{pgfonlayer}{edgelayer}
		\draw (0) to (1);
		\draw (12) to (19);
		\draw (43) to (42);
		\draw (42) to (44);
		\draw (1) to (48.center);
		\draw (49.center) to (43);
		\draw (44) to (50.center);
		\draw (51.center) to (12);
		\draw (55) to (42);
	\end{pgfonlayer}
\end{tikzpicture}
}
\end{tabular}	& 
  \begin{tabular}{c} $\mathrm{SO}(4k+6)\times \mathrm{U}(1)$ \; $(k \geq 1)$\end{tabular}&
  \begin{tabular}{c}{
  $ \sum\limits_{i=1}^{k} \mu_{2i} t^{2i} + t^2 + \left(\mu_{2k+2}^2 q^2+\mu_{2k+2}\mu_{2k+3}  + \mu_{2k+3}^2 \frac{1}{q^2} \right) t^{2k+2}  -\mu_{2k+2}^2\mu_{2k+3}^2t^{4k+4}$} 
  \end{tabular}\\ 
 \hline	
  	\begin{tabular}{c}
  	\scalebox{0.80}{	
\begin{tikzpicture}
	\begin{pgfonlayer}{nodelayer}
		\node [style=miniU] (0) at (-3.7, 0) {};
		\node [style=miniBlue] (1) at (-2.7, 0) {};
		\node [style=miniBlue] (12) at (2.675, 0) {};
		\node [style=miniU] (19) at (3.675, 0) {};
		\node [style=none] (21) at (-3.7, -0.5) {2};
		\node [style=none] (22) at (-2.7, -0.5) {2};
		\node [style=none] (27) at (0, -0.5) {\small$2k+2$};
		\node [style=none] (28) at (1.15, -0.5) {\small$2k$};
		\node [style=none] (32) at (2.675, -0.5) {2};
		\node [style=none] (33) at (3.675, -0.5) {2};
		\node [style=miniU] (42) at (0, 0) {};
		\node [style=miniBlue] (43) at (-1.15, 0) {};
		\node [style=miniBlue] (44) at (1.15, 0) {};
		\node [style=none] (46) at (-1.9, 0) {$\cdots$};
		\node [style=none] (47) at (1.9, 0) {$\cdots$};
		\node [style=none] (48) at (-2.25, 0) {};
		\node [style=none] (49) at (-1.5, 0) {};
		\node [style=none] (50) at (1.5, 0) {};
		\node [style=none] (51) at (2.25, 0) {};
		\node [style=none] (54) at (-1.15, -0.5) {\small$2k$};
		\node [style=gauge3] (55) at (0, 1) {};
		\node [style=none] (56) at (0, 1.5) {1};
	\end{pgfonlayer}
	\begin{pgfonlayer}{edgelayer}
		\draw (0) to (1);
		\draw (12) to (19);
		\draw (43) to (42);
		\draw (42) to (44);
		\draw (1) to (48.center);
		\draw (49.center) to (43);
		\draw (44) to (50.center);
		\draw (51.center) to (12);
		\draw (55) to (42);
	\end{pgfonlayer}
\end{tikzpicture}
}
\end{tabular}	& 
  \begin{tabular}{c}     $\mathrm{PSO}(4k+4)\times \mathrm{U}(1)$ \; $(k\geq 1)$\end{tabular}&
  \begin{tabular}{c}{
  $ \sum\limits_{i=1}^{k} \mu_{2i} t^{2i} + t^2 +\mu_{2k+2} ^2 \left( q^2 + 1+\frac{1}{q^2} \right) t^{2k+2}-\mu_{2k+2} ^4t^{4k+4}$} 
  \end{tabular} \\ 
\hline
  	\begin{tabular}{c}
  	\scalebox{0.80}{	
\begin{tikzpicture}
	\begin{pgfonlayer}{nodelayer}
		\node [style=miniU] (0) at (-3.725, 0) {};
		\node [style=miniBlue] (1) at (-2.725, 0) {};
		\node [style=miniBlue] (12) at (3.05, 0) {};
		\node [style=miniU] (19) at (4.05, 0) {};
		\node [style=none] (21) at (-3.725, -0.5) {2};
		\node [style=none] (22) at (-2.725, -0.5) {2};
		\node [style=none] (26) at (-1, -0.5) {\small $2k-2l$};
		\node [style=none] (27) at (0.15, -0.5) {\small $2k-2l$};
		\node [style=none] (28) at (1.3, -0.5) {\small $2k-2l$};
		\node [style=none] (32) at (3.05, -0.5) {2};
		\node [style=none] (33) at (4.05, -0.5) {2};
		\node [style=miniU] (34) at (-1, 0) {};
		\node [style=miniU] (35) at (1.3, 0) {};
		\node [style=gauge3] (36) at (0.15, 1) {};
		\node [style=miniBlue] (38) at (0.15, 0) {};
		\node [style=none] (41) at (0.8, 1) {1};
		\node [style=gauge3] (41) at (0.15, 1) {};
		\node [style=none] (44) at (1, 2) {$l+1$};
		\node [style=flavor2] (48) at (0.15, 2) {};
		\node [style=none] (49) at (-1.8, 0) {$\cdots$};
		\node [style=none] (50) at (2.125, 0) {$\cdots$};
		\node [style=none] (51) at (2.5, 0) {};
		\node [style=none] (52) at (1.725, 0) {};
		\node [style=none] (53) at (-1.4, 0) {};
		\node [style=none] (54) at (-2.225, 0) {};
	\end{pgfonlayer}
	\begin{pgfonlayer}{edgelayer}
		\draw (0) to (1);
		\draw (12) to (19);
		\draw (34) to (38);
		\draw (38) to (36);
		\draw (38) to (35);
		\draw (53.center) to (34);
		\draw (1) to (54.center);
		\draw (35) to (52.center);
		\draw (51.center) to (12);
		\draw [line join=round,decorate, decoration={zigzag, segment length=4,amplitude=.9,post=lineto,post length=2pt}]  (48) -- (41);
	\end{pgfonlayer}
\end{tikzpicture}
}
\end{tabular}	& 
  \begin{tabular}{c}   $\mathrm{U}(1)$\; $(k=l)$ \\ $\mathrm{SO}(4k-4l+2)\times \mathrm{U}(1)$ \; otherwise
  \end{tabular}&	
  \begin{tabular}{c}
    \parbox{7cm}{
    $ \sum\limits_{i=1}^{k-l-1} \mu_{2i} t^{2i} + t^2 + \mu_{2k-2l}\mu_{2k-2l+1}t^{2k-2l} \\- \mu_{2k-2l}\mu_{2k-2l+1}t^{2k+2} \\+ \left( \mu_{2k-2l}^2 q^2 + \mu_{2k-2l+1}^2 \frac{1}{q^2} \right) t^{2k+2} -\mu_{2k-2l}\mu_{2k-2l+1}t^{4k+4}$}
  \end{tabular}\\ 
\hline	
  	\begin{tabular}{c}
  	\scalebox{0.80}{	
\begin{tikzpicture}
	\begin{pgfonlayer}{nodelayer}
		\node [style=miniU] (0) at (-3.725, 0) {};
		\node [style=miniBlue] (1) at (-2.725, 0) {};
		\node [style=miniBlue] (12) at (3.05, 0) {};
		\node [style=miniU] (19) at (4.05, 0) {};
		\node [style=none] (21) at (-3.725, -0.5) {2};
		\node [style=none] (22) at (-2.725, -0.5) {2};
		\node [style=none] (26) at (-1, -1) {\small $2k-2l-2$};
		\node [style=none] (27) at (0.15, -0.5) {\small $2k-2l$};
		\node [style=none] (28) at (1.3, -1) {\small $2k-2l-2$};
		\node [style=none] (32) at (3.05, -0.5) {2};
		\node [style=none] (33) at (4.05, -0.5) {2};
		\node [style=miniU] (34) at (-1, 0) {};
		\node [style=miniU] (35) at (1.3, 0) {};
		\node [style=gauge3] (36) at (0.15, 1) {};
		\node [style=miniBlue] (38) at (0.15, 0) {};
		\node [style=none] (410) at (0.8, 1) {1};
		\node [style=gauge3] (41) at (0.15, 1) {};
		\node [style=none] (44) at (1, 2) {$l+1$};
		\node [style=flavor2] (48) at (0.15, 2) {};
		\node [style=none] (49) at (-1.8, 0) {$\cdots$};
		\node [style=none] (50) at (2.125, 0) {$\cdots$};
		\node [style=none] (51) at (2.5, 0) {};
		\node [style=none] (52) at (1.725, 0) {};
		\node [style=none] (53) at (-1.4, 0) {};
		\node [style=none] (54) at (-2.225, 0) {};
		\node [style=miniU] (55) at (0.15, 0) {};
		\node [style=miniBlue] (56) at (1.3, 0) {};
		\node [style=miniBlue] (57) at (-1, 0) {};
	\end{pgfonlayer}
	\begin{pgfonlayer}{edgelayer}
		\draw (0) to (1);
		\draw (12) to (19);
		\draw (34) to (38);
		\draw (38) to (36);
		\draw (38) to (35);
		\draw (53.center) to (34);
		\draw (1) to (54.center);
		\draw (35) to (52.center);
		\draw (51.center) to (12);
		\draw [line join=round,decorate, decoration={zigzag, segment length=4,amplitude=.9,post=lineto,post length=2pt}]  (48) -- (41);
	\end{pgfonlayer}
\end{tikzpicture}
}
\end{tabular}	& 
  \begin{tabular}{c}   $\mathrm{U}(1)$\; $(k=l )$\\$\mathrm{PSO}(4k-4l)\times \mathrm{U}(1)$ $\;$ otherwise
  \end{tabular}&	
  \begin{tabular}{c}
     \parbox{7cm}{
     $\sum\limits_{i=1}^{k-l-1} \mu_{2i}t^{2i} + t^2 +\mu_{2k-2l}^2t^{2k-2l} -\mu_{2k-2l}^2t^{2k+2} +\mu_{2k-2l}^2 \left( q^2+\dfrac{1}{q^2} \right) t^{2k+2}-\mu_{2k-2l}^4t^{4k+4}$}
  \end{tabular}\\ 
  \hline
	\end{tabular}
\end{adjustbox}
\caption[$E_n$ unitary-orthsymplectic quivers]{The unitary-orthosymplectic $E_n$ families. The discrete group here is chosen to be $H=\{1\}$. The Coulomb branches are $\mathbb{Z}_2$ orbifolds of those in Figure \ref{generalOrtho}. We also provide the Hilbert series in the form of Highest Weight Generating (HWG) functions where $\mu_i$, $\nu$ and $q$ are the Dynkin fugacities of the (non-exceptional) global symmetry. The wiggly line represents a hypermultiplet of charge 2. The last two rows are the $E_{4-2l}$ and $E_{3-2l}   (\textrm{larger cone})$ families.  }
\label{Enorbifolds}
\end{figure}
\end{landscape}
\section{Applications in 4d and 6d}\label{applications}

In this section, we apply the monopole formula to unframed orthosymplectic quivers that arise in the study of certain 4d and 6d SCFTs. The choice of ungauging is  $H=\mathrm{ker}\phi=\mathbb{Z}_2^{\mathrm{diag}}$ for all cases in this section. The goal of this section is to demonstrate that the choice of $H$ confirms existing predictions of the 4d and 6d physics, without getting into a detailed study of these topics, while providing quantitative data on the corresponding moduli spaces.
As illustrations, Appendices \ref{classS} and \ref{6dapp} tabulate numerous explicit computations.  

\subsection{4d \texorpdfstring{$\mathcal{N}=2$}{N=2} class \texorpdfstring{$\mathcal{S}$}{S} theories }
{\label{classSguys}}
4d $\mathcal{N}=2$ class $\mathcal{S}$ theories can be constructed from elementary building blocks called \emph{fixtures} \cite{DistlerA} (or triskelions in \cite{BeniniSicilian}). These are $6d$ $\mathcal{N}=(2,0)$ theories with gauge algebra $J$ on a three-punctured Riemann sphere.
By computing the $3$d $\mathcal{N}=4$ Coulomb branch of the star-shaped magnetic quiver that is derived from the class $\mathcal{S}$ data along the lines proposed in \cite{Benini:2010uu}, we obtain the Higgs branch of the corresponding 4d class $\mathcal{S}$ theories:
\begin{equation}
   \mathcal{H}^{4d} \left( \raisebox{-.5\height}{\scalebox{.6}{\begin{tikzpicture}
        \filldraw[color=blue!60, fill=blue!5, very thick](0,0) circle (2);
        \filldraw[fill=black!100] (1.4,0) circle (.1);
        \filldraw[fill=black!100] (-.7,1) circle (.1);
        \filldraw[fill=black!100] (-.7,-1) circle (.1);
        \node at (1.4,.4) {\large $\rho_1$};
        \node at (-.7,1.4) {\large$\rho_2$};
        \node at (-.7,-.6) {\large$\rho_3$};
    \end{tikzpicture}}} \right) = \mathcal{C}^{3d} \left(\begin{tabular}{c}\text{Three-legged magnetic quiver} \\ \text{ with} $T_{\rho_1}(J)$, $T_{\rho_2}(J)$ , $T_{\rho_3}(J)$ \\ \text{joined at central node} $J$   \end{tabular} \right) \,. 
\end{equation}
where the quivers  $T_{\rho}(J)$ are elaborated in more detail in Appendix \ref{punctures}.
As examples, we consider $A_{2r-1}$ twisted fixtures, which have a central $\mathrm{USp(2r)}$ gauge group, and  (untwisted) $D_r$-type fixtures, which have a central $\mathrm{SO(2r)}$ gauge group. The corresponding three-legged magnetic quivers are unitary-orthosymplectic.

\subsubsection{Twisted $A_{\mathrm{odd}}$ fixtures}
For $A_3$,  the linear quivers $T_{\rho}(\mathrm{USp}(4))$ are tabulated in Table \ref{A3puncture}  and $T_{\rho}(\mathrm{SU}(4))$  are tabulated in Table \ref{A3punctureUntwisted}.
The magnetic quivers are constructed by gluing such quivers. For example, the magnetic quiver for the fixture containing two maximal $A_3$ twisted punctures $T_{(1^5)}(\mathrm{USp}(4))$ and one maximal $A_3$ untwisted puncture $T_{(1^5)}(\mathrm{SU}(4))$ can be glued together in the following way:
\begin{equation}
\raisebox{-.5\height}{
\scalebox{0.7}{
    \begin{tikzpicture}
	\begin{pgfonlayer}{nodelayer}
		\node [style=gauge3] (0) at (-4, 2.5) {};
		\node [style=gauge3] (1) at (-4, 1.5) {};
		\node [style=none] (3) at (-4.75, 2.5) {1};
		\node [style=none] (4) at (-4.75, 1.5) {2};
		\node [style=gauge3] (7) at (-4, 0.5) {};
		\node [style=none] (8) at (-4.75, 0.5) {3};
		\node [style=flavor2] (9) at (-4, -0.5) {};
		\node [style=none] (10) at (-4.75, -0.5) {4};
		\node [style=none] (11) at (-4, 3.25) {$T_{(1^4)}(\mathrm{SU}(4))$};
		\node [style=flavorBlue] (17) at (-4.75, -1.5) {};
		\node [style=miniU] (18) at (-2.25, -1.5) {};
		\node [style=bluegauge] (20) at (-1.25, -1.5) {};
		\node [style=miniU] (21) at (-0.25, -1.5) {};
		\node [style=flavorBlue] (22) at (-3.25, -1.5) {};
		\node [style=none] (23) at (-4.75, -2) {4};
		\node [style=none] (24) at (-3.25, -2) {4};
		\node [style=none] (25) at (-2.25, -2) {4};
		\node [style=none] (26) at (-1.25, -2) {2};
		\node [style=none] (27) at (-0.25, -2) {2};
		\node [style=none] (31) at (1, 0.25) {};
		\node [style=none] (32) at (3, 0.25) {};
		\node [style=gauge3] (33) at (7.75, 1.5) {};
		\node [style=gauge3] (34) at (7.75, 0.5) {};
		\node [style=none] (35) at (7, 1.5) {1};
		\node [style=none] (36) at (7, 0.5) {2};
		\node [style=gauge3] (38) at (7.75, -0.5) {};
		\node [style=none] (39) at (7, -0.5) {$3$};
		\node [style=miniU] (43) at (6.75, -1.5) {};
		\node [style=bluegauge] (45) at (5.75, -1.5) {};
		\node [style=miniU] (46) at (4.75, -1.5) {};
		\node [style=miniU] (48) at (8.75, -1.5) {};
		\node [style=bluegauge] (50) at (9.75, -1.5) {};
		\node [style=miniU] (51) at (10.75, -1.5) {};
		\node [style=none] (53) at (7.75, -2) {4};
		\node [style=none] (55) at (8.75, -2) {4};
		\node [style=none] (56) at (9.75, -2) {2};
		\node [style=none] (57) at (10.75, -2) {2};
		\node [style=none] (58) at (6.75, -2) {4};
		\node [style=none] (59) at (5.75, -2) {2};
		\node [style=none] (60) at (4.75, -2) {2};
		\node [style=bluegauge] (61) at (7.75, -1.5) {};
		\node [style=none] (62) at (2, 0.75) {Gluing};
		\node [style=none] (63) at (-6.25, -2.75) {$T_{(1^5)}(\mathrm{USp}(4))$};
		\node [style=none] (64) at (-1.5, -2.75) {$T_{(1^5)}(\mathrm{USp}(4))$};
		\node [style=miniU] (65) at (-5.75, -1.5) {};
		\node [style=bluegauge] (66) at (-6.75, -1.5) {};
		\node [style=miniU] (67) at (-7.75, -1.5) {};
		\node [style=none] (68) at (-5.75, -2) {4};
		\node [style=none] (69) at (-6.75, -2) {2};
		\node [style=none] (70) at (-7.75, -2) {2};
	\end{pgfonlayer}
	\begin{pgfonlayer}{edgelayer}
		\draw (0) to (1);
		\draw (7) to (9);
		\draw (18) to (22);
		\draw (20) to (21);
		\draw [style=->] (31.center) to (32.center);
		\draw (33) to (34);
		\draw (46) to (45);
		\draw (50) to (51);
		\draw (61) to (48);
		\draw (43) to (61);
		\draw (38) to (61);
		\draw (1) to (7);
		\draw (18) to (20);
		\draw (45) to (43);
		\draw (48) to (50);
		\draw (34) to (38);
		\draw (67) to (66);
		\draw (66) to (65);
		\draw (65) to (17);
	\end{pgfonlayer}
\end{tikzpicture}}
}
\end{equation}
where the quiver after gluing is the three-legged magnetic quiver. The results for $A_3$ twisted fixtures are tabulated in Table \ref{A3fixtures}.

\subsubsection{Untwisted D fixtures}
The magnetic quiver of untwisted D-type fixtures are obtained by gluing together  $T_\rho(\mathrm{SO}(2r))$ quivers. The linear quivers $T_\rho(\mathrm{SO}(2r))$ are listed in Figure \ref{D3puncture} for $r=3$ and Figure \ref{D4puncture} for $r=4$. 
The magnetic quiver for a fixture with three maximal $D_3$ punctures is obtained by gluing three maximal legs of $T_\rho(\mathrm{SO}(6))$ \cite{BeniniSicilian}:
\begin{equation}
\scalebox{0.7}{\begin{tikzpicture}
	\begin{pgfonlayer}{nodelayer}
		\node [style=none] (11) at (-4, 4) {$T_{(1^6)}(\mathrm{SO}(6))$};
		\node [style=miniU] (18) at (-1.25, -2) {};
		\node [style=bluegauge] (20) at (-0.25, -2) {};
		\node [style=miniU] (21) at (0.75, -2) {};
		\node [style=none] (23) at (-5.75, -2.5) {4};
		\node [style=none] (24) at (-2.25, -2.5) {4};
		\node [style=none] (25) at (-1.25, -2.5) {4};
		\node [style=none] (26) at (-0.25, -2.5) {2};
		\node [style=none] (27) at (0.75, -2.5) {2};
		\node [style=none] (31) at (1, -0.25) {};
		\node [style=none] (32) at (3, -0.25) {};
		\node [style=none] (62) at (2, 0.25) {Gluing};
		\node [style=miniU] (65) at (-6.75, -2) {};
		\node [style=bluegauge] (66) at (-7.75, -2) {};
		\node [style=miniU] (67) at (-8.75, -2) {};
		\node [style=none] (68) at (-6.75, -2.5) {4};
		\node [style=none] (69) at (-7.75, -2.5) {2};
		\node [style=none] (70) at (-8.75, -2.5) {2};
		\node [style=miniU] (71) at (-4, 3) {};
		\node [style=bluegauge] (72) at (-4, 2) {};
		\node [style=bluegauge] (73) at (-4, 0) {};
		\node [style=miniU] (75) at (-4, 1) {};
		\node [style=flavorRed] (76) at (-4, -1) {};
		\node [style=miniBlue] (77) at (-2.25, -2) {};
		\node [style=miniBlue] (78) at (-5.75, -2) {};
		\node [style=flavorRed] (79) at (-4.75, -2) {};
		\node [style=flavorRed] (80) at (-3.25, -2) {};
		\node [style=none] (81) at (-4.75, -2.5) {6};
		\node [style=none] (82) at (-3.25, -2.5) {6};
		\node [style=none] (83) at (-4.75, 3) {2};
		\node [style=none] (84) at (-4.75, 2) {2};
		\node [style=none] (85) at (-4.75, 1) {4};
		\node [style=none] (86) at (-4.75, 0) {4};
		\node [style=none] (87) at (-4.75, -1) {6};
		\node [style=miniU] (89) at (10, -2) {};
		\node [style=bluegauge] (90) at (11, -2) {};
		\node [style=miniU] (91) at (12, -2) {};
		\node [style=none] (92) at (7, -2.5) {4};
		\node [style=none] (93) at (9, -2.5) {4};
		\node [style=none] (94) at (10, -2.5) {4};
		\node [style=none] (95) at (11, -2.5) {2};
		\node [style=none] (96) at (12, -2.5) {2};
		\node [style=miniU] (99) at (6, -2) {};
		\node [style=bluegauge] (100) at (5, -2) {};
		\node [style=miniU] (101) at (4, -2) {};
		\node [style=none] (102) at (6, -2.5) {4};
		\node [style=none] (103) at (5, -2.5) {2};
		\node [style=none] (104) at (4, -2.5) {2};
		\node [style=miniU] (105) at (8, 2) {};
		\node [style=bluegauge] (106) at (8, 1) {};
		\node [style=bluegauge] (107) at (8, -1) {};
		\node [style=miniU] (108) at (8, 0) {};
		\node [style=miniBlue] (110) at (9, -2) {};
		\node [style=miniBlue] (111) at (7, -2) {};
		\node [style=none] (116) at (7.25, 2) {2};
		\node [style=none] (117) at (7.25, 1) {2};
		\node [style=none] (118) at (7.25, 0) {4};
		\node [style=none] (119) at (7.25, -1) {4};
		\node [style=miniU] (120) at (8, -2) {};
		\node [style=none] (121) at (8, -2.5) {6};
		\node [style=none] (122) at (-1, -3.25) {$T_{(1^6)}(\mathrm{SO}(6))$};
		\node [style=none] (123) at (-7, -3.25) {$T_{(1^6)}(\mathrm{SO}(6))$};
	\end{pgfonlayer}
	\begin{pgfonlayer}{edgelayer}
		\draw (20) to (21);
		\draw [style=->] (31.center) to (32.center);
		\draw (18) to (20);
		\draw (67) to (66);
		\draw (66) to (65);
		\draw (71) to (72);
		\draw (72) to (75);
		\draw (75) to (73);
		\draw (73) to (76);
		\draw (65) to (78);
		\draw (78) to (79);
		\draw (80) to (77);
		\draw (77) to (18);
		\draw (90) to (91);
		\draw (89) to (90);
		\draw (101) to (100);
		\draw (100) to (99);
		\draw (105) to (106);
		\draw (106) to (108);
		\draw (108) to (107);
		\draw (99) to (111);
		\draw (110) to (89);
		\draw (107) to (120);
		\draw (120) to (110);
		\draw (120) to (111);
	\end{pgfonlayer}
\end{tikzpicture}
}
\end{equation}
Computations for untwisted $D_3$ fixtures are given in Table \ref{D3fixtures} and for $D_4$ fixtures in Tables \ref{D4fixture1} and \ref{D4fixture2}. 
The Coulomb branch Hilbert series of the magnetic quivers of the twisted $A$ and the untwisted $D$ fixtures are consistent with the expectations in \cite{Chacaltana:2012ch,DistlerD}. 
\subsection{D-type 6d \texorpdfstring{$\mathcal{N}=(1,0)$}{N=1} theories  }{\label{bouquets}}
We now turn to examples of orthosymplectic magnetic quivers for the Higgs branch of some $6$d $\mathcal{N}=(1,0)$ SCFTs. Specifically, we consider magnetic quivers associated to the $6$d $\mathcal{N}=(1,0)$ theories living on the world-volume of $m$ M5 branes on a $\mathbb{C}^2/D_n$ orbifold, with $n\geq4$ and $m\geq1$. 
The electric quiver for the low energy theory of this brane system is \cite{Hanany:1997gh,Brunner:1997gk,Brunner:1997gf,Intriligator:1997kq,Blum:1997mm,Intriligator:1997dh,Ferrara:1998vf,DelZotto:2014hpa}:
\begin{equation}
\scalebox{.8}{\raisebox{-.5\height}{
\begin{tikzpicture}
	\begin{pgfonlayer}{nodelayer}
		\node [style=bluegauge] (1) at (-5.75, 0) {};
		\node [style=redgauge] (2) at (-4.75, 0) {};
		\node [style=bluegauge] (3) at (-3.75, 0) {};
		\node [style=bluegauge] (8) at (-2.25, 0) {};
		\node [style=redgauge] (9) at (-1.25, 0) {};
		\node [style=bluegauge] (14) at (-0.25, 0) {};
		\node [style=none] (29) at (-1.25, -0.5) {$2n$};
		\node [style=none] (30) at (-2.25, -0.5) {$2n-8$};
		\node [style=none] (33) at (-0.25, -0.5) {$2n-8$};
		\node [style=none] (37) at (-5.75, -0.5) {$2n-8$};
		\node [style=none] (38) at (-4.75, -0.5) {$2n$};
		\node [style=none] (39) at (-3.75, -0.5) {$2n-8$};
		\node [style=flavorRed] (40) at (-5.75, 1.5) {};
		\node [style=flavorRed] (41) at (-0.25, 1.5) {};
		\node [style=none] (42) at (-5.75, 2) {$2n$};
		\node [style=none] (43) at (-0.25, 2) {$2n$};
		\node [style=none] (44) at (-0.25, -0.75) {};
		\node [style=none] (45) at (-5.75, -0.75) {};
		\node [style=none] (46) at (-3, -1.75) {$2m-1$};
		\node [style=none] (47) at (-3, 0) {$\cdots$};
		\node [style=none] (48) at (-3.425, 0) {};
		\node [style=none] (49) at (-2.575, 0) {};
	\end{pgfonlayer}
	\begin{pgfonlayer}{edgelayer}
		\draw (8) to (9);
		\draw (3) to (2);
		\draw (2) to (1);
		\draw (40) to (1);
		\draw (41) to (14);
		\draw (14) to (9);
		\draw [style=brace] (44.center) to (45.center);
		\draw (48.center) to (3);
		\draw (49.center) to (8);
	\end{pgfonlayer}
\end{tikzpicture}
 \,.
}}
\label{electric6d}
\end{equation}
Magnetic quivers for this brane system were addressed in depth in \cite{newMarcus}. If the $m$ M5 branes are away from the singularity, and do not pairwise coincide, the magnetic quiver is:
\begin{equation}
\scalebox{.8}{\raisebox{-.5\height}{
\begin{tikzpicture}
	\begin{pgfonlayer}{nodelayer}
		\node [style=redgauge] (0) at (-6.5, 0) {};
		\node [style=bluegauge] (1) at (-5.5, 0) {};
		\node [style=redgauge] (2) at (-4.5, 0) {};
		\node [style=bluegauge] (3) at (-3.5, 0) {};
		\node [style=bluegauge] (8) at (-2, 0) {};
		\node [style=redgauge] (9) at (-1, 0) {};
		\node [style=redgauge] (13) at (2.5, 0) {};
		\node [style=bluegauge] (14) at (3.5, 0) {};
		\node [style=redgauge] (15) at (4.5, 0) {};
		\node [style=bluegauge] (16) at (1.5, 0) {};
		\node [style=bluegauge] (17) at (0, 0) {};
		\node [style=bluegauge] (18) at (0.5, 1) {};
		\node [style=bluegauge] (19) at (-0.25, 1) {};
		\node [style=dotsize] (20) at (-1, 1) {};
		\node [style=dotsize] (21) at (-1.25, 1) {};
		\node [style=dotsize] (22) at (-0.75, 1) {};
		\node [style=bluegauge] (23) at (-1.75, 1) {};
		\node [style=bluegauge] (24) at (-2.5, 1) {};
		\node [style=none] (25) at (-2.5, 1.5) {2};
		\node [style=none] (26) at (-1.75, 1.5) {2};
		\node [style=none] (27) at (-0.25, 1.5) {2};
		\node [style=none] (28) at (0.5, 1.5) {2};
		\node [style=none] (29) at (-1, -0.5) {$2n$};
		\node [style=none] (30) at (-2, -0.5) {$2n-2$};
		\node [style=none] (31) at (0, -0.5) {$2n-2$};
		\node [style=none] (32) at (4.5, -0.5) {2};
		\node [style=none] (33) at (3.5, -0.5) {2};
		\node [style=none] (34) at (2.5, -0.5) {4};
		\node [style=none] (35) at (1.5, -0.5) {4};
		\node [style=none] (36) at (-6.5, -0.5) {2};
		\node [style=none] (37) at (-5.5, -0.5) {2};
		\node [style=none] (38) at (-4.5, -0.5) {4};
		\node [style=none] (39) at (-3.5, -0.5) {4};
		\node [style=none] (40) at (-2.5, 2) {};
		\node [style=none] (41) at (0.5, 2) {};
		\node [style=none] (42) at (-1, 2.5) {$m$};
		\node [style=none] (43) at (-2.75, 0) {$\cdots$};
		\node [style=none] (44) at (0.75, 0) {$\cdots$};
		\node [style=none] (45) at (-2.45, 0) {};
		\node [style=none] (46) at (-3.05, 0) {};
		\node [style=none] (47) at (0.45, 0) {};
		\node [style=none] (48) at (1.05, 0) {};
	\end{pgfonlayer}
	\begin{pgfonlayer}{edgelayer}
		\draw (24) to (9);
		\draw (23) to (9);
		\draw (19) to (9);
		\draw (9) to (18);
		\draw (9) to (17);
		\draw (8) to (9);
		\draw (3) to (2);
		\draw (2) to (1);
		\draw (0) to (1);
		\draw (13) to (14);
		\draw (14) to (15);
		\draw (16) to (13);
		\draw [style=brace] (40.center) to (41.center);
		\draw (3) to (46.center);
		\draw (45.center) to (8);
		\draw (47.center) to (17);
		\draw (48.center) to (16);
	\end{pgfonlayer}
\end{tikzpicture}
\,.
}} 
\label{fullbouquet}
\end{equation}
If the M5 branes become coincident, we can encode the subsets of coinciding M5 branes using partitions $[m_1,m_2,...,m_l]$, with $\sum_{i=1}^l m_i = m $. 
For a fixed partition the magnetic quiver takes the form:
\begin{equation}
\scalebox{.8}{\raisebox{-.5\height}{
\begin{tikzpicture}
	\begin{pgfonlayer}{nodelayer}
		\node [style=redgauge] (0) at (-6.5, 0) {};
		\node [style=bluegauge] (1) at (-5.5, 0) {};
		\node [style=redgauge] (2) at (-4.5, 0) {};
		\node [style=bluegauge] (3) at (-3.5, 0) {};
		\node [style=bluegauge] (8) at (-2, 0) {};
		\node [style=redgauge] (9) at (-1, 0) {};
		\node [style=redgauge] (13) at (2.5, 0) {};
		\node [style=bluegauge] (14) at (3.5, 0) {};
		\node [style=redgauge] (15) at (4.5, 0) {};
		\node [style=bluegauge] (16) at (1.5, 0) {};
		\node [style=bluegauge] (17) at (0, 0) {};
		\node [style=none] (29) at (-1, -0.5) {$2n$};
		\node [style=none] (30) at (-2, -0.5) {$2n-2$};
		\node [style=none] (31) at (0, -0.5) {$2n-2$};
		\node [style=none] (32) at (4.5, -0.5) {2};
		\node [style=none] (33) at (3.5, -0.5) {2};
		\node [style=none] (34) at (2.5, -0.5) {4};
		\node [style=none] (35) at (1.5, -0.5) {4};
		\node [style=none] (36) at (-6.5, -0.5) {2};
		\node [style=none] (37) at (-5.5, -0.5) {2};
		\node [style=none] (38) at (-4.5, -0.5) {4};
		\node [style=none] (39) at (-3.5, -0.5) {4};
		\node [style=bluegauge] (40) at (-2, 1) {};
		\node [style=none] (41) at (-2.7, 1) {$2m_1$};
		\node [style=none] (42) at (-1.9, 1.9) {$\Lambda^2$};
		\node [style=dotsize] (43) at (-1.5, 1) {};
		\node [style=dotsize] (44) at (-1, 1) {};
		\node [style=dotsize] (45) at (-0.5, 1) {};
		\node [style=bluegauge] (46) at (0, 1) {};
		\node [style=none] (47) at (0.7, 1) {$2m_l$};
		\node [style=none] (48) at (0.1, 1.9) {$\Lambda^2$};
		\node [style=none] (49) at (-2.75, 0) {$\cdots$};
		\node [style=none] (50) at (0.75, 0) {$\cdots$};
		\node [style=none] (51) at (-2.45, 0) {};
		\node [style=none] (52) at (-3.05, 0) {};
		\node [style=none] (53) at (0.45, 0) {};
		\node [style=none] (54) at (1.05, 0) {};
	\end{pgfonlayer}
	\begin{pgfonlayer}{edgelayer}
		\draw (9) to (17);
		\draw (8) to (9);
		\draw (3) to (2);
		\draw (2) to (1);
		\draw (0) to (1);
		\draw (13) to (14);
		\draw (14) to (15);
		\draw (16) to (13);
		\draw (40) to (9);
		\draw [in=135, out=45, loop] (40) to ();
		\draw [in=135, out=45, loop] (46) to ();
		\draw (46) to (9);
		\draw (52.center) to (3);
		\draw (51.center) to (8);
		\draw (53.center) to (17);
		\draw (54.center) to (16);
	\end{pgfonlayer}
\end{tikzpicture}
 \,.
}}
\label{fullanti}
\end{equation}
On the field theory side, this corresponds to taking gauge couplings to infinity \cite{newMarcus}. It is expected that the effect on the moduli space is that of a discrete quotient \cite{Hanany:2018vph,AntonBouquet,MarcusSymmetric,6dmagnetic,newMarcus}. For the choice of $H=\mathbb{Z}_2$ this can be explicitly checked using the summation over integer and integer-plus-half lattice in the monopole formula. 
The global symmetry for all members of \eqref{fullbouquet} and \eqref{fullanti} (with $n \geq 4 $) is 
\begin{equation}
    G_{\text{global}}=\mathfrak{so}(2n)\times \mathfrak{so}(2n) \,.
\end{equation}
The claim in \cite{MarcusSymmetric,newMarcus} is then the Coulomb branch moduli spaces $\mathcal{M}_{\mathcal{C}}(\ref{fullanti})$ and $\mathcal{M}_{\mathcal{C}}(\ref{fullbouquet})$ satisfy:
\begin{equation}
\mathcal{M}_{\mathcal{C}}(\ref{fullanti}) = \frac{\mathcal{M}_{\mathcal{C}}(\ref{fullbouquet})}{\prod_i S_{m_i}} \,.
\label{symmetriceqn}
\end{equation}
A necessary condition whether moduli spaces are related by discrete quotients is given by computing their respective Coulomb branch Hilbert series $\text{HS}_{\mathcal{C}}(\ref{fullanti})$ and $\text{HS}_{\mathcal{C}}(\ref{fullbouquet})$ and taking their ratio \cite{AntonBouquet}. In the limit $t\to 1$, the ratio should evaluate to the order of the symmetry groups $\prod_i |S_{m_i}|$:
\begin{equation}
    \frac{\text{HS}_{\mathcal{C}}(\ref{fullbouquet})}{\text{HS}_{\mathcal{C}}(\ref{fullanti})}\bigg \rvert_{t \rightarrow 1}  =  \prod_i \left| S_{m_i} \right| \,.
    \label{quotientHS}
\end{equation}
One need not worry about any divergences when we set $t \rightarrow 1$ as the term in the denominator $(1-t^2)^d$, where $d$ is the complex dimension of the Coulomb branch, cancels out. This is because discrete gauging does  not change the dimension of the Coulomb branch.

The results are tabulated in Appendix \ref{6dapp} for $n=2,3,4$. And for all cases that we can check, the relation \eqref{quotientHS} is satisfied. 

\section{Orthosymplectic quivers with SO(odd) gauge groups}
\label{sodd}
So far, only unframed unitary-orthosymplectic quivers containing $\mathrm{SO}(\mathrm{even})$ gauge groups have been discussed.  As $\mathrm{SO}(\mathrm{odd})$ gauge groups have trivial centers, the quivers containing them have $H=\mathrm{ker} \, \phi=\{1\}$ in \eqref{defGaugeGroup}. Henceforth, we do not expect to mod out an overall $\mathbb{Z}_2^{\textrm{diag}}$ such that the magnetic lattice should only contain integer magnetic charges. 

The dearth of physically motivated unframed orthosymplectic quivers with $\mathrm{SO}(\mathrm{odd})$ gauge groups in the literature makes it difficult to compare results. To our knowledge, the known examples are:
\begin{itemize}
    \item  Magnetic quivers of class $\mathcal{S}$ theories with twisted $A_{2r}$ and twisted $D_{r}$ punctures \cite{Benini:2010uu,TwistedA2n,twisted2n}.
    \item  Magnetic quivers for the Higgs branch of $6$d $\mathcal{N}=(1,0)$  $\mathrm{O}(2k+8)$ gauge theory with $N_f=2k$ at infinite coupling \cite{newMarcus}. 
    \item  Magnetic quivers for the Higgs branch of $5$d $\mathcal{N}={1}$ theories with 
    \begin{itemize}
        \item $\mathrm{O}(2k)$ gauge group and $N_f \leq 2k-3$
        \item $\mathrm{O}(2k+1)$ gauge group and $N_f\leq 2k-2$ 
    \end{itemize} at infinite gauge coupling \cite{5dweb}. 
\end{itemize}
Unfortunately, the monopole formula diverges for these examples. One reason for this is that most of the above cases contain  $ T_{}^{}(\mathrm{SO}(\mathrm{odd}))$ legs where the USp nodes are bad, see \cite{tsigmarho}, leading to the divergence in the Hilbert series. 

We must therefore look beyond these theories to construct quivers with $\mathrm{SO}(\mathrm{odd})$ gauge groups for which the monopole formula does not diverge in order to test our prescription. We start by considering legs that do not contain bad nodes. These legs can be found in \cite{RudolphSlodowy} indicated by $\Delta \neq 0$. One such leg is $(\mathrm{SO}(3))-(\mathrm{USp}(4))-[\mathrm{SO}(7)]$. We can choose to glue several identical legs together to obtain an unframed orthosymplectic quiver. It turns out that the minimal number of legs, to ensure a converging monopole formula, is four:
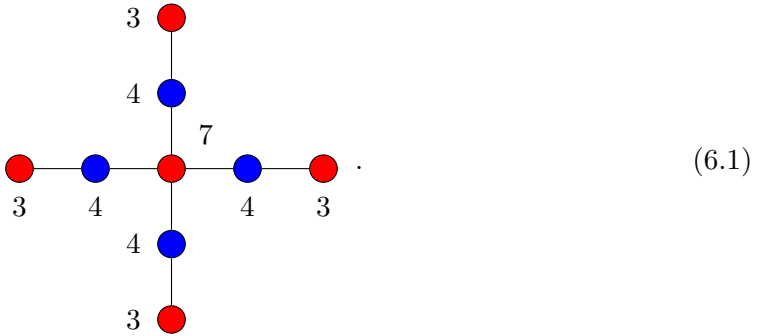
\begin{equation}
\raisebox{-.5\height}{
  \begin{tikzpicture}
	\begin{pgfonlayer}{nodelayer}
		\node [style=miniU] (0) at (0, 0) {};
		\node [style=miniBlue] (1) at (0, 1) {};
		\node [style=miniBlue] (2) at (1, 0) {};
		\node [style=miniBlue] (3) at (-1, 0) {};
		\node [style=miniBlue] (4) at (0, -1) {};
		\node [style=miniU] (5) at (0, 2) {};
		\node [style=miniU] (6) at (-2, 0) {};
		\node [style=miniU] (7) at (2, 0) {};
		\node [style=miniU] (8) at (0, -2) {};
		\node [style=none] (9) at (0.45, 0.45) {7};
		\node [style=none] (10) at (2, -0.5) {3};
		\node [style=none] (11) at (-0.5, -2) {3};
		\node [style=none] (12) at (-2, -0.5) {3};
		\node [style=none] (13) at (-0.5, 2) {3};
		\node [style=none] (14) at (-0.5, 1) {4};
		\node [style=none] (15) at (-1, -0.5) {4};
		\node [style=none] (16) at (1, -0.5) {4};
		\node [style=none] (17) at (-0.5, -1) {4};
	\end{pgfonlayer}
	\begin{pgfonlayer}{edgelayer}
		\draw (5) to (1);
		\draw (1) to (0);
		\draw (0) to (3);
		\draw (3) to (6);
		\draw (0) to (4);
		\draw (4) to (8);
		\draw (0) to (2);
		\draw (2) to (7);
	\end{pgfonlayer}
\end{tikzpicture}
} \,.
\label{soddguy}
\end{equation}
 An explicit computation of the Hilbert series using the monopole formula summing over \emph{only integer charges} gives:
\begin{eqnarray}
\begin{aligned}
    \mathrm{HS}_{\mathcal{C}}(\ref{soddguy}) = & {\scriptsize{\dfrac
    {(1-t^2)^4{~}{\cal P}_{276}}{(1-t^4)^2 (1-t^6)^6 (1-t^8)^9 (1-t^{10})^8 (1-t^{12})^4 (1-t^{14})^5} }} \\
    =  & 1 + 12 t^2 + 99 t^4 + 698 t^6 + 4398 t^8 + 25611 t^{10} + 138706 t^{12} +  700384 t^{14}\\ & + 3309803 t^{16}+ 14687048 t^{18} + 61402972 t^{20} + \mathcal{O}(t^{22})
\end{aligned}
\end{eqnarray}
which is consistent with an independent computation using slices and Hall-Littlewood methods. The palindromic numerator is detailed in Appendix \ref{apx:PP}. The computation for orthosymplectic quivers containing $\mathrm{SO}(\mathrm{odd})$ gauge groups is therefore consistent with our prescription with the magnetic lattice being just the integer lattice. 

\subsection{\texorpdfstring{$E_6$}{E6} quiver with \texorpdfstring{$\mathrm{SO}(\mathrm{odd})$}{SO(odd)} groups}
Using the isomorphism $\mathfrak{so}(5)\cong \mathfrak{usp}(4)$, we can use the $E_6$ quiver in Section \ref{5dexamples} to create another orthosymplectic quiver. The quiver takes the form:
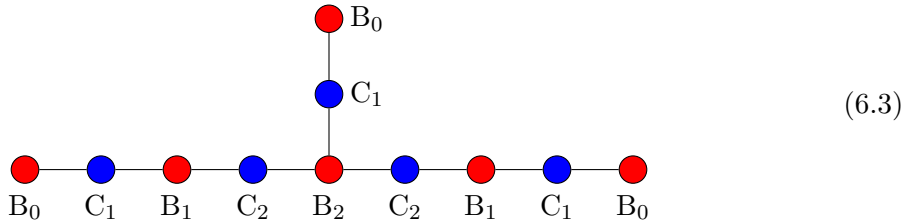
\begin{equation}
\raisebox{-.5\height}{
    \begin{tikzpicture}
	\begin{pgfonlayer}{nodelayer}
		\node [style=miniU] (0) at (0, 4) {};
		\node [style=miniBlue] (1) at (0, 3) {};
		\node [style=miniU] (2) at (0, 2) {};
		\node [style=miniU] (3) at (2, 2) {};
		\node [style=miniU] (4) at (4, 2) {};
		\node [style=miniU] (5) at (-2, 2) {};
		\node [style=miniU] (6) at (-4, 2) {};
		\node [style=miniBlue] (7) at (1, 2) {};
		\node [style=miniBlue] (8) at (3, 2) {};
		\node [style=miniBlue] (9) at (-1, 2) {};
		\node [style=miniBlue] (10) at (-3, 2) {};
		\node [style=none] (11) at (0.5, 4) {$\mathrm{B}_0$};
		\node [style=none] (12) at (0.5, 3) {$\mathrm{C}_1$};
		\node [style=none] (13) at (0, 1.5) {$\mathrm{B}_2$};
		\node [style=none] (14) at (1, 1.5) {$\mathrm{C}_2$};
		\node [style=none] (15) at (-1, 1.5) {$\mathrm{C}_2$};
		\node [style=none] (16) at (2, 1.5) {$\mathrm{B}_1$};
		\node [style=none] (17) at (3, 1.5) {$\mathrm{C}_1$};
		\node [style=none] (18) at (4, 1.5) {$\mathrm{B}_0$};
		\node [style=none] (19) at (-2, 1.5) {$\mathrm{B}_1$};
		\node [style=none] (20) at (-3, 1.5) {$\mathrm{C}_1$};
		\node [style=none] (21) at (-4, 1.5) {$\mathrm{B}_0$};
	\end{pgfonlayer}
	\begin{pgfonlayer}{edgelayer}
		\draw (0) to (1);
		\draw (1) to (2);
		\draw (2) to (7);
		\draw (7) to (3);
		\draw (3) to (8);
		\draw (8) to (4);
		\draw (2) to (9);
		\draw (9) to (5);
		\draw (5) to (10);
		\draw (10) to (6);
	\end{pgfonlayer}
\end{tikzpicture}
}
\label{E6odd}
\end{equation}
The Coulomb branch Hilbert series cannot be computed using the monopole formula as it diverges. However, one can compute the Coulomb branch using the Hall-Littlewood formula (see Appendix \ref{app:HL}) and the Hilbert series gives $\overline{\mathcal{O}}^{\mathfrak{e}_6}_{\text{min}} $. However, the Hall-Littlewood formula does not specify whether the groups are special orthogonal or orthogonal. Therefore, in \eqref{E6odd} we restrict the labeling of the (special) orthogonal nodes to just the algebras. 

We can generalize \eqref{E6odd} to the following familiy:
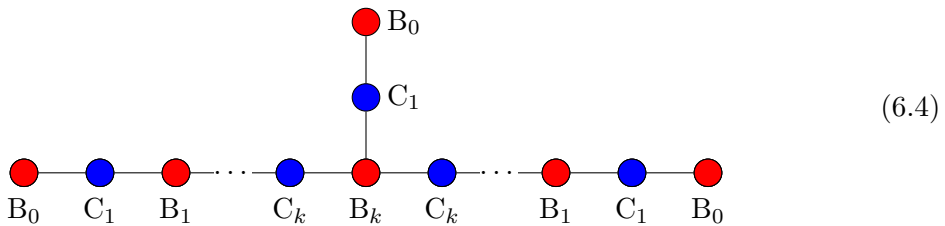
\begin{equation}
\raisebox{-.5\height}{
 \begin{tikzpicture}
	\begin{pgfonlayer}{nodelayer}
		\node [style=miniU] (0) at (0, 1) {};
		\node [style=miniBlue] (1) at (1, 1) {};
		\node [style=miniBlue] (2) at (-1, 1) {};
		\node [style=none] (3) at (1.75, 1) {\dots};
		\node [style=none] (4) at (-1.75, 1) {\dots};
		\node [style=miniU] (5) at (2.5, 1) {};
		\node [style=miniBlue] (10) at (0, 2) {};
		\node [style=miniBlue] (6) at (3.5, 1) {};
		\node [style=miniBlue] (7) at (-3.5, 1) {};
		\node [style=miniU] (8) at (-2.5, 1) {};
		\node [style=miniU] (9) at (-4.5, 1) {};
		\node [style=miniU] (11) at (4.5, 1) {};
		\node [style=none] (12) at (1.5, 1) {};
		\node [style=none] (13) at (2, 1) {};
		\node [style=none] (14) at (-1.5, 1) {};
		\node [style=none] (15) at (-2, 1) {};
		\node [style=none] (16) at (0.5, 2) {$\mathrm{C}_1$};
		\node [style=none] (17) at (0, 0.5) {$\mathrm{B}_k$};
		\node [style=none] (18) at (1, 0.5) {$\mathrm{C}_k$};
		\node [style=none] (19) at (-1, 0.5) {$\mathrm{C}_k$};
		\node [style=none] (20) at (-2.5, 0.5) {$\mathrm{B}_1$};
		\node [style=none] (21) at (-3.5, 0.5) {$\mathrm{C}_1$};
		\node [style=none] (22) at (-4.5, 0.5) {$\mathrm{B}_0$};
		\node [style=none] (23) at (2.5, 0.5) {$\mathrm{B}_1$};
		\node [style=none] (24) at (3.5, 0.5) {$\mathrm{C}_1$};
		\node [style=none] (25) at (4.5, 0.5) {$\mathrm{B}_0$};
		\node [style=miniBlue] (26) at (0, 1) {};
		\node [style=miniBlue] (27) at (2.5, 1) {};
		\node [style=miniBlue] (28) at (4.5, 1) {};
		\node [style=miniBlue] (29) at (-2.5, 1) {};
		\node [style=miniBlue] (30) at (-4.5, 1) {};
		\node [style=miniU] (32) at (-1, 1) {};
		\node [style=miniU] (33) at (1, 1) {};
		\node [style=miniU] (34) at (3.5, 1) {};
		\node [style=miniU] (35) at (-3.5, 1) {};
		\node [style=miniU] (36) at (0, 3) {};
		\node [style=none] (37) at (0.5, 3) {$\mathrm{B}_0$};
		\node [style=miniU] (38) at (0, 1) {};
		\node [style=miniBlue] (39) at (1, 1) {};
		\node [style=miniBlue] (40) at (-1, 1) {};
		\node [style=miniU] (41) at (2.5, 1) {};
		\node [style=miniU] (42) at (-2.5, 1) {};
		\node [style=miniBlue] (43) at (3.5, 1) {};
		\node [style=miniBlue] (44) at (-3.5, 1) {};
		\node [style=miniU] (45) at (4.5, 1) {};
		\node [style=miniU] (46) at (-4.5, 1) {};
	\end{pgfonlayer}
	\begin{pgfonlayer}{edgelayer}
		\draw (2) to (0);
		\draw (0) to (10);
		\draw (0) to (1);
		\draw (5) to (6);
		\draw (8) to (7);
		\draw (7) to (9);
		\draw (6) to (11);
		\draw (15.center) to (8);
		\draw (14.center) to (2);
		\draw (1) to (12.center);
		\draw (13.center) to (5);
		\draw (36) to (10);
	\end{pgfonlayer}
\end{tikzpicture}
}
\label{generalB}
\end{equation}
and the Coulomb branch of \eqref{generalB} is the same as the Coulomb branch of \eqref{E6non-simply-laced2} for $n=2k$. The HWG is therefore the same as \eqref{foldedE6HWG}. The result here is consistent with the expectation in \cite{Bourget:2020asf} that the generalized sequence for $C_5$ rank 1 SCFT can be described by both a non-simply laced unitary quiver \eqref{E6non-simply-laced2} and the magnetic quiver for twisted D-type fixture \eqref{generalB} for $n$ even. 

\section{Conclusions and outlook}
\label{sec:conclusion}
The main point of this paper is the precise identification of the magnetic lattice for a given quiver $\mathsf{Q}$, which depends on the choice of the gauge group $G_H$. Focusing on the case where $\mathrm{ker} \phi$ is discrete, the different choices of $G_H=G\slash H$ for any $H \trianglelefteq \mathrm{ker} \phi$ lead to different Coulomb branches associated to the pair $(\mathsf{Q},H)$. For $H=\mathrm{ker}\phi$, the Coulomb branch is denoted as $\mathcal{C}_{\mathrm{ker} \phi}$. For any choice $H$, one computes $N_H = \frac{\mathrm{ker}\phi}{ H} $.
By means of the Coulomb branch Hilbert series, the effects of ungauging different $H$ choices are identified as $N_H$ quotients of the respective moduli spaces, i.e.\ $\mathcal{C}_H = \mathcal{C}_{\mathrm{ker} \phi} \slash N_H$. Hence, a single quiver can give rise to various different moduli spaces, related as orbifolds.

For the exceptional sequences $E_n$, the orbifold relation is realised by the $\mathbb{Z}_2$ action on the spinors of the global symmetry. In other words, for the choice $H=\mathbb{Z}_2$ the moduli space $\mathcal{C}_{\mathbb{Z}_2}$ has the largest global symmetry, which contains the spinor representations. Transitioning to the $H=\{1\}$ choice, the moduli space global symmetry of  $\mathcal{C}_{\{1\}}$ is reduced 
due to the $\mathbb{Z}_2$ action, to a group without spinor representations. Computationally, the magnetic lattice for $H=\{1\}$ is the so-called integer lattice $\mathbb{Z}$, whereas the choice $H=\mathbb{Z}_2$ requires the integer lattice plus the interger-plus-half magnetic lattice $\mathbb{Z}+\frac{1}{2}$. The Hilbert series $\mathrm{HS}_{\mathbb{Z}}$ equals the  Molien sum of $\mathrm{HS}_{\mathbb{Z}} + \mathrm{HS}_{\mathbb{Z}+\frac{1}{2}} $ over the $\mathbb{Z}_2$ acting on the spinors of the global symmetry. 

In detail, the main point of this paper is demonstrated for unframed \emph{unitary-orthosymplectic quivers} arising as magnetic quivers of:
\begin{itemize}
    \item The exceptional $E_n$ sequences with choices $H = \mathbb{Z}_2^{\mathrm{diag}} $ and $\{1\}$ in Section \ref{5dexamples}.
    \item Some moduli spaces of $5$d $\mathcal{N}=1$ theories with choices $H = \mathbb{Z}_2^{\mathrm{diag}} $ and $\{1\}$ in Section \ref{general5dfamilysection}.
    \item For Higgs branches of $4$d $\mathcal{N}=2$ class $\mathcal{S}$ theories with $H = \mathbb{Z}_2^{\mathrm{diag}} $ in Section \ref{classSguys}.
    \item For moduli spaces of $6$d $\mathcal{N}=(1,0)$ theories with  $H = \mathbb{Z}_2^{\mathrm{diag}} $ in Section \ref{bouquets}.
\end{itemize}
Several non-trivial consistency checks are preformed to validate the results: in case of known Coulomb branch Hilbert series, the results of this paper are in agreement.\\

In the case of a simply laced, unframed unitary quiver with a single $\mathrm{SU}(N)$ node one has $\mathrm{ker} \phi=\mathbb{Z}_N^{\rm diag}$. This provides an opportunity to compute many discrete quotients, as one can take an unframed unitary quiver and replace any $\mathrm{U}(N)$ node by a $\mathrm{SU}(N)$ node and then pick different choices of $H\trianglelefteq\mathrm{Z}_{N}
^{\rm diag}$, producing discrete quotients of the Coulomb branch of the unitary quiver. In the cases where Hall-Littlewood methods can be applied, for instance in Section \ref{applications}, and indeed for many loop-free unframed quivers, the results agree with those from the $\mathbb{Z}_N^{\textrm{diag}}$ analysis.\\

Besides the realisation of the different orbifolds of moduli spaces, the second important conclusion of this paper is that a single moduli space, or a single symplectic singularity, can admit different magnetic quiver representations.  In other words, there are different magnetic quivers that describe the same moduli space. The examples considered here are particularly explicit for the minimal nilpotent orbit of $E_6$ for which the following \emph{four} magnetic quivers are known:
\begin{itemize}
\item A simply-laced unitary quiver \eqref{E6_simply_laced}.
\item A non-simply laced unitary quiver \eqref{E6non-simply-laced}.
\item An orthosymplectic quiver \eqref{quiverE6} with D-type nodes.
\item An orthosymplectic quiver \eqref{E6odd} with B-type nodes.
\end{itemize}
All of the quivers have different properties, as for instance unitary quivers allow for full refinement; nevertheless, they all describe the same moduli space.
In view of the analysis of hyper-K\"ahler spaces via Hasse diagrams, it is an imperative task to systematise all possible magnetic quiver representations of a given moduli space.\\

In light of the results presented, future directions include the following:

\paragraph{Orbifolds.}
For a given quiver $\mathsf{Q}$, the different choices of $H$ lead to a multitude of possible orbifold moduli spaces to be explored. In particular, one can construct orbifolds of spaces with exceptional symmetries or any other space of interest.

\paragraph{O vs SO.} The question of orthogonal versus special orthogonal gauge groups on the Coulomb branch of flavored orthosymplectic quivers is first tackled in \cite{tsigmarho} and expanded upon in \cite{CabreraZhong,Antoinefirstquiverpaper}. For unframed quivers, it is possible that the extra $\mathbb{Z}_2$ factor in orthogonal gauge groups interferes with decoupling of an overall $\mathbb{Z}_2^{\textrm{diag}}$ resulting in a non-trivial effect on the magnetic lattice. A more detailed study of the magnetic lattice of orthogonal gauge groups is desirable. This would turn out useful for exploring quivers with a bouquet of $\mathrm{O}(2)$ nodes.

\paragraph{Non-simply laced orthosymplectic quivers.}
So far we only discussed orthosymplectic quivers that are simply laced. The possibility of constructing non-simply laced orthosymplectic quivers is interesting because these might lead to new moduli spaces or new realizations of known moduli spaces. 

\paragraph{Acknowledgement.}
We are grateful to Ofer Aharony, Mohammad Akhond, Santiago Cabrera, Jacques Distler, Matt Glis, Noppadol Mekareeya, Dominik Miketa, Sakura Sch\"afer-Nameki, Gabi Zafrir, and Anton Zajac for helpful discussions.
The work of A.B., J.F.G., A.H., R.K. and Z.Z. is supported by STFC grant ST/P000762/1. The work of M.S. is supported by the National Thousand-Young-Talents Program of China, the National Natural Science Foundation of China (grant no.\ 11950410497), and the China Postdoctoral Science Foundation (grant no.\ 2019M650616). We thank the Simons Center for Geometry and Physics, Stony Brook University for the hospitality and the partial support during an early stage of this work at the Simons Summer workshop 2019. We thank the 11th Joburg Workshop on String Theory from 8-13 December 2019 for hospitality, and the MIT-Imperial College London Seed Fund for support.
\appendix
\section{Global symmetry -- group versus algebra}
\label{originalGF}
\begin{figure}[!t]
    \centering
    \begin{tabular}{|c|c|}
    \hline 
        Group  & Representations \\ \hline 
        $\mathrm{Spin}(4n)$  & adjoint, vector, spinor, conjugate spinor  \\  
        $\mathrm{Spin}(4n)/\mathbb{Z}_2^V = \mathrm{SO}(4n)$  & adjoint, vector  \\  
        $\mathrm{Spin}(4n)/\mathbb{Z}_2^S = \mathrm{Ss}(4n)$  & adjoint, spinor \\  
        $\mathrm{Spin}(4n)/\mathbb{Z}_2^C = \mathrm{Sc}(4n)$  & adjoint, conjugate spinor  \\  
        $\mathrm{Spin}(4n)/\mathbb{Z}_2^2 = \mathrm{PSO}(4n)$  & adjoint \\  \hline 
        $\mathrm{Spin}(4n+2)$  & adjoint, vector, spinor, conjugate spinor \\  
        $\mathrm{Spin}(4n+2)/\mathbb{Z}_2 = \mathrm{SO}(4n+2)$  &  adjoint, vector \\  
        $\mathrm{Spin}(4n+2)/\mathbb{Z}_4 = \mathrm{PSO}(4n+2)$  & adjoint \\  \hline 
    \end{tabular}
    \caption[$D_N$ representations of some groups]{$D_N$ representations of some groups. $\mathrm{Ss}(4n)$ and $\mathrm{Sc}(4n)$  are the semispin groups. We have the particular cases $\mathrm{Ss}(4)=\mathrm{SU}(2) \times \mathrm{SO}(3)$ and $\mathrm{Ss}(8)\simeq \mathrm{SO}(8)$ by triality. $\mathrm{PSO}(16)\subset \mathrm{Ss}(16)\subset E_8$ (while $\mathrm{SO}(16)\not\subset E_8$)}
    \label{tableGroupsD}
\end{figure}

The global symmetry group, as opposed to the global symmetry algebra, can be determined using the expression of the highest weight generating function (HWG). 

Let $\mathfrak{g}$ be the  
Lie algebra of the global symmetry group. We denote by $\tilde{G}$ the compact simply connected group with Lie algebra $\mathfrak{g}$, and let $Z(\tilde{G})$ be its center, which is a finite Abelian group. Then the other compact connected groups with Lie algebra $\mathfrak{g}$ are the groups $\tilde{G} / H$ where $H$ is a subgroup of $Z(\tilde{G})$. 

To a given highest weight representation $\phi$ of $\mathfrak{g}$ one can associate its congruency class $c(\phi) \in Z(\tilde{G})$. The congruency classes are given, for instance, in \cite{lemire1980congruence,slansky1981group}.
Given the HWG, we first compute the subgroup $N_H$ of $Z(\tilde{G})$ generated by the $c(\phi_i)$, where the $\phi_i$ are the representations appearing in the HWG. $N_H$ is a normal subgroup of $Z(\tilde{G})$ since $Z(\tilde{G})$ is Abelian. Then we set $H = Z(\tilde{G})/N_H$, and the global symmetry group of the corresponding moduli space is $\tilde{G} / H$. In this paper, the global symmetry often involves type $D_N$ where the center has order 4. The corresponding groups $\tilde{G} / H$ are listed in Figure \ref{tableGroupsD}.

\section{Class $\mathcal{S}$ magnetic quivers}\label{classS}
Here we study the magnetic quivers associated with $A_{\mathrm{odd}}$ and $D$ type class $\mathcal{S}$ theories in more detail. We discuss how the magnetic quiver is constructed by gluing together $T_{\rho}^{}(J)$ theories and tabulate the Coulomb branch Hilbert series. The specific class $\mathcal{S}$ fixtures are a subset of those tabulated in \cite{Chacaltana:2012ch, DistlerD} where the monopole formula converges.
\subsection{Punctures and Slodowy slices}\label{punctures}
\begin{figure}[tp]
\centering
\begin{displaymath}
    \xymatrix{
  & 
  T_{\bv \sigma}^{} (J)  \ar[dl]|{\mathop {Higgs}\limits_{(O)}} \ar@{.>}[dr]|{\mathop {Coulomb}\limits_{(SO)} } 
  & \\
 {\orbit{\sigma}}  
 & 
 \text{\scriptsize \it Special~Duality} \ar@{-->}[l] \ar@{-->}[r] 
 &  
 {{{{\cal S}}}_{{\cal N},\bv{\sigma}}}  \\
   &   
   T_{}^{\bv {\sigma}} (J^\vee)  \ar@{.>}[ul]|{\mathop {Coulomb}\limits_{(O/SO)} } \ar[ur]|{\mathop {Higgs}\limits_{(O)} } 
   & }
\end{displaymath}
\caption[BV Duality of Special $BCD$ Orbits]{BV Duality of Special $BCD$ Orbits. The closure $\orbit{\sigma}$ of a \emph{special} nilpotent orbit of $J$ is dual under the Barbasch Vogan map to a Slodowy slice $\slice{\bv \sigma}$ of the GNO dual group $J^\vee$. $T_{\bv \sigma}(J)$ and $T_{}^{\bv {\sigma}}(J^\vee)$ are related by $3$d-mirror duality. For a fuller explanation see \cite{RudolphSlodowy}.}
\label{fig:TSRBV}
\end{figure}
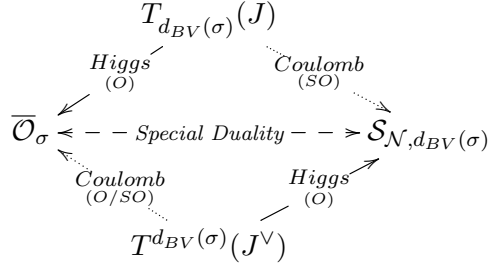

 In this section we consider fixtures of type $A$ or $D$. The punctures are labeled by a class of integer partitions. More precisely, the magnetic quivers for these theories are constructed by identifying each puncture with a linear quiver $T_{\rho}(J)$ where $J$ is the algebra of the flavor node \cite{BeniniSicilian} and $\rho$ is a partition of the GNO dual group $J^\vee$ \cite{GaiottoWitten}. Upon identifying such linear quivers, one can glue them together by gauging the central $J$ node, to form a star-shaped quiver. 
There are important relationships between $T_{\rho}^{}(J)$ theories, nilpotent orbits, and Slodowy slices, which draw on the Barbasch Vogan (BV) map \cite{barbasch_vogan_1985}, as summarized in Figure \ref{fig:TSRBV}. Here, a partition $\rho$ is termed \emph{special} if the BV map acts on it as an involution, $\bv {\bv \rho}=\rho$.

Now, the quivers $T_{\bv{\sigma}}(J)$ have the property that their Higgs branches are closures of nilpotent orbits $\orbit \sigma$ in the Lie algebra of $J$. On the other hand, their Coulomb branches have interpretations as the intersections ${{{{\cal S}}}_{{\cal N},\rho}}$ of the nilpotent cone ${\cal N}$ with Slodowy slices ${{{{\cal S}}}_{\rho}}$ to the orbits $\orbit \rho$ of the GNO dual group $J^\vee$, where $\rho = \bv{\sigma}$ \cite{Benini:2010uu}. The status of whether the monopole formula for these slices  diverges is given in \cite{RudolphSlodowy}.\footnote{We follow the convention where the partitions $\rho$ correspond to the Nahm poles or nilpotent orbits of $J^\vee$, as in \cite{DistlerA,Chacaltana:2011ze,tsigmarho,twistedGaiotto}.}

While identifying the $T_{\rho}(J)$ quiver, whose Coulomb branch is a particular slice, is straightforward when $J = \mathrm{SU}(r+1)$, there are many complications for orthosymplectic types. First, the partitions are subject to selection rules. Second, while there is a direct map from $B/C/D$ partitions to quivers whose $3$d Higgs branches are nilpotent orbits, the map to orthosymplectic quivers whose $3$d Coulomb branches are slices involves node shifting and is incomplete \cite{RudolphSlodowy}. Consequently, not all orthosymplectic slices can be represented by $T_{\rho}(J)$ quivers.

In order to construct the magnetic quiver of a class $\mathcal{S}$ theory, we first need to specify the linear quivers corresponding to the punctures, which we label by the group $J$. Figures \ref{A3puncture} and \ref{A4puncture} give quivers for twisted $A_3$ and $A_5$ punctures, which have $J=\mathrm{USp}(4)$ and $J=\mathrm{USp}(6)$ flavor nodes. Figures \ref{D3puncture} and \ref{D4puncture} give quivers for $D_3$ and $D_4$ punctures, respectively, which have $J=\mathrm{SO}(6)$ and $J=\mathrm{SO}(8)$ flavor nodes. In all cases, we provide the data for the corresponding orbits and slices, along with the global symmetry of the Coulomb branch. Such tables can in principle be compiled for any classical group $J$ using methods outlined in \cite{RudolphSlodowy}. The global symmetry of the entire magnetic quiver is typically the product of the linear quiver (i.e.\ GNO dual slice) global symmetry groups, but in notable cases this symmetry is enhanced.

The identification of punctures as Slodowy slices allow natural computations of exact Hilbert series for the Coulomb branches of star-shaped (unitary)-orthosymplectic quivers, using  Hall-Littlewood related functions, as described in Appendix \ref{app:HL}. This method has been used throughout the paper, where star-shaped magnetic quivers without loops\footnote{Here, loops refer to quivers containing hypermultiplets transforming in the adjoint, symmetric, or antisymmetric representations.} appear. Such independent calculations confirm the results obtained by $\mathbb{Z}_2^{\textrm{diag}}$ ungauging.


\begin{figure}[t]
\small
\begin{adjustbox}{center}
	\begin{tabular}{|c|c|c|c|c|}
		\hline
		  	\begin{tabular}{c} Orbit \\ $\sigma$ \end{tabular}
		  	& \begin{tabular}{c} Dual Orbit\\ $\bv {\sigma}$ \end{tabular}
		  	&\begin{tabular}{c} Dual Slice\\ Dimension \end{tabular}&
		  	\begin{tabular}{c} Dual Slice \\ Symmetry \end{tabular}  & 
			Quiver 
			\\ \hline
$(1^4)$ & $(5)$ & 0 &
 $\emptyset$ & 
Trivial \\ \hline
$(2^2) $	& $(3,1^2)$ & 1 &
$\mathrm{SO}(2)$ &
\begin{tabular}{c}
\begin{tikzpicture}
	\begin{pgfonlayer}{nodelayer}
		\node [style=flavorBlue] (0) at (0, 0) {};
		\node [style=redgauge] (1) at (-1, 0) {};
		\node [style=none] (6) at (-1, -0.5) {2};
		\node [style=none] (7) at (0, -0.5) {4};
				\node [style=none] (5) at (0, 0.8) {};
	\end{pgfonlayer}
	\begin{pgfonlayer}{edgelayer}
		\draw (1) to (0);
	\end{pgfonlayer}
\end{tikzpicture}
\end{tabular}
\\  
\hline	
$(4)$ &$(1^5)$ & 4 &
$\mathrm{SO}(5)$ &
\begin{tabular}{l}
\begin{tikzpicture}
	\begin{pgfonlayer}{nodelayer}
		\node [style=flavorBlue] (0) at (0, 0) {};
		\node [style=redgauge] (1) at (-1, 0) {};
		\node [style=bluegauge] (2) at (-2, 0) {};
		\node [style=redgauge] (3) at (-3, 0) {};
		\node [style=none] (4) at (-3, -0.5) {2};
		\node [style=none] (5) at (-2, -0.5) {2};
		\node [style=none] (6) at (-1, -0.5) {4};
				\node [style=none] (5) at (0, 0.8) {};
		\node [style=none] (7) at (0, -0.5) {4};
	\end{pgfonlayer}
	\begin{pgfonlayer}{edgelayer}
		\draw (3) to (2);
		\draw (2) to (1);
		\draw (1) to (0);
	\end{pgfonlayer}
\end{tikzpicture}
\end{tabular}
\\ \hline	
	\end{tabular}
\end{adjustbox}
\caption[$T(\mathrm{USp}(4))$ linear quivers]{$T(\mathrm{USp}(4))$ linear quivers. These are used in twisted $A_3$ fixtures. The quivers $T_{\bv {\sigma}}(\mathrm{USp}(4))$ have non-diverging monopole formula. The partitions identify special orbits of $\mathrm{USp}(4)$ and its GNO dual $\mathrm{SO}(5)$.}
\label{A3puncture}
\end{figure}


 \begin{figure}[!ht]
 \small
 \begin{adjustbox}{center}
 	\begin{tabular}{|c|c|c|c|c|}
 		\hline
 		  	\begin{tabular}{c} Orbit \\ $\sigma$ \end{tabular}
		  	& \begin{tabular}{c} Dual Orbit\\ $\bv {\sigma}$ \end{tabular}
		  	&\begin{tabular}{c} Dual Slice\\ Dimension \end{tabular}&
 		  	\begin{tabular}{c} Dual Slice \\ Symmetry \end{tabular}  & 
 			Quiver 
 			\\ \hline
 $(1^6)$ & (7) & 0 &
  $\emptyset$ &
 Trivial \\ \hline
 \begin{tabular}{l} $(2^2,1^2)$ \end{tabular} & $(5,1^2)$ & 1 &
 $\mathrm{SO}(2)$&
 \begin{tabular}{l}\begin{tikzpicture}
 	\begin{pgfonlayer}{nodelayer}
 		\node [style=flavorBlue] (0) at (0, 0) {};
 		\node [style=redgauge] (1) at (-1, 0) {};
 		\node [style=none] (6) at (-1, -0.5) {2};
 		\node [style=none] (7) at (0, -0.5) {6};
 						\node [style=none] (5) at (0, 0.8) {};
 	\end{pgfonlayer}
 	\begin{pgfonlayer}{edgelayer}
 		\draw (1) to (0);
 	\end{pgfonlayer}
 \end{tikzpicture}
 \end{tabular} \\ \hline
 $(2^3) $	& $(3^2,1)$ & 2 &
 $\mathrm{SO}(2)$&
 \begin{tabular}{l}
 \begin{tikzpicture}
 	\begin{pgfonlayer}{nodelayer}
 		\node [style=flavorBlue] (0) at (0, 0) {};
 		\node [style=redgauge] (1) at (-1, 0) {};
 		\node [style=none] (6) at (-1, -0.5) {4};
 		\node [style=none] (7) at (0, -0.5) {6};
 		\node [style=none] (5) at (0, 0.8) {};
 	\end{pgfonlayer}
 	\begin{pgfonlayer}{edgelayer}
 		\draw (1) to (0);
 	\end{pgfonlayer}
 \end{tikzpicture}
 \end{tabular}
 \\  \hline		
 $(3,3)$ & $(3,2^2)$ & 3 &
  $\mathrm{USp}(2)$ &
 Bad quiver \\ \hline
 $(4,2)$ & $(3,1^4)$ & 4 &
 $\mathrm{SO}(4)$&
 \begin{tabular}{l}
 \begin{tikzpicture}
 	\begin{pgfonlayer}{nodelayer}
 		\node [style=flavorBlue] (0) at (0, 0) {};
 		\node [style=redgauge] (1) at (-1, 0) {};
 		\node [style=none] (6) at (-1, -0.5) {4};
 		\node [style=none] (7) at (0, -0.5) {6};
 		\node [style=bluegauge] (8) at (-2, 0) {};
 		\node [style=redgauge] (9) at (-3, 0) {};
 		\node [style=none] (10) at (-2, -0.5) {2};
 		\node [style=none] (11) at (-3, -0.5) {2};
 		\node [style=none] (5) at (0, 0.8) {};
 	\end{pgfonlayer}
 	\begin{pgfonlayer}{edgelayer}
 		\draw (1) to (0);
 		\draw (9) to (8);
 		\draw (8) to (1);
 	\end{pgfonlayer}
 \end{tikzpicture}
 \end{tabular}
 \\ \hline
 $(6)$ & $(1^7)$ & 9 &
 $\mathrm{SO}(7)$&
 \begin{tabular}{l}
 \begin{tikzpicture}
 	\begin{pgfonlayer}{nodelayer}
 		\node [style=flavorBlue] (0) at (0, 0) {};
 		\node [style=redgauge] (1) at (-1, 0) {};
 		\node [style=none] (6) at (-1, -0.5) {6};
 		\node [style=none] (7) at (0, -0.5) {6};
 		\node [style=bluegauge] (8) at (-2, 0) {};
 		\node [style=redgauge] (9) at (-3, 0) {};
 		\node [style=none] (10) at (-2, -0.5) {4};
 		\node [style=none] (11) at (-3, -0.5) {4};
 		\node [style=bluegauge] (12) at (-4, 0) {};
 		\node [style=redgauge] (13) at (-5, 0) {};
 		\node [style=none] (14) at (-4, -0.5) {2};
 		\node [style=none] (15) at (-5, -0.5) {2};
 		\node [style=none] (5) at (0, 0.8) {};
 	\end{pgfonlayer}
 	\begin{pgfonlayer}{edgelayer}
 		\draw (1) to (0);
 		\draw (9) to (8);
 		\draw (8) to (1);
		\draw (13) to (12);
 		\draw (12) to (9);
 	\end{pgfonlayer}
 \end{tikzpicture}
 \end{tabular}
 \\ \hline	
 	\end{tabular}
 \end{adjustbox}
 \caption[$T(\mathrm{USp}(6))$ linear quivers]{$T(\mathrm{USp}(6))$ linear quivers.  These are used in twisted $A_5$ fixtures. The quivers $T_{\bv {\sigma}}(\mathrm{USp}(6))$  have non-diverging monopole formula; the bad quiver with zero conformal dimension is not shown. The partitions identify special orbits of $\mathrm{USp}(6)$ and its GNO dual $\mathrm{SO}(7)$.}
 \label{A4puncture}
 \end{figure}


\begin{figure}[!ht]
\small
\begin{adjustbox}{center}
		\begin{tabular}{|c|c|c|c|c|c|}
		\hline
		  	\begin{tabular}{c} Orbit \\ $\sigma$ \end{tabular}
		  	&\begin{tabular}{c} Dual Orbit\\
		  	$\bv {\sigma}$
		  	\end{tabular}
		  	&\begin{tabular}{c} Dual Slice\\ Dimension \end{tabular} &	
		  	\begin{tabular}{c} Dual Slice \\ Symmetry \end{tabular}  &
			Quiver 
			\\ 
		  	\hline 
$(1^4)$   &$(4)$   & 0 & $\emptyset$ &  Trivial
\\ \hline	
$(2,1^2)$   & $(3,1)$   & 1 & $\mathrm{U}(1)$ & 
\begin{tabular}{l}
\scalebox{1}{
\begin{tikzpicture}
	\begin{pgfonlayer}{nodelayer}
		\node [style=flavor0] (0) at (0, 0) {};
		\node [style=gauge3] (1) at (-1, 0) {};
		\node [style=none] (2) at (-1, -0.5) {1};
		\node [style=none] (3) at (0, -0.5) {4};
				\node [style=none] (5) at (0, 0.8) {};
	\end{pgfonlayer}
	\begin{pgfonlayer}{edgelayer}
		\draw (1) to (0);
	\end{pgfonlayer}
\end{tikzpicture}}
\end{tabular}
\\ \hline	
$(2^2)$   & $(2^2)$   & 2 & $\mathrm{SU}(2)$ & 
\begin{tabular}{l}
\scalebox{1}{
\begin{tikzpicture}
	\begin{pgfonlayer}{nodelayer}
		\node [style=flavor0] (0) at (0, 0) {};
		\node [style=gauge3] (1) at (-1, 0) {};
		\node [style=none] (2) at (-1, -0.5) {2};
		\node [style=none] (3) at (0, -0.5) {4};
				\node [style=none] (5) at (0, 0.8) {};
	\end{pgfonlayer}
	\begin{pgfonlayer}{edgelayer}
		\draw (1) to (0);
	\end{pgfonlayer}
\end{tikzpicture}}
\end{tabular}
\\ \hline	
$(3,1)$   &$(2,1^2)$   & 3 & $\mathrm{SU}(2) \times \mathrm{U}(1)$ & 
\begin{tabular}{l}
\scalebox{1}{
\begin{tikzpicture}
	\begin{pgfonlayer}{nodelayer}
		\node [style=flavor0] (0) at (0, 0) {};
		\node [style=gauge3] (1) at (-1, 0) {};
		\node [style=gauge3] (6) at (-2, 0) {};
		\node [style=none] (4) at (-2, -0.5) {1};
		\node [style=none] (2) at (-1, -0.5) {2};
		\node [style=none] (3) at (0, -0.5) {4};
				\node [style=none] (5) at (0, 0.8) {};
	\end{pgfonlayer}
	\begin{pgfonlayer}{edgelayer}
		\draw (1) to (0);
		\draw (1) to (6);
	\end{pgfonlayer}
\end{tikzpicture}}
\end{tabular}
\\ \hline	
$(4)$   &$(1^4)$   & 6 & $\mathrm{SU}(4)$ & 
\begin{tabular}{l}
\scalebox{1}{
\begin{tikzpicture}
	\begin{pgfonlayer}{nodelayer}
		\node [style=flavor0] (0) at (0, 0) {};
		\node [style=gauge3] (1) at (-1, 0) {};
		\node [style=gauge3] (2) at (-2, 0) {};
		\node [style=gauge3] (3) at (-3, 0) {};
		\node [style=none]  at (-3, -0.5) {1};
		\node [style=none]  at (-2, -0.5) {2};
		\node [style=none]  at (-1, -0.5) {3};
		\node [style=none]  at (0, -0.5) {4};
				\node [style=none] (5) at (0, 0.8) {};
	\end{pgfonlayer}
	\begin{pgfonlayer}{edgelayer}
		\draw (1) to (0);
		\draw (1) to (2);
		\draw (2) to (3);
	\end{pgfonlayer}
\end{tikzpicture}}
\end{tabular}
\\ \hline	
	\end{tabular}
\end{adjustbox}
\caption[$T(\mathrm{SU}(4))$ linear quivers]{$T(\mathrm{SU}(4))$ linear quivers. These quivers correspond to untwisted $A_3$ punctures. All the quivers $T_{\bv {\sigma}}(\mathrm{SU}(4))=T_{ {\sigma}^T}(\mathrm{SU}(4))$  have non-diverging monopole formula;  }
\label{A3punctureUntwisted}
\end{figure}


\begin{figure}[!ht]
\small
\begin{adjustbox}{center}
		\begin{tabular}{|c|c|c|c|c|c|}
		\hline
		  	\begin{tabular}{c} Orbit \\ $\sigma$ \end{tabular}
		  	& \begin{tabular}{c} Dual Orbit\\ $\bv {\sigma}$ \end{tabular}
		  	&\begin{tabular}{c} Dual Slice\\ Dimension \end{tabular} &	
		  	\begin{tabular}{c} Dual Slice \\ Symmetry \end{tabular}  &
			Quiver 
			\\ 
		  	\hline
$(1^6)$ & $(5,1)$ & 0 & 
$\emptyset$ &
Trivial \\ \hline
$(2^2,1^2)$ &$(3^2)$ & 1 & 
$\mathrm{SO}(2)$&
\begin{tabular}{l}
\scalebox{1}{
\begin{tikzpicture}
	\begin{pgfonlayer}{nodelayer}
		\node [style=flavorRed] (0) at (0, 0) {};
		\node [style=bluegauge] (1) at (-1, 0) {};
		\node [style=none] (2) at (-1, -0.5) {2};
		\node [style=none] (3) at (0, -0.5) {6};
				\node [style=none] (5) at (0, 0.8) {};
	\end{pgfonlayer}
	\begin{pgfonlayer}{edgelayer}
		\draw (1) to (0);
	\end{pgfonlayer}
\end{tikzpicture}}
\end{tabular}
 \\ \hline
$(3,1^3) $	& $(3,1^3) $ & 2 &
$\mathrm{SO}(3)$&
\begin{tabular}{l}
\begin{tikzpicture}
	\begin{pgfonlayer}{nodelayer}
		\node [style=flavorRed] (0) at (0, 0) {};
		\node [style=bluegauge] (1) at (-1, 0) {};
		\node [style=none] (2) at (-1, -0.5) {2};
		\node [style=none] (3) at (0, -0.5) {6};
		\node [style=redgauge] (4) at (-2, 0) {};
		\node [style=none] (5) at (-2, -0.5) {2};
				\node [style=none] (5) at (0, 0.8) {};
	\end{pgfonlayer}
	\begin{pgfonlayer}{edgelayer}
		\draw (1) to (0);
		\draw (4) to (1);
	\end{pgfonlayer}
\end{tikzpicture}
\end{tabular}
\\  \hline		
$(3^2) $	& $(2^2,1^2)$ & 3  &
$\mathrm{USp}(2) \times \mathrm{SO}(2)$&
Bad quiver
\\  \hline		
$(5,1)$ & $(1^6)$ & 6 & 
$\mathrm{SO}(6)$&
\begin{tabular}{l}
\scalebox{1}{
\begin{tikzpicture}
	\begin{pgfonlayer}{nodelayer}
		\node [style=flavorRed] (8) at (1, 0) {};
		\node [style=bluegauge] (0) at (0, 0) {};
		\node [style=redgauge] (1) at (-1, 0) {};
		\node [style=bluegauge] (2) at (-2, 0) {};
		\node [style=redgauge] (3) at (-3, 0) {};
		\node [style=none] (4) at (-3, -0.5) {2};
		\node [style=none] (5) at (-2, -0.5) {2};
		\node [style=none] (6) at (-1, -0.5) {4};
		\node [style=none] (9) at (1, -0.5) {6};
		\node [style=none] (7) at (0, -0.5) {4};
		\node [style=none] (5) at (0, 0.8) {};
	\end{pgfonlayer}
	\begin{pgfonlayer}{edgelayer}
		\draw (3) to (2);
		\draw (2) to (1);
		\draw (1) to (0);
			\draw (8) to (3);
	\end{pgfonlayer}
\end{tikzpicture}}
\end{tabular}
\\ \hline	
	\end{tabular}
\end{adjustbox}
\caption[$T(\mathrm{SO}(6))$ linear quivers]{$T(\mathrm{SO}(6))$ linear quivers. These are used in untwisted $D_3$ fixtures. The quivers $T_{\bv {\sigma}}(\mathrm{SO}(6))$  have non-diverging monopole formula; the bad quiver with zero conformal dimension is not shown. The partitions identify special orbits of $\mathrm{SO}(6)$.}
\label{D3puncture}
\end{figure}

\begin{figure}[!ht]
\small	
\begin{adjustbox}{center}
		\begin{tabular}{|c|c|c|c|c|}
\hline
		  	\begin{tabular}{c} Orbit \\ $\sigma$ \end{tabular}
		  	& \begin{tabular}{c} Dual Orbit\\ $\bv {\sigma}$ \end{tabular}
		  	&\begin{tabular}{c} Dual Slice\\ Dimension \end{tabular}&
		  	\begin{tabular}{c} Dual Slice \\ Symmetry \end{tabular}  &
		 	Quiver 
		 	\\ 
\hline
$(1^8) $	& $(7,1)$ & 0 &
$\emptyset$ &
Trivial \\ 
\hline
$(2^2,1^4)$ & $(5,3)$ & 1 &
$\mathrm{O}(1)$&
\begin{tabular}{l}
\scalebox{0.8}{
\begin{tikzpicture}
	\begin{pgfonlayer}{nodelayer}
		\node [style=flavorRed] (0) at (0, 0) {};
		\node [style=bluegauge] (1) at (-1, 0) {};
		\node [style=none] (2) at (-1, -0.5) {2};
		\node [style=none] (3) at (0, -0.5) {8};
		\node [style=none] (5) at (0, 0.8) {};
	\end{pgfonlayer}
	\begin{pgfonlayer}{edgelayer}
		\draw (1) to (0);
	\end{pgfonlayer}
\end{tikzpicture}}
\end{tabular}\\ 
\hline
$(2^4)$ & $(4^2)$ & 2 &
$\mathrm{USp}(2)$&
Bad quiver \\ 
\hline
$(3,1^5)$ & $(5,1^3)$ & 2 &
$\mathrm{SO}(3)$&
\begin{tabular}{l}
\scalebox{0.8}{
\begin{tikzpicture}
	\begin{pgfonlayer}{nodelayer}
		\node [style=flavorRed] (0) at (0, 0) {};
		\node [style=none] (1) at (0, -0.5) {8};
		\node [style=bluegauge] (2) at (-1, 0) {};
		\node [style=redgauge] (3) at (-2, 0) {};
		\node [style=none] (4) at (-1, -0.5) {2};
		\node [style=none] (5) at (-2, -0.5) {2};
		\node [style=none] (5) at (0, 0.8) {};
	\end{pgfonlayer}
	\begin{pgfonlayer}{edgelayer}
		\draw (3) to (2);
		\draw (2) to (0);
	\end{pgfonlayer}
\end{tikzpicture}}
\end{tabular} \\ 
\hline
$(3^2,1^2)$ & $(3^2,1^2)$ & 3 &
\begin{tabular}{c}$\mathrm{SO}(2)$ \\ $\times$ \\ $\mathrm{SO}(2)$ \end{tabular}&
\begin{tabular}{l}
\scalebox{0.8}{
\begin{tikzpicture}
	\begin{pgfonlayer}{nodelayer}
		\node [style=flavorRed] (0) at (0, 0) {};
		\node [style=none] (1) at (0, -0.5) {8};
		\node [style=bluegauge] (2) at (-1, 0) {};
		\node [style=redgauge] (3) at (-2, 0) {};
		\node [style=none] (4) at (-1, -0.5) {4};
		\node [style=none] (5) at (-2, -0.5) {2};
		\node [style=none] (5) at (0, 0.8) {};
	\end{pgfonlayer}
	\begin{pgfonlayer}{edgelayer}
		\draw (3) to (2);
		\draw (2) to (0);
	\end{pgfonlayer}
\end{tikzpicture}}
\end{tabular} \\ 
\hline
$(4^2)$ & $(2^4)$ & 6 &
$\mathrm{USp}(4)$&
Bad quiver \\ 
\hline
$(5,1^3)$ & $(3,1^5)$ & 6 &
$\mathrm{SO}(5)$&
\begin{tabular}{l}
\scalebox{0.8}{
\begin{tikzpicture}
	\begin{pgfonlayer}{nodelayer}
		\node [style=flavorRed] (0) at (0, 0) {};
		\node [style=none] (1) at (0, -0.5) {8};
		\node [style=bluegauge] (2) at (-1, 0) {};
		\node [style=redgauge] (3) at (-2, 0) {};
		\node [style=none] (4) at (-1, -0.5) {4};
		\node [style=none] (5) at (-2, -0.5) {4};
		\node [style=bluegauge] (6) at (-3, 0) {};
		\node [style=redgauge] (7) at (-4, 0) {};
		\node [style=none] (8) at (-3, -0.5) {2};
		\node [style=none] (9) at (-4, -0.5) {2};
		\node [style=none] (5) at (0, 0.8) {};
	\end{pgfonlayer}
	\begin{pgfonlayer}{edgelayer}
		\draw (3) to (2);
		\draw (2) to (0);
		\draw (7) to (6);
		\draw (6) to (3);
	\end{pgfonlayer}
\end{tikzpicture}}
\end{tabular}
\\ \hline
$(5,3)$ & $(2^2,1^4)$ & 7 &
\begin{tabular}{c} $\mathrm{SO}(4)$ \\ $\times$ \\ $\mathrm{USp}(2)$ \end{tabular}&
Bad quiver
\\ \hline
$(7,1)$ & $(1^8)$ & 12 &
$\mathrm{SO}(8)$&
\begin{tabular}{c}
\scalebox{0.8}{
\begin{tikzpicture}
	\begin{pgfonlayer}{nodelayer}
		\node [style=flavorRed] (0) at (0, 0) {};
		\node [style=none] (1) at (0, -0.5) {8};
		\node [style=bluegauge] (2) at (-1, 0) {};
		\node [style=redgauge] (3) at (-2, 0) {};
		\node [style=none] (4) at (-3, -0.5) {4};
		\node [style=none] (5) at (-4, -0.5) {4};
		\node [style=none] (5) at (0, 0.8) {};
		\node [style=bluegauge] (6) at (-3, 0) {};
		\node [style=redgauge] (7) at (-4, 0) {};
		\node [style=none] (8) at (-5, -0.5) {2};
		\node [style=none] (9) at (-6, -0.5) {2};
		\node [style=bluegauge] (10) at (-5, 0) {};
		\node [style=redgauge] (11) at (-6, 0) {};
		\node [style=none] (12) at (-2, -0.5) {6};
		\node [style=none] (13) at (-1, -0.5) {6};
	\end{pgfonlayer}
	\begin{pgfonlayer}{edgelayer}
		\draw (3) to (2);
		\draw (2) to (0);
		\draw (7) to (6);
		\draw (6) to (3);
		\draw (7) to (10);
		\draw (10) to (11);
	\end{pgfonlayer}
\end{tikzpicture}}
\end{tabular} \\ 
\hline
	\end{tabular}
\end{adjustbox}
\caption[$T(\mathrm{SO}(8))$ linear quivers]{$T(\mathrm{SO}(8))$ linear quivers. These are used in untwisted $D_4$ fixtures. The quivers $T_{\bv {\sigma}}(\mathrm{SO}(8))$  have non-diverging monopole formula; bad quivers with zero conformal dimension are not shown. The partitions identify special orbits of $\mathrm{SO}(8)$.}
\label{D4puncture}
\end{figure}

\subsection{Twisted \texorpdfstring{$A_3$}{A3} fixtures}
The quivers investigated here are known as $A_{2r-1}$ twisted fixtures. These include two twisted $A_{2r-1}$ punctures and an untwisted $A_{2r-1}$ puncture. The twisting map folds $A_{2r-1}$ onto $C_r$ via outer automorphism. As a result, when we study the three-legged magnetic quiver, the two twisted $A_{2r-1}$ punctures have $\mathrm{USp}(2r)$ flavor nodes, and the fixture acquires a $J=\mathrm{USp}(2r)$ central gauge node.  The untwisted puncture has a $\mathrm{SU}(2r)$ flavor node. During gluing, only the $\mathrm{USp}(2r) \subset \mathrm{SU}(2r)$ subgroup of the untwisted $A_{2r-1}$ puncture is gauged\footnote{This reproduces a similar procedure in \cite{Benini:2010uu} when treating twisted $D_r$ punctures. The quiver corresponding to twisted $D_r$ punctures have $\mathrm{SO}(2r-1)$ flavor nodes whereas untwisted $D_r$ have $\mathrm{SO}(2r)$ flavor nodes. As a result, when constructing the three-legged magnetic quiver, only the $\mathrm{SO}(2r-1) \subset \mathrm{SO}(2r)$ subgroup is gauged in the untwisted puncture.}. This produces an unitary-orthosymplectic quiver.  For $r=2$, the linear quivers corresponding to different twisted punctures are tabulated in Figure \ref{A3puncture} and for untwisted punctures in Figure \ref{A3punctureUntwisted}.

We tabulate twisted fixtures of $A_3$ in Figure \ref{A3fixtures}. The global symmetries match those in \cite{twistedGaiotto}. From now on, we only provide the algebra of the global symmetry group as we often do not have the HWG required to precisely identify the group. These quivers are a subset of an exhaustive list in \cite{twistedGaiotto} as we only list the magnetic quivers whose Coulomb branch Hilbert series does not diverge when evaluated with the monopole formula. With the monopole formula, we are unable to refine a Hilbert series with (special) orthogonal and symplectic gauge nodes. However, when there are unitary gauge nodes, we may be able to partially refine the Hilbert series by assigning the usual root fugacities to the unitary gauge nodes. 

For $A_{2n}$ twisted punctures we obtain $\mathrm{SO}(2n+1)$ flavor nodes and therefore the central node of our star-shaped magnetic quiver will be $\mathrm{SO}(2n+1)$ \cite{twisted2n,TwistedA2n}. We will not look at this family of quivers in this paper.

\begin{landscape}
\begin{figure}[t]
\small
\begin{adjustbox}{center}
	\begin{tabular}{|c|c|c|c|c|}
		\hline
		 Quiver &
		 \begin{tabular}{c} Global\\ Symmetry \end{tabular}   &
		 $\text{dim}_{\mathbb{H}}(\mathcal{C})$ &
		\begin{tabular}{c} Hilbert Series \end{tabular}&
		 	\begin{tabular}{c}Plethystic \\Logarithm\end{tabular} \\ 
		 \hline
\begin{tabular}{c}
\begin{tikzpicture}
	\begin{pgfonlayer}{nodelayer}
		\node [style=miniU] (3) at (3, 0) {};
		\node [style=miniU] (7) at (-3, 0) {};
		\node [style=none] (12) at (-3, -0.5) {2};
		\node [style=none] (13) at (-1, -0.5) {4};
		\node [style=none] (14) at (0, -0.5) {4};
		\node [style=none] (15) at (1, -0.5) {4};
		\node [style=none] (16) at (3, -0.5) {2};
		\node [style=none] (19) at (0, 1.5) {1};
		\node [style=miniBlue] (20) at (0, 0) {};
		\node [style=miniU] (21) at (-1, 0) {};
		\node [style=miniU] (22) at (1, 0) {};
		\node [style=miniBlue] (23) at (-2, 0) {};
		\node [style=miniBlue] (24) at (2, 0) {};
		\node [style=none] (25) at (-2, -0.5) {2};
		\node [style=none] (26) at (2, -0.5) {2};
		\node [style=gauge3] (27) at (0, 1) {};
	\end{pgfonlayer}
	\begin{pgfonlayer}{edgelayer}
		\draw (7) to (23);
		\draw (23) to (21);
		\draw (21) to (20);
		\draw (20) to (27);
		\draw (20) to (22);
		\draw (22) to (24);
		\draw (24) to (3);
	\end{pgfonlayer}
\end{tikzpicture}
\end{tabular}
& 	$\mathfrak{e}_6$ &	11 &
		\begin{tabular}{l}
		\parbox[t]{6cm}{
	$\dfrac
    {(1 + t^2){~}{\cal P}_{20}(t)}
    {(1 - t^2)^{22}}$ \\
    \\
    $= 1 + 78 t^2 + 2430 t^4 + 43758 t^6 + 537966 t^8 + 4969107 t^{10} + O\left(t^{12}\right)$}
		\end{tabular}	
		& 
        \begin{tabular}{l}
        \parbox[t]{3cm}
        {$78 t^2 - 651 t^4 + 12376 t^6 - 296946 t^8 + 7755189 t^{10}+ O\left(t^{12}\right)$}   
        \end{tabular}\\ 
        \hline		
\begin{tabular}{c}	
\begin{tikzpicture}
	\begin{pgfonlayer}{nodelayer}
		\node [style=miniU] (3) at (3, 0) {};
		\node [style=miniU] (7) at (-3, 0) {};
		\node [style=none] (12) at (-3, -0.5) {2};
		\node [style=none] (13) at (-1, -0.5) {4};
		\node [style=none] (14) at (0, -0.5) {4};
		\node [style=none] (15) at (1, -0.5) {4};
		\node [style=none] (16) at (3, -0.5) {2};
		\node [style=none] (19) at (0, 1.5) {2};
		\node [style=miniBlue] (20) at (0, 0) {};
		\node [style=miniU] (21) at (-1, 0) {};
		\node [style=miniU] (22) at (1, 0) {};
		\node [style=miniBlue] (23) at (-2, 0) {};
		\node [style=miniBlue] (24) at (2, 0) {};
		\node [style=none] (25) at (-2, -0.5) {2};
		\node [style=none] (26) at (2, -0.5) {2};
		\node [style=gauge3] (27) at (0, 1) {};
	\end{pgfonlayer}
	\begin{pgfonlayer}{edgelayer}
		\draw (7) to (23);
		\draw (23) to (21);
		\draw (21) to (20);
		\draw (20) to (27);
		\draw (20) to (22);
		\draw (22) to (24);
		\draw (24) to (3);
	\end{pgfonlayer}
\end{tikzpicture}
\end{tabular}
	& 		
	\centering	\begin{tabular}{c}	$\mathfrak{usp}(8)$\\ $\times$ \\ $\mathfrak{su}(2)$	\end{tabular}& 12&	
	\begin{tabular}{c} 
	\begin{tabular}{l}
	\parbox[t]{6cm}{
	$\dfrac
 {{\cal P}_{48}(t)}
 {(1 - t^2)^{24} (1 + t^2)^{12}} $ \\
 \\
 $= 1 + 39 t^2 + 878 t^4 + 13396 t^6 + 152412 t^8 + 1370975 t^{10} +O\left(t^{12}\right)$} 
		\end{tabular} 
 \end{tabular}
	&
	\begin{tabular}{c}
	\parbox[t]{3cm}{$39 t^2 + 98 t^4 - 1086 t^6 + 1545 t^8 + 67761 t^{10}+O\left(t^{12}\right)$}\end{tabular}\\ 
	\hline			
	\begin{tabular}{c}
\begin{tikzpicture}
	\begin{pgfonlayer}{nodelayer}
		\node [style=miniU] (3) at (3, 0) {};
		\node [style=miniU] (7) at (-3, 0) {};
		\node [style=none] (12) at (-3, -0.5) {2};
		\node [style=none] (13) at (-1, -0.5) {4};
		\node [style=none] (14) at (0, -0.5) {4};
		\node [style=none] (15) at (1, -0.5) {4};
		\node [style=none] (16) at (3, -0.5) {2};
		\node [style=none] (19) at (0.75, 1) {2};
		\node [style=miniBlue] (20) at (0, 0) {};
		\node [style=miniU] (21) at (-1, 0) {};
		\node [style=miniU] (22) at (1, 0) {};
		\node [style=miniBlue] (23) at (-2, 0) {};
		\node [style=miniBlue] (24) at (2, 0) {};
		\node [style=none] (25) at (-2, -0.5) {2};
		\node [style=none] (26) at (2, -0.5) {2};
		\node [style=gauge3] (27) at (0, 1) {};
		\node [style=gauge3] (28) at (0, 2) {};
		\node [style=none] (29) at (0, 2.5) {1};
	\end{pgfonlayer}
	\begin{pgfonlayer}{edgelayer}
		\draw (7) to (23);
		\draw (23) to (21);
		\draw (21) to (20);
		\draw (20) to (27);
		\draw (20) to (22);
		\draw (22) to (24);
		\draw (24) to (3);
		\draw (28) to (27);
	\end{pgfonlayer}
\end{tikzpicture}
\end{tabular}	& 		
		\begin{tabular}{c} 	$\mathfrak{so}(5)^2$\\
		$\times$ \\ $\mathfrak{su}(2)$ \\ $\times$ \\ $\mathfrak{u}(1)$	\end{tabular}
		& 13 &
  \begin{tabular}{c}
  \parbox[t]{6cm}{
  	$\dfrac
 {(1-t)^9 {~}{\cal P}_{86,c_2}(t)}
 {(1-t^2)^7 (1-t^3)^{13} (1-t^4)^7 (1-t^5)^8} $ \\
 \\
  {$= 1 + 24 t^2 + 36 t^3 + 356 t^4 + 932 t^5 + 4367 t^6 + 13272 t^7 +  46189 t^8 + 137468 t^9 + 413087 t^{10}+O\left(t^{12}\right)$}}
 \end{tabular}&
 \begin{tabular}{c}
 \parbox[t]{3cm}
 {$24 t^2 + 36 t^3 + 56 t^4 + 68 t^5 - 243 t^6 - 1176 t^7 - 2357 t^8 -  188 t^9 + 18121 t^{10}+ O\left(t^{12}\right)$}
 \end{tabular}\\ 
 \hline	
 	\begin{tabular}{c}
\begin{tikzpicture}
	\begin{pgfonlayer}{nodelayer}
		\node [style=miniU] (3) at (3, 0) {};
		\node [style=miniU] (7) at (-3, 0) {};
		\node [style=none] (12) at (-3, -0.5) {2};
		\node [style=none] (13) at (-1, -0.5) {4};
		\node [style=none] (14) at (0, -0.5) {4};
		\node [style=none] (15) at (1, -0.5) {4};
		\node [style=none] (16) at (3, -0.5) {2};
		\node [style=none] (19) at (0.75, 1) {3};
		\node [style=miniBlue] (20) at (0, 0) {};
		\node [style=miniU] (21) at (-1, 0) {};
		\node [style=miniU] (22) at (1, 0) {};
		\node [style=miniBlue] (23) at (-2, 0) {};
		\node [style=miniBlue] (24) at (2, 0) {};
		\node [style=none] (25) at (-2, -0.5) {2};
		\node [style=none] (26) at (2, -0.5) {2};
		\node [style=gauge3] (27) at (0, 1) {};
		\node [style=gauge3] (28) at (0, 2) {};
		\node [style=none] (29) at (0.75, 2) {2};
		\node [style=gauge3] (30) at (0, 3) {};
		\node [style=none] (31) at (0, 3.5) {1};
	\end{pgfonlayer}
	\begin{pgfonlayer}{edgelayer}
		\draw (7) to (23);
		\draw (23) to (21);
		\draw (21) to (20);
		\draw (20) to (27);
		\draw (20) to (22);
		\draw (22) to (24);
		\draw (24) to (3);
		\draw (28) to (27);
		\draw (30) to (28);
	\end{pgfonlayer}
\end{tikzpicture}
\end{tabular}	& 		
	\begin{tabular}{c} 	$\mathfrak{so}(5)^2$ \\ $\times$ \\ $\mathfrak{su}(4)$	\end{tabular}
	& 16 & 
  \begin{tabular}{c}
  \parbox[t]{6cm}{
  $\dfrac
 {{\cal P}_{100}(t)}
 {(1-t^2)^{10} (1-t^4)^{10} (1-t^6)^{12}} $ \\
 \\
  {$ = 1 + 35 t^2 + 724 t^4 + 11242 t^6 + 140062 t^8 + 1453129 t^{10} +  12880215 t^{12} + 99473971 t^{14} + 680140044 t^{16} + 4172259667 t^{18} +  23223084225 t^{20}+ O\left(t^{22}\right)$}}
  \end{tabular}
  &	
  \begin{tabular}{c}
  \parbox[t]{3cm}
  {$35 t^2 + 94 t^4 + 182 t^6 - 3808 t^8 - 7771 t^{10}+ O\left(t^{12}\right)$}
  \end{tabular}\\ 
  \hline	
\end{tabular}
\end{adjustbox}
\caption[Magnetic quivers for twisted $A_3$ fixtures]{Magnetic quivers for twisted $A_3$  fixtures. We show the subset of orthosymplectic quivers whose Coulomb branch is not bad. The choice of the discrete group is $H=\ker(\phi)=\mathbb{Z}_2$. The palindromic numerator terms ${\cal P}_k(t)$ are given in Appendix \ref{apx:PP}.  }
\label{A3fixtures}
\end{figure}
\end{landscape}

The first row of Figure \ref{A3fixtures} is the $E_6$ quiver already investigated in Section \ref{5dexamples}. The unitary-orthosymplectic quiver in the second row has a  non-simply laced unitary quiver counterpart with the same Coulomb branch and is a member of the $C_{n+1}\times A_1$ rank 1 $4d$ $\mathcal{N}=1$ SCFT sequence \cite{Bourget:2020asf}. 

\subsection{\texorpdfstring{$D_3$}{D3} and \texorpdfstring{$D_4$}{D4} fixtures}
\begin{figure}[pt]
\centering
    \begin{subtable}[t]{1\textwidth}
    \centering
	\scalebox{0.80}{	
\begin{adjustbox}{center}
	\begin{tabular}{|l|l|l|l|}
		\hline
		$2\Delta$ & Generators & Irreps & Dimensions \\ \hline
  $t^2$ &\begin{tabular}{c}$\mu_A, \mu_B, \mu_C$\end{tabular}		&\begin{tabular}{c}$+[1,0,1],+[1,0,1],+[1,0,1] $\end{tabular}	& \begin{tabular}{c}+45	\end{tabular}	 \\ \hline
    $t^3$ &	\begin{tabular}{c}$Q^{i_Ai_Bi_C}$\\$Q_{i_Ai_Bi_C}$\end{tabular} 	&	\begin{tabular}{c}$+[1,0,0]\otimes[1,0,0]\otimes [1,0,0] $\\$+[0,0,1]\otimes[0,0,1]\otimes [0,0,1] $\end{tabular}  &\begin{tabular}{c}+64\\+64\end{tabular} 	\\ \hline
      $t^4$ &\begin{tabular}{c}$Q^{[i_Aj_A][i_Bj_B][i_Cj_C]}$	\end{tabular}		&\begin{tabular}{c} $+[0,1,0]\otimes[0,1,0]\otimes[0,1,0]$\end{tabular}&\begin{tabular}{c} +216	\end{tabular}	 \\ \hline
	\end{tabular}
\end{adjustbox}}
\caption{Generators of the $T_4$ theory.}
\label{T4gen}
\end{subtable} 

 \vspace{0.5cm}
    \begin{subtable}[t]{1\textwidth}
    \centering
	\scalebox{0.80}{	
\begin{adjustbox}{center}
	\begin{tabular}{|l|l|l|l|}
		\hline
		$2\Delta$ & Relations & Irreps & Dimensions   \\ \hline
      $t^4$  &\begin{tabular}{c}$\text{tr}(\mu_{A}^k)=\text{tr}(\mu_{B}^k)=\text{tr}(\mu_{C}^k)$\end{tabular} &\begin{tabular}{c}$-2[0,0,0]$ \end{tabular}&\begin{tabular}{c}$-2$\end{tabular}\\ \hline
        $t^5$	 &\begin{tabular}{c}	$(\mu_A)^{i_A}_{j_A}Q^{j_Ai_Bi_C}=(\mu_B)^{i_B}_{j_B}Q^{i_Aj_Bi_C}=(\mu_C)^{i_C}_{j_C}Q^{i_Ai_Bj_C}$\\$(\mu_A)^{j_A}_{i_A}Q_{j_Ai_Bi_C}=(\mu_B)^{j_B}_{i_B}Q_{i_Aj_Bi_C}=(\mu_C)^{j_C}_{i_C}Q_{i_Ai_Bj_C}$ \end{tabular} &\begin{tabular}{c}$-2\times [1,0,0]\otimes[1,0,0]\otimes [1,0,0]$\\ $-2\times [0,0,1]\otimes[0,0,1]\otimes [0,0,1] $\end{tabular} &\begin{tabular}{c} $-128$ \\$-128$ \end{tabular} \\ \hline
	\end{tabular}
\end{adjustbox}}
\caption[Relations of $T_4$ theory]{An incomplete list of the relations of the $T_4$ theory.}
\label{T4rel}
\end{subtable}
\caption[Generators and relations of $T_4$ theory]{The generators and some relations  of the $T_4$ theory, ordered by their conformal dimension $\Delta$ along with the irreps they transform under and their dimensions, are summarised in \subref{T4gen} and \subref{T4rel}, respectively. $[n_1,n_2,n_3]$ is the Dynkin label for $\mathfrak{su}(4)$.
}
\label{fig:T4}
\end{figure}
We present a list of fixtures that can be constructed using the $D_3$ and $D_4$ punctures outlined in  \crefrange{D3puncture}{D4fixture2}. The global symmetry is given in \cite{DistlerA,Chacaltana:2011ze}. The quivers listed here form a subset of an exhaustive list in \cite{DistlerA,Chacaltana:2011ze} as we only list the magnetic quivers whose monopole formula is convergent. 
We would like to point out:
\begin{itemize}
    \item The quivers listed in \crefrange{D3fixtures}{D4fixture2} contain the $E_7$ and $E_8$ theory from Section \ref{5dexamples}. 
    \item The Coulomb branch of the quiver in the last row of Figure \ref{D3fixtures} is the $T_4$ theory. The generators and some of the relations of the Coulomb branch are explicitly listed in \cite{T4}. We can compute the plethystic logarithm (PL), where the first few positive terms are the dimension of the generators and negative terms are the dimension of the relations, as a check. The results up to $t^5$ are consistent with what we expect from \cite{T4} and are tabulated in Figure \ref{fig:T4}. Higher order relations are obscured by higher syzygies, making it difficult to extract from the PL. 
    \item The Coulomb branch of the quiver in the last row of Figure \ref{D4fixture2} is $\overline{\mathcal{O}}^{\mathfrak{e}_6}_{\text{min}}\times \overline{\mathcal{O}}^{\mathfrak{e}_6}_{\text{min}}$ which is the product of two minimal nilpotent orbit closures of $E_6$ \cite{DistlerD}. The HWG is $\text{PE}\left [\mu_2t^2 + \eta_2t^2 \right]$ where $\mu_i$ and $\eta_i$ are the highest weight fugacities of the two $E_6$'s. This is clear as the HWG here is just the product of the HWG of each individual $\overline{\mathcal{O}}^{\mathfrak{e}_6}_{\text{min}}$. 
\end{itemize}

\begin{landscape}
\begin{figure}[!ht]
\small
\begin{adjustbox}{center}

\end{adjustbox}
\caption[Magnetic quivers for $D_3$ fixtures]{$D_3$ fixtures. We show the subset of orthosymplectic quivers whose Coulomb branch is not bad. The choice of the discrete group is $H=\ker(\phi)=\mathbb{Z}_2$. The palindromic numerator terms ${\cal P}_k(t)$ are given in Appendix \ref{apx:PP}.}
\label{D3fixtures}
\end{figure}

\begin{figure}[!ht]
\small
\begin{adjustbox}{center}
		%
\end{adjustbox}
\caption[Magnetic quivers for $D_4$ fixtures I]{Magnetic quivers for $D_4$ fixtures. We provide only the subset of orthosymplectic quivers whose Coulomb branch is not bad. The choice of the discrete group is $H=\ker(\phi)=\mathbb{Z}_2$. The palindromic numerator terms ${\cal P}_k(t)$ are given in Appendix \ref{apx:PP}.}
\label{D4fixture1}
\end{figure}

\begin{figure}[!ht]
\small
\begin{adjustbox}{center}
	%
\\
 \hline	
\end{tabular}
\end{adjustbox}
\caption[Magnetic quivers for $D_4$ fixtures II]{Magnetic quivers for $D_4$ fixtures. We provide only the subset of orthosymplectic quivers whose Coulomb branch is not bad. The choice of the discrete group is $H=\ker(\phi)=\mathbb{Z}_2$. The palindromic numerator terms ${\cal P}_k(t)$ are given in Appendix \ref{apx:PP}. In the last line, the Hilbert series is the square of the Hilbert series for the closure of the minimal nilpotent orbit of $\mathfrak{e}_6$. }
\label{D4fixture2}
\end{figure}
\end{landscape}
\section{6d magnetic quiver}
\label{6dapp}
In this appendix, we tabulate the magnetic quivers for $6d$ $\mathcal{N}=(1,0)$ theories discussed in Section \ref{applications}. 
\subsection{\texorpdfstring{$D_3 \cong A_3$}{D3=A3} singularity }
We start by considering the first case of \eqref{fullbouquet} where the monopole formula converges. This magnetic quiver does not have a corresponding electric quiver in the form of (\ref{electric6d}) as it would contain $\mathrm{USp}(-2)$ gauge groups. Nonetheless, one can construct a Type IIA brane configuration similar to those constructed in \cite{newMarcus}. Furtermore, since it is a special case of $D_3$ singularity, we can equivalently use $A_3$ singularities and express the results with unitary quivers. The electric quiver is given in \cite{Hanany:2018vph,MarcusSymmetric,6dmagnetic} and takes the form:
\begin{equation}
\scalebox{.8}{\raisebox{-.5\height}{
\begin{tikzpicture}
	\begin{pgfonlayer}{nodelayer}
		\node [style=gauge3] (1) at (-5.75, 0) {};
		\node [style=gauge3] (2) at (-4.75, 0) {};
		\node [style=gauge3] (3) at (-3.75, 0) {};
		\node [style=gauge3] (8) at (-2.25, 0) {};
		\node [style=gauge3] (9) at (-1.25, 0) {};
		\node [style=gauge3] (14) at (-0.25, 0) {};
		\node [style=none] (29) at (-1.25, -0.5) {\scriptsize{$\mathrm{SU}(4)$}};
		\node [style=none] (30) at (-2.25, -0.5) {\scriptsize{$\mathrm{SU}(4)$}};
		\node [style=none] (33) at (-0.25, -0.5) {\scriptsize$\mathrm{SU}(4)$};
		\node [style=none] (37) at (-5.75, -0.5) {\scriptsize$\mathrm{SU}(4)$};
		\node [style=none] (38) at (-4.75, -0.5) {\scriptsize$\mathrm{SU}(4)$};
		\node [style=none] (39) at (-3.75, -0.5) {\scriptsize$\mathrm{SU}(4)$};
		\node [style=flavor2] (40) at (-5.75, 1.5) {};
		\node [style=flavor2] (41) at (-0.25, 1.5) {};
		\node [style=none] (42) at (-5.75, 2) {\scriptsize $\mathrm{SU}(4)$};
		\node [style=none] (43) at (-0.25, 2) {\scriptsize$\mathrm{SU}(4)$};
		\node [style=none] (44) at (-0.25, -0.75) {};
		\node [style=none] (45) at (-5.75, -0.75) {};
		\node [style=none] (46) at (-3, -1.75) {$m-1$};
		\node [style=none] (47) at (-3, 0) {$\cdots$};
		\node [style=none] (48) at (-3.425, 0) {};
		\node [style=none] (49) at (-2.575, 0) {};
	\end{pgfonlayer}
	\begin{pgfonlayer}{edgelayer}
		\draw (8) to (9);
		\draw (3) to (2);
		\draw (2) to (1);
		\draw (40) to (1);
		\draw (41) to (14);
		\draw (14) to (9);
		\draw [style=brace] (44.center) to (45.center);
		\draw (48.center) to (3);
		\draw (49.center) to (8);
	\end{pgfonlayer}
\end{tikzpicture}
}}
\label{unitaryelectric}
\end{equation}
\paragraph{Case $\mathbf{  m=1}$.}
In terms of the orthosymplectic quiver, for $m=1$ there is only a single $\mathrm{USp}(2)$ gauge node on top of \eqref{fullbouquet}, so this is a trivial bouquet. The magnetic quiver describes a free theory, which is displayed in the first row of Figure \ref{free} and we reproduce it here:
\begin{equation}
\raisebox{-.5\height}{
    \begin{tikzpicture}
	\begin{pgfonlayer}{nodelayer}
		\node [style=miniBlue] (0) at (0, 1) {};
		\node [style=miniU] (1) at (0, 0) {};
		\node [style=miniBlue] (2) at (1, 0) {};
		\node [style=miniBlue] (3) at (-1, 0) {};
		\node [style=miniBlue] (4) at (3, 0) {};
		\node [style=miniBlue] (5) at (-3, 0) {};
		\node [style=miniU] (6) at (2, 0) {};
		\node [style=miniU] (7) at (4, 0) {};
		\node [style=miniU] (8) at (-2, 0) {};
		\node [style=miniU] (9) at (-4, 0) {};
		\node [style=none] (10) at (0, 1.5) {2};
		\node [style=none] (11) at (0, -0.5) {6};
		\node [style=none] (12) at (1, -0.5) {4};
		\node [style=none] (13) at (-1, -0.5) {4};
		\node [style=none] (14) at (-2, -0.5) {4};
		\node [style=none] (15) at (2, -0.5) {4};
		\node [style=none] (16) at (3, -0.5) {2};
		\node [style=none] (17) at (4, -0.5) {2};
		\node [style=none] (18) at (-3, -0.5) {2};
		\node [style=none] (19) at (-4, -0.5) {2};
	\end{pgfonlayer}
	\begin{pgfonlayer}{edgelayer}
		\draw (9) to (5);
		\draw (5) to (8);
		\draw (8) to (3);
		\draw (3) to (1);
		\draw (1) to (0);
		\draw (1) to (2);
		\draw (2) to (6);
		\draw (6) to (4);
		\draw (4) to (7);
	\end{pgfonlayer}
\end{tikzpicture}
}
\end{equation}
If we look at the unitary quiver counterpart, the electric quiver in \eqref{unitaryelectric}, $m=1$ case is the theory of 16 free hypers where the unitary magnetic quiver takes the form:
\begin{equation}
 \scalebox{.8}{ 
 \raisebox{-.5\height}{
 \begin{tikzpicture}
	\begin{pgfonlayer}{nodelayer}
		\node [style=gauge3] (1) at (-4, 0) {};
		\node [style=gauge3] (2) at (-3, 0) {};
		\node [style=gauge3] (3) at (-2, 0) {};
		\node [style=gauge3] (9) at (-1, 0) {};
		\node [style=gauge3] (13) at (1, 0) {};
		\node [style=gauge3] (14) at (2, 0) {};
		\node [style=gauge3] (16) at (0, 0) {};
		\node [style=none] (29) at (-1, -0.5) {4};
		\node [style=none] (33) at (2, -0.5) {1};
		\node [style=none] (34) at (1, -0.5) {2};
		\node [style=none] (35) at (0, -0.5) {3};
		\node [style=none] (37) at (-4, -0.5) {1};
		\node [style=none] (38) at (-3, -0.5) {2};
		\node [style=none] (39) at (-2, -0.5) {3};
		\node [style=gauge3] (40) at (-1, 1) {};
		\node [style=none] (42) at (-1, 1.5) {1};
	\end{pgfonlayer}
	\begin{pgfonlayer}{edgelayer}
		\draw (3) to (2);
		\draw (2) to (1);
		\draw (13) to (14);
		\draw (16) to (13);
		\draw (3) to (9);
		\draw (9) to (40);
		\draw (9) to (16);
	\end{pgfonlayer}
\end{tikzpicture}}
}
\end{equation}
\paragraph{Case $\mathbf{  m=2}$.}
For $m=2$ the bouqueted quiver takes the form:
\begin{equation}
\raisebox{-.5\height}{
  \scalebox{.8}{  \begin{tikzpicture}
	\begin{pgfonlayer}{nodelayer}
		\node [style=redgauge] (0) at (-5, 0) {};
		\node [style=bluegauge] (1) at (-4, 0) {};
		\node [style=redgauge] (2) at (-3, 0) {};
		\node [style=bluegauge] (3) at (-2, 0) {};
		\node [style=redgauge] (9) at (-1, 0) {};
		\node [style=redgauge] (13) at (1, 0) {};
		\node [style=bluegauge] (14) at (2, 0) {};
		\node [style=redgauge] (15) at (3, 0) {};
		\node [style=bluegauge] (16) at (0, 0) {};
		\node [style=none] (29) at (-1, -0.5) {6};
		\node [style=none] (32) at (3, -0.5) {2};
		\node [style=none] (33) at (2, -0.5) {2};
		\node [style=none] (34) at (1, -0.5) {4};
		\node [style=none] (35) at (0, -0.5) {4};
		\node [style=none] (36) at (-5, -0.5) {2};
		\node [style=none] (37) at (-4, -0.5) {2};
		\node [style=none] (38) at (-3, -0.5) {4};
		\node [style=none] (39) at (-2, -0.5) {4};
		\node [style=bluegauge] (40) at (-1.75, 1) {};
		\node [style=bluegauge] (41) at (-0.25, 1) {};
		\node [style=none] (42) at (-1.75, 1.5) {2};
		\node [style=none] (43) at (-0.25, 1.5) {2};
					\node [style=none] (45) at (6, 0.5) {$F = \mathfrak{su}(8)\times \mathfrak{u}(1)$};
	\end{pgfonlayer}
	\begin{pgfonlayer}{edgelayer}
		\draw (3) to (2);
		\draw (2) to (1);
		\draw (0) to (1);
		\draw (13) to (14);
		\draw (14) to (15);
		\draw (16) to (13);
		\draw (3) to (9);
		\draw (9) to (40);
		\draw (9) to (41);
		\draw (9) to (16);
	\end{pgfonlayer}
\end{tikzpicture}}
}
\label{D3bouquet}
\end{equation}
and coinciding the two $\mathrm{USp}(2)$ nodes:
\begin{equation}
\scalebox{.8}{\raisebox{-.5\height}{
    \begin{tikzpicture}
	\begin{pgfonlayer}{nodelayer}
		\node [style=redgauge] (0) at (-5, 0) {};
		\node [style=bluegauge] (1) at (-4, 0) {};
		\node [style=redgauge] (2) at (-3, 0) {};
		\node [style=bluegauge] (3) at (-2, 0) {};
		\node [style=redgauge] (9) at (-1, 0) {};
		\node [style=redgauge] (13) at (1, 0) {};
		\node [style=bluegauge] (14) at (2, 0) {};
		\node [style=redgauge] (15) at (3, 0) {};
		\node [style=bluegauge] (16) at (0, 0) {};
		\node [style=none] (29) at (-1, -0.5) {6};
		\node [style=none] (32) at (3, -0.5) {2};
		\node [style=none] (33) at (2, -0.5) {2};
		\node [style=none] (34) at (1, -0.5) {4};
		\node [style=none] (35) at (0, -0.5) {4};
		\node [style=none] (36) at (-5, -0.5) {2};
		\node [style=none] (37) at (-4, -0.5) {2};
		\node [style=none] (38) at (-3, -0.5) {4};
		\node [style=none] (39) at (-2, -0.5) {4};
		\node [style=bluegauge] (41) at (-1, 1) {};
		\node [style=none] (43) at (-0.5, 1) {4};
		\node [style=none] (44) at (-1, 2) {$\Lambda^2$};
					\node [style=none] (45) at (6, 0.5) {$F = \mathfrak{su}(8)$};
	\end{pgfonlayer}
	\begin{pgfonlayer}{edgelayer}
		\draw (3) to (2);
		\draw (2) to (1);
		\draw (0) to (1);
		\draw (13) to (14);
		\draw (14) to (15);
		\draw (16) to (13);
		\draw (3) to (9);
		\draw (9) to (41);
		\draw (9) to (16);
		\draw [in=135, out=45, loop] (41) to ();
	\end{pgfonlayer}
\end{tikzpicture}}
}
\label{D3anti}
\end{equation}
Checking the ratio \eqref{quotientHS} is straightforward with the exact computations:
\begin{equation}
\begin{split}
\text{HS}_{\mathcal{C}}(\ref{D3bouquet})= &\\
& \scriptsize
    \frac{
    \left(
    \begin{array}{c}1 + 47 t^2 + 1187 t^4 + 19608 t^6 + 230525 t^8 + 2028432 t^{10} \\+ 
 13846924 t^{12} + 75329630 t^{14} + 333435759 t^{16} + 1220534386 t^{18}+ 
 3742539039 t^{20}\\+9711761582 t^{22} + 21501330350 t^{24} + 
 40872803196 t^{26} + 67040503765 t^{28} + 95225971156 t^{30}\\+ 
 117430324170 t^{32} + 125907479318 t^{34} + \pal + t^{68}
 \end{array}
 \right) }{
 (1-t^2)^{34}  (1 + t^2)^{17}
 }
 \end{split}
  \label{exactD3}
\end{equation}

\begin{equation}
\begin{split}
\text{HS}_{\mathcal{C}}(\ref{D3anti}) = &\\
& \scriptsize
    \frac{
    \left(
    \begin{array}{c}1 + 46 t^2 + 1071 t^4 + 16143 t^6 + 174118 t^8 + 1419714 t^{10}\\ + 9073146 t^{12} + 46643305 t^{14} + 196711367 t^{16} + 691043361 t^{18} +  2046661987 t^{20} \\+ 5159425991 t^{22} + 11154844150 t^{24} + 20807528268 t^{26} + 33640921000 t^{28} + 47302761853 t^{30}\\ + 57982719020 t^{32} + 62044510334 t^{34} \pal + t^{68}
 \end{array}
 \right) }{
 (1-t^2)^{34}  (1 + t^2)^{17}
 }.
  \end{split}
 \label{exactD3B}
\end{equation}
Dividing their volumes yields:
\begin{equation}
    \frac{\text{HS}_{\mathcal{C}}(\ref{exactD3})}{\text{HS}_{\mathcal{C}}(\ref{exactD3B})}\bigg \rvert_{t \rightarrow 1}  = 2=\left| \mathbb{Z}_2 \right| \,.
\end{equation}
This satisfies relation \eqref{quotientHS}, where the discrete group that we quotient by is $\mathbb{Z}_2$.
Here, we emphasise that this $\mathbb{Z}_2$ is \textit{different} to the overall $\mathbb{Z}_2^{\textrm{diag}}$ discrete group which is our discrete group $H$.

\paragraph{Case $\mathbf{m\geq3}$.}
When there are more than two $\mathrm{USp}(2)$ nodes in the bouquet of the magnetic quiver, we have more than one choice of coinciding the nodes (corresponding to multiple ways of coinciding M5 branes). 
For instance, $m=3$ there are three inequivalent choices labelled by partitions:
\begin{itemize}
    \item $[1,1,1]$: Trivial partition where the gauge nodes in the bouquet are all separated.
    \item $[2,1]$: Two of the gauge nodes in the bouquet coincide.
    \item $[3]$: All three gauge nodes in the bouquet coincide. 
\end{itemize}
We tabulate the Hilbert series for a bouquet of three $\mathrm{USp}(2)$ nodes in Figure \ref{D3quotientm3} and a bouquet of four $\mathrm{USp}(2)$ nodes in Figure \ref{D3quotientm4}.

\paragraph{Padé approximation.}
Verifying (\ref{quotientHS}) is easily achieved when the Hilbert series of the two moduli spaces are exact (in a rational form). But obtaining the exact unrefined Hilbert series is not always possible and becomes increasingly difficult when the quivers are more involved. As a result, for larger quivers we can only obtain a perturbative Hilbert series. If we compute the two Hilbert series perturbatively and divide them, we should expect the resulting series to converge to $|S_m|$ as $t \rightarrow 1$. However, with a finite number of terms, it is hopeless to get the ration of Taylor series to converge to $|S_m|$. 

We provide a solution via the Padé approximation \cite{pade1892representation,bender2013advanced} which is much superior in approximating the exact rational function than a Taylor series. As a result, it allows us to get the ratio of the volumes very close to the desired integer as we set $t \rightarrow 1$ by computing a finite number of terms in the Hilbert series. Naturally, the higher the number of terms, the more accurate the approximation: 
\begin{equation}
\begin{split}
    \text{HS}_{\mathcal{C}}(\ref{D3bouquet})=&1 + 64 t^2 + 2156 t^4 + 49035 t^6 + 832376 t^8 + 11170539 t^{10} + 
 123240632 t^{12}\\ &+ 1150637073 t^{14} + 9298241055 t^{16} + 66216202892 t^{18} + 
 421718414235 t^{20}\\& + 2431558547462 t^{22} + 12823770808393 t^{24} + 
 62403806228172 t^{26}\\& + 282310404939196 t^{28} + 1195031191158327 t^{30} + 
 4760155076556964 t^{32}\\& + 17930932092838988 t^{34}+64153544161616080 t^{36} + 218854365637620864 t^{38}\\& + 
 714337732919816234 t^{40} +O\left(t^{42}\right)
\end{split}
\end{equation}
\begin{equation}
    \begin{split}
    \text{HS}_{\mathcal{C}}(\ref{D3anti})  =&  1 + 63 t^2 + 2023 t^4 + 43428 t^6 + 696086 t^8 + 8860325 t^{10} + 
 93249581 t^{12}\\& + 835269497 t^{14} + 6509509030 t^{16} + 44911956047 t^{18} + 
 278222007609 t^{20}\\& + 1565674022207 t^{22} + 8082467996707 t^{24} + 
 38595225758861 t^{26}\\& + 171701062128429 t^{28} + 716059431853538 t^{30} + 
 2814530485234069 t^{32} \\&+ 10476161772590996 t^{34} + 37081407192725272 t^{36} + 125280303807253313 t^{38}\\& + 
 405343172433875147 t^{40} + O\left(t^{42}\right) \,.
    \end{split}
\end{equation}
Dividing the two series and applying the Padé approximation for $t \rightarrow 1$ gives $2.0001$. This is approximately two, the expected order of the discrete group $\mathbb{Z}_2$. 
We use the Padé approximation to verify (\ref{quotientHS}) whenever an exact Hilbert series cannot be obtained. 

\subsection{\texorpdfstring{$D_4$}{D4} singularity}
We now study the smallest member of \eqref{electric6d} that has a corresponding electric quiver. 

\paragraph{Case $\mathbf{m=1}$.}
In the case $m=1$, the bouquet only has one node and the quiver is identical to  \eqref{quiverE8}, with enhanced symmetry $E_8$. This is consistent with the 6d theory: for $n=4$ and $m=1$, the quiver (\ref{electric6d}) becomes
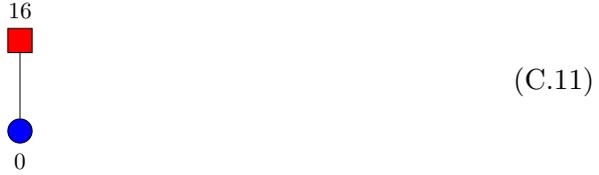
\begin{equation}
\scalebox{.8}{\raisebox{-.5\height}{
\begin{tikzpicture}
	\begin{pgfonlayer}{nodelayer}
		\node [style=bluegauge] (14) at (0, 0) {};
		\node [style=none] (33) at (0, -0.5) {$0$};
		\node [style=flavorRed] (41) at (0, 1.5) {};
		\node [style=none] (43) at (0, 2) {$16$};
		\node [style=none] (44) at (0, -0.75) {};
	\end{pgfonlayer}
	\begin{pgfonlayer}{edgelayer}
		\draw (41) to (14);
	\end{pgfonlayer}
\end{tikzpicture}
}}
\label{electric6dEstring}
\end{equation}
This indeed corresponds, at the UV fixed point, to the E-string theory with $E_8$ symmetry.

\paragraph{Case $\mathbf{m \geq 2}$.}
Formally, the electric quiver becomes
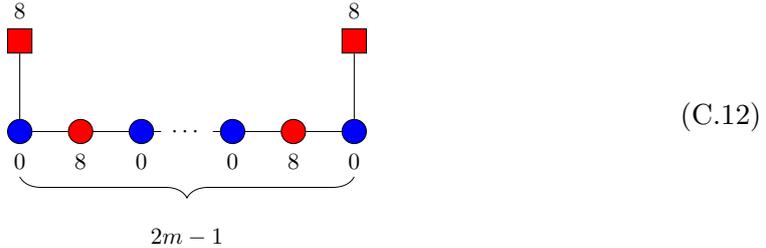
\begin{equation}
\scalebox{.8}{\raisebox{-.5\height}{
\begin{tikzpicture}
	\begin{pgfonlayer}{nodelayer}
		\node [style=bluegauge] (1) at (-5.75, 0) {};
		\node [style=redgauge] (2) at (-4.75, 0) {};
		\node [style=bluegauge] (3) at (-3.75, 0) {};
		\node [style=bluegauge] (8) at (-2.25, 0) {};
		\node [style=redgauge] (9) at (-1.25, 0) {};
		\node [style=bluegauge] (14) at (-0.25, 0) {};
		\node [style=none] (29) at (-1.25, -0.5) {$8$};
		\node [style=none] (30) at (-2.25, -0.5) {$0$};
		\node [style=none] (33) at (-0.25, -0.5) {$0$};
		\node [style=none] (37) at (-5.75, -0.5) {$0$};
		\node [style=none] (38) at (-4.75, -0.5) {$8$};
		\node [style=none] (39) at (-3.75, -0.5) {$0$};
		\node [style=flavorRed] (40) at (-5.75, 1.5) {};
		\node [style=flavorRed] (41) at (-0.25, 1.5) {};
		\node [style=none] (42) at (-5.75, 2) {$8$};
		\node [style=none] (43) at (-0.25, 2) {$8$};
		\node [style=none] (44) at (-0.25, -0.75) {};
		\node [style=none] (45) at (-5.75, -0.75) {};
		\node [style=none] (46) at (-3, -1.75) {$2m-1$};
		\node [style=none] (47) at (-3, 0) {$\cdots$};
		\node [style=none] (48) at (-3.425, 0) {};
		\node [style=none] (49) at (-2.575, 0) {};
	\end{pgfonlayer}
	\begin{pgfonlayer}{edgelayer}
		\draw (8) to (9);
		\draw (3) to (2);
		\draw (2) to (1);
		\draw (40) to (1);
		\draw (41) to (14);
		\draw (14) to (9);
		\draw [style=brace] (44.center) to (45.center);
		\draw (48.center) to (3);
		\draw (49.center) to (8);
	\end{pgfonlayer}
\end{tikzpicture}}
}
\label{electric6dd}
\end{equation}
such that, in this minimal case, the Higgs branch at finite gauge coupling is trivial. Nonetheless, as we  tune gauge couplings to infinity, massless degrees of freedom arise from tensionless strings yielding a non-trivial Higgs branch. The resulting magnetic quiver therefore belongs to (\ref{fullanti}) with $n=4$. We tabulate the results for $m=2,3,4$ in  \crefrange{D4quotientm2}{D4quotientm4} along with the Hilbert series and the associated plethystic logarithm. The global symmetry of all the tabulated quivers can be read off the Hilbert series and is the expected $\mathfrak{so}(8) \times \mathfrak{so}(8)$. 
\subsection{\texorpdfstring{$D_5$}{D5} singularity}
Next we consider the first member of (\ref{electric6d}) that has a non-trivial Higgs branch at finite gauge coupling: 
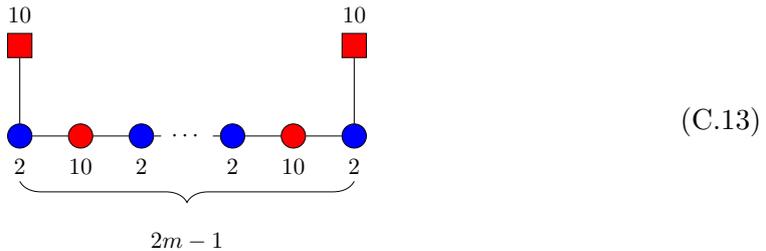
\begin{equation}
\scalebox{.8}{\raisebox{-.5\height}{
\begin{tikzpicture}
	\begin{pgfonlayer}{nodelayer}
		\node [style=bluegauge] (1) at (-5.75, 0) {};
		\node [style=redgauge] (2) at (-4.75, 0) {};
		\node [style=bluegauge] (3) at (-3.75, 0) {};
		\node [style=bluegauge] (8) at (-2.25, 0) {};
		\node [style=redgauge] (9) at (-1.25, 0) {};
		\node [style=bluegauge] (14) at (-0.25, 0) {};
		\node [style=none] (29) at (-1.25, -0.5) {$10$};
		\node [style=none] (30) at (-2.25, -0.5) {$2$};
		\node [style=none] (33) at (-0.25, -0.5) {$2$};
		\node [style=none] (37) at (-5.75, -0.5) {$2$};
		\node [style=none] (38) at (-4.75, -0.5) {$10$};
		\node [style=none] (39) at (-3.75, -0.5) {$2$};
		\node [style=flavorRed] (40) at (-5.75, 1.5) {};
		\node [style=flavorRed] (41) at (-0.25, 1.5) {};
		\node [style=none] (42) at (-5.75, 2) {$10$};
		\node [style=none] (43) at (-0.25, 2) {$10$};
		\node [style=none] (44) at (-0.25, -0.75) {};
		\node [style=none] (45) at (-5.75, -0.75) {};
		\node [style=none] (46) at (-3, -1.75) {$2m-1$};
		\node [style=none] (47) at (-3, 0) {$\cdots$};
		\node [style=none] (48) at (-3.425, 0) {};
		\node [style=none] (49) at (-2.575, 0) {};
	\end{pgfonlayer}
	\begin{pgfonlayer}{edgelayer}
		\draw (8) to (9);
		\draw (3) to (2);
		\draw (2) to (1);
		\draw (40) to (1);
		\draw (41) to (14);
		\draw (14) to (9);
		\draw [style=brace] (44.center) to (45.center);
		\draw (48.center) to (3);
		\draw (49.center) to (8);
	\end{pgfonlayer}
\end{tikzpicture}
}}
\label{electric6dn5}
\end{equation}
We tabulate the Hilbert series for the magnetic quivers with bouquets of size $m=2,3$ in Figures \ref{D5quotientm2} and \ref{D5quotientm3}. The global symmetry is again consistent with the expected $\mathfrak{so}(10) \times \mathfrak{so}(10)$. Furthermore, the quiver theory (\ref{electric6dn5}) implies that there are generators transforming in the $[1,0,0,0,0]_{\mathrm{SO}(10)} \otimes [1,0,0,0,0]_{\mathrm{SO}(10)}$ representation which should appear at order ${2m}$. 
Indeed the computed plethystic logarithm, shows these field theoretic generators along with additional generators that are suspected to arise from tensionless strings.
Unfortunately, as we can only compute unrefined Hilbert series for orthosymplectic quivers, their irreducible representation content under the global symmetry group remains elusive. 

\subsection{Global symmetry}
Whenever star shaped quivers, such as the quivers with a bouquet, have no enhancement the global symmetry can be obtained as the product of the global symmetries of the individual legs \cite{RudolphSlodowy}. The two long legs
\begin{equation}
\scalebox{.8}{\raisebox{-.5\height}{\begin{tikzpicture}
		\node [style=redgauge] (0) at (0, 0) [label=below:2] {};
		\node [style=bluegauge] (1) at (1, 0) [label=below:2] {};
		\node [style=redgauge] (2) at (2, 0) [label=below:4] {};
		\node [style=bluegauge] (3) at (3, 0) [label=below:4] {};
		\node [style=flavorRed] (4) at (4, 0) [label=below:6] {};
		\draw (0)--(1)--(2)--(3)--(4);
\end{tikzpicture}}}
\end{equation}
both contribute a global symmetry of $\mathfrak{so}(6)$ and the bouquet legs of 
\begin{equation}
\scalebox{.8}{\raisebox{-.5\height}{\begin{tikzpicture}
		\node [style=bluegauge] (3) at (3, 0) [label=below:2] {};
		\node [style=flavorRed] (4) at (4, 0) [label=below:6] {};
		\draw (3)--(4);
\end{tikzpicture}}}
\end{equation}
are $D_3$ Kleinian singularities, which have $\mathrm{SO}(2)\cong \mathfrak{u}(1)$ global symmetry. Taking their product gives us the expected results. 
For cases with $n\geq 4$,  the two long legs 
\begin{equation}
\scalebox{.8}{\raisebox{-.5\height}{\begin{tikzpicture}
		\node [style=redgauge] (0) at (0, 0) [label=below:2] {};
		\node [style=bluegauge] (1) at (1, 0) [label=below:2] {};
		\node (2) at (2, 0) {$\cdots$};
		\node [style=bluegauge] (3) at (3, 0) [label=below:$2n-2$] {};
		\node [style=flavorRed] (4) at (4, 0) [label=below:$2n$] {};
		\draw (0)--(1)--(2)--(3)--(4);
\end{tikzpicture}}}
\end{equation}
each contribute a $\mathfrak{so}(2n)$ factor. The bouquet legs of 
\begin{equation}
\scalebox{.8}{\raisebox{-.5\height}{\begin{tikzpicture}
		\node [style=bluegauge] (3) at (3, 0) [label=below:2] {};
		\node [style=flavorRed] (4) at (4, 0) [label=below:$2n$] {};
		\draw (3)--(4);
\end{tikzpicture}}}
\end{equation}
are $D_{n\geq 4} $ Kleinian singularity with trivial global symmetry. This explains why the magnetic quivers have global symmetry $\mathfrak{so}(2n)\times \mathfrak{so}(2n)$ which is consistent with expectation from the electric quiver (\ref{electric6d}). 

Furthermore, if $m_i$ legs of the type 
\begin{equation}
\scalebox{.8}{\raisebox{-.5\height}{\begin{tikzpicture}
		\node [style=bluegauge] (3) at (3, 0) [label=below:2] {};
		\node [style=flavorRed] (4) at (4, 0) [label=below:$2n$] {};
		\draw (3)--(4);
\end{tikzpicture}}}
\end{equation}
coincide to give an 
\begin{equation}
\scalebox{.8}{\raisebox{-.5\height}{\begin{tikzpicture}
		\node at (2, 0) {$\Lambda^2$};
		\node [style=bluegauge] (3) at (3, 0) [label=below:$2m_i$] {};
		\node [style=flavorRed] (4) at (4, 0) [label=below:$2n$] {};
		\draw (3)--(4);
		\draw [in=135, out=-135, loop] (3) to ();
\end{tikzpicture}}}
\end{equation}
then the global symmetry of this leg is $\mathfrak{u}(1)$ for $n=3$ and trivial for any $n \geq 4$. Therefore, we observe the global symmetry becoming smaller whenever we coincide the bouquet of nodes.

\begin{landscape}
\begin{figure}[!ht]
\small
\begin{adjustbox}{center}
	\begin{tabular}{|c|c|c|c|c|}
		\hline
		Quiver &
		\begin{tabular}{c}
		Global\\Symmetry
		\end{tabular} &
		Hilbert Series &
		\begin{tabular}{c}
		Plethystic\\Logarithm
		\end{tabular}&
		\begin{tabular}{c}
		Discrete\\Quotient
		\end{tabular}\\ 
		\hline
\begin{tabular}{c}
\centering
\scalebox{0.8}{
\begin{tikzpicture}
	\begin{pgfonlayer}{nodelayer}
		\node [style=redgauge] (0) at (-5, 0) {};
		\node [style=bluegauge] (1) at (-4, 0) {};
		\node [style=redgauge] (2) at (-3, 0) {};
		\node [style=bluegauge] (3) at (-2, 0) {};
		\node [style=redgauge] (9) at (-1, 0) {};
		\node [style=redgauge] (13) at (1, 0) {};
		\node [style=bluegauge] (14) at (2, 0) {};
		\node [style=redgauge] (15) at (3, 0) {};
		\node [style=bluegauge] (16) at (0, 0) {};
		\node [style=none] (29) at (-1, -0.5) {6};
		\node [style=none] (32) at (3, -0.5) {2};
		\node [style=none] (33) at (2, -0.5) {2};
		\node [style=none] (34) at (1, -0.5) {4};
		\node [style=none] (35) at (0, -0.5) {4};
		\node [style=none] (36) at (-5, -0.5) {2};
		\node [style=none] (37) at (-4, -0.5) {2};
		\node [style=none] (38) at (-3, -0.5) {4};
		\node [style=none] (39) at (-2, -0.5) {4};
		\node [style=bluegauge] (41) at (-2, 1) {};
		\node [style=none] (43) at (-2, 1.5) {2};
		\node [style=bluegauge] (44) at (-1, 1) {};
		\node [style=bluegauge] (45) at (0, 1) {};
		\node [style=none] (46) at (-1, 1.5) {2};
		\node [style=none] (47) at (0, 1.5) {2};
	\end{pgfonlayer}
	\begin{pgfonlayer}{edgelayer}
		\draw (3) to (2);
		\draw (2) to (1);
		\draw (0) to (1);
		\draw (13) to (14);
		\draw (14) to (15);
		\draw (16) to (13);
		\draw (3) to (9);
		\draw (9) to (41);
		\draw (9) to (16);
		\draw (44) to (9);
		\draw (9) to (45);
	\end{pgfonlayer}
\end{tikzpicture}
}
\end{tabular}
& 		
\centering
\begin{tabular}{c} $\mathfrak{so}(6)^2$ \\ $\times$ \\ $\mathfrak{u}(1)^3$\end{tabular}
&
\begin{tabular}{l}
       \parbox[t]{5cm}{
       {
       {\footnotesize ${ \dfrac{(1-t)^{11}(1-t^2)^{-8} (1-t^3)^{-8} {~}{\cal P}_{170}(t) }
       {(1-t^4)^2 (1-t^5)^{12} (1-t^6)^{10} (1-t^7)^7}}$}\\
       \\
       $ = 1 + 33 t^2 + 32 t^3 + 566 t^4 + 1120 t^5 + 7422 t^6 + 20128 t^7 + 85959 t^8 + 257280 t^9 + 893629 t^{10} + O\left(t^{12}\right)$
       } }\\
       \end{tabular}
       &
       \begin{tabular}{c}
       \parbox{3cm}
       {$33 t^2 + 32 t^3 + 5 t^4 + 64 t^5 + 184 t^6 - 96 t^7 - 1310 t^8 - 2368 t^9 - 1842 t^{10}+ O\left(t^{12}\right)$}
       \end{tabular}
       &\begin{tabular}{c} 1\\ \\ $[1,1,1]$ \end{tabular}\\  
       \hline		
\begin{tabular}{c}	
\centering
\scalebox{0.8}{
\begin{tikzpicture}
	\begin{pgfonlayer}{nodelayer}
		\node [style=redgauge] (0) at (-5, 0) {};
		\node [style=bluegauge] (1) at (-4, 0) {};
		\node [style=redgauge] (2) at (-3, 0) {};
		\node [style=bluegauge] (3) at (-2, 0) {};
		\node [style=redgauge] (9) at (-1, 0) {};
		\node [style=redgauge] (13) at (1, 0) {};
		\node [style=bluegauge] (14) at (2, 0) {};
		\node [style=redgauge] (15) at (3, 0) {};
		\node [style=bluegauge] (16) at (0, 0) {};
		\node [style=none] (29) at (-1, -0.5) {6};
		\node [style=none] (32) at (3, -0.5) {2};
		\node [style=none] (33) at (2, -0.5) {2};
		\node [style=none] (34) at (1, -0.5) {4};
		\node [style=none] (35) at (0, -0.5) {4};
		\node [style=none] (36) at (-5, -0.5) {2};
		\node [style=none] (37) at (-4, -0.5) {2};
		\node [style=none] (38) at (-3, -0.5) {4};
		\node [style=none] (39) at (-2, -0.5) {4};
		\node [style=bluegauge] (41) at (-2, 1) {};
		\node [style=none] (43) at (-1.5, 1) {4};
		\node [style=bluegauge] (45) at (0, 1) {};
		\node [style=none] (47) at (0, 1.5) {2};
		\node [style=none] (48) at (-2, 1.875) {$\Lambda^2$};
	\end{pgfonlayer}
	\begin{pgfonlayer}{edgelayer}
		\draw (3) to (2);
		\draw (2) to (1);
		\draw (0) to (1);
		\draw (13) to (14);
		\draw (14) to (15);
		\draw (16) to (13);
		\draw (3) to (9);
		\draw (9) to (41);
		\draw (9) to (16);
		\draw (9) to (45);
		\draw [in=135, out=45, loop] (41) to ();
	\end{pgfonlayer}
\end{tikzpicture}
}
\end{tabular}
	& 		
	\begin{tabular}{c}$\mathfrak{so}(6)^2$ \\ $\times$ \\ $\mathfrak{u}(1)^2$	\end{tabular}
	&	
	\begin{tabular}{c}
	\parbox[t]{5cm}
	{$1 + 32 t^2 + 32 t^3 + 532 t^4 + 1056 t^5 + 6754 t^6 + 18016 t^7 + 75262 t^8 + 219712 t^9 + 749188 t^{10}+ O\left(t^{12}\right)$}
	\end{tabular}
	&
	\begin{tabular}{c}
	\parbox[t]{3cm}
	{$32 t^2 + 32 t^3 + 4 t^4 + 32 t^5 + 114 t^6 - 32 t^7 - 788 t^8 - 1504 t^9 - 660 t^{10}+ O\left(t^{12}\right)$}
	\end{tabular}
	& 
	\begin{tabular}{c}
	$\mathbb{Z}_2$ \\ \\ $[2,1]$
	\end{tabular}\\
	\hline
\begin{tabular}{c}
		\centering
\scalebox{0.8}{
\begin{tikzpicture}
	\begin{pgfonlayer}{nodelayer}
		\node [style=redgauge] (0) at (-5, 0) {};
		\node [style=bluegauge] (1) at (-4, 0) {};
		\node [style=redgauge] (2) at (-3, 0) {};
		\node [style=bluegauge] (3) at (-2, 0) {};
		\node [style=redgauge] (9) at (-1, 0) {};
		\node [style=redgauge] (13) at (1, 0) {};
		\node [style=bluegauge] (14) at (2, 0) {};
		\node [style=redgauge] (15) at (3, 0) {};
		\node [style=bluegauge] (16) at (0, 0) {};
		\node [style=none] (29) at (-1, -0.5) {6};
		\node [style=none] (32) at (3, -0.5) {2};
		\node [style=none] (33) at (2, -0.5) {2};
		\node [style=none] (34) at (1, -0.5) {4};
		\node [style=none] (35) at (0, -0.5) {4};
		\node [style=none] (36) at (-5, -0.5) {2};
		\node [style=none] (37) at (-4, -0.5) {2};
		\node [style=none] (38) at (-3, -0.5) {4};
		\node [style=none] (39) at (-2, -0.5) {4};
		\node [style=bluegauge] (41) at (-1, 1) {};
		\node [style=none] (43) at (-0.5, 1) {6};
		\node [style=none] (48) at (-1, 1.875) {$\Lambda^2$};
	\end{pgfonlayer}
	\begin{pgfonlayer}{edgelayer}
		\draw (3) to (2);
		\draw (2) to (1);
		\draw (0) to (1);
		\draw (13) to (14);
		\draw (14) to (15);
		\draw (16) to (13);
		\draw (3) to (9);
		\draw (9) to (41);
		\draw (9) to (16);
		\draw [in=135, out=45, loop] (41) to ();
	\end{pgfonlayer}
\end{tikzpicture}
}
\end{tabular}
& 		
		\begin{tabular}{c}
		$\mathfrak{so}(6)^2$ \\ $\times$ \\ $\mathfrak{u}(1)$	
		\end{tabular}
&
  \begin{tabular}{c}  
  \parbox[t]{5cm}
  {$1 + 31 t^2 + 32 t^3 + 498 t^4 + 992 t^5 + 6089 t^6 + 15936 t^7 + 64733 t^8 + 183264 t^9 + 609515 t^{10}+O\left(t^{14}\right)$}
 \end{tabular}
 &	
 \begin{tabular}{c}
 \parbox[t]{3cm}
 {$31 t^2 + 32 t^3 + 2 t^4 + 43 t^6 - 339 t^8 - 672 t^9 + 29 t^{10} + O \left(t^{12}\right)$}
 \end{tabular}
 & \begin{tabular}{c} $S_3$ \\ \\$[3]$ \end{tabular}\\
 \hline	
	\end{tabular}
\end{adjustbox}
\caption[Bouquet quivers of 3 USp(2) nodes and SO(6) central node]{Magnetic quivers corresponding to the $n=3$, $m=3$ cases of (\ref{fullanti}), with different discrete quotients fixed by the choice of partition. In all cases $\text{dim}_{\mathbb{H}} (\mathcal{C})=18$. The Coulomb branch Hilbert series and their plethystic logarithms are shown along with their global symmetries. The discrete symmetry gauge groups are also shown. Palindromic numerator terms ${\cal P}_k(t)$ are given in Appendix \ref{apx:PP}.
}
\label{D3quotientm3}
\end{figure}

\begin{figure}[!ht]
\small
\begin{adjustbox}{center}
	\begin{tabular}{|c|c|c|c|c|}
		\hline
		Quiver & \begin{tabular}{c}Global\\Symmetry \end{tabular} &
		Hilbert Series  &
		\begin{tabular}{c}	Plethystic\\Logarithm \end{tabular}&
		\begin{tabular}{c} Discrete\\ Quotient \end{tabular}\\ 
		\hline
\begin{tabular}{c}
\centering
\scalebox{0.8}{
\begin{tikzpicture}
	\begin{pgfonlayer}{nodelayer}
		\node [style=redgauge] (0) at (-5, 0) {};
		\node [style=bluegauge] (1) at (-4, 0) {};
		\node [style=redgauge] (2) at (-3, 0) {};
		\node [style=bluegauge] (3) at (-2, 0) {};
		\node [style=redgauge] (9) at (-1, 0) {};
		\node [style=redgauge] (13) at (1, 0) {};
		\node [style=bluegauge] (14) at (2, 0) {};
		\node [style=redgauge] (15) at (3, 0) {};
		\node [style=bluegauge] (16) at (0, 0) {};
		\node [style=none] (29) at (-1, -0.5) {6};
		\node [style=none] (32) at (3, -0.5) {2};
		\node [style=none] (33) at (2, -0.5) {2};
		\node [style=none] (34) at (1, -0.5) {4};
		\node [style=none] (35) at (0, -0.5) {4};
		\node [style=none] (36) at (-5, -0.5) {2};
		\node [style=none] (37) at (-4, -0.5) {2};
		\node [style=none] (38) at (-3, -0.5) {4};
		\node [style=none] (39) at (-2, -0.5) {4};
		\node [style=bluegauge] (40) at (-2.25, 1) {};
		\node [style=bluegauge] (41) at (-1.425, 1) {};
		\node [style=bluegauge] (42) at (-0.575, 1) {};
		\node [style=bluegauge] (43) at (0.25, 1) {};
		\node [style=none] (44) at (-2.25, 1.5) {2};
		\node [style=none] (45) at (-1.425, 1.5) {2};
		\node [style=none] (46) at (-0.55, 1.5) {2};
		\node [style=none] (47) at (0.25, 1.5) {2};
	\end{pgfonlayer}
	\begin{pgfonlayer}{edgelayer}
		\draw (3) to (2);
		\draw (2) to (1);
		\draw (0) to (1);
		\draw (13) to (14);
		\draw (14) to (15);
		\draw (16) to (13);
		\draw (3) to (9);
		\draw (9) to (16);
		\draw (40) to (9);
		\draw (9) to (41);
		\draw (42) to (9);
		\draw (9) to (43);
	\end{pgfonlayer}
\end{tikzpicture}}
\end{tabular}
& 		
\begin{tabular}{c}$\mathfrak{so}(6)^2$ \\ $\times$\\ $\mathfrak{u}(1)^4$\end{tabular}	
&
\begin{tabular}{l}
       \parbox[t]{5cm}{{
       {\footnotesize {
       $\dfrac{{\cal P}_{240}(t)}{(1 - t^4)^4 (1 - t^6)^{12} (1 - t^8)^{15} (1 - t^{10})^7}$}}\\
       \\
       $= 1 + 34 t^2 + 634 t^4 + 8561 t^6 + 93664 t^8 + 878472 t^{10} + 7285312 t^{12} + 54410497 t^{14} + 370331950 t^{16} + 2316834748 t^{18} + 13413526606 t^{20} + O\left(t^{22}\right)$
       }}\\
       \end{tabular}
       & 
       \begin{tabular}{c}
       \parbox{3cm}{$34 t^2 + 39 t^4 + 95 t^6 + 404 t^8 - 2416 t^{10}+O\left(t^{12}\right)$} \end{tabular} &\begin{tabular}{c} 1\\ \\$[1,1,1,1]$\end{tabular} \\
       \hline		
\begin{tabular}{c}	
\centering
\scalebox{0.8}{
\begin{tikzpicture}
	\begin{pgfonlayer}{nodelayer}
		\node [style=redgauge] (0) at (-5, 0) {};
		\node [style=bluegauge] (1) at (-4, 0) {};
		\node [style=redgauge] (2) at (-3, 0) {};
		\node [style=bluegauge] (3) at (-2, 0) {};
		\node [style=redgauge] (9) at (-1, 0) {};
		\node [style=redgauge] (13) at (1, 0) {};
		\node [style=bluegauge] (14) at (2, 0) {};
		\node [style=redgauge] (15) at (3, 0) {};
		\node [style=bluegauge] (16) at (0, 0) {};
		\node [style=none] (29) at (-1, -0.5) {6};
		\node [style=none] (32) at (3, -0.5) {2};
		\node [style=none] (33) at (2, -0.5) {2};
		\node [style=none] (34) at (1, -0.5) {4};
		\node [style=none] (35) at (0, -0.5) {4};
		\node [style=none] (36) at (-5, -0.5) {2};
		\node [style=none] (37) at (-4, -0.5) {2};
		\node [style=none] (38) at (-3, -0.5) {4};
		\node [style=none] (39) at (-2, -0.5) {4};
		\node [style=bluegauge] (40) at (-1.75, 1) {};
		\node [style=bluegauge] (42) at (-0.575, 1) {};
		\node [style=bluegauge] (43) at (0.25, 1) {};
		\node [style=none] (44) at (-1.25, 1) {4};
		\node [style=none] (46) at (-0.55, 1.5) {2};
		\node [style=none] (47) at (0.25, 1.5) {2};
		\node [style=none] (48) at (-1.75, 1.875) {$\Lambda^2$};
	\end{pgfonlayer}
	\begin{pgfonlayer}{edgelayer}
		\draw (3) to (2);
		\draw (2) to (1);
		\draw (0) to (1);
		\draw (13) to (14);
		\draw (14) to (15);
		\draw (16) to (13);
		\draw (3) to (9);
		\draw (9) to (16);
		\draw (40) to (9);
		\draw (42) to (9);
		\draw (9) to (43);
		\draw [in=135, out=45, loop] (40) to ();
	\end{pgfonlayer}
\end{tikzpicture}
}
\end{tabular}
	& 		
	\centering	\begin{tabular}{c}	$\mathfrak{so}(6)^2$ \\ $\times$ \\$\mathfrak{u}(1)^3$	\end{tabular}&	\begin{tabular}{c}  \parbox[t]{5cm}
	{$1 + 33 t^2 + 599 t^4 + 7864 t^6 + 83405 t^8 + 756264 t^{10} + 6054943 t^{12} + 43659582 t^{14} + 287234433 t^{16} + 1740270395 t^{18} + 9779780797 t^{20}+O\left(t^{22}\right)$}\end{tabular}
	&
	\begin{tabular}{c}
	\parbox[t]{3cm}
	{$33 t^2 + 38 t^4 + 65 t^6 + 296 t^8 - 1499 t^{10}+O\left(t^{12}\right)$}
	\end{tabular} 
	&
	\begin{tabular}{c}
	$\mathbb{Z}_2$\\ \\ $[2,1,1]$
	\end{tabular} \\
	\hline			 
	\begin{tabular}{c}
		\centering
		\scalebox{0.8}{
\begin{tikzpicture}
	\begin{pgfonlayer}{nodelayer}
		\node [style=redgauge] (0) at (-5, 0) {};
		\node [style=bluegauge] (1) at (-4, 0) {};
		\node [style=redgauge] (2) at (-3, 0) {};
		\node [style=bluegauge] (3) at (-2, 0) {};
		\node [style=redgauge] (9) at (-1, 0) {};
		\node [style=redgauge] (13) at (1, 0) {};
		\node [style=bluegauge] (14) at (2, 0) {};
		\node [style=redgauge] (15) at (3, 0) {};
		\node [style=bluegauge] (16) at (0, 0) {};
		\node [style=none] (29) at (-1, -0.5) {6};
		\node [style=none] (32) at (3, -0.5) {2};
		\node [style=none] (33) at (2, -0.5) {2};
		\node [style=none] (34) at (1, -0.5) {4};
		\node [style=none] (35) at (0, -0.5) {4};
		\node [style=none] (36) at (-5, -0.5) {2};
		\node [style=none] (37) at (-4, -0.5) {2};
		\node [style=none] (38) at (-3, -0.5) {4};
		\node [style=none] (39) at (-2, -0.5) {4};
		\node [style=bluegauge] (40) at (-1.75, 1) {};
		\node [style=bluegauge] (42) at (-0.325, 1) {};
		\node [style=none] (44) at (-1.25, 1) {4};
		\node [style=none] (46) at (0.2, 1) {4};
		\node [style=none] (48) at (-1.75, 1.875) {$\Lambda^2$};
		\node [style=none] (49) at (-0.25, 1.875) {$\Lambda^2$};
	\end{pgfonlayer}
	\begin{pgfonlayer}{edgelayer}
		\draw (3) to (2);
		\draw (2) to (1);
		\draw (0) to (1);
		\draw (13) to (14);
		\draw (14) to (15);
		\draw (16) to (13);
		\draw (3) to (9);
		\draw (9) to (16);
		\draw (40) to (9);
		\draw (42) to (9);
		\draw [in=135, out=45, loop] (40) to ();
		\draw [in=135, out=45, loop] (42) to ();
	\end{pgfonlayer}
\end{tikzpicture}
}
\end{tabular}	& 		
		\centering	\begin{tabular}{c} 	$\mathfrak{so}(6)^2$\\ $\times$ \\$\mathfrak{u}(1)^2$	\end{tabular}	& 
  \begin{tabular}{c}
  \parbox[t]{5cm}{$1 + 32 t^2 + 565 t^4 + 7203 t^6 + 73975 t^8 + 648013 t^{10} + 
 5007177 t^{12} + 34855370 t^{14} + 221675879 t^{16} + 1300945624 t^{18} + 
 7098111346 t^{20}+ O\left(t^{22}\right)$}
 \end{tabular}&	
 \begin{tabular}{c}
 \parbox[t]{3cm}{$32 t^2 + 37 t^4 + 35 t^6 + 256 t^8 - 850 t^{10}+O\left(t^{12}\right)$}
 \end{tabular}
 &
 \begin{tabular}{c}
 $\mathbb{Z}_2 \times \mathbb{Z}_2$ \\ \\$[2,2]$ 
  \end{tabular}\\
 \hline	
 \begin{tabular}{c}
 \scalebox{0.8}{
		\centering
\begin{tikzpicture}
	\begin{pgfonlayer}{nodelayer}
		\node [style=redgauge] (0) at (-5, 0) {};
		\node [style=bluegauge] (1) at (-4, 0) {};
		\node [style=redgauge] (2) at (-3, 0) {};
		\node [style=bluegauge] (3) at (-2, 0) {};
		\node [style=redgauge] (9) at (-1, 0) {};
		\node [style=redgauge] (13) at (1, 0) {};
		\node [style=bluegauge] (14) at (2, 0) {};
		\node [style=redgauge] (15) at (3, 0) {};
		\node [style=bluegauge] (16) at (0, 0) {};
		\node [style=none] (29) at (-1, -0.5) {6};
		\node [style=none] (32) at (3, -0.5) {2};
		\node [style=none] (33) at (2, -0.5) {2};
		\node [style=none] (34) at (1, -0.5) {4};
		\node [style=none] (35) at (0, -0.5) {4};
		\node [style=none] (36) at (-5, -0.5) {2};
		\node [style=none] (37) at (-4, -0.5) {2};
		\node [style=none] (38) at (-3, -0.5) {4};
		\node [style=none] (39) at (-2, -0.5) {4};
		\node [style=bluegauge] (40) at (-1.75, 1) {};
		\node [style=bluegauge] (42) at (-0.325, 1) {};
		\node [style=none] (44) at (-1.25, 1) {6};
		\node [style=none] (46) at (-0.3, 1.5) {2};
		\node [style=none] (48) at (-1.75, 1.875) {$\Lambda^2$};
	\end{pgfonlayer}
	\begin{pgfonlayer}{edgelayer}
		\draw (3) to (2);
		\draw (2) to (1);
		\draw (0) to (1);
		\draw (13) to (14);
		\draw (14) to (15);
		\draw (16) to (13);
		\draw (3) to (9);
		\draw (9) to (16);
		\draw (40) to (9);
		\draw (42) to (9);
		\draw [in=135, out=45, loop] (40) to ();
	\end{pgfonlayer}
\end{tikzpicture}}
\end{tabular}	& 		
		\centering	\begin{tabular}{c} 	$\mathfrak{so}(6)^2$\\ $\times$ \\$\mathfrak{u}(1)^2$	\end{tabular}	& 
  \begin{tabular}{c}  \parbox[t]{5cm}
  {$1 + 32 t^2 + 564 t^4 + 7170 t^6 + 73277 t^8 + 637078 t^{10} + 4872441 t^{12} + 33491820 t^{14} + 209952106 t^{16} + 1213058254 t^{18} +  6512119045 t^{20}+ O\left(t^{22}\right)$}
  \end{tabular}&
  \begin{tabular}{c}
    \parbox[t]{3cm}
    {$32 t^2 + 36 t^4 + 34 t^6 + 155 t^8 - 786 t^{10}+ O\left(t^{12}\right)$}
  \end{tabular}
  & \begin{tabular}{c} $S_3$\\ \\  $[3,1]$ \end{tabular}\\
  \hline	
 \begin{tabular}{c}
		\centering
\scalebox{0.8}{
\begin{tikzpicture}
	\begin{pgfonlayer}{nodelayer}
		\node [style=redgauge] (0) at (-5, 0) {};
		\node [style=bluegauge] (1) at (-4, 0) {};
		\node [style=redgauge] (2) at (-3, 0) {};
		\node [style=bluegauge] (3) at (-2, 0) {};
		\node [style=redgauge] (9) at (-1, 0) {};
		\node [style=redgauge] (13) at (1, 0) {};
		\node [style=bluegauge] (14) at (2, 0) {};
		\node [style=redgauge] (15) at (3, 0) {};
		\node [style=bluegauge] (16) at (0, 0) {};
		\node [style=none] (29) at (-1, -0.5) {6};
		\node [style=none] (32) at (3, -0.5) {2};
		\node [style=none] (33) at (2, -0.5) {2};
		\node [style=none] (34) at (1, -0.5) {4};
		\node [style=none] (35) at (0, -0.5) {4};
		\node [style=none] (36) at (-5, -0.5) {2};
		\node [style=none] (37) at (-4, -0.5) {2};
		\node [style=none] (38) at (-3, -0.5) {4};
		\node [style=none] (39) at (-2, -0.5) {4};
		\node [style=bluegauge] (41) at (-1, 1) {};
		\node [style=none] (43) at (-0.5, 1) {8};
		\node [style=none] (48) at (-1, 1.875) {$\Lambda^2$};
	\end{pgfonlayer}
	\begin{pgfonlayer}{edgelayer}
		\draw (3) to (2);
		\draw (2) to (1);
		\draw (0) to (1);
		\draw (13) to (14);
		\draw (14) to (15);
		\draw (16) to (13);
		\draw (3) to (9);
		\draw (9) to (41);
		\draw (9) to (16);
		\draw [in=135, out=45, loop] (41) to ();
	\end{pgfonlayer}
\end{tikzpicture}}
\end{tabular}	& 		
		\centering
		\begin{tabular}{c}
		$\mathfrak{so}(6)^2$\\ $\times$ \\$\mathfrak{u}(1)$	\end{tabular}	& 
  \begin{tabular}{c}  \parbox[t]{5cm}
  {$1 + 31 t^2 + 530 t^4 + 7145 t^6 + 82749 t^8 + 821163 t^{10} +  7016522 t^{12} + 52556792 t^{14} + 351654724 t^{16} + 2134233829 t^{18} +  11889073429 t^{20}+ O\left(t^{22}\right)$}
 \end{tabular}&	
 \begin{tabular}{c}
 \parbox[t]{3cm}
 {$31 t^2 + 34 t^4 + 635 t^6 - 771 t^8 - 20067 t^{10}+O\left(t^{12}\right)$}
 \end{tabular}&
  \begin{tabular}{c}
 {$S_4$}\\ \\ $[4]$
  \end{tabular}\\
 \hline	
	\end{tabular}
\end{adjustbox}
\caption[Bouquet quivers of 4 USp(2) nodes and SO(6) central node]{Magnetic quivers corresponding to the $n=3,m=4$ cases of (\ref{fullanti}), with different discrete quotients fixed by the choice of partition. In all cases $\text{dim}_{\mathbb{H}} (\mathcal{C})=19$. The Coulomb branch Hilbert series and their plethystic logarithms are shown along with their global symmetries. The discrete symmetry gauge groups are also shown. Palindromic numerator terms ${\cal P}_k(t)$ are given in Appendix \ref{apx:PP}.
}
\label{D3quotientm4}
\end{figure}
\end{landscape}

\begin{landscape}

\begin{figure}[!ht]
\small
\begin{adjustbox}{center}
	\begin{tabular}{|c|c|c|c|c|}
		\hline
		Quiver	& 
		\begin{tabular}{c} 	Global\\Symmetry \end{tabular}&
		Hilbert Series &
		\begin{tabular}{c} Plethystic \\Logarithm \end{tabular}&
		\begin{tabular}{c} Discrete\\ Quotient \end{tabular}\\
		\hline
\begin{tabular}{c}
\centering
\scalebox{0.7}{
\begin{tikzpicture}
	\begin{pgfonlayer}{nodelayer}
		\node [style=redgauge] (0) at (-5, 0) {};
		\node [style=bluegauge] (1) at (-4, 0) {};
		\node [style=redgauge] (2) at (-3, 0) {};
		\node [style=bluegauge] (3) at (-2, 0) {};
		\node [style=redgauge] (9) at (-1, 0) {};
		\node [style=redgauge] (13) at (1, 0) {};
		\node [style=bluegauge] (14) at (2, 0) {};
		\node [style=redgauge] (15) at (3, 0) {};
		\node [style=bluegauge] (16) at (0, 0) {};
		\node [style=none] (29) at (-1, -0.5) {8};
		\node [style=none] (32) at (3, -0.5) {4};
		\node [style=none] (33) at (2, -0.5) {4};
		\node [style=none] (34) at (1, -0.5) {6};
		\node [style=none] (35) at (0, -0.5) {6};
		\node [style=none] (36) at (-5, -0.5) {4};
		\node [style=none] (37) at (-4, -0.5) {4};
		\node [style=none] (38) at (-3, -0.5) {6};
		\node [style=none] (39) at (-2, -0.5) {6};
		\node [style=bluegauge] (40) at (4, 0) {};
		\node [style=redgauge] (41) at (5, 0) {};
		\node [style=none] (42) at (5, -0.5) {2};
		\node [style=none] (43) at (4, -0.5) {2};
		\node [style=redgauge] (44) at (-7, 0) {};
		\node [style=bluegauge] (45) at (-6, 0) {};
		\node [style=none] (46) at (-7, -0.5) {2};
		\node [style=none] (47) at (-6, -0.5) {2};
		\node [style=bluegauge] (48) at (-2, 1) {};
		\node [style=bluegauge] (49) at (0, 1) {};
		\node [style=none] (50) at (-2, 1.5) {2};
		\node [style=none] (51) at (0, 1.5) {2};
	\end{pgfonlayer}
	\begin{pgfonlayer}{edgelayer}
		\draw (3) to (2);
		\draw (2) to (1);
		\draw (0) to (1);
		\draw (13) to (14);
		\draw (14) to (15);
		\draw (16) to (13);
		\draw (3) to (9);
		\draw (9) to (16);
		\draw (40) to (41);
		\draw (15) to (40);
		\draw (44) to (45);
		\draw (45) to (0);
		\draw (49) to (9);
		\draw (9) to (48);
	\end{pgfonlayer}
\end{tikzpicture}}
\end{tabular}
& 		
\centering	
\begin{tabular}{c}
$\mathfrak{so}(8)^2$
\end{tabular}
&
\begin{tabular}{l}
       \parbox[t]{4.5cm}
      { { 
      {
     $\dfrac
     {{\cal P}_{172}(t)}
      {(1 - t^2)^{17} (1 - t^4)^{30} (1 - t^6)^{13}}$}}\\
      \\
    {$= 1 + 56 t^2 + 1791 t^4 + 41970 t^6 + 793817 t^8 + 12731945 t^{10} + 178313546 t^{12} + 2222731227 t^{14} + 24994138263 t^{16} +  256098078865 t^{18} + 2410109564419 t^{20} + O\left(t^{22}\right)$
    }}\\
       \end{tabular}
       & 
\begin{tabular}{c}
       \parbox{2cm}
       {$56 t^2 + 195 t^4 + 194 t^6 - 2503 t^8 - 23933 t^{10}+O\left(t^{12}\right)$} 
       \end{tabular} 
       &\begin{tabular}{c} 1\\ \\ $[1,1]$ \end{tabular} \\
       \hline		
\begin{tabular}{c}	
\centering
\scalebox{0.7}{
\begin{tikzpicture}
	\begin{pgfonlayer}{nodelayer}
		\node [style=redgauge] (0) at (-5, 0) {};
		\node [style=bluegauge] (1) at (-4, 0) {};
		\node [style=redgauge] (2) at (-3, 0) {};
		\node [style=bluegauge] (3) at (-2, 0) {};
		\node [style=redgauge] (9) at (-1, 0) {};
		\node [style=redgauge] (13) at (1, 0) {};
		\node [style=bluegauge] (14) at (2, 0) {};
		\node [style=redgauge] (15) at (3, 0) {};
		\node [style=bluegauge] (16) at (0, 0) {};
		\node [style=none] (29) at (-1, -0.5) {8};
		\node [style=none] (32) at (3, -0.5) {4};
		\node [style=none] (33) at (2, -0.5) {4};
		\node [style=none] (34) at (1, -0.5) {6};
		\node [style=none] (35) at (0, -0.5) {6};
		\node [style=none] (36) at (-5, -0.5) {4};
		\node [style=none] (37) at (-4, -0.5) {4};
		\node [style=none] (38) at (-3, -0.5) {6};
		\node [style=none] (39) at (-2, -0.5) {6};
		\node [style=bluegauge] (40) at (4, 0) {};
		\node [style=redgauge] (41) at (5, 0) {};
		\node [style=none] (42) at (5, -0.5) {2};
		\node [style=none] (43) at (4, -0.5) {2};
		\node [style=redgauge] (44) at (-7, 0) {};
		\node [style=bluegauge] (45) at (-6, 0) {};
		\node [style=none] (46) at (-7, -0.5) {2};
		\node [style=none] (47) at (-6, -0.5) {2};
		\node [style=bluegauge] (48) at (-1, 1) {};
		\node [style=none] (50) at (-0.5, 1) {4};
		\node [style=none] (51) at (-1, 1.85) {$\Lambda^2$};
	\end{pgfonlayer}
	\begin{pgfonlayer}{edgelayer}
		\draw (3) to (2);
		\draw (2) to (1);
		\draw (0) to (1);
		\draw (13) to (14);
		\draw (14) to (15);
		\draw (16) to (13);
		\draw (3) to (9);
		\draw (9) to (16);
		\draw (40) to (41);
		\draw (15) to (40);
		\draw (44) to (45);
		\draw (45) to (0);
		\draw (9) to (48);
		\draw [in=135, out=45, loop] (48) to ();
	\end{pgfonlayer}
\end{tikzpicture}}
\end{tabular}
	& 		
	\centering	\begin{tabular}{c}	$\mathfrak{so}(8)^2 $
	\end{tabular}&	
	\begin{tabular}{c}  
	\parbox[t]{4.5cm}
	{$1 + 56 t^2 + 1789 t^4 + 41665 t^6 + 778839 t^8 + 12284023 t^{10} + 168694837 t^{12} + 2061512225 t^{14} + 22768693238 t^{16} + 229816006542 t^{18} + 2137060265760 t^{20}+ O\left(t^{22}\right)$}
	\end{tabular}
	&
	\begin{tabular}{c}
	\parbox[t]{2cm}
	{$56 t^2 + 193 t^4 + t^6 - 3092 t^8 - 9710 t^{10}+ O\left(t^{12}\right)$}
	\end{tabular} 
	& 
	\begin{tabular}{c}
	$\mathbb{Z}_2$\\ \\$[2]$
		\end{tabular} \\
	\hline
	\end{tabular}
\end{adjustbox}
\caption[Bouquet quivers of 2 USp(2) nodes and SO(8) central node]{Magnetic quivers corresponding to the $n=4$, $m=2$ cases of (\ref{fullanti}), with different discrete quotients fixed by the choice of partition. In all cases $\text{dim}_{\mathbb{H}} (\mathcal{C})=30$. The Coulomb branch Hilbert series and their plethystic logarithms are shown along with their global symmetries. The discrete symmetry gauge groups are also shown. Palindromic numerator terms ${\cal P}_k(t)$ are given in Appendix \ref{apx:PP}.
}
\label{D4quotientm2}
\end{figure}


\begin{figure}[!ht]
\small
{\begin{adjustbox}{center}
	\begin{tabular}{|c|c|c|c|c|}
		\hline
		Quiver &
		\begin{tabular}{c}	Global\\Symmetry \end{tabular} & 
		Hilbert Series &
		\begin{tabular}{c}	Plethystic\\Logarithm \end{tabular} &
		\begin{tabular}{c}	Discrete\\Quotient \end{tabular}\\ 
		\hline
\begin{tabular}{c}
\centering
\scalebox{0.7}{
\begin{tikzpicture}
	\begin{pgfonlayer}{nodelayer}
		\node [style=redgauge] (0) at (-5, 0) {};
		\node [style=bluegauge] (1) at (-4, 0) {};
		\node [style=redgauge] (2) at (-3, 0) {};
		\node [style=bluegauge] (3) at (-2, 0) {};
		\node [style=redgauge] (9) at (-1, 0) {};
		\node [style=redgauge] (13) at (1, 0) {};
		\node [style=bluegauge] (14) at (2, 0) {};
		\node [style=redgauge] (15) at (3, 0) {};
		\node [style=bluegauge] (16) at (0, 0) {};
		\node [style=none] (29) at (-1, -0.5) {8};
		\node [style=none] (32) at (3, -0.5) {4};
		\node [style=none] (33) at (2, -0.5) {4};
		\node [style=none] (34) at (1, -0.5) {6};
		\node [style=none] (35) at (0, -0.5) {6};
		\node [style=none] (36) at (-5, -0.5) {4};
		\node [style=none] (37) at (-4, -0.5) {4};
		\node [style=none] (38) at (-3, -0.5) {6};
		\node [style=none] (39) at (-2, -0.5) {6};
		\node [style=bluegauge] (40) at (4, 0) {};
		\node [style=redgauge] (41) at (5, 0) {};
		\node [style=none] (42) at (5, -0.5) {2};
		\node [style=none] (43) at (4, -0.5) {2};
		\node [style=redgauge] (44) at (-7, 0) {};
		\node [style=bluegauge] (45) at (-6, 0) {};
		\node [style=none] (46) at (-7, -0.5) {2};
		\node [style=none] (47) at (-6, -0.5) {2};
		\node [style=bluegauge] (48) at (-2, 1) {};
		\node [style=bluegauge] (49) at (-1, 1) {};
		\node [style=bluegauge] (50) at (0, 1) {};
		\node [style=none] (51) at (0, 1.5) {2};
		\node [style=none] (52) at (-1, 1.5) {2};
		\node [style=none] (53) at (-2, 1.5) {2};
	\end{pgfonlayer}
	\begin{pgfonlayer}{edgelayer}
		\draw (3) to (2);
		\draw (2) to (1);
		\draw (0) to (1);
		\draw (13) to (14);
		\draw (14) to (15);
		\draw (16) to (13);
		\draw (3) to (9);
		\draw (9) to (16);
		\draw (40) to (41);
		\draw (15) to (40);
		\draw (44) to (45);
		\draw (45) to (0);
		\draw (48) to (9);
		\draw (9) to (49);
		\draw (50) to (9);
	\end{pgfonlayer}
\end{tikzpicture}}
\end{tabular}
& 
\begin{tabular}{c}	$\mathfrak{so}(8)^2$ \end{tabular}	
&
\begin{tabular}{l}
       \parbox[t]{4.5cm}
       {{$1 + 56 t^2 + 1601 t^4 + 31331 t^6 + 474423 t^8 + 5950219 t^{10} + 64526860 t^{12} + 622861123 t^{14} + 5461745110 t^{16} + 44149380513 t^{18} + 332528337809 t^{20} + O\left(t^{22}\right)$} }\\
       \end{tabular}& 
       \begin{tabular}{c}
       \parbox{2cm}{$56t^2 + 5 t^4 + 195 t^6 + 382 t^8 - 329 t^{12}+ O\left(t^{12}\right)$} 
       \end{tabular} & 
       \begin{tabular}{c}
       1\\  \\$[1,1,1]$
        \end{tabular}\\
       \hline		
\begin{tabular}{c}	
\centering
\scalebox{0.7}{
\begin{tikzpicture}
	\begin{pgfonlayer}{nodelayer}
		\node [style=redgauge] (0) at (-5, 0) {};
		\node [style=bluegauge] (1) at (-4, 0) {};
		\node [style=redgauge] (2) at (-3, 0) {};
		\node [style=bluegauge] (3) at (-2, 0) {};
		\node [style=redgauge] (9) at (-1, 0) {};
		\node [style=redgauge] (13) at (1, 0) {};
		\node [style=bluegauge] (14) at (2, 0) {};
		\node [style=redgauge] (15) at (3, 0) {};
		\node [style=bluegauge] (16) at (0, 0) {};
		\node [style=none] (29) at (-1, -0.5) {8};
		\node [style=none] (32) at (3, -0.5) {4};
		\node [style=none] (33) at (2, -0.5) {4};
		\node [style=none] (34) at (1, -0.5) {6};
		\node [style=none] (35) at (0, -0.5) {6};
		\node [style=none] (36) at (-5, -0.5) {4};
		\node [style=none] (37) at (-4, -0.5) {4};
		\node [style=none] (38) at (-3, -0.5) {6};
		\node [style=none] (39) at (-2, -0.5) {6};
		\node [style=bluegauge] (40) at (4, 0) {};
		\node [style=redgauge] (41) at (5, 0) {};
		\node [style=none] (42) at (5, -0.5) {2};
		\node [style=none] (43) at (4, -0.5) {2};
		\node [style=redgauge] (44) at (-7, 0) {};
		\node [style=bluegauge] (45) at (-6, 0) {};
		\node [style=none] (46) at (-7, -0.5) {2};
		\node [style=none] (47) at (-6, -0.5) {2};
		\node [style=bluegauge] (48) at (-2, 1) {};
		\node [style=none] (50) at (-1.5, 1) {4};
		\node [style=none] (51) at (-2, 1.85) {$\Lambda^2$};
		\node [style=miniBlue] (52) at (0, 1) {};
		\node [style=none] (53) at (0, 1.5) {2};
	\end{pgfonlayer}
	\begin{pgfonlayer}{edgelayer}
		\draw (3) to (2);
		\draw (2) to (1);
		\draw (0) to (1);
		\draw (13) to (14);
		\draw (14) to (15);
		\draw (16) to (13);
		\draw (3) to (9);
		\draw (9) to (16);
		\draw (40) to (41);
		\draw (15) to (40);
		\draw (44) to (45);
		\draw (45) to (0);
		\draw (9) to (48);
		\draw [in=135, out=45, loop] (48) to ();
		\draw (52) to (9);
	\end{pgfonlayer}
\end{tikzpicture}}
\end{tabular}
	& 		
	\centering	\begin{tabular}{c}	$\mathfrak{so}(8)^2$ \end{tabular}
	&
	\begin{tabular}{c} 
	\parbox[t]{4.5cm}{$1 + 56 t^2 + 1599 t^4 + 31218 t^6 + 470977 t^8 + 5875432 t^{10} + 63245499 t^{12} + 604592195 t^{14} + 5237830546 t^{16} + 41735338778 t^{18} + 309233426478 t^{20}+ O\left(t^{22}\right)$}
	\end{tabular}
	&
	\begin{tabular}{c}
	\parbox[t]{2cm}
	{$56 t^2 + 3 t^4 + 194 t^6 + 193 t^8 + 2 t^{10} + O\left(t^{12}\right)$}
	\end{tabular} &
	\begin{tabular}{c}
	$\mathbb{Z}_2$\\ \\$[2,1]$
	\end{tabular} \\
	\hline
\begin{tabular}{c}	
\centering
\scalebox{0.7}{
\begin{tikzpicture}
	\begin{pgfonlayer}{nodelayer}
		\node [style=redgauge] (0) at (-5, 0) {};
		\node [style=bluegauge] (1) at (-4, 0) {};
		\node [style=redgauge] (2) at (-3, 0) {};
		\node [style=bluegauge] (3) at (-2, 0) {};
		\node [style=redgauge] (9) at (-1, 0) {};
		\node [style=redgauge] (13) at (1, 0) {};
		\node [style=bluegauge] (14) at (2, 0) {};
		\node [style=redgauge] (15) at (3, 0) {};
		\node [style=bluegauge] (16) at (0, 0) {};
		\node [style=none] (29) at (-1, -0.5) {8};
		\node [style=none] (32) at (3, -0.5) {4};
		\node [style=none] (33) at (2, -0.5) {4};
		\node [style=none] (34) at (1, -0.5) {6};
		\node [style=none] (35) at (0, -0.5) {6};
		\node [style=none] (36) at (-5, -0.5) {4};
		\node [style=none] (37) at (-4, -0.5) {4};
		\node [style=none] (38) at (-3, -0.5) {6};
		\node [style=none] (39) at (-2, -0.5) {6};
		\node [style=bluegauge] (40) at (4, 0) {};
		\node [style=redgauge] (41) at (5, 0) {};
		\node [style=none] (42) at (5, -0.5) {2};
		\node [style=none] (43) at (4, -0.5) {2};
		\node [style=redgauge] (44) at (-7, 0) {};
		\node [style=bluegauge] (45) at (-6, 0) {};
		\node [style=none] (46) at (-7, -0.5) {2};
		\node [style=none] (47) at (-6, -0.5) {2};
		\node [style=bluegauge] (48) at (-1, 1) {};
		\node [style=none] (50) at (-0.5, 1) {6};
		\node [style=none] (51) at (-1, 1.85) {$\Lambda^2$};
	\end{pgfonlayer}
	\begin{pgfonlayer}{edgelayer}
		\draw (3) to (2);
		\draw (2) to (1);
		\draw (0) to (1);
		\draw (13) to (14);
		\draw (14) to (15);
		\draw (16) to (13);
		\draw (3) to (9);
		\draw (9) to (16);
		\draw (40) to (41);
		\draw (15) to (40);
		\draw (44) to (45);
		\draw (45) to (0);
		\draw (9) to (48);
		\draw [in=135, out=45, loop] (48) to ();
	\end{pgfonlayer}
\end{tikzpicture}}
\end{tabular}
	& 		
	\centering	
	\begin{tabular}{c}	$\mathfrak{so}(8)^2$ \end{tabular}
	&
	\begin{tabular}{c}  
	\parbox[t]{4.5cm}
	{$1 + 56 t^2 + 1597 t^4 + 31105 t^6 + 467532 t^8 + 5800703 t^{10} + 61966235 t^{12} + 586380081 t^{14} + 5015120160 t^{16} + 39341912485 t^{18} + 286232448834 t^{20}+ O\left(t^{22}\right)$}
	\end{tabular}
	&
	\begin{tabular}{c}
	\parbox[t]{2cm}
	{$56 t^2 + t^4 + 193 t^6 + t^8 + 2 t^{10} +O\left(t^{12}\right)$}
	\end{tabular} 
	&
	\begin{tabular}{c}
	    $S_3$\\ \\$[3]$
	\end{tabular} \\
	\hline
	\end{tabular}
\end{adjustbox}}
\caption[Bouquet quivers of 3 USp(2) nodes and SO(8) central node]{Magnetic quivers corresponding to the $n=4$, $m=3$ cases of (\ref{fullanti}), with different discrete quotients fixed by the choice of partition. In all cases $\text{dim}_{\mathbb{H}} (\mathcal{C})=31$. The Coulomb branch Hilbert series and their plethystic logarithms are shown along with their global symmetries. The discrete symmetry gauge groups are also shown.}
\label{D4quotientm3}
\end{figure}


\begin{figure}[!ht]
\small
\begin{adjustbox}{center}
	\begin{tabular}{|c|c|c|c|c|}
		\hline
		Quiver &
		\begin{tabular}{c}	Global\\Symmetry \end{tabular} & 
		Hilbert Series &
		\begin{tabular}{c}Plethystic\\Logarithm\end{tabular} &
		\begin{tabular}{c}Discrete\\Quotient \end{tabular}\\ 
		\hline
\begin{tabular}{c}
\centering
\scalebox{0.7}{
\begin{tikzpicture}
	\begin{pgfonlayer}{nodelayer}
		\node [style=redgauge] (0) at (-5, 0) {};
		\node [style=bluegauge] (1) at (-4, 0) {};
		\node [style=redgauge] (2) at (-3, 0) {};
		\node [style=bluegauge] (3) at (-2, 0) {};
		\node [style=redgauge] (9) at (-1, 0) {};
		\node [style=redgauge] (13) at (1, 0) {};
		\node [style=bluegauge] (14) at (2, 0) {};
		\node [style=redgauge] (15) at (3, 0) {};
		\node [style=bluegauge] (16) at (0, 0) {};
		\node [style=none] (29) at (-1, -0.5) {8};
		\node [style=none] (32) at (3, -0.5) {4};
		\node [style=none] (33) at (2, -0.5) {4};
		\node [style=none] (34) at (1, -0.5) {6};
		\node [style=none] (35) at (0, -0.5) {6};
		\node [style=none] (36) at (-5, -0.5) {4};
		\node [style=none] (37) at (-4, -0.5) {4};
		\node [style=none] (38) at (-3, -0.5) {6};
		\node [style=none] (39) at (-2, -0.5) {6};
		\node [style=bluegauge] (40) at (4, 0) {};
		\node [style=redgauge] (41) at (5, 0) {};
		\node [style=none] (42) at (5, -0.5) {2};
		\node [style=none] (43) at (4, -0.5) {2};
		\node [style=redgauge] (44) at (-7, 0) {};
		\node [style=bluegauge] (45) at (-6, 0) {};
		\node [style=none] (46) at (-7, -0.5) {2};
		\node [style=none] (47) at (-6, -0.5) {2};
		\node [style=bluegauge] (48) at (-2-0.25, 1) {};
		\node [style=bluegauge] (49) at (-1+0.5, 1) {};
		\node [style=bluegauge] (50) at (0.25, 1) {};
		\node [style=bluegauge] (500) at (-1.5, 1) {};
		\node [style=none] (51) at (0.25, 1.5) {2};
		\node [style=none] (52) at (-1+0.5, 1.5) {2};
		\node [style=none] (53) at (-2-0.25, 1.5) {2};
			\node [style=none] (54) at (-1.5, 1.5) {2};
	\end{pgfonlayer}
	\begin{pgfonlayer}{edgelayer}
		\draw (3) to (2);
		\draw (2) to (1);
		\draw (0) to (1);
		\draw (13) to (14);
		\draw (14) to (15);
		\draw (16) to (13);
		\draw (3) to (9);
		\draw (9) to (16);
		\draw (40) to (41);
		\draw (15) to (40);
		\draw (44) to (45);
		\draw (45) to (0);
		\draw (48) to (9);
		\draw (9) to (49);
		\draw (50) to (9);
		\draw (500) to (9);
	\end{pgfonlayer}
\end{tikzpicture}}
\end{tabular}
& 		$\mathfrak{so}(8)^2$ &
\begin{tabular}{l}
       \parbox[t]{4.5cm}{{$1 + 56 t^2 + 1603 t^4 + 31252 t^6 + 466740 t^8 + 5696700 t^{10} + 59218233 t^{12} + 539642360 t^{14} + 4404382326 t^{16} + 32733983761 t^{18} + 224481692035 t^{20} + O\left(t^{22}\right)$} }\\
       \end{tabular}&
       \begin{tabular}{c}
       \parbox{2cm}{$56 t^2 + 7 t^4 + 4 t^6 + 190 t^8 + 576 t^{10} ++O\left(t^{12}\right)$}
       \end{tabular} &
       \begin{tabular}{c}
       1 \\ \\$[1,1,1,1]$ 
       \end{tabular}\\
       \hline		
\begin{tabular}{c}	
\centering
\scalebox{0.7}{
\begin{tikzpicture}
	\begin{pgfonlayer}{nodelayer}
		\node [style=redgauge] (0) at (-5, 0) {};
		\node [style=bluegauge] (1) at (-4, 0) {};
		\node [style=redgauge] (2) at (-3, 0) {};
		\node [style=bluegauge] (3) at (-2, 0) {};
		\node [style=redgauge] (9) at (-1, 0) {};
		\node [style=redgauge] (13) at (1, 0) {};
		\node [style=bluegauge] (14) at (2, 0) {};
		\node [style=redgauge] (15) at (3, 0) {};
		\node [style=bluegauge] (16) at (0, 0) {};
		\node [style=none] (29) at (-1, -0.5) {8};
		\node [style=none] (32) at (3, -0.5) {4};
		\node [style=none] (33) at (2, -0.5) {4};
		\node [style=none] (34) at (1, -0.5) {6};
		\node [style=none] (35) at (0, -0.5) {6};
		\node [style=none] (36) at (-5, -0.5) {4};
		\node [style=none] (37) at (-4, -0.5) {4};
		\node [style=none] (38) at (-3, -0.5) {6};
		\node [style=none] (39) at (-2, -0.5) {6};
		\node [style=bluegauge] (40) at (4, 0) {};
		\node [style=redgauge] (41) at (5, 0) {};
		\node [style=none] (42) at (5, -0.5) {2};
		\node [style=none] (43) at (4, -0.5) {2};
		\node [style=redgauge] (44) at (-7, 0) {};
		\node [style=bluegauge] (45) at (-6, 0) {};
		\node [style=none] (46) at (-7, -0.5) {2};
		\node [style=none] (47) at (-6, -0.5) {2};
		\node [style=bluegauge] (48) at (-2, 1) {};
		\node [style=none] (50) at (-2.5, 1) {4};
		\node [style=none] (51) at (-2, 1.85) {$\Lambda^2$};
		\node [style=miniBlue] (52) at (0, 1) {};
		\node [style=miniBlue] (54) at (-1, 1) {};
		\node [style=none] (55) at (-1, 1.5) {2};
		\node [style=none] (56) at (0, 1.5) {2};
	\end{pgfonlayer}
	\begin{pgfonlayer}{edgelayer}
		\draw (3) to (2);
		\draw (2) to (1);
		\draw (0) to (1);
		\draw (13) to (14);
		\draw (14) to (15);
		\draw (16) to (13);
		\draw (3) to (9);
		\draw (9) to (16);
		\draw (40) to (41);
		\draw (15) to (40);
		\draw (44) to (45);
		\draw (45) to (0);
		\draw (9) to (48);
		\draw [in=135, out=45, loop] (48) to ();
		\draw (9) to (54);
		\draw (9) to (52);
	\end{pgfonlayer}
\end{tikzpicture}}
\end{tabular}
	& 	$\mathfrak{so}(8)^2$	&	
	\begin{tabular}{c}  
	\parbox[t]{4.5cm}{$1 + 56 t^2 + 1601 t^4 + 31139 t^6 + 463482 t^8 + 5632629 t^{10} + 58249189 t^{12} + 527596032 t^{14} + 4275989980 t^{16} + 31526199502 t^{18} + 214242906887 t^{20}+O\left(t^{22}\right)$}
	\end{tabular}
	&
	\begin{tabular}{c}
	\parbox[t]{2cm}{$56 t^2 + 5 t^4 + 3 t^6 + 193 t^8 + 386 t^{10} +O\left(t^{12}\right)$}
	\end{tabular} &
	\begin{tabular}{c}
	$\mathbb{Z}_2$\\ \\$[2,1,1]$
	\end{tabular} \\
	\hline
\begin{tabular}{c}	
\centering
\scalebox{0.7}{
\begin{tikzpicture}
	\begin{pgfonlayer}{nodelayer}
		\node [style=redgauge] (0) at (-5, 0) {};
		\node [style=bluegauge] (1) at (-4, 0) {};
		\node [style=redgauge] (2) at (-3, 0) {};
		\node [style=bluegauge] (3) at (-2, 0) {};
		\node [style=redgauge] (9) at (-1, 0) {};
		\node [style=redgauge] (13) at (1, 0) {};
		\node [style=bluegauge] (14) at (2, 0) {};
		\node [style=redgauge] (15) at (3, 0) {};
		\node [style=bluegauge] (16) at (0, 0) {};
		\node [style=none] (29) at (-1, -0.5) {8};
		\node [style=none] (32) at (3, -0.5) {4};
		\node [style=none] (33) at (2, -0.5) {4};
		\node [style=none] (34) at (1, -0.5) {6};
		\node [style=none] (35) at (0, -0.5) {6};
		\node [style=none] (36) at (-5, -0.5) {4};
		\node [style=none] (37) at (-4, -0.5) {4};
		\node [style=none] (38) at (-3, -0.5) {6};
		\node [style=none] (39) at (-2, -0.5) {6};
		\node [style=bluegauge] (40) at (4, 0) {};
		\node [style=redgauge] (41) at (5, 0) {};
		\node [style=none] (42) at (5, -0.5) {2};
		\node [style=none] (43) at (4, -0.5) {2};
		\node [style=redgauge] (44) at (-7, 0) {};
		\node [style=bluegauge] (45) at (-6, 0) {};
		\node [style=none] (46) at (-7, -0.5) {2};
		\node [style=none] (47) at (-6, -0.5) {2};
		\node [style=bluegauge] (48) at (-2, 1) {};
		\node [style=none] (50) at (-2.5, 1) {4};
		\node [style=none] (51) at (-2, 1.85) {$\Lambda^2$};
		\node [style=bluegauge] (52) at (0, 1) {};
		\node [style=none] (53) at (0.5, 1) {4};
		\node [style=none] (54) at (0, 1.85) {$\Lambda^2$};
	\end{pgfonlayer}
	\begin{pgfonlayer}{edgelayer}
		\draw (3) to (2);
		\draw (2) to (1);
		\draw (0) to (1);
		\draw (13) to (14);
		\draw (14) to (15);
		\draw (16) to (13);
		\draw (3) to (9);
		\draw (9) to (16);
		\draw (40) to (41);
		\draw (15) to (40);
		\draw (44) to (45);
		\draw (45) to (0);
		\draw (9) to (48);
		\draw [in=135, out=45, loop] (48) to ();
		\draw [in=135, out=45, loop] (52) to ();
		\draw (52) to (9);
	\end{pgfonlayer}
\end{tikzpicture}}
\end{tabular}
	& 			$\mathfrak{so}(8)^2$ &	
	\begin{tabular}{c}  
	\parbox[t]{4.5cm}{$1 + 56 t^2 + 1599 t^4 + 31026 t^6 + 460228 t^8 + 5568786 t^{10} + 57286958 t^{12} + 515691588 t^{14} + 4149904285 t^{16} + 30349484503 t^{18} + 204363531903 t^{20}+O\left(t^{22}\right)$}
	\end{tabular}
	&
	\begin{tabular}{c}
	\parbox[t]{2cm}{$56 t^2 + 3 t^4 + 2 t^6 + 196 t^8 + 196 t^{10}+ O\left(t^{12}\right)$}
	\end{tabular} &
	\begin{tabular}{c}
	    $\mathbb{Z}_2\times \mathbb{Z}_2$\\ \\$[2,2]$
	\end{tabular}\\
	\hline
	\begin{tabular}{c}	
\centering
\scalebox{0.7}{
\begin{tikzpicture}
	\begin{pgfonlayer}{nodelayer}
		\node [style=redgauge] (0) at (-5, 0) {};
		\node [style=bluegauge] (1) at (-4, 0) {};
		\node [style=redgauge] (2) at (-3, 0) {};
		\node [style=bluegauge] (3) at (-2, 0) {};
		\node [style=redgauge] (9) at (-1, 0) {};
		\node [style=redgauge] (13) at (1, 0) {};
		\node [style=bluegauge] (14) at (2, 0) {};
		\node [style=redgauge] (15) at (3, 0) {};
		\node [style=bluegauge] (16) at (0, 0) {};
		\node [style=none] (29) at (-1, -0.5) {8};
		\node [style=none] (32) at (3, -0.5) {4};
		\node [style=none] (33) at (2, -0.5) {4};
		\node [style=none] (34) at (1, -0.5) {6};
		\node [style=none] (35) at (0, -0.5) {6};
		\node [style=none] (36) at (-5, -0.5) {4};
		\node [style=none] (37) at (-4, -0.5) {4};
		\node [style=none] (38) at (-3, -0.5) {6};
		\node [style=none] (39) at (-2, -0.5) {6};
		\node [style=bluegauge] (40) at (4, 0) {};
		\node [style=redgauge] (41) at (5, 0) {};
		\node [style=none] (42) at (5, -0.5) {2};
		\node [style=none] (43) at (4, -0.5) {2};
		\node [style=redgauge] (44) at (-7, 0) {};
		\node [style=bluegauge] (45) at (-6, 0) {};
		\node [style=none] (46) at (-7, -0.5) {2};
		\node [style=none] (47) at (-6, -0.5) {2};
		\node [style=bluegauge] (48) at (-2, 1) {};
		\node [style=none] (50) at (-2.5, 1) {6};
		\node [style=none] (51) at (-2, 1.85) {$\Lambda^2$};
		\node [style=miniBlue] (52) at (0, 1) {};
		\node [style=none] (53) at (0, 1.5) {2};
	\end{pgfonlayer}
	\begin{pgfonlayer}{edgelayer}
		\draw (3) to (2);
		\draw (2) to (1);
		\draw (0) to (1);
		\draw (13) to (14);
		\draw (14) to (15);
		\draw (16) to (13);
		\draw (3) to (9);
		\draw (9) to (16);
		\draw (40) to (41);
		\draw (15) to (40);
		\draw (44) to (45);
		\draw (45) to (0);
		\draw (9) to (48);
		\draw [in=135, out=45, loop] (48) to ();
		\draw (52) to (9);
	\end{pgfonlayer}
\end{tikzpicture}}
\end{tabular}
	& 		
	$\mathfrak{so}(8)^2$	&
\begin{tabular}{c}  
\parbox[t]{4.5cm}{$1 + 56 t^2 + 1599 t^4 + 31026 t^6 + 460225 t^8 + 5568616 t^{10} + 57281860 t^{12} + 515584539 t^{14} + 4148148363 t^{16} + 30325670104 t^{18} + 204087091615 t^{20}+O \left(t^{22}\right)$}
\end{tabular}
	&
\begin{tabular}{c}
\parbox[t]{2cm}{$56 t^2 + 3 t^4 + 2 t^6 + 193 t^8 + 194 t^{10} +O\left(t^{12}\right)$}
\end{tabular} &
\begin{tabular}{c}
$S_3$\\ \\ $[3,1]$
\end{tabular}\\
\hline
\begin{tabular}{c}	
\centering
\scalebox{0.7}{
\begin{tikzpicture}
	\begin{pgfonlayer}{nodelayer}
		\node [style=redgauge] (0) at (-5, 0) {};
		\node [style=bluegauge] (1) at (-4, 0) {};
		\node [style=redgauge] (2) at (-3, 0) {};
		\node [style=bluegauge] (3) at (-2, 0) {};
		\node [style=redgauge] (9) at (-1, 0) {};
		\node [style=redgauge] (13) at (1, 0) {};
		\node [style=bluegauge] (14) at (2, 0) {};
		\node [style=redgauge] (15) at (3, 0) {};
		\node [style=bluegauge] (16) at (0, 0) {};
		\node [style=none] (29) at (-1, -0.5) {8};
		\node [style=none] (32) at (3, -0.5) {4};
		\node [style=none] (33) at (2, -0.5) {4};
		\node [style=none] (34) at (1, -0.5) {6};
		\node [style=none] (35) at (0, -0.5) {6};
		\node [style=none] (36) at (-5, -0.5) {4};
		\node [style=none] (37) at (-4, -0.5) {4};
		\node [style=none] (38) at (-3, -0.5) {6};
		\node [style=none] (39) at (-2, -0.5) {6};
		\node [style=bluegauge] (40) at (4, 0) {};
		\node [style=redgauge] (41) at (5, 0) {};
		\node [style=none] (42) at (5, -0.5) {2};
		\node [style=none] (43) at (4, -0.5) {2};
		\node [style=redgauge] (44) at (-7, 0) {};
		\node [style=bluegauge] (45) at (-6, 0) {};
		\node [style=none] (46) at (-7, -0.5) {2};
		\node [style=none] (47) at (-6, -0.5) {2};
		\node [style=bluegauge] (48) at (-1, 1) {};
		\node [style=none] (50) at (-1.5, 1) {8};
		\node [style=none] (51) at (-1, 1.85) {$\Lambda^2$};
	\end{pgfonlayer}
	\begin{pgfonlayer}{edgelayer}
		\draw (3) to (2);
		\draw (2) to (1);
		\draw (0) to (1);
		\draw (13) to (14);
		\draw (14) to (15);
		\draw (16) to (13);
		\draw (3) to (9);
		\draw (9) to (16);
		\draw (40) to (41);
		\draw (15) to (40);
		\draw (44) to (45);
		\draw (45) to (0);
		\draw (9) to (48);
		\draw [in=135, out=45, loop] (48) to ();
	\end{pgfonlayer}
\end{tikzpicture}}
\end{tabular}
	& 	$\mathfrak{so}(8)^2$&
	\begin{tabular}{c}  
	\parbox[t]{4.5cm}{$1 + 56 t^2 + 1597 t^4 + 30913 t^6 + 456972 t^8 + 5504831 t^{10} +  56321344 t^{12} + 503714929 t^{14} + 4022613338 t^{16} + 29156208003 t^{18} + 194290639132 t^{20}+ O\left(t^{22}\right)$}
	\end{tabular}
	&\begin{tabular}{c}
	\parbox[t]{2cm}{$56 t^2 + t^4 + t^6 + 193 t^8 + 2 t^{10} +O\left(t^{12}\right)$}
	\end{tabular} &
	\begin{tabular}{c}  
	$S_4$\\ \\$[4]$
	\end{tabular}\\
	\hline
	\end{tabular}
\end{adjustbox}
\caption[Bouquet quivers of 4 USp(2) nodes and SO(8) central node]{Magnetic quivers corresponding to the $n=4$, $m=4$ cases of (\ref{fullanti}), with different discrete quotients fixed by the choice of partition. In all cases $\text{dim}_{\mathbb{H}} (\mathcal{C})=32$. The Coulomb branch Hilbert series and their plethystic logarithms are shown along with their global symmetries. The discrete symmetry gauge groups are also shown.}
\label{D4quotientm4}
\end{figure}


\begin{figure}[!ht]
\small
\begin{adjustbox}{center}
	\begin{tabular}{|c|c|c|c|c|}
		\hline
		Quiver &
		\begin{tabular}{c}	Global\\Symmetry \end{tabular} & 
		Hilbert Series &
		\begin{tabular}{c}Plethystic\\Logarithm\end{tabular} &
		\begin{tabular}{c}Discrete\\Quotient \end{tabular}\\ 
		\hline
\begin{tabular}{c}
\centering
	\scalebox{0.8}{
\begin{tikzpicture}
	\begin{pgfonlayer}{nodelayer}
		\node [style=bluegauge] (0) at (-5, 0) {};
		\node [style=none] (1) at (-4, 0) {$\cdots$};
		\node [style=redgauge] (2) at (-3, 0) {};
		\node [style=bluegauge] (3) at (-2, 0) {};
		\node [style=redgauge] (9) at (-1, 0) {};
		\node [style=redgauge] (13) at (1, 0) {};
		\node [style=none] (14) at (2, 0) {$\cdots$};
		\node [style=bluegauge] (15) at (3, 0) {};
		\node [style=bluegauge] (16) at (0, 0) {};
		\node [style=none] (29) at (-3, -0.5) {8};
		\node [style=none] (33) at (4, -0.5) {2};
		\node [style=none] (34) at (3, -0.5) {2};
		\node [style=none] (37) at (-6, -0.5) {2};
		\node [style=none] (38) at (-5, -0.5) {2};
		\node [style=redgauge] (40) at (4, 0) {};
		\node [style=redgauge] (45) at (-6, 0) {};
		\node [style=bluegauge] (48) at (-2, 1) {};
		\node [style=none] (58) at (1, -0.5) {8};
		\node [style=none] (59) at (-2, -0.5) {8};
		\node [style=none] (60) at (-1, -0.5) {10};
		\node [style=none] (61) at (0, -0.5) {8};
		\node [style=none] (62) at (-2, 1.5) {2};
		\node [style=miniBlue] (63) at (0, 1) {};
		\node [style=none] (64) at (0, 1.5) {2};
	\end{pgfonlayer}
	\begin{pgfonlayer}{edgelayer}
		\draw (3) to (2);
		\draw (2) to (1);
		\draw (0) to (1);
		\draw (13) to (14);
		\draw (14) to (15);
		\draw (16) to (13);
		\draw (3) to (9);
		\draw (9) to (16);
		\draw (15) to (40);
		\draw (45) to (0);
		\draw (48) to (9);
		\draw (63) to (9);
	\end{pgfonlayer}
\end{tikzpicture}}
\end{tabular}
& 		$\mathfrak{so}(10)^2$ 	&
\begin{tabular}{l}
       \parbox[t]{4.5cm}{{$1 + 90 t^2 + 4196 t^4 + 135084 t^6 + 3376345 t^8 + 69823410 t^{10} + 1243044566 t^{12} + 19567608611 t^{14} + 277613510257 t^{16} + 3600143050467 t^{18} + 43134857718712 t^{20} + O\left(t^{22}\right)$} }\\
       \end{tabular}& 
       \begin{tabular}{c}
       \parbox{2cm}{$90 t^2 + 101 t^4 + 414 t^6 + 604 t^8 - 6282 t^{10}+ O\left(t^{12}\right)$} 
       \end{tabular} &
       \begin{tabular}{c}
       1 \\  \\ $[1,1]$
    \end{tabular}\\
       \hline	
\begin{tabular}{c}	
\centering
	\scalebox{0.8}{
\begin{tikzpicture}
	\begin{pgfonlayer}{nodelayer}
		\node [style=bluegauge] (0) at (-5, 0) {};
		\node [style=none] (1) at (-4, 0) {$\cdots$};
		\node [style=redgauge] (2) at (-3, 0) {};
		\node [style=bluegauge] (3) at (-2, 0) {};
		\node [style=redgauge] (9) at (-1, 0) {};
		\node [style=redgauge] (13) at (1, 0) {};
		\node [style=none] (14) at (2, 0) {$\cdots$};
		\node [style=bluegauge] (15) at (3, 0) {};
		\node [style=bluegauge] (16) at (0, 0) {};
		\node [style=none] (29) at (-3, -0.5) {8};
		\node [style=none] (33) at (4, -0.5) {2};
		\node [style=none] (34) at (3, -0.5) {2};
		\node [style=none] (37) at (-6, -0.5) {2};
		\node [style=none] (38) at (-5, -0.5) {2};
		\node [style=redgauge] (40) at (4, 0) {};
		\node [style=redgauge] (45) at (-6, 0) {};
		\node [style=bluegauge] (48) at (-1, 1) {};
		\node [style=none] (53) at (-1, 2) {$\Lambda^2$};
		\node [style=none] (58) at (1, -0.5) {8};
		\node [style=none] (59) at (-2, -0.5) {8};
		\node [style=none] (60) at (-1, -0.5) {10};
		\node [style=none] (61) at (0, -0.5) {8};
		\node [style=none] (62) at (-1.75, 1) {4};
	\end{pgfonlayer}
	\begin{pgfonlayer}{edgelayer}
		\draw (3) to (2);
		\draw (2) to (1);
		\draw (0) to (1);
		\draw (13) to (14);
		\draw (14) to (15);
		\draw (16) to (13);
		\draw (3) to (9);
		\draw (9) to (16);
		\draw (15) to (40);
		\draw (45) to (0);
		\draw (48) to (9);
	\draw [in=135, out=45, loop] (48) to ();
	\end{pgfonlayer}
\end{tikzpicture}}
\end{tabular}
	& 		
	$\mathfrak{so}(10)^2 $	&	
	\begin{tabular}{c}  
	\parbox[t]{4.5cm}{$1 + 90 t^2 + 4195 t^4 + 134993 t^6 + 3371447 t^8 + 69627639 t^{10} + 1236842023 t^{12} + 19406387991 t^{14} + 274081195871 t^{16} + 3533394567656 t^{18} + 42025998937738 t^{20}+ O\left(t^{22}\right)$}
	\end{tabular}
	&
	\begin{tabular}{c}
	\parbox[t]{2cm}{$90 t^2 + 100 t^4 + 413 t^6 - 8 t^8 - 7694 t^{10}+ O\left(t^{12}\right)$}
	\end{tabular} &
		\begin{tabular}{c}
	$\mathbb{Z}_2$\\ \\$[2]$
	\end{tabular}\\
	\hline
	\end{tabular}
\end{adjustbox}
\caption[Bouquet quivers of 2 USp(2) nodes and SO(10) central node]{Magnetic quivers corresponding to the $n=5$, $m=2$ cases of (\ref{fullanti}), with different discrete quotients fixed by the choice of partition. In all cases $\text{dim}_{\mathbb{H}} (\mathcal{C})=47$. The Coulomb branch Hilbert series and their plethystic logarithms are shown along with their global symmetries. The discrete symmetry gauge groups are also shown.}
\label{D5quotientm2}
\end{figure}


\begin{figure}[!ht]
\small
\adjustbox{center}{ 
	\begin{tabular}{|c|c|c|c|c|}
		\hline
		Quiver &
		\begin{tabular}{c}	Global\\Symmetry \end{tabular} & 
		Hilbert Series &
		\begin{tabular}{c}Plethystic\\Logarithm\end{tabular} &
		\begin{tabular}{c}Discrete\\Quotient \end{tabular}\\ 
		\hline
\begin{tabular}{c}
\centering
\scalebox{0.8}{
\begin{tikzpicture}
	\begin{pgfonlayer}{nodelayer}
		\node [style=bluegauge] (0) at (-5, 0) {};
		\node [style=none] (1) at (-4, 0) {$\cdots$};
		\node [style=redgauge] (2) at (-3, 0) {};
		\node [style=bluegauge] (3) at (-2, 0) {};
		\node [style=redgauge] (9) at (-1, 0) {};
		\node [style=redgauge] (13) at (1, 0) {};
		\node [style=none] (14) at (2, 0) {$\cdots$};
		\node [style=bluegauge] (15) at (3, 0) {};
		\node [style=bluegauge] (16) at (0, 0) {};
		\node [style=none] (29) at (-3, -0.5) {8};
		\node [style=none] (33) at (4, -0.5) {2};
		\node [style=none] (34) at (3, -0.5) {2};
		\node [style=none] (37) at (-6, -0.5) {2};
		\node [style=none] (38) at (-5, -0.5) {2};
		\node [style=redgauge] (40) at (4, 0) {};
		\node [style=redgauge] (45) at (-6, 0) {};
		\node [style=bluegauge] (48) at (-2, 1) {};
		\node [style=bluegauge] (49) at (-1, 1) {};
		\node [style=bluegauge] (50) at (0, 1) {};
		\node [style=none] (51) at (0, 1.5) {2};
		\node [style=none] (52) at (-1, 1.5) {2};
		\node [style=none] (53) at (-2, 1.5) {2};
		\node [style=none] (58) at (1, -0.5) {8};
		\node [style=none] (59) at (-2, -0.5) {8};
		\node [style=none] (60) at (-1, -0.5) {10};
		\node [style=none] (61) at (0, -0.5) {8};
	\end{pgfonlayer}
	\begin{pgfonlayer}{edgelayer}
		\draw (3) to (2);
		\draw (2) to (1);
		\draw (0) to (1);
		\draw (13) to (14);
		\draw (14) to (15);
		\draw (16) to (13);
		\draw (3) to (9);
		\draw (9) to (16);
		\draw (15) to (40);
		\draw (45) to (0);
		\draw (48) to (9);
		\draw (9) to (49);
		\draw (50) to (9);
	\end{pgfonlayer}
\end{tikzpicture}}
\end{tabular}
& 			$\mathfrak{so}(10)^2$	&
\begin{tabular}{l}
       \parbox[t]{4.5cm}{{$1 + 90 t^2 + 4097 t^4 + 125863 t^6 + 2937100 t^8 + 512 t^9 + 55555918 t^{10} + O\left(t^{11}\right)$} }\\
       \end{tabular}&
       \begin{tabular}{c}
       \parbox{2.5cm}{$90 t^2 + 2 t^4 + 103 t^6 - 98 t^8 + 512 t^9 + 299 t^{10}+O\left(t^{11}\right)$} 
       \end{tabular} &
        \begin{tabular}{c}
       1 \\ \\$[1,1,1]$
        \end{tabular}\\
       \hline	
\begin{tabular}{c}	
\centering
\scalebox{0.8}{
\begin{tikzpicture}
	\begin{pgfonlayer}{nodelayer}
		\node [style=bluegauge] (0) at (-5, 0) {};
		\node [style=none] (1) at (-4, 0) {$\cdots$};
		\node [style=redgauge] (2) at (-3, 0) {};
		\node [style=bluegauge] (3) at (-2, 0) {};
		\node [style=redgauge] (9) at (-1, 0) {};
		\node [style=redgauge] (13) at (1, 0) {};
		\node [style=none] (14) at (2, 0) {$\cdots$};
		\node [style=bluegauge] (15) at (3, 0) {};
		\node [style=bluegauge] (16) at (0, 0) {};
		\node [style=none] (29) at (-3, -0.5) {8};
		\node [style=none] (33) at (4, -0.5) {2};
		\node [style=none] (34) at (3, -0.5) {2};
		\node [style=none] (37) at (-6, -0.5) {2};
		\node [style=none] (38) at (-5, -0.5) {2};
		\node [style=redgauge] (40) at (4, 0) {};
		\node [style=redgauge] (45) at (-6, 0) {};
		\node [style=bluegauge] (48) at (-2, 1) {};
		\node [style=bluegauge] (49) at (0, 1) {};
		\node [style=none] (52) at (0, 1.5) {2};
		\node [style=none] (53) at (-2, 2) {$\Lambda^2$};
		\node [style=none] (58) at (1, -0.5) {8};
		\node [style=none] (59) at (-2, -0.5) {8};
		\node [style=none] (60) at (-1, -0.5) {10};
		\node [style=none] (61) at (0, -0.5) {8};
		\node [style=none] (62) at (-2.75, 1) {4};
	\end{pgfonlayer}
	\begin{pgfonlayer}{edgelayer}
		\draw (3) to (2);
		\draw (2) to (1);
		\draw (0) to (1);
		\draw (13) to (14);
		\draw (14) to (15);
		\draw (16) to (13);
		\draw (3) to (9);
		\draw (9) to (16);
		\draw (15) to (40);
		\draw (45) to (0);
		\draw (48) to (9);
		\draw (9) to (49);
		\draw [in=135, out=45, loop] (48) to ();
	\end{pgfonlayer}
\end{tikzpicture}}
\end{tabular}
	& 			$\mathfrak{so}(10)^2$	&
	\begin{tabular}{c}  
	\parbox[t]{4.5cm}{$1 + 90 t^2 + 4096 t^4 + 125772 t^6 + 2932913 t^8 + 512 t^9 + 55425860 t^{10}+ O\left(t^{11}\right)$}
	\end{tabular}
	&
	\begin{tabular}{c}
	\parbox[t]{2.5cm}{$90 t^2 + t^4 + 102 t^6 - 98 t^8 + 512 t^9 + 200 t^{10} +O\left(t^{11}\right)$}
	\end{tabular} &
	\begin{tabular}{c}
	$\mathbb{Z}_2$\\ \\$[2,1]$
	\end{tabular}\\
	\hline
\begin{tabular}{c}	
\centering
\scalebox{0.8}{
\begin{tikzpicture}
	\begin{pgfonlayer}{nodelayer}
		\node [style=bluegauge] (0) at (-5, 0) {};
		\node [style=none] (1) at (-4, 0) {$\cdots$};
		\node [style=redgauge] (2) at (-3, 0) {};
		\node [style=bluegauge] (3) at (-2, 0) {};
		\node [style=redgauge] (9) at (-1, 0) {};
		\node [style=redgauge] (13) at (1, 0) {};
		\node [style=none] (14) at (2, 0) {$\cdots$};
		\node [style=bluegauge] (15) at (3, 0) {};
		\node [style=bluegauge] (16) at (0, 0) {};
		\node [style=none] (29) at (-3, -0.5) {8};
		\node [style=none] (33) at (4, -0.5) {2};
		\node [style=none] (34) at (3, -0.5) {2};
		\node [style=none] (37) at (-6, -0.5) {2};
		\node [style=none] (38) at (-5, -0.5) {2};
		\node [style=redgauge] (40) at (4, 0) {};
		\node [style=redgauge] (45) at (-6, 0) {};
		\node [style=bluegauge] (48) at (-1, 1) {};
		\node [style=none] (53) at (-1, 2) {$\Lambda^2$};
		\node [style=none] (58) at (1, -0.5) {8};
		\node [style=none] (59) at (-2, -0.5) {8};
		\node [style=none] (60) at (-1, -0.5) {10};
		\node [style=none] (61) at (0, -0.5) {8};
		\node [style=none] (62) at (-1.75, 1) {6};
		\draw [in=135, out=45, loop] (48) to ();
	\end{pgfonlayer}
	\begin{pgfonlayer}{edgelayer}
		\draw (3) to (2);
		\draw (2) to (1);
		\draw (0) to (1);
		\draw (13) to (14);
		\draw (14) to (15);
		\draw (16) to (13);
		\draw (3) to (9);
		\draw (9) to (16);
		\draw (15) to (40);
		\draw (45) to (0);
		\draw (48) to (9);
	\end{pgfonlayer}
\end{tikzpicture}}
\end{tabular}
	& 			$\mathfrak{so}(10)^2$	&
	\begin{tabular}{c}  
	\parbox[t]{4.5cm}{$1 + 90 t^2 + 4095 t^4 + 125681 t^6 + 2928726 t^8 + 512 t^9 + 55295803 t^{10}+ O\left(t^{11}\right)$}
	\end{tabular}
	&
	\begin{tabular}{c}
	\parbox[t]{2.5cm}{$90 t^2 + 101 t^6 - 99 t^8 + 512 t^9 + 100 t^{10} + O\left(t^{11}\right)$}
	\end{tabular} &
	\begin{tabular}{c} 
	$S_3$\\ \\$[3]$ 
	\end{tabular}\\
	\hline
	\end{tabular}}
\caption[Bouquet quivers of 3 USp(2) nodes and SO(10) central node]{Magnetic quivers corresponding to the $n=5$, $m=3$ cases of (\ref{fullanti}), with different discrete quotients fixed by the choice of partition. In all cases $\text{dim}_{\mathbb{H}} (\mathcal{C})=48$. The Coulomb branch Hilbert series and their plethystic logarithms are shown along with their global symmetries. The discrete symmetry gauge groups are also shown. }
\label{D5quotientm3}
\end{figure}
\end{landscape}


\section{Hall Littlewood polynomials and star shaped quivers}
\label{app:HL}

We can calculate the Coulomb branch Hilbert series of many $3d$ ${\cal N}=4$ unframed (unitary)- orthosymplectic star shaped quivers ${\mathsf{Q}}(J)$, that are characterised by a central node with gauge group $J$, by gluing the resolved Slodowy slices of the GNO dual group $J^\vee$. The procedure depends upon being able to identify the linear quivers that form the legs of the star shaped quiver as Slodowy slices of  $J^\vee$. This correspondence between linear quivers and Slodowy slices draws on the Barbasch Vogan map and related dualities, as elaborated in \cite{RudolphSlodowy, RudolphIntersections}. The pairings for low rank Classical groups are tabulated in \cite{RudolphSlodowy}, and can in principle be extended to higher rank.

Recall that, for any group $G$, the Slodowy slice $\slice{\rho}^G$ transverse to the closure of a nilpotent orbit $\orbit{\rho}$ is defined by the $SU(2)$ homomorphism $\rho: G \to F(\rho) \otimes SU(2)$, identified by the partition $\rho$. The partition $ \rho $ determines a fugacity map ${{\bf x} \to ({\bf y}},t) $,  where the $ {\bf x}$ and $ {\bf y}$ are CSA fugacities for $G$ and the subgroup $  F \subseteq  {{G}}$, respectively, and $t$ is a CSA fugacity for the $SU(2)$. The Slodowy slice transforms under $F$.

The Hilbert series for a Slodowy slice can be found from the  branching of the adjoint representation of $G$ under $\rho$: 

\begin{equation}
\label{eq:eq:HL1}
\begin{aligned}
\chi _{[adjoint]} ^G ({\bf x})&= \mathop  \bigoplus \limits_{[n],{\bf [m]}} {a_{n,{\bf m}}}\left( {\chi _{ [n]}^{SU(2)} \otimes \chi _{\bf [m]}^F} ({\bf y}) \right),
\end{aligned}
\end{equation}
where the $ {a_{n,{\bf m}}}$ are branching coefficients. The HS is obtained from \eqref{eq:eq:HL1} by (i) replacing the $SU(2)$ characters by highest weight fugacities \cite{RudolphHWG}, $\chi_{[n]}^{SU(2)} \to t^n$, (ii) symmetrising the representations of $F(\rho)$ under a grading by $t^2$, and (iii) taking a quotient by the Casimirs of $G$. This leads to the refined and unrefined HS:
\begin{equation}
\label{eq:eq:HL2}
\begin{aligned}
g_{HS}^{{{\cal S}_{{\cal N},\rho} ^G}}({\bf y},t) &= PE\left[ {\mathop  \bigoplus \limits_{[n], \bf [m]}  a_{n,\bf m} {~} \chi _{\bf [m]}^F({\bf y}){t^{n + 2}} - \sum\limits_{i = 1}^r {{t^{2{d_i}}}} } \right],\\
g_{HS}^{{{{\cal S}}_{{{\cal N},\rho}} ^G}}(1,t) &= PE\left[ {\sum\limits_n {{a_n}{t^{n + 2}} - \sum\limits_{i = 1}^r {{t^{2{d_i}}}} } } \right],
\end{aligned}
\end{equation}
where the $d_i$ are the degrees of the Casimirs of $G$. Further details on the construction of slices can be found in \cite{RudolphSlodowy}.

We now introduce \emph{resolved} Slodowy slices, $ \slice{\rho,{\bf [n]}}^{{G}} $, which carry background charges ${\bf [n]}$ parameterised by the Dynkin labels of ${{G}}$, as described in Appendix B of \cite{RudolphIntersections}. These are related to uncharged slices (which effectively carry singlet charges) by a quotient of Hall Littlewood polynomials \cite{RudolphInstantons} under the fugacity map for $\rho$:

\begin{equation}
\ghs{ \slice{\rho,{\bf [n]}}^{{G}}}({\bf y},t)
\equiv {\ghs{ \slice{\rho}^{{G}}}}    ({\bf y},t) {~}
{\left. { \frac{ HL_{{\bf{[n]}}}^{{G}}({\bf x},t)}{ HL_{{\bf{[0]}}}^{{G}}({\bf x},t)} } \right|_{\rho:{~} {\bf x} \to ( {\bf y}, t )    }}.
\label{eq:HL3}
\end{equation}
The above considerations are quite general and apply to any Lie group $G$.

Returning to our star shaped quiver ${\mathsf{Q}}(J)$, let us assume that we can identify its $k$ legs with a set of Slodowy slices $ \slice{\rho(i)}^{{J}^\vee} $, where $i \in \{1,\ldots, k \}$. The partitions $ \rho (i) $ determine fugacity maps ${{\bf x} \to ({\bf y}(i)},t) $. Given such a set of slices, we can apply charges $\bf [n]$ to these slices and carry out gluing \cite{CremonesiHall} by summation over the weight lattice of $J^\vee$. We obtain the following Hilbert series:

\begin{equation}
\ghs{\Sigma}({\bf y},t)= 
\sum\limits_{\left[ \bf n \right] \in { \Gamma_{J^\vee / W^\vee}}}
\underbrace{
 {P_{\left[ \bf n \right]}^{J ^\vee}} {~}
{t^{2 \Delta [ {\bf [n]}]}}
 }_{\text{central term}}
 {~}
 \prod\limits_{i=1}^{k}
 {t^{ - \Delta [ {\bf [n]}]}}
 \underbrace{\ghs{ \slice{{\rho (i),[\bf n]}}^{J^\vee }}({\bf y}(i),t)} _{\text{slices}}.
\label{eq:HL4}
\end{equation}
Here, the symmetry factors ${P_{\left[ \bf n \right]}^{J ^\vee}}$ match the ${P_J ({\bf m})}$ terms that appear in the monopole formula \eqref{mono}, being related by the map between the magnetic weight lattice charges $\bf m$ in the orthogonal basis of ${J ^\vee}$, and the weight lattice charges $\bf [n]$ in the Dynkin label or $\omega$-basis \cite{Feger:2012bs,Feger:2019tvk}. The term $\Delta[\bf[n]]$ equals the conformal dimension contribution $\Delta_{vec}(\bf m)$, as in table \ref{tableLattices}.\footnote{Alternatively, $\Delta[\bf[n]]$ can be found directly, either from the weight map \cite{rudolph1601} associated to the nilcone of $J ^{\vee}$, as ${\Delta [\bf[ n]]}=- [\bf n] \cdot \omega ({\cal N}) $, or from the Cartan matrix ${\bf A}^{\vee}$ and Weyl vector ${\bf 1}$, as $\Delta[{\bf[n]}] = - 2 {\bf [n]} \cdot {{\bf A} ^{\vee} }^{-1} \cdot {\bf 1}$.} Each leg, described by the remaining terms, corresponds to the Coulomb branch of a theory of type $T(J)$ carrying external charges,
\begin{equation}
T_{\rho(i),[\bf n]}(J) = {t^{- \Delta [ {\bf [n]}]}}{~} \ghs{ \slice{{\rho (i),[\bf n]}}^{J^\vee }}({\bf y}(i),t).
\label{eq:HL4a}
\end{equation}
The Hilbert series $g_{HS}^{\Sigma}$ transforms in the product group $\mathop  \otimes \limits_i F( \rho(i) )$ (before possible symmetry enhancement) and matches that of the Coulomb branch of the star shaped quiver $\mathsf{Q}(J)$.

\paragraph{Selection Rule.}
Not every collection of slices yields a well formed Hilbert series wherein all the fields (other than the singlet at its origin) have positive conformal dimension, $ \forall {\bf{n}} \ne \left[ {\bf{0}} \right]:\Delta \left[ {\left[ {\bf{n}} \right]} \right] > 0$. The selection rule can be formulated in terms of the weights associated with each slice by the partitions $\rho(i)$ and their contribution via \eqref{eq:HL3} and \eqref{eq:HL4} to the overall charges carried by $t$. Thus, each fugacity map ${{\bf x} \to ({\bf y}(i)},t) $ incorporates a weight map $\omega(i)$ that assigns R-charges to the CSA fugacities $\bf{x}$, viz $\omega(i):{~}\{x_1,\ldots,x_r\} \to \{t^{\omega_1(i)},\ldots,t^{\omega_r(i)}\}$ \cite{rudolph1601}. The selection rule requires:


\begin{equation}
    \omega(\mathsf{Q}) \equiv  - 2{~} \omega (\text{reg.}) + \sum\limits_{i = 1}^k\left( {\omega (\text{reg.}) - \omega (i)}\right)  \mathop  > \limits_{\text{strict}} {\bf{0}},
    \label{eq:puncselection}
\end{equation}
where $\omega (\text{reg.})$ is the weight map associated with the zero dimensional slice to the regular orbit, and the inequality requires that all the entries of the weight vector $ \omega(\mathsf{Q})$ should be greater than zero.

For example, consider the quiver with $E_7$ global symmetry in table \ref{D3fixtures}. We can read from table \ref{D3puncture} that this comprises slices to orbits with $D$-partitions $\rho=(1^6),(1^6)$ and $(3,1^3)$. Using the nilpotent orbit data in Appendix B of \cite{rudolph1601}, we find that these correspond to weights $[0,0,0],[0,0,0] \text{ and } [2,1,1]$, and that the weight map for the regular slice is $[4,3,3]$. We obtain,
\begin{equation}
\begin{aligned}
\omega(\mathsf{Q})& =-2[4,3,3]+ 2([4,3,3]-[0,0,0])+([4,3,3]-[2,1,1])\\
& =[2,2,2].
\end{aligned}
\end{equation}
This is strictly greater than $[0,0,0]$, so $\omega(\mathsf{Q})$ is the weight vector of a good quiver.

\paragraph{Integer and Fractional Integer Lattices.}
The regular summation over resolved slices in \eqref{eq:HL4} is carried out over the entire weight or Dynkin label lattice ${\left[ \bf n \right] \in { \Gamma_{J^\vee / W^\vee}}}$ of the GNO dual group $J^\vee$ of the cental quiver node. This corresponds to the combination of integer and half integer lattices as in \eqref{e}. The summation can be restricted to correspond to the integer lattice \eqref{a} by the simple expedient of restricting the summation over the special orthogonal $J^\vee$ weight lattice to exclude irreps from spinor lattices, i.e. to exclude those ${\left[ \bf n \right]}$ where the sum of spinor Dynkin labels is odd.

Similarly, in the case of the $\rm{SU(N)}/{\mathbb Z}_N^{\rm{diag}}$ split magnetic lattices discussed in sections \ref{sec:D4ex} and \ref{sec:E6ex}, the corresponding summations over resolved slices are restricted to the sub-lattice(s) of $\rm{SU(N)}$ Dynkin labels ${\left[\bf n \right]}$ with the relevant $N$-ality.

\paragraph{Twisted Fixtures.}
The foregoing applies to magnetic quivers for regular fixtures, where all the slices are transverse to orbits from the same symmetry group $J^\vee$. Quivers can also be constructed for ``twisted" fixtures, where some of the punctures are represented by slices from a different symmetry group ${J'}$ (generally of different rank), and are ``twisted" to fit the (Dynkin label) lattice of the $J^\vee$ symmetry group. Compatibility requirements between the lattices of $J^\vee$ and ${J'}$ mean that only certain pairs of groups can be related by such twists.

To accommodate twisted quivers, equations \eqref{eq:HL4} and \eqref{eq:HL4a} require modification. We set the terms for the twisted puncture to contain a conformal dimension contribution and slice drawn from ${{J'}}$:
\begin{equation}
\begin{aligned}
T_{\rho(i),[\bf n']}({{J'}^\vee}) = {t^{- \Delta' [ {\bf [n']}]}}{~} \ghs{ \slice{{\rho (i),[\bf n']}}^{{{J'}}}}({\bf y}(i),t).
\end{aligned}
\end{equation}
\begin{enumerate}
    \item We identify the partition $\rho (i)$ from amongst the orbits of ${J'}$, based on the quiver diagram, so that the resolved slices $\slice{{\rho (i),[\bf n']}}^{{{J'}}}$ belong to the twisted group.
\item We also need to identify the ``twisted" map $[\bf n] \to [\bf n']$ between the Dynkin labels of $J^\vee$ and those of ${J'}$, 
and use this to express the $\Delta' [ {\bf [n']}]$ contribution to conformal dimension in terms of $[\bf n]$.
\end{enumerate}

The twisted $A_{2r-1}$ fixtures analysed in section \ref{applications} contain a $J'=A_{2r-1}$ linear quiver affixed to a star shaped quiver with $ J = C_r$ and $ J^\vee = B_r$. The map between the Dynkin labels of $B_r$ and $A_{2r-1}$ under the \emph{twisted} map is $[n_1,n_2,\ldots, n_r]  \Leftrightarrow [n_1,n_2,\ldots,n_r,n_{r-1},\ldots,n_1]$. 

For example, consider the quiver in table \ref{A3fixtures} with $E_6$ global symmetry. We can read from table \ref{A3puncture} that this comprises two slices to orbits with the $B$-partitions $\rho=(1^5)$ and a slice to the orbit with $A_3$ partition $(3,1)$. Using the nilpotent orbit data in Appendix B of \cite{rudolph1601}, we find these correspond to $B_2$ weights $[0,0]$ and $A_3$ weights $[2,2,2]$, respectively. The weight map for the regular slice of $B_2$ is $[4,3]$ and that for $A_3$ is $[3,4,3]$. So,
\begin{equation}
\begin{aligned}
\omega(\mathsf{Q})& =-2[4,3]+ 2([4,3]-[0,0])+([3,4,3]-[2,2,2])\\
& =[1,2,1].
\end{aligned}
\end{equation}
Application of the selection rule shows $\omega(\mathsf{Q})$ is strictly positive and therefore the weight vector is that of a good star shaped quiver. The summation is carried out over the $B_2$ Dynkin label lattice $[n_1,n_2]$. Only $A_3$ slices with Dynkin labels of the form $[n_1,n_2,n_1]$ are involved. Further discussion of twisted fixtures is left for future work.
\clearpage
\section{Palindromic polynomials}
\label{apx:PP}
\begin{figure}[h]
\small
\scalebox{0.90}{
\begin{tabular}{|c|c|}
		\hline
		  Label & Polynomial\\
		  \hline
${\cal P}_{20}(t) $ & 
   \parbox[t]{13cm}{
$ 1 + 55 t^2 + 890 t^4 + 5886 t^6 + 17929 t^8 + 26060 t^{10} + \pal + t^{20} $}\\
\hline
${\cal P}_{32}(t) $ & 
          \parbox[t]{13cm}{
$ 1 + 98t^2 + 3312t^4 + 53305t^6 + 468612t^8+2421286t^{10} + 7664780t^{12} + 15203076t^{14}\\ + 19086400t^{16}+ \pal +t^{32} $}\\
\hline
 ${\cal P}_{34}(t) $ & 
	  \parbox[t]{13cm}{
	  $ 1 - t + 13 t^{2} + 12 t^{3} + 96 t^{4} + 172 t^{5} + 572 t^{6} + 1072 t^{7} + 2479 t^{8} + 4265 t^{9} + 7813 t^{10} + 11874 t^{11} + 18146 t^{12} + 24124 t^{13} + 31540 t^{14} + 36640 t^{15} + 41456 t^{16} + 42064 t^{17} + \pal +t^{34} $}\\
	  \hline
${\cal P}_{36}(t) $ & 
	  \parbox[t]{13cm}{
	  $ 1 + 10 t^{2} + 134 t^{4} + 1036 t^{6} + 6086 t^{8} + 25084 t^{10} + 76145 t^{12} + 167014 t^{14} + 268598 t^{16} + 313912 t^{18} + \pal + t^{36}$}\\
	  \hline
${\cal P}_{40}(t) $ & 
  \parbox[t]{13cm}{
$1 + 25 t^2 + 320 t^4 + 2497 t^6 + 13343 t^8 + 51447 t^{10} + 
  149336 t^{12} + 334179 t^{14} + 587816 t^{16} + 820972 t^{18} + 917376 t^{20} + \pal + t^{40} $}\\
\hline
 ${\cal P}_{44}(t) $ & 
         \parbox[t]{13cm}{
$ 1 + 35 t^2 + 816 t^4 + 10176 t^6 + 79992 t^8 + 426525 t^{10} + 1647613 t^{12} + 4776684 t^{14} + 10697290 t^{16}\\ + 18803936 t^{18} + 26278416 t^{20} + 29348968 t^{22} + \pal + t^{44} $}\\
    \hline
${\cal P}_{48}(t) $ & 
   \parbox[t]{13cm}
   {$  1 + 27 t^2 + 464 t^4 + 4890 t^6 + 35877 t^8 + 190734 t^{10} + 768953 t^{12} + 2408906 t^{14} + 5996581 t^{16} \\+ 12020149 t^{18} +  19625479 t^{20} + 26259134 t^{22} + 28929866 t^{24} + \pal + t^{48} $}\\
\hline
	\end{tabular}}
\caption[Palindromic polynomials I]{Palindromic polynomials I. These palindromic polynomials appear in the numerators of Hilbert series. They are identified by the degree of the fugacity $t$, augmented by further labels as necessary.}
\label{tab:PPI}
\end{figure}

\begin{figure}[h]
\small
	\begin{tabular}{|c|c|}
		\hline
		  Label & Polynomial\\
		  \hline
${\cal P}_{100}(t) $ & 
	   \parbox[t]{13cm}{$   1 + 25 t^{2} + 409 t^{4} + 5195 t^{6} + 51787 t^{8} + 420264 t^{10} +  2840102 t^{12} + 16273789 t^{14} \\+ 80306700 t^{16} + 345692590 t^{18} +  1312122037 t^{20} + 4431518098 t^{22} + 13420420448 t^{24} \\+  36681742342 t^{26} + 90995317496 t^{28} + 205840630198 t^{30} +  426323091615 t^{32}\\ + 811209029112 t^{34} + 1422246527224 t^{36} +  2303168695267 t^{38} + 3451967941502 t^{40} \\+ 4796423730026 t^{42} +  6186566897809 t^{44} + 7414766944277 t^{46} + 8263532776006 t^{48}\\ +  8567127089506 t^{50} + \pal  + t^{100}	   $}\\
\hline
	\end{tabular}
\caption[Palindromic polynomials II]{Palindromic polynomials II.}
\label{tab:PPII}
\end{figure}
\begin{figure}[h]
\small
	\begin{tabular}{|c|c|}
		\hline
		  Label & Polynomial\\
\hline
		  \end{tabular}
\caption[Palindromic polynomials III]{Palindromic polynomials III.}
\label{tab:PPIII}
\end{figure}
\begin{figure}[h]
\small
	\begin{tabular}{|c|c|}
		\hline
		  Label & Polynomial\\
\hline
	${\cal P}_{176}(t) $ & 
	  \parbox[t]{13cm}{ $1 + 24 t^{2} + 305 t^{4} + 2860 t^{6} + 22411 t^{8} + 153603 t^{10} +  936878 t^{12} + 5134768 t^{14} \\+ 25486046 t^{16} + 115373699 t^{18} +  479492068 t^{20} + 1840476253 t^{22} + 6560029731 t^{24}\\ +  21818202154 t^{26} + 68007524835 t^{28} + 199436887333 t^{30} +  552167369135 t^{32} \\+ 1447772374047 t^{34} + 3604970823988 t^{36} +  8545930534516 t^{38} + 19330733192922 t^{40}\\ + 41806847646416 t^{42} +  86606358190228 t^{44} + 172136222958565 t^{46} + 328748388774696 t^{48} \\+  604103685282463 t^{50} + 1069420301216580 t^{52} +  1825817343019054 t^{54} \\+ 3009372615128096 t^{56} +  4792929174120706 t^{58} + 7382297817372560 t^{60}\\ +  11004547427977587 t^{62} + 15886770100322027 t^{64} +  22225167817786249 t^{66} \\+ 30146454564251645 t^{68} +  39666076108671224 t^{70} + 50650056441438755 t^{72}\\ +  62788685722476841 t^{74} + 75590115928073512 t^{76} +  88399840242104967 t^{78}\\ + 100448071397392135 t^{80} +  110921815558614130 t^{82} + 119053041604350681 t^{84} \\+  124210076994665741 t^{86} + 125977403206155610 t^{88} +  \pal + t^{176} $ }\\
\hline
		  \end{tabular}
\caption[Palindromic polynomials IV]{Palindromic polynomials IV.}
\label{tab:PPIV}
\end{figure}

\begin{figure}[h]
\small
	\begin{tabular}{|c|c|}
		\hline
		  Label & Polynomial\\
\hline
	\end{tabular}
\caption[Palindromic polynomials V]{Palindromic polynomials V.}
\label{tab:PPV}
\end{figure}

\begin{figure}[h]
\small
	\begin{tabular}{|c|c|}
		\hline
		  Label & Polynomial\\
\hline
	  		  \end{tabular}
\caption[Palindromic polynomials VI]{Palindromic polynomials VI.}
\label{tab:PPVI}
\end{figure}

\begin{figure}[h]
\small
\scalebox{0.9}{
	\begin{tabular}{|l|l|}
		\hline
		  Label & Polynomial\\
\hline
 \end{tabular}}
\caption[Palindromic polynomials VII]{Palindromic polynomials VII.}
\label{tab:PPVII}
\end{figure}

\begin{figure}[h]
\small
	\begin{tabular}{|l|m{13cm}|}
		\hline
		  Label & Polynomial\\
\hline
 \end{tabular}
\caption[Palindromic polynomials VIII]{Palindromic polynomials VIII.}
\label{tab:PPVIII}
\end{figure}

\clearpage
{\footnotesize{
\listoffigures
}}

\bibliographystyle{JHEP2}
\bibliography{bibli.bib}

\end{document}